\documentclass[english, twoside]{sapthesis}
\usepackage[utf8]{inputenx}
\usepackage{indentfirst}
\usepackage{microtype}
\usepackage{csquotes}
\usepackage{comment}
\usepackage{lettrine}
\usepackage{comment}
\usepackage{color, colortbl}
\usepackage[T1]{fontenc} 
\usepackage{braket}
\usepackage{placeins}
\usepackage{tabularx}
\usepackage{ulem}
\usepackage{url}
\usepackage{xcolor}
\usepackage{newverbs}
\usepackage{siunitx}
\usepackage[multiple]{footmisc}
\usepackage{bbold}

\usepackage[nottoc, notlof, notlot]{tocbibind}

\usepackage{graphicx}
\usepackage{adjustbox}
\usepackage{doi}
\usepackage{subcaption}
\usepackage{epsfig}
\usepackage{booktabs}
\usepackage{float}
\usepackage{relsize}
\usepackage[innercaption]{sidecap}
\usepackage{lmodern}
\usepackage{enumitem}
\usepackage{textcomp}
\usepackage{amsmath}
\usepackage[authoryear,round]{natbib}
\usepackage{amssymb}
\usepackage{euclid}
\usepackage{aas_macros}
\usepackage{epstopdf}
\usepackage{array}   
\usepackage{hyperref}

\newcolumntype{L}{>{$}l<{$}}
\newcolumntype{C}{>{$}c<{$}}

\defcitealias{ISTF2020}{EC20}
\defcitealias{Lacasa_2019}{LG19}
\defcitealias{Barreira2018response_approach}{B18}

\def\be{\begin{equation}} 
\def\ee{\end{equation}}

\newcommand{\ML}[1]{#1}

\newcommand{\FB}[1]{#1}

\newcommand{\kmax}{k_{\rm max}}

\newcommand{\gK}{g_k}

\graphicspath{{./}{graphics/}}

\hypersetup{
			colorlinks=true,
			linkcolor=sapred,
                linktoc=page,
			anchorcolor=black,
			citecolor=sapred,
			urlcolor=blue,
			pdftitle={Deriving Cosmological Parameters from the \Euclid mission},
			pdfauthor={Davide Sciotti},
			pdfkeywords={thesis, sapienza, roma, university}
}

\newcommand{\chap}[1]{\hyperref[#1]{Chapter~\ref*{#1}}}

\title{Deriving Cosmological Parameters from the \Euclid mission}
\author{Davide Sciotti}
\IDnumber{1570331}
\course{PhD program in Astronomy, Astrophysics and Space Science}
\courseorganizer{Sapienza Università di Roma}
\cycle{XXXVI}
\AcademicYear{2023/2024}
\advisor{Prof. Roberto Maoli}
\advisor{Prof. Roberto Scaramella}
\advisor{Dr. Vincenzo Fabrizio Cardone}
\customdirectorlabel{Coordinator:} 
\director{Prof. Francesco Piacentini}
\authoremail{davide.sciotti@uniroma1.it}
\copyyear{2023}

\examdate{22 December 2023}
\examiner[]{Prof. Stefano Borgani, University of Trieste}
\examiner[]{Prof. Salvatore Capozziello, University of Napoli Federico II}
\examiner[]{Prof. Francesco Rosario Ferraro, University of Bologna}

\thesistype{PhD thesis}

\begin{document}

\frontmatter
\maketitle

\begin{acknowledgements}
I would like to express my deep gratitude to all the people who kindly assisted me in this work, through their comments, suggestions, encouragement, and fruitful discussion. In particular, I would like to thank Sylvain Gouyou Beauchamps, Marco Bonici, Santiago Casas, Guadalupe Ca\~{n}as Herrera, Vincenzo Cardone, Matteo Martinelli, Roberto Maoli, Roberto Scaramella, Stefano Camera, Isaac Tutusauus, Fabien Lacasa, Alex Barreira, Peter Taylor, Martin Crocce, Alkistis Pourtsidou, Carlo Giocoli, Robin Upham, Louis Legrand, Massimiliano Lattanzi, Riccardo La Placa, and the many others who made this journey possible. 
\end{acknowledgements}

\dedication{``A Giò, ancora e sempre.\\E a Guido, Fabrizio, Milo, Cecilia, Francesco, Andrea, Yari, e Simona. \\ }


\tableofcontents

\mainmatter



\addcontentsline{toc}{chapter}{Introduction}
\chapter*{Introduction}\label{chap:introduction} 

\lettrine[lines=2, findent=3pt, nindent=0pt]{T}{he} cosmological parameters are fundamental quantities of the cosmological model, the theoretical framework that tries to predict and explain the dynamics, geometry, and evolutionary course of the Universe. From the rate of expansion, encapsulated by the Hubble constant, to the density parameters, describing the relative proportions of the different cosmic species, these quantities allow us, amongst other things, to reconstruct the Universe's evolution from a homogeneous, hot, and dense early state to the complex and beautiful structures we observe at low redshift. \\

We are now living in the epoch of precision cosmology, characterized by unprecedented amounts of high-resolution data. Amongst the most important milestones in the direction of this new era of Cosmology is the accurate measurement of the anisotropies in Cosmic Microwave Background (CMB) by the \Planck Satellite, giving us a wealth of information about the early Universe. However, the quest for unravelling the cosmic mysteries calls for independent and complementary routes to improve the precision of the parameters' estimates, hence opening the possibility of selecting between competing cosmological models. \\

In this regard, the \Euclid mission, a project undertaken by the European Space Agency (ESA), emerges as a promising tool to probe the low-redshift Universe, and hence study its expansion history and geometry. By surveying the position and morphology of billions of galaxies, \Euclid will construct the largest and most accurate map of the large-scale structure (LSS) of the Universe, allowing for unique insights into the nature of dark matter and dark energy.\\

This thesis revolves around the central theme of quantifying the constraining power of the upcoming \Euclid photometric survey, by taking into account several factors which have been neglected to this date in the official forecasts, especially more subtle sources of uncertainty which need to be included in the forecast (and data) analysis due to the precision of the observations.\\

The present work is organized as follows: Chapter~\ref{chap:cosmology} will introduce the current concordance cosmological model and the aspects that will be best investigated by \Euclid. Chapters~\ref{chap:statistics} and~\ref{chap:observables} will illustrate the statistical framework and cosmological probes used in our forecasts (and often in the literature). Chapters~\ref{chap:SSC} and~\ref{chap:SSC_for_Euclid} will discuss super-sample covariance, its analytical treatment, the approximations used to study it in this context and the results obtained for the photometric survey; this is the result of one of the Key Papers of the Galaxy Clustering Science Working Group (GC SWG), \citet{Sciotti2023}, submitted to Astronomy \& Astrophysics after having successfully undergone the internal review of the Euclid Collaboration. Finally, Chapter~\ref{chap:scalecuts} will give some details about the work done in the context of the latest Science Performance Verification (SPV) of the survey, the last before the first data release, regarding the inclusion of scale cuts in the forecast analysis through the use of the Bernardeau-Nishimichi-Taruya (BNT) transform.
\chapter[Cosmology]{Cosmology}\label{chap:cosmology}
Cosmology is the domain of physics that investigates the global characteristics of the Universe. This includes an exploration of the types, densities, and distributions of matter-energy components within the cosmos, along with its age, expansion history and geometric properties. Such a study, as with all physics, is based on the continuous feedback loop between data and model: we formulate a model based on key insights and existing observations, which motivates further experiments, which corroborate or \FB{conforte} the existing models, and so on. This Chapter will introduce the key principles and models at the heart of modern cosmological analyses, laying the foundation for the concepts presented in the rest of the work.

\section{The Cosmological Principle}
At the basis of our current understanding of the Universe is the cosmological principle, the observation that on sufficiently large scales  ($\gtrsim 10^2  \mathrm{Mpc}$, \citealt{Peebles1993}) the Universe manifests as homogeneous and isotropic. Mathematically, this means that on sufficiently large scales the cosmological fields -- and the metric tensor, describing the geometry of spacetime -- possess translational and rotational symmetry. This principle has in turn its roots in the Copernican principle, stating that we are not special observers of the Universe: there is no privileged point or direction.
It is interesting to note that these two assumptions are not independent: if the Universe appears the same in every direction, and if this is true for any observer - as a natural consequence of the Copernican principle - it must also be homogeneous. Isotropy for every observer is the stronger condition:
\begin{align}
    &\text{isotropy for every observer $\Rightarrow$ homogeneity} \nonumber \\
    &\text{homogeneity $\nRightarrow$ isotropy for every observer.} \nonumber
\end{align}
Such a principle implies the following general form for the metric tensor, the mathematical object which allows defining distances and angles on a manifold, expressed through the line element $\mathrm{d}s^2$
\begin{equation}
\label{eq:FLRW_metric}
    \mathrm{d}s^2 = -c^2\mathrm{d}t^2 + a(t)^2[\mathrm{d}r^2 + S_{k}^{2}(r)\mathrm{d}\Omega^2]
\end{equation}
in spherical coordinates $(r,\theta , \phi)$, with $\mathrm{d}\Omega^2 = \mathrm{d}\theta ^2 + \sin^2 \theta \mathrm{d} \phi ^2$. This is called the Friedmann-Lemaître-Robertson-Walker (FLRW) metric. The isotropy condition forces the mixed terms $g_{0i} = g_{i0}$ to vanish. The homogeneity condition primarily governs the spatial part of the metric tensor $ g_{ij} $, making the spatial components maximally symmetric. This is expressed by the fact that $ g_{ij} $ can be written as a function of a single spatial coordinate $ r $ and a time-dependent scale factor $ a(t) $, along with $ S_k(r) $ to account for possible (Gaussian) curvature $k$, which has to be constant due to, again, homogeneity and isotropy. This constant curvature can be positive, negative or null, and has units of $[L^{-2}]$: 
\begin{equation} \label{eq:S_k(r)}
  S_k(r)=\begin{cases}
    \frac{\sin(r\sqrt{k})}{\sqrt{k}}        & (k > 0) \\
    r                                       & (k = 0) \\
    \frac{\sinh(r\sqrt{|k|})}{\sqrt{|k|}}   & (k < 0) \; ,
  \end{cases}
\end{equation}
where $r$ is the radial coordinate. The dimensionless quantity $a(t)$ is called the \textit{scale factor}. It encodes the time dependence of the spatial degrees of freedom, therefore allowing the possibility of an expanding or contracting space (again, an isotropic model needs just one such quantity - all spatial directions scale at the same rate). It is defined as
\begin{equation}
    a(t) = d_{\rm pr}(t)/d_{{\rm pr}, 0} \; ,
\end{equation}
where $d_{\rm pr}(t)$ is the \textit{proper distance} at some time $t$, compared to the proper distance at some reference time $d_{\rm pr}(t_0) = d_{{\rm pr}, 0}$. This definition ensures that $a(t_0) = a_0 = 1$. It is customary to take as a reference time the present epoch, so $a=1$ today. The time coordinate is called \textit{cosmic time}, and it represents the time measured by a clock moving along (\textit{comoving}) with the expanding Universe (with the so-called \enquote{Hubble flow}). \\
An expanding Universe naturally leads to define different types of distance indicators. Two points whose distance changes in time only due to the expansion of space are said to be at a constant comoving distance $\chi$. This is the distance measured in a coordinate system which follows the expansion of space. The proper distance, as seen above, is measured instead with a coordinate system that does not track the space expansion, hence the distance between the two points will change just by virtue of the scale factor variation. Proper distance is therefore the distance between two regions of space at a constant cosmological time. It is not directly observable, but can be calculated by factoring in the scale factor: $d_{\rm pr}(t) = a(t)\chi$. Since $a(t_0) = 1$, today the two distances correspond.

\section{Hubble's law and Friedmann equations}

One of the first direct observations of the expansion of the Universe was performed by Edwin Hubble in the late 1920s. In the law that now carries his name, he stated that distant galaxies tend to recede from us, with a velocity directly proportional to their distance:
\begin{equation}
    v = H_0d \; .
\end{equation}
$H_0$ is called Hubble (more recently\footnote{\url{https://www.iau.org/news/pressreleases/detail/iau1812/}} Hubble-Lemaître) constant, and is often expressed in $\mathrm{km \, s^{-1}\, Mpc^{-1}}$. Its measurement constitutes one of the most important open problems in Cosmology, since observations from early-time and late-time probes give statistically incompatible results (the so-called \enquote{Hubble tension}: see \citealt{DiValentino2021_hubble_tension} for a recent comprehensive review). Cosmologists often use the reduced dimensionless Hubble constant $h = H_0 /(100 \mathrm{\, km \, s^{-1} \, Mpc^{-1}}) $. In this work we will set $h=0.67$, following results from the \Planck mission \cite{Planck2020_params}. \\
Hubble's law can be recast in terms of the scale factor by introducing one of the most important quantities in Astrophysics: the \textit{redshift} $z$, defined as
\begin{equation}
    z = \frac{\lambda_{\rm obs}-\lambda_{\rm em}}{\lambda_{\rm em}}.
\end{equation}
which is the relative differential shift between the observed and emitted wavelength of the radiation source. This can be caused by different factors, such as the relative peculiar velocity of the source, or the presence of a gravitational potential. In the context of Cosmology, we are mainly interested in the redshift caused by the expansion of the Universe, which increases the radiation wavelength (hence decreasing its energy) as it follows its geodesic. For small values of $v/c$ - or small distances \textit{d} - we have, from special relativity (see e.g. \citealt{Hogg1999}):
\begin{equation}
    z=\sqrt{\frac{1+v/c}{1-v/c}} \simeq v/c \; ,
\end{equation}   
from which
\begin{equation}
    z \simeq \frac{H_0d}{c} \; .
\end{equation}
We can connect the redshift to the scale factor via the following equation:
\begin{equation}
    \frac{a_{\rm obs}}{a_{\rm em}} = \frac{1}{a} = \frac{\lambda_{\rm obs}}{\lambda_{\rm em}} = 1+z \; .
\end{equation}
This makes redshift, which is directly observable, a key quantity in our analysis: light coming from distant galaxies carries an imprint of the value of the scale factor at the time of the signal emission. This allows reconstructing the expansion history $a(t)$ of the Universe, one of its most fundamental properties. We can predict the functional form of $a(t)$ and connect it to other fundamental quantities by using the Friedmann equations. These can be derived by using the metric \eqref{eq:FLRW_metric} and assuming General Relativity (GR) to be the correct theory of gravity on cosmological scales. \\

We begin by introducing the Einstein field equations, a set of differential equations relating the Einstein tensor $G_{\mu\nu}$ to the stress-energy tensor $T_{\mu\nu}$:
\begin{equation} \label{eq:EFE}
    G_{\mu\nu} + \Lambda g_{\mu\nu}= \frac{8\pi G}{c^4} T_{\mu\nu} \; ,
\end{equation}
where $G$ is the gravitational constant, $\Lambda$ is the cosmological constant (with a positive value) and the second term on the l.h.s. was originally introduced to achieve a static Universe, which was thought to be the correct model at the beginning of the 20th century -- before the observations by Hubble. The Einstein tensor encodes the geometric properties (i.e., the curvature) of spacetime, while the stress-energy tensor characterizes the distribution and flow of energy-momentum in that spacetime. The stress-energy tensor $T_{\mu\nu}$ acts as the source term in Eq.~\eqref{eq:EFE}, dictating how energy and momentum contribute to the curvature. The Einstein tensor, in turn, governs the geodesic motion of particles and radiation in the curved spacetime. Hence, this equation captures the dynamical interplay between the distribution of matter-energy in the Universe and the geometry of spacetime. \\
The Einstein tensor can be expressed as a combination of the Ricci tensor $R_{\mu\nu}$, the Ricci scalar $R = g^{\mu\nu}R_{\mu\nu}$ and the metric tensor $g_{\mu\nu}$:
\begin{equation}\label{eq:Gmunu_vs_metric}
    G_{\mu\nu} = R_{\mu\nu} -\frac{1}{2}R g_{\mu\nu} \; .
\end{equation}
Since the Ricci tensor only depends on the metric tensor and its derivatives, the l.h.s. is purely geometrical. Evaluating it for the FLRW metric~\eqref{eq:FLRW_metric}, we get for the time-time component $G_{00}$:
\begin{equation}
    G_{00} = 3\left(  \frac{\dot{a}}{a}\right)^2 \; ,
\end{equation}
with the dot denoting the time derivative; we note that $\dot{a}/a$, the relative rate of change of the scale factor, is just the Hubble parameter $H(t)$. We can then move the cosmological constant term to the r.h.s. of Eq.~\eqref{eq:EFE}, interpreting it as a matter-energy component (\textit{dark energy}) rather than a property of space, which makes it easier to generalize it to dynamical forms. Considering the different components of the Universe as non-interacting perfect fluids with no anisotropic stresses \ML{(as implied by the requirement of homogeneity and isotropy)}, the stress-energy tensor will be ${\rm diag}(\rho_i, P_i, P_i, P_i)$) where $\rho_i$ is the density and $P_i$ the pressure of the $i$-th component and the \enquote{diag} notation indicates a diagonal matrix with entries specified by the arguments. The $\Lambda$ component has a stress-energy tensor
\begin{equation}
T_{(\Lambda)\nu}^\mu = -\frac{\Lambda}{8\pi G}\delta^\mu_{\;\;\nu} = {\rm diag}(-\rho_\Lambda, -\rho_\Lambda, -\rho_\Lambda, -\rho_\Lambda) \; ; \; \rho_\Lambda = \frac{\Lambda}{8\pi G} \; 
\end{equation}
with a time- and position-independent energy density (hence the name cosmological \textit{constant}).
We can now write the first Friedmann equation, describing the evolution of the homogeneous Universe, in the form of a differential equation for the scale factor:
\begin{equation}
    \label{friedmann1_initial}
    H^2 = \left(\frac{\dot{a}}{a}\right)^2 = \frac{8 \pi G}{3}\rho =  \frac{8 \pi G}{3} \sum_i \rho_i
\end{equation}
where $i$ runs over the different (non-interacting) energy components and $\rho(t)$ is the total energy density. This derivation holds for a flat universe; in the most general case, there is an additional term accounting for curvature:
\begin{equation}
    \label{eq:friedmann1}
    H^2 = \left(\frac{\dot{a}}{a}\right)^2 = \frac{8 \pi G}{3}\rho - \frac{kc^2}{a^2} 
\end{equation}
which makes it natural to define the \textit{critical density} $\rho_{\rm cr}$ as the density which makes the curvature vanish: its value today is $\rho_{\rm cr} = 3H_0^2/8\pi G$.\\
Taking the time derivative of Eq.~\eqref{eq:friedmann1}, we get 
\begin{equation}\label{eq:friedmann1_der}
2H\dot{H} = \frac{8\pi G}{3}\dot{\rho} + \frac{2kc^2}{a^3}\dot{a} \; .
\end{equation}
We now need to relate $\dot{\rho}$ to other known quantities. The continuity equation for perfect fluids in an expanding universe fluid is given by the conservation equation $T^{\mu\nu}_{;\nu}=0$ (more specifically, from the $\mu=0$ term), where the semicolon indicates the covariant derivative :
\begin{equation}\label{eq:continuity}
\dot{\rho} + 3H(\rho + \frac{P}{c^2}) = 0 \; .
\end{equation}
Solving for $\dot{\rho}$ and substituting into Eq.~\eqref{eq:friedmann1_der}, we arrive at the second Friedmann equation:
\begin{equation}\label{eq:friedmann2}
\frac{\ddot{a}}{a} = \dot{H} + H^2 = -\frac{4\pi G}{3} \left(\rho + \frac{3P}{c^2}\right) \; .
\end{equation}
To solve the two Friedmann equations (Eqs.~\ref{eq:friedmann1} and~\ref{eq:friedmann2}) we need an equation relating the density and pressure of the perfect fluids involved, i.e., an equation of state (EoS): we assume cosmological fluids to be described by a simple EoS, in the form 
\begin{equation} \label{eq:eos}
    P = w\rho c^2 \; ,
\end{equation}
parameterized by $w$. Combining this with the continuity equation~\eqref{eq:continuity}, we get
\begin{equation}
    \rho \propto a^{-3(1+w)}
\end{equation}
which gives the evolution of the density as a function of the scale factor under the assumption of constant $w$. \\

The perfect fluids which make up the Universe described by our model are essentially of three types: matter (baryonic matter, cold dark matter, non-relativistic neutrinos), radiation, and cosmological constant. \ML{Concerning its impact on the time evolution of the scale factor}, curvature can be treated as a further matter-energy component; the EoS parameter and scale factor dependence of these different species are given in Table~\ref{tab:eos_parameters}.
\begin{table}[h]
\centering
\caption{The equation of state parameters and the dependence of the energy densities with the scale factor for the different cosmic species.}
\label{tab:eos_parameters}
\begin{tabular}{ |c|c|c| }
\hline
radiation & $ w=1/3 $ & $ \rho_{\rm r} \propto a^{-4} $ \\
matter & $ w=0 $ & $ \rho_{\rm m} \propto a^{-3} $ \\
curvature & $ w=-1/3 $ & $ \rho_k \propto a^{-2} $ \\
$ \Lambda $ & $ w=-1 $ & $ \rho_\Lambda = \mathrm{const} $ \\
\hline
\end{tabular}
\end{table}
The energy densities $\rho_i$ can be normalized by the critical density to obtain the \textit{density parameters} $\Omega_i = \rho_i/\rho_{\rm cr}$. Curvature and $\Lambda$ will respectively have $\Omega_{\Lambda, 0} = \Omega_\Lambda = \Lambda c^2/3H_0^2$ and $\Omega_{k, 0} = -k/H_0^2$. As usual, the 0 subscript indicates the value today. The different scaling with $a$ shows how different components were prevalent at different times, allowing distinguishing between \textit{epochs} when they were dominant over the others. These different components will now be briefly introduced.\\

Recent observations \citep{Perlmutter1999, Riess1998} have shown that the Universe is currently accelerating its expansion ($\ddot{a}>0$); in the current concordance model, this means that we now live in the $\Lambda$-dominated epoch. Explaining the cause of this accelerated expansion is arguably the central problem in modern Cosmology. After being rejected by Einstein because of the rise of the expanding Universe model, the cosmological constant was reintroduced after this discovery; in fact, from the second Friedmann equation:
\begin{equation}
\frac{\ddot{a}}{a}=-\frac{4 \pi G}{3}\left(\rho_{\rm m}+\rho_{\Lambda}+\frac{3 P_{\rm m}}{c^2}+\frac{3 P_{\Lambda}}{c^2}\right)=-\frac{4 \pi G}{3}\left(\rho_{\rm m}-2 \rho_{\Lambda}\right) \, ,
\end{equation}
from which we get $\ddot{a} > 0$ if $\rho_\Lambda > \rho_{\rm m}/2$. As mentioned earlier, the $\Lambda$ term in Eq.~\eqref{eq:EFE} can be interpreted as an additional component, a form of a yet-to-be-understood type of energy (Dark Energy, DE, or as the energy of vacuum), or as a property of space itself, depending on whether we choose to include it in right- or left-hand side (l.h.s.) of the Einstein equations. In the first case, the DE EoS can be allowed to have time dependence by letting $w = w(t)$; one of the simplest time-dependent parameterizations of the DE EoS is the Chevallier-Polarski-Linder (CPL) parametrization \citep{Chevallier2001, Linder2005}: 
\begin{equation}\label{eq:CPL}
    w_{\rm DE}(z) = w_0 + w_a \frac{z}{1+z} = w_0 + w_a(1-a).
\end{equation}
where $w_0$ is the current time value and $w_a$ controls the time dependence of $w_{\rm DE}$. Any observation of $w_a \neq 0$ would allow discriminating between different DE theories; a key goal of the \Euclid mission is to place tight bounds on $w_0$ and $w_a$, hence selecting between the cosmological constant and dynamical DE models with high statistical power \citep{laureijs2011euclid}. Of course, another possibility to explain the accelerated expansion is that GR is not the correct theory of gravity (not on all scales, at least) and must be modified: this class of solutions is referred to as \textit{Modified Gravity} (MG) models.\\

The second component is matter, which is pressureless; it is composed of baryonic (protons, neutrons and electrons; in Cosmology, the term \enquote{baryons} is used more loosely than in particle physics) and Dark matter (DM). This second type is of unknown nature and origin and exhibits very low electromagnetic interaction (hence the name \enquote{dark}) but does interact gravitationally. It makes up $\sim 27\%$ of the total matter-energy content, and its existence has been proposed to explain several observations, such as the statistical properties of the large-scale matter distribution (Sect.~\ref{sec:angular_ps_theory}), Baryon Acoustic Oscillations (Sect.~\ref{sec:GC_spectro}), galaxy rotation curves \citep{Rubin1982} and the Bullet Cluster \citep{Clowe2006_bullet}. \ML{In the concordance model, this form of matter is cold -- i.e., has $v_{\rm DM}/c \ll  1$; a possible DM candidate are Weakly Interacting Massive Particles (WIMPs, \citealt{Jungman1996_WIMP}), with masses ranging from a few GeV to several TeV}. Massive neutrinos are another potential form of Dark Matter, although their large velocity dispersion makes them a kind of \textit{hot} DM: they too are weakly interacting particles, but their energy density is far too low to be the only actor at play in the DM sector. They are treated as radiation at early times and matter when they become non-relativistic, so their equation of state varies as well.\\
Our current standard cosmological model is called Lambda - Cold Dark Matter ($\Lambda$CDM), reflecting our current most accepted hypotheses on the nature of the dark components. \\ 

We can recast the first Friedmann equation as a function of the different density parameters:
\begin{equation}
    H^2 = H_0^2(\Omega_{\rm r} + \Omega_{\rm m} + \Omega_k + \Omega_\Lambda) \; , 
\end{equation}
or:
\begin{equation}
    H^2 = H_0^2\left( \frac{\Omega_{{\rm r}, 0}}{a^4} + \frac{\Omega_{{\rm m}, 0}}{a^3} + \frac{\Omega_{k, 0}}{a^2} + \Omega_{\rm DE} a^{-3(1+w_{\rm DE})}    \right) \; ,
\end{equation}
The sum of the density parameters gives information about the curvature:
\begin{align} 
\Omega_k = 0 & \rightarrow \Omega_{\rm tot} = 1 \rightarrow \text{flat geometry} \\ 
\Omega_k > 0 & \rightarrow \Omega_{\rm tot} < 1 \rightarrow \text{open geometry} \\ 
\Omega_k < 0 & \rightarrow \Omega_{\rm tot} > 1 \rightarrow \text{closed geometry} \; ,
\end{align}
having defined $\Omega_{\rm tot} = 1-\Omega_{k, 0} = \Omega_{{\rm m}, 0} + \Omega_{{\rm r}, 0} + \Omega_{\Lambda}$. Current observations and theoretical models (in particular, inflation: \citealt{Starobinsky1980_inflation}, \citealt{Guth1981_inflation}) tend to favour a flat geometry, with $\Omega_{k}=0$. \\
In this work we will mainly deal with low-redshift probes; because of this, we can consider the contribution from the radiation component as negligible (setting $\Omega_{{\rm r}, 0} = 0$) and the massive neutrinos as non-relativistic, splitting $\Omega_{{\rm m}, 0}$ in
\begin{equation}
    \Omega_{{\rm m}, 0} = \Omega_{{\rm b}, 0} + \Omega_{{\rm c}, 0} + \Omega_{{\nu}, 0} 
\end{equation}
with subscripts $\nu$ for neutrinos, ${\rm b}$ for baryons and ${\rm c}$ for Cold Dark Matter.

\section{Distances in Cosmology} \label{sec:distances}
Until now, we have encountered two possible distance measures in Cosmology, the proper and comoving distance. A distinction between the two was made necessary by the dynamic nature of space. Other distance indicators are needed to connect the comoving distance to other observable quantities, such as the angular scale or luminosity of an object. This section will briefly introduce these.\\
To begin with, the comoving distance can be expressed as a function of the cosmological fluids by integrating the first Friedmann equation:
\begin{equation} \label{chiz}
    \chi(z) = \frac{c}{H_0}\int^z_0 \frac{\diff z}{E(z)} \; ,
\end{equation}
with $H(z) = H_0E(z)$ and $E(z) =  \sqrt{ \Omega_{{\rm m}, 0}(1+z)^3 + \Omega_{k, 0}(1+z)^2 + \Omega_{0\Lambda} }$ (for a $\Lambda$CDM model, having set $\Omega_{{\rm r}, 0}$ = 0). \\

This can be used to compute another useful quantity, the \textit{comoving volume}, used for example to get the volume of a region subtending a solid angle $\Omega$ between two redshifts $z_1$ and $z_2$:
\begin{equation}
    V(z_1, z_2) = \Omega\int_{\chi(z_1)}^{\chi(z_2)}\chi^2\diff \chi = \frac{\Omega}{3}[\chi^3(z_2)-\chi^3(z_1)]
\end{equation}
Another important distance indicator is the \textit{angular diameter distance}, defined as the ratio of an object’s physical transverse size to its angular size in radians \citep{Hogg1999}:
\begin{equation}
    D_A(z)=\frac{c}{(1+z) H_0 \sqrt{\left|\Omega_{k 0}\right|}} S_k\left[\sqrt{\left|\Omega_{k 0}\right|} \int_0^z \frac{\mathrm{d} z^{\prime}}{E\left(z^{\prime}\right)}\right] \; ,
\end{equation}
or, writing explicitly the different cases of Eq.~\eqref{eq:S_k(r)}:
\begin{equation} \label{eq:ADD}
  D_A(z)=\begin{cases}
    (1+z)^{-1}\frac{c}{H_0}\frac{1}{\sqrt{|\Omega_{k, 0}|}}\sin\left(\sqrt{|\Omega_{k, 0}|}\frac{H_0}{c}\chi(z)\right) & \Omega_{k, 0} < 0\\ \\
    
    (1+z)^{-1}\chi(z) & \Omega_{k, 0} = 0\\ \\
    
    (1+z)^{-1}\frac{c}{H_0}\frac{1}{\sqrt{\Omega_{k, 0}}}\sinh\left(\sqrt{\Omega_{k, 0}}\frac{H_0}{c}\chi(z)\right) & \Omega_{k, 0} > 0\\
  \end{cases}
\end{equation}
which depends on the density parameters through $\chi(z)$.\\
Finally, the \textit{luminosity distance} $D_L$ is given by the bolometric (i.e., integrated over all frequencies) luminosity $L$  and the bolometric flux $S$:
\begin{equation}
D_L = \sqrt{\frac{L}{4\pi S}}
\end{equation}
and it is related to the angular diameter distance by 
\begin{equation}
D_L = (1+z)^2 D_A \;
\end{equation}

\section{Inhomogeneities}
While the Cosmological principle posits the homogeneity and isotropy of the Universe at large scales, observations clearly show this not to be the case when restricting our field of view (FoV) to smaller regions. Large-scale galaxy surveys have uncovered a rich and complex structure of galaxies and matter, organized in clusters, voids and filaments (the \textit{cosmic web}), and which act as a proxy for the underlying dark matter distribution. Hence, the large-scale matter distribution breaks the assumptions of homogeneity and isotropy, creating the need to go beyond the treatment seen so far to describe the evolution of such inhomogeneities. \\

Our starting point in this treatment, leading to the equations of the (linear) growth of perturbations in an expanding Universe for the different species, is the perturbed FLRW metric. In the following, we will rely on the approximation of small perturbations around the smooth background, and rewrite the metric as:
\begin{align}\label{eq:pert_FLRW_metric}
\begin{aligned}
    & g_{00}(\vec{x}, t)=-1-2 \Psi(\vec{x}, t) \\
    & g_{0 i}(\vec{x}, t)=0 \\
    & g_{i j}(\vec{x}, t)=a^2(t) \delta_{i j}\left[1+2 \Phi(\vec{x}, t)\right] \: ,
\end{aligned}
\end{align}
where the space- and time-dependent metric perturbations $\Psi$ and $\Phi$ parameterize deviation from the FLRW metric, and describe the gravitational (or temporal) potential and the curvature (or spatial) potential. Since they are small ($\mathcal{O}(10^{-4})$, \citealt{Dodelson2020}), we can treat them at linear order to a good approximation. We note that Eq.~\eqref{eq:pert_FLRW_metric} describes only scalar perturbations, the most relevant to structure formation, whereas tensor perturbations capture gravitational waves; moreover, the above expression is in the \textit{conformal Newtonian} gauge \FB{\citep{Mukhanov1992}}, a choice which does not influence our results since GR possesses gauge freedom. \\
\FB{The framework for the perturbative description of inhomogeneities in the Universe was laid out in \citet{Lifshitz1946}, and used to study photon and baryon perturbations in \citet{PeeblesYu1970}; neutrinos were added to the picture in \citet{BondSzalay1983}, and a complete treatment including dark energy, CDM and massive and massless neutrinos in a flat universe was presented in \citet{MaBertschinger1995}, which solved numerically the linearized Boltzmann, Einstein, and fluid equations for the evolution of the metric perturbations in both the synchronous and the conformal Newtonian gauges. A gauge-invariant approach to perturbation theory, on the other hand, was introduced in \citet{Bardeen1980}; one of the advantages of the conformal Newtonian gauge is precisely that the fields $\Psi$ and $\Phi$ are equal (modulo a $-1$ factor) to the gauge-invariant variables thereby identified.}

\subsection{The Boltzmann equation}
Until this point, we have derived the equations governing the evolution, or expansion history, of the smooth background, as a function of its various components. Now we turn our attention to the evolution of the species inhabiting the Universe; in the context of Cosmology, a statistical, average description of the collective behaviour of such species (luckily) suffices. This evolution is regulated by the central equation in statistical mechanics, the Boltzmann equation, These differential equations describe the time evolution of the phase-space distribution function $f(\vec{x}, \vec{p}, t)$ of a given set of particles, also taking into account their couplings to other species. The conjugate variables $\vec{x}$ and $\vec{p}$, respectively the position and momentum of the particles, completely characterize the system. This distribution is defined by:
\begin{equation}
N(\vec{x}, \vec{p}, t)=f(\vec{x}, \vec{p}, t)\left(\frac{\Delta x \Delta p}{2 \pi}\right)^3 \; ,
\end{equation}
where $N(\vec{x}, \vec{p}, t)$ is the number of particles in the phase-space volume around $(\vec{x}, \vec{p})$. The Boltzmann equation essentially expresses the conservation of \ML{phase space volume}, which implies e null total derivative ($\diff /\diff t$) of the distribution function:
\begin{equation}
\frac{\diff f}{\diff t}=0 \quad \text{with} \quad \frac{\diff }{\diff t} = \frac{\partial }{\partial t} + \dot{\vec{x}}\vec{\nabla}_x + \dot{\vec{x
p}}\vec{\nabla}_p \; ,
\end{equation}
where $\vec{\nabla}$ is the gradient.
We can account for particle-particle interactions, including particle annihilation and creation, by including a source term on the r.h.s., the \textit{collision term}:
\begin{equation}
\frac{\diff f}{\diff t}=C[f] \; .
\end{equation}
%
Having obtained the distribution function for a given species by solving the Boltzmann equation, we can derive macroscopic quantities such as the stress-energy tensor
\begin{equation}
T_{\;\; \nu}^\mu(\vec{x}, t)=\frac{g_s}{\sqrt{-\operatorname{det}\left[g_{\alpha \beta}\right]}} \int \frac{\diff P_1 \diff P_2 \diff P_3}{(2 \pi)^3} \frac{P^\mu P_\nu}{P^0} f(\vec{x}, \vec{p}, t)
\end{equation}
%
where $g_s$ represents the quantum degeneracy factor of the species considered, counting the number of particle states described by $f$, typically given by the number of spin (or helicity) states. We will not delve into the details of the solution of the Boltzmann equation for the different species here. \\

As seen above, the connection between the perturbed metric \eqref{eq:pert_FLRW_metric} and the stress-energy tensor derived through the Boltzmann equations can be made via the Einstein equations \eqref{eq:EFE}. Inserting the perturbed metric into the Einstein tensor \eqref{eq:Gmunu_vs_metric} we get a 0-order term (the Friedmann equations) and a first-order term, which we are interested in in the context of the present discussion. The time-time component of the perturbation to the Einstein tensor is 
\begin{align}
     \delta G^0_{\;\; 0} & =8 \pi G \delta T^0_{\;\; 0} \\
     \delta G^0_{\;\; 0} & =-6 H \Phi_{, 0}+6 \Psi H^2-2 \frac{k^2 \Phi}{a^2} \; ,
\end{align}
where the first line comes from Eq.~\eqref{eq:EFE} and the second from plugging the perturbed metric into Eq.~\eqref{eq:Gmunu_vs_metric}, at linear order in Fourier-space (with $k$ indicating the wavenumber) and the comma stands for the partial derivative:  $_{,\alpha} \equiv \partial/\partial x^\alpha$. \\
The Boltzmann equation then gives us the form of the $\delta T^0_{\;\;0}$, allowing us to couple the gravitational potentials $\Phi$ and $\Psi$ to the various components of matter and energy in the Universe. Here, we consider cold dark matter (CDM), baryons, photons, and neutrinos:
\begin{equation}\label{eq:einstein_eq_1}
    k^2 \Phi + 3 \frac{a^{\prime}}{a} \left( \Phi^{\prime} - \Psi \frac{a^{\prime}}{a} \right) = 4 \pi G a^2 \left[ \rho_{\mathrm{c}} \delta_{\mathrm{c}} + \rho_{\mathrm{b}} \delta_{\mathrm{b}} + 4 \rho_\gamma \Theta_0 + 4 \rho_\nu \mathcal{N}_0 \right] \; ,
\end{equation}
where the derivative w.r.t. the conformal time, defined as
\begin{equation}\label{eq:conformal_time}
    \eta(t) = \int_0^t\frac{\diff t'}{a(t')} \; ,
\end{equation}
is denoted by the prime symbol (${}'$), $a^{\prime}/a$ represents the Hubble parameter in terms of conformal time, and we have defined the \textit{density contrast} for species $i$ as the relative difference between the local and average density at a given redshift $z$:
\begin{equation}
    \delta_i(\vec{x},z) = 
    \frac{\rho_i(\vec{x},z) - \bar{\rho_i}(z)}{\bar{\rho_i}(z)} 
    \in [-1, \infty] \; ,
\end{equation}
where $\vec{x}$ is the comoving coordinate vector and $\bar{\rho}_i$ the average density. The meaning of the different contributions on the r.h.s. is listed below:
\begin{itemize}
    \item $\rho_{\mathrm{c}} \delta_{\mathrm{c}}$ represents the contribution from cold dark matter, where $\rho_{\mathrm{c}}$ is its energy density.

    \item $\rho_{\mathrm{b}} \delta_{\mathrm{b}}$ is the analogous for baryonic matter.
    
    \item $4 \rho_\gamma \Theta_0$ represents the contribution from photons. $\rho_\gamma$ is the photon energy density and $\Theta_0$ is the monopole term of the photon temperature perturbation in a multipole expansion.
    
    \item $4 \rho_\nu \mathcal{N}_0$ is the analogous for \ML{massless} neutrinos, with $\mathcal{N}_0$ the monopole term of the neutrino distribution function perturbation. \ML{In principle, one such term would be required for each mass eigenstate; neutrino masses can however be neglected at least until recombination \citep{Dodelson2020}}.
\end{itemize}
The second equation, from the spatial part of the Einstein equations, gives instead
\begin{equation}\label{eq:einstein_eq_2}
k^2(\Phi+\Psi)=-32 \pi G a^2\left[\rho_\gamma \Theta_2+\rho_\nu \mathcal{N}_2\right] \; ,
\end{equation}
which tells us that the two gravitational potentials $\Phi$ and $\Psi$ are equal and opposite unless the photons or neutrinos exhibit a non-negligible quadrupole moment, which is true only for collisionless neutrinos during radiation domination. \\
To solve this system of equations, we need some initial conditions; these will be discussed in the next section.
\subsection{Initial conditions}
In order to solve the system of equations for the evolution of perturbations, the initial conditions for all fields should be set in principle; we will see how, under specific assumptions, the initial conditions for just one field will suffice. At early times, such that $k \eta \ll 1$, the Boltzmann equations for radiation and cold dark matter become:
\begin{align}
    \Theta_0^{\prime}+\Phi^{\prime}=0, \nonumber \\
    \mathcal{N}_0^{\prime}+\Phi^{\prime}=0, \nonumber \\
    \delta_b^{\prime}= - 3 \Phi^{\prime}, \nonumber \\
    \delta_c^{\prime}= - 3 \Phi^{\prime} \nonumber \; .
\end{align}
In this epoch radiation is the dominant energy component, so we can neglect the matter terms in the r.h.s. of Eq.~\eqref{eq:einstein_eq_1} to get:
\begin{equation}\label{eq:time_time_early_times}
    3 \frac{a'}{a}\left(\Phi^{\prime} - \frac{a'}{a} \Psi\right)=
    16\pi G a^2 \left(\rho_\gamma \Theta_0+\rho_\nu \mathcal{N}_0\right) \; .
\end{equation}
Combining this with the above early-time equations for photons and neutrinos and \eqref{eq:einstein_eq_2} and setting $\Phi \simeq-\Psi$ (i.e. neglecting the neutrinos and photons quadrupole moment) we obtain:
\begin{equation}
\Phi^{\prime \prime} \eta+4 \Phi^{\prime}=0 \; .
\end{equation}
To set the initial conditions, we just need to specify the properties of the perturbations. Inflationary models predict that at early times the Universe contains a stochastic background of Gaussian, adiabatic, and nearly scale-invariant perturbations. The adiabatic perturbation hypothesis has been tested with great accuracy by CMB experiments, such as \Planck \citep{Planck2020_inflation}. For such perturbations, different points of the Universe have different overdensities, but the relative density perturbations are the same for all species:
\begin{equation}
\frac{\delta \rho_i}{\rho_i}=\frac{\delta \rho}{\rho} \; ,
\end{equation}
for $i$ running over the different components. With this choice, the evolution equations become:
\begin{equation}
\Phi\left(\vec{k}, \eta_i\right)=2 \Theta_0\left(\vec{k}, \eta_i\right)=2 \mathcal{N}_0\left(\vec{k}, \eta_i\right)
\end{equation}
\begin{equation}
\delta_{\mathrm{c}}(\vec{k}, \eta_i)=\delta_{\mathrm{b}}(\vec{k}, \eta_i)=3 \Theta_0(\vec{k}, \eta_i) = \frac{3}{2}\Phi(\vec{k}, \eta_i)
\end{equation}
where $\eta_i$ is the conformal time (Eq.~\ref{eq:conformal_time}) of the initial conditions.
The last two equations have a profound consequence: it is sufficient to set the initial conditions just for one field, for example, the metric perturbation $\Phi$.

It is commonly assumed that the primordial fluctuations of the field $\Phi_{\rm p}$ are Gaussian with zero mean $\langle \Phi_{\rm p}(k) \rangle = 0$ and with a two-point function given in Fourier space by 
\begin{equation}
\left\langle\Phi_{\rm p}(\vec{k}) \Phi_{\rm p}^*\left(\vec{k}^{\prime}\right)\right\rangle=P_{\Phi_{\rm p}}(k)(2 \pi)^3 \delta^3_{\rm D}\left(\vec{k}-\vec{k}^{\prime}\right) .
\end{equation}
where $P_{\Phi_{\rm p}}(k)$ is the power spectrum (PS) of the field $\Phi_{\rm p}$, which gives its variance as a function of scale (see Sect.~\ref{sec:2_point_stat} for a formal introduction of two-point statistics and the PS), and $\delta^3_{\rm D}$ the three-dimensional Dirac delta. The initial perturbation power spectrum is usually set to a power law with the Harrison-Zel'dovich-Peebles parametrization:
\begin{equation}\label{eq:Harrison_pk}
P_{\Phi}(k)=A_{\rm s}\left(\frac{k}{k_{\rm p}}\right)^{n_{\rm s}-1} k^{-3}
\end{equation}
where the dimensionless parameter $A_{\rm s}$ sets the amplitude of the initial perturbations, $n_{\rm s}$, the \textit{scalar spectral index}, their scale dependence, and $k_{\rm p}$ is the pivot scale, which has to be fixed conventionally. Inflationary models predict $n_{\rm s} \simeq 1$, i.e., nearly scale-invariant perturbations.
Given the $A_{\rm s}$ and $n_{\rm s}$ parameters, the initial conditions are completely characterized.
\subsection{Linear structure formation}\label{sec:lin_struct_formation}
In this section, we want to find a form for the PS of the matter density contrast field $\delta_{\rm m}$ (the \textit{matter} PS) as a function of scale and time (expressed by $z$ or $a$). In linear theory, density modes with different wavenumbers $ k $ evolve independently, since the governing evolution equations for each $ k $ are decoupled.\\
This linear evolution can be divided into different phases, or regimes, based on the scale of the perturbation ($k$) relative to the scale of the horizon (which equals $\eta$ in natural units): the super-horizon regime, horizon crossing, and sub-horizon regime. The horizon grows with time, becoming larger than a broader and broader range of scales $k$. In particular, modes evolve in different ways depending on the dominant species at the epoch they \enquote{cross the horizon} -- whether it is during radiation ($ a \ll a_{\text{eq}} $) or matter ($ a \gg a_{\text{eq}} $) domination. This scale-dependent behavior can be factorized as a function of $ k $ and $ a $ in the potential $ \Phi $: the \textit{transfer function} $ T(k) $ and the \textit{growth function} $ D(a) $:
\begin{equation}
\Phi(\vec{k}, a) = \frac{3}{5} \Phi_{\rm p}(\vec{k}) T(k) \frac{D(a)}{a} \; .
\end{equation}
During the radiation-dominated era, smaller-scale modes of $ \Phi $ decay as they enter the horizon, and the growth of perturbations is suppressed. Conversely, larger scale modes remain constant, entering the horizon during the matter-dominated epoch. Intermediate scales exhibit modest decay as they cross the horizon near the matter-radiation equality.\\
Analytical models for the transfer function exist in the literature (e.g. \citealt{Eisenstein1998_transfer}), but are nowadays less relevant thanks to the accuracy and speed reached by numerical solvers of the coupled Boltzmann equations such as \texttt{CLASS}\footnote{\texttt{\url{https://lesgourg.github.io/class\_public/class.html}}} \citep{Blas2011} and \texttt{CAMB}\footnote{\texttt{\url{https://camb.info/}}} \citep{Lewis2011}. Additionally, these codes can go beyond the linear perturbative treatment expounded in this section and compute the power spectrum even with nonlinear effects, for example using $N$-body simulation-based fitting formulas like \texttt{halofit} \citep{Smith2003_halofit, Takahashi2012_halofit}, introduced in the next section.\\
\begin{figure}
  \begin{minipage}{.6\textwidth}
    \includegraphics[width=\linewidth]{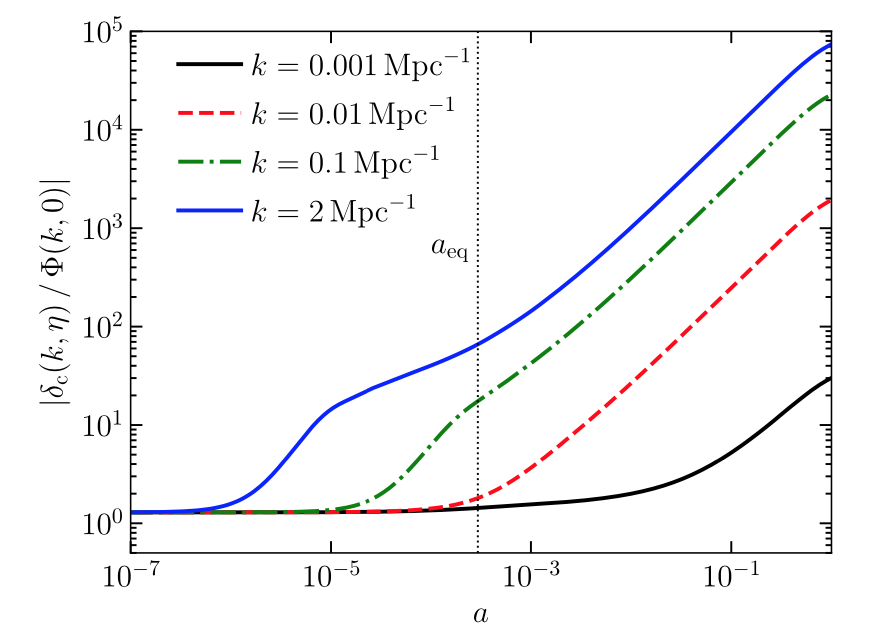}
  \end{minipage}%
  \begin{minipage}{.4\textwidth}
    \caption{Evolution of CDM perturbations in the concordance cosmological model, normalized by the potential at early times. During matter domination, all scales evolve identically, proportionally to $a$. At very late times, $\Lambda$ domination starts suppressing the growth again. Source: \citealt{Dodelson2020}}
    \label{fig:deltac_vs_a}
  \end{minipage}
\end{figure}
The dark matter overdensity evolves with $ \Phi $ as illustrated in Fig.~\ref{fig:deltac_vs_a}. Applying the late-time, no-radiation constraint of Eq.~\eqref{eq:einstein_eq_1} yields:
\begin{equation}
k^2 \Phi(\vec{k}, a) = 4 \pi G \rho_{\mathrm{m}}(a) a^2 \delta_{\mathrm{m}}(\vec{k}, a) \quad (k \gg a H) \; ,
\end{equation}
and substituting the zeroth-order Friedmann equation provides an expression for $\delta_{\mathrm{m}} $:
\begin{equation}
\delta_{\mathrm{m}}(\vec{k}, a) = \frac{2}{5} \frac{k^2}{\Omega_{\mathrm{m}, 0} H_0^2} \Phi_{\rm p}(\vec{k}) T(k) D(a) \; .
\end{equation}
The parameterization of the linear power spectrum of matter density fluctuations is then given by:
\begin{equation}
P_{\delta \delta}^{\mathrm{L}}(k, a) = \frac{8 \pi^2}{25} \frac{A_{\mathrm{s}}}{\Omega_{\mathrm{m}, 0}^2} D^2(a) T^2(k) \frac{k^{n_{\mathrm{s}}}}{H_0^4 k_{\mathrm{p}}^{n_{\mathrm{s}}-1}}.
\end{equation}
Traditionally, $ \sigma_8 $, the standard deviation of cold matter perturbations at scale $ 8 h^{-1} \text{Mpc} $, is used instead of the scalar amplitude $ A_{\mathrm{s}} $. The relationship between them is:
\begin{equation}
\sigma_8^2 = \int \frac{\mathrm{d} k}{k} \frac{k^3 P_{\mathrm{L}}(k)}{2 \pi^2} |W_{\mathrm{TH}}(k R_8)|^2 \; ,
\end{equation}
where 
\begin{equation}
W_{\mathrm{TH}}(x) = \frac{3(\sin x - x \cos x)}{x^3} 
\end{equation}
is the Fourier-transformed top-hat filter. The evolution equation for the growth function $ D(a) $ can now be derived:
\begin{equation}\label{eq:delta_m_linear}
    \frac{\diff^2 \delta_{\mathrm{m}}}{\partial a^2} + \frac{\partial \ln (a^3 H)}{\partial a} \frac{\partial \delta_{\mathrm{m}}}{\partial a} - \frac{3 \Omega_{\mathrm{m}, 0} H_0^2}{2 a^5 H^2} \delta_{\rm{m}} = 0.
\end{equation}
In the late Universe, this equation simplifies to an integral form:
\begin{equation}
D(a) = \frac{5 \Omega_{\rm{m}, 0}}{2} \frac{H(a)}{H_0} \int_0^a \frac{\diff a'}{(a' H(a') / H_0)^3} \; .
\end{equation}
A useful empirical fit exists for the logarithmic growth rate $ f(a) $ \citep{Lahav1991_gamma055, Linder2005}:
\begin{equation}\label{eq:growth_rate}
f(a) = \ML{\frac{\diff \ln D(a)}{\diff \ln a}} \approx [\Omega_{\rm{m}, 0}(a)]^\gamma \; ,
\end{equation}
where $ \gamma $ is the \textit{growth parameter}, approximately 0.55 in $ \Lambda \text{CDM} $. This parameterization is useful to detect deviations from GR, which would result in a different observed value for the $\gamma$ parameter. In fact, placing tight constraints on $\gamma$ is one of the main aims of the \Euclid mission \citep{laureijs2011euclid}.\\

A more complete description of the growth of structures, allowing, for example, a scale-dependent growth function, can be obtained by introducing the $\mu_{\rm MG}$ and $\Sigma_{\rm MG}$ free functions \citep{Planck2016_MG}:
\begin{align}
k^2\Psi &= -\mu_{\rm MG}(a,k) 4\pi Ga^2\sum_i[\rho_i\delta_i
+ 3(\rho_i + p_i)\sigma_i]\\
k^2(\Psi + \Phi) &= -\Sigma_{\rm MG}(a,k) 4\pi Ga^2\sum_i[2\rho_i\delta_i
- 3(\rho_i + p_i)\sigma_i].
\end{align}
$\mu_{\rm MG}$ and $\Sigma_{\rm MG}$ parametrize respectively the deviations of the growth of structures and light deflection from the GR case. $\sigma_i$ is the anisotropic stress, non-vanishing for relativistic species; it is negligible at low redshifts. In the  $(\mu_{\rm MG},\Sigma_{\rm MG}) = (1,1)$ case we re-obtain the $\Lambda$CDM model; deviations from these values can be due either to modified gravity or to the presence of extra relativistic species. \\
We can relate $\mu_{\rm MG}$ to $\gamma$ via \citep{ISTF2020} -- hereafter \citetalias{ISTF2020}:
\begin{equation}
\mu_{\rm MG}(a,\gamma) = \frac{2}{3}\Omega_{\rm m}^{\gamma-1}
\left[ \Omega_{\rm m}^\gamma + 2 + \frac{H'}{H}+ \gamma\frac{\Omega'_m}{\Omega_{\rm m}} + \gamma'\ln\Omega_{\rm m}
\right]
\end{equation}
where $\Omega_{\rm m} = \Omega_{\rm m}(a)$ and the prime indicates differentiation with respect to $\ln a$. For a constant $\gamma$ we still have to choose the form of the second free function, $\Sigma_{\rm MG}$: we can pick $\Sigma_{\rm MG} = 1$ to recover the same light deflection as in $\Lambda$CDM. In this way, a $\gamma\neq0.55$ value would only affect the growth of structures through $\mu_{\rm MG} \neq 1$, and the resultant different dependence of the matter PS from the growth rate \eqref{eq:growth_rate}. 
\subsection{Nonlinear scales}\label{sec:nonlinear_scales}
The unprecedented resolution of \Euclid observations will allow probing the galaxy distribution and the weak lensing signal up to very small angular scales, where structure formation enters the nonlinear regime. This means that the linear treatment of perturbations seen so far, which allowed us to derive an analytical expression for the linear matter power spectrum, is no longer valid. Density perturbations are larger on smaller scales, hence structures collapse on small scales first; the large-scale structures are then built hierarchically \citep{Dodelson2020}. Therefore, to begin with, we can look for a more accurate description of the clustering of DM on small scales, which is much simpler to model being collisionless, with no baryonic effects such as radiative cooling of gas, AGN feedback and so forth, which become important only deeply in the nonlinear regime ($k \gtrsim 1 \, h \, {\rm Mpc}^{-1}$) anyways. The following discussion will introduce the perturbative and halo-model approach to the problem.
\subsubsection{Perturbation theory}
Perturbation theory is a possible (analytical) approach to the solution of this problem, at least up to scales at which the perturbative correction to the leading order term (the linear PS, in this case) is small ($k\lesssim 0.2 \, h \, {\rm Mpc}^{-1}$, \citealt{Dodelson2020}). The starting point of this treatment is the collisionless Boltzmann (Vlasov) equation and the Poisson equation, governing the dynamics of the (dark) matter density and velocity fields. By taking and truncating moments of this equation we obtain the simpler fluid (continuity and Euler) equations \citep{Bernardeau2002_pert, Desjacques_2018}:
\begin{equation*}
\begin{aligned}
& \frac{\partial}{\partial \eta} \delta(\vec{x}, \eta)+\vec{\nabla} \cdot\{[1+\delta(\vec{x}, \eta)] \vec{v}(\vec{x}, \eta)\}=0 \\
& \frac{\partial}{\partial \eta} \vec{v}(\vec{x}, \eta)+[\vec{v}(\vec{x}, \eta) \cdot \vec{\nabla}] \vec{v}(\vec{x}, \eta)+\mathcal{H}(\eta) \vec{v}(\vec{x}, \eta)=-\vec{\nabla} \Phi(\vec{x}, \eta) \\
& \nabla^2 \Phi(\vec{x}, \eta)=\frac{3}{2} \mathcal{H}^2 \Omega_{\rm m}(\eta) \delta(\vec{x}, \eta) \; .
\end{aligned}
\end{equation*}
Having introduced $\mathcal{H} = a'/a = aH$ and writing $\Omega_{\rm m}(\eta)$ to highlight that we are referring to the density parameter as a function of (conformal) time. In Fourier space these equations become, introducing the velocity divergence $\theta \equiv \vec{\nabla} \cdot \vec{v}(\vec{x}, \eta)$ and using the Fourier transform of the Poisson equation to eliminate $\Phi$:
\begin{align}
    & \frac{\partial \delta(\vec{k}, \eta)}{\partial \eta}+\theta(\vec{k}, \eta)=-\int_{\vec{k}_1} \int_{\vec{k}_2}(2 \pi)^3 \delta_{\rm D}\left(\vec{k}-\vec{k}_{12}\right) \alpha\left(\vec{k}_1, \vec{k}_2\right) \theta\left(\vec{k}_1, \eta\right) \delta\left(\vec{k}_2, \eta\right) \label{eq:ddelta_deta}\\
    & \frac{\partial \theta(\vec{k}, \eta)}{\partial \eta}+\mathcal{H}(\eta) \theta(\vec{k}, \eta)+\frac{3}{2} \mathcal{H}^2(\eta) \Omega_{\rm m}(\eta) \delta(\vec{k}, \eta)=\nonumber \\
    & \quad\quad -\int_{\vec{k}_1} \int_{\vec{k}_2}(2 \pi)^3 \delta_{\rm D}\left(\vec{k}-\vec{k}_{12}\right) \beta\left(\vec{k}_1, \vec{k}_2\right) \theta\left(\vec{k}_1, \eta\right) \theta\left(\vec{k}_2, \eta\right) \label{eq:dtheta_deta}\; .
\end{align}
with $\vec{k}_{i j \ldots} \equiv \vec{k}_i+\vec{k}_j+\cdots$, and
\begin{equation}
    \alpha\left(\vec{k}_1, \vec{k}_2\right)=\frac{\vec{k}_{12} \cdot \vec{k}_1}{k_1^2}, \quad \beta\left(\vec{k}_1, \vec{k}_2\right)=\frac{k_{12}^2\left(\vec{k}_1 \cdot \vec{k}_2\right)}{2 k_1^2 k_2^2} \; ,
\end{equation}
These can be solved using a perturbative approach:
\begin{equation}\label{eq:perturbative_expansion}
    \delta(\vec{k}, \eta)=\sum_{n=1}^{\infty} \delta^{(n)}(\vec{k}, \eta), \quad \theta(\vec{k}, \eta)=\sum_{n=1}^{\infty} \theta^{(n)}(\vec{k}, \eta),
\end{equation}
with $n$ indicating the order of the perturbative expansion. \\
On large scales ($k \rightarrow 0, \delta^{(1)} \ll 1$), the quadratic source term on the right-hand side (r.h.s.) of Eqs.~\eqref{eq:ddelta_deta} and~\eqref{eq:dtheta_deta} can be neglected. This is equivalent to assuming that the interactions between different Fourier modes are negligible: we obtain then the differential equation for the evolution of the linear density contrast (which is Eq.~\ref{eq:delta_m_linear}):
\begin{equation}
    \frac{\partial^2}{\partial \eta^2} \delta^{(1)}(\vec{k}, \eta)+\mathcal{H}(\eta) \frac{\partial}{\partial \eta} \delta^{(1)}(\vec{k}, \eta)-\frac{3}{2} \Omega_{\rm m}(\eta) \mathcal{H}^2(\eta) \delta^{(1)}(\vec{k}, \eta)=0
\end{equation}

The equations of motion~\eqref{eq:ddelta_deta} and~\eqref{eq:dtheta_deta} suggest the ansatz of writing the $n$-th order solution in Eq.~\eqref{eq:perturbative_expansion} as:
\begin{align}
    \delta^{(n)}(\vec{k}, \eta) & =\int_{\vec{k}_{1}} \cdots \int_{\vec{k}_{n}}(2 \pi)^{3} \delta_{\rm D} \left(\vec{k}-\vec{k}_{12 \cdots n}\right) \nonumber \\
    & \quad \times F_{n}\left(\vec{k}_{1}, \cdots, \vec{k}_{n}, \eta\right)  \delta^{(1)}\left(\vec{k}_{1}, \eta\right) \cdots \delta^{(1)}\left(\vec{k}_{n}, \eta\right) \\
    \theta^{(n)}(\vec{k}, \eta) & =-\mathcal{H}(\eta) f(\eta) \int_{\vec{k}_{1}} \cdots \int_{\vec{k}_{n}}(2 \pi)^{3} \delta_{\rm D}\left(\vec{k}-\vec{k}_{12 \cdots n}\right) \nonumber \\
    & \quad \times G_{n}\left(\vec{k}_{1}, \cdots, \vec{k}_{n}, \eta\right) \delta^{(1)}\left(\vec{k}_{1}, \eta\right) \cdots \delta^{(1)}\left(\vec{k}_{n}, \eta\right),
\end{align}
with $F_{n}$ and $G_{n}$ being respectively the density and velocity divergence kernels, which are time-independent in Einstein-de Sitter cosmologies ($\Omega_{{\rm m}, 0} = 1$) -- and to good approximation in $\Lambda$CDM --, where they take the form:
\begin{align}
F_{2}\left(\vec{k}_{1}, \vec{k}_{2}\right) & =\frac{5}{7}+\frac{2}{7} \frac{\left(\vec{k}_{1} \cdot \vec{k}_{2}\right)^{2}}{k_{1}^{2} k_{2}^{2}}+\frac{\vec{k}_{1} \cdot \vec{k}_{2}}{2 k_{1} k_{2}}\left(\frac{k_{1}}{k_{2}}+\frac{k_{2}}{k_{1}}\right) \label{eq:F2_kernel}\\
G_{2}\left(\vec{k}_{1}, \vec{k}_{2}\right) & =\frac{3}{7}+\frac{4}{7} \frac{\left(\vec{k}_{1} \cdot \vec{k}_{2}\right)^{2}}{k_{1}^{2} k_{2}^{2}}+\frac{\vec{k}_{1} \cdot \vec{k}_{2}}{2 k_{1} k_{2}}\left(\frac{k_{1}}{k_{2}}+\frac{k_{2}}{k_{1}}\right) \label{eq:G2_kernel} .
\end{align}
We finally arrive at the solution for the \enquote{next-to-leading order} (NLO) contribution to the matter PS (and velocity PS, which we do not write explicitly), which allows expressing the nonlinear PS as
\begin{equation}
    P_{\rm mm}(k)=P_{\rm L}(k)+P_{\rm m m}^{\rm{NLO}}(k) \; ,
\end{equation}
with
\begin{align}
    & P_{\rm m m}^{\rm{NLO}}(k) = P_{\rm m m}^{(22)}(k)+2 P_{\rm m m}^{(13)}(k) \\
    & P_{\rm m m}^{(22)}(k) \equiv\left\langle\delta^{(2)}(\vec{k}) \delta^{(2)}\left(\vec{k}^{\prime}\right)\right\rangle=2 \int \frac{\diff^3\vec{p}}{(2\pi)^3}\left[F_{2}(\vec{p}, \vec{k}-\vec{p})\right]^{2} P_{\rm L}(p) P_{\rm L}(|\vec{k}-\vec{p}|) \\
    & P_{\rm m m}^{(13)}(k) \equiv\left\langle\delta^{(1)}(\vec{k}) \delta^{(3)}\left(\vec{k}^{\prime}\right)\right\rangle=3 P_{\rm L}(k) \int \frac{\diff^3\vec{p}}{(2\pi)^3} F_{3}(\vec{p},-\vec{p}, \vec{k}) P_{\rm L}(p) \; .
\end{align}
In the above equation, we indicated with $P_{\rm L}$ the linear matter PS and introduced the third-order mode-coupling density kernel $F_3$:
\begin{align}\label{eq:F3_kernel}
    F_3\left(\vec{k}_1, \vec{k}_2, \vec{k}_3\right) & \equiv \frac{7}{18} \frac{\vec{k}_{12} \cdot \vec{k}_1}{k_1^2}\left[F_2\left(\vec{k}_2, \vec{k}_3\right)+G_2\left(\vec{k}_1, \vec{k}_2\right)\right] \nonumber \\
    & \quad + \frac{1}{18} \frac{k_{12}^2\left(\vec{k}_1 \cdot \vec{k}_2\right)}{k_1^2 k_2^2}\left[G_2\left(\vec{k}_2, \vec{k}_3\right)+G_2\left(\vec{k}_1, \vec{k}_2\right)\right] \; .
\end{align}
Perturbation theory will also be used in Sect.~\ref{sec:hm_responses} to derive some of the terms needed to compute the matter power spectrum response to a shift (perturbation) in the background density. 
\subsubsection{Halo Model}\label{sec:pk_nonlin_halomodel}
\begin{figure}
    \centering
    \includegraphics[width=0.8\textwidth]{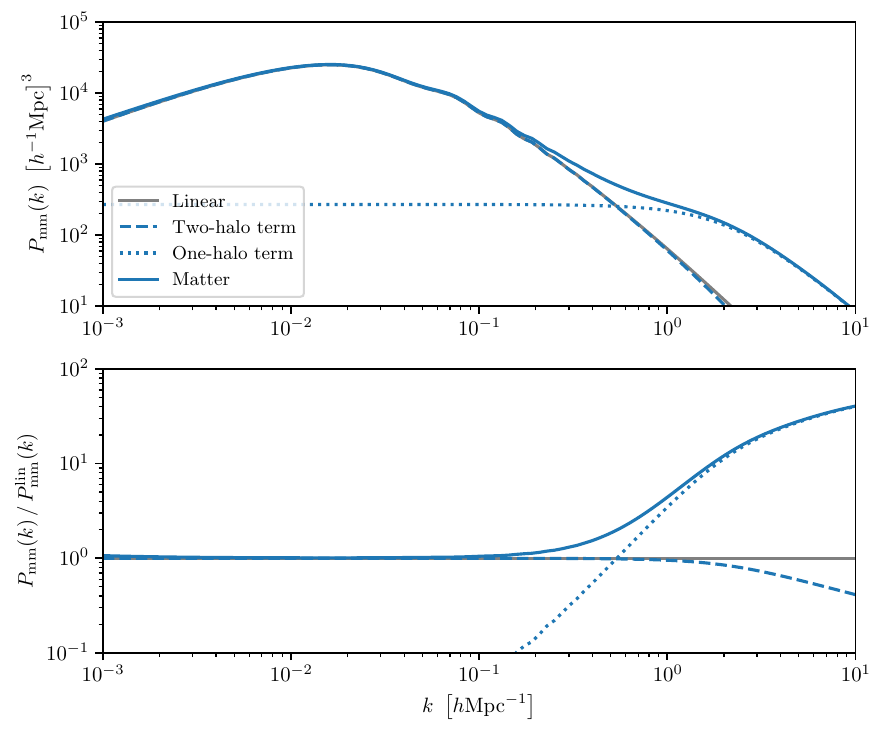}
    \caption{The one- and two-halo terms in the matter PS computed with the HM. The two-halo term dominates on large scales and captures well the linear PS (see bottom plot) up to $k\sim 1 h \, {\rm Mpc}^{-1}$, where the one-halo term is already dominant. The one-halo term acts as a shot-noise contribution on large scales -- unphysical but of very little effect --, becoming important at the nonlinear threshold $k \sim 0.1 h \, {\rm Mpc}^{-1}$. Plot produced with the \texttt{pyhalomodel} code \citep{Asgari2023_halomodel_review}, adapted from one of the notebooks available at \texttt{\url{https://github.com/alexander-mead/pyhalomodel/tree/main}}}
    \label{fig:pk_hm_terms}
\end{figure}
Another powerful framework to model the nonlinear clustering of matter is the halo model (HM, \citet{Seljak2000_halomodel, Ma2000_halomodel, Peacock2000_halomodel}; see also \cite{Asgari2023_halomodel_review} for a recent comprehensive review). This model relies on the main assumption that most of the matter in our Universe is concentrated in \textit{halos}, regions with $\sim 200$ times the critical density (although different definitions are possible). The power spectrum is then seen as the sum of two contributions:
\begin{equation}
    P(k) = P_\mathrm{1h}(k) + P_\mathrm{2h}(k) \; ,
\end{equation}
where $P_\mathrm{1h}(k)$ describes the two-point correlations amongst the same halo (\enquote{intra-halo}, or 1-halo), which become relevant at small (\enquote{sub-halo}) scales, and $P_\mathrm{2h}(k)$ captures the correlations amongst two different halos (\enquote{inter-halo}, or 2-halo), which describe the large-scale PS. These are shown in Fig.~\ref{fig:pk_hm_terms}. A key assumption of this approach is that the growth of halos should depend only on the local physics at the scale of the halo, and not on the large-scale distribution of matter \citep{Bird2012_nu}.\\ 
The expressions for these two terms are given by:
\begin{align}
P_\mathrm{1h}(k) & = I_2^0(k, k) \\
P_\mathrm{2h}(k) & = \left[I_1^1(k)\right]^2 P_\mathrm{L}(k) \; ,
\end{align}
having defined the auxiliary halo model functions $I_\mu^\beta(k)$ \citep{Cooray2001_halomodel_integral}, consisting essentially of integrals over the halo mass:
\begin{equation}
    I_\mu^\beta\left(k_1, \ldots, k_\mu\right) \equiv \int \diff M \Phi_{\rm MF} \left(\frac{M}{\bar{\rho}_\mathrm{m}}\right)^\mu b^\mathrm{h}_\beta \prod_{i=1}^\mu \tilde{u}\left(k_i\right | M) \; ,
    \label{eq:halomodel_auxiliary}
\end{equation}
$M$ being the halo mass, $\Phi_{\rm MF}$ the halo mass function, $\bar{\rho}_\mathrm{m}$ the background matter density, $b_0^\mathrm{h}=1$, $b_1^\mathrm{h}=b^\mathrm{h}(M)$ the linear halo bias, and $\tilde{u}_M(k | M)$ the Fourier transform of the halo density profile normalized so that $\tilde{u}(k=0 | M)=1$.\\
The halo mass $M$ is usually defined as the mass enclosed within a sphere of radius $R$ where the average density is $\Delta$ times the critical (or mean matter) density of the Universe, with e.g. $\Delta=200$ (as mentioned above). 
The halo mass function $\Phi_{\rm MF} = \diff n/\diff M$ describes the number density $n$ of halos as a function of their mass, commonly described through the  Press-Schechter formalism \citep{Press1974_PSformalism} and its extensions.
Analogously to the linear galaxy bias, the linear halo bias $b^\mathrm{h}(M)$ quantifies how the distribution of halos of a given mass $M$ traces the underlying matter distribution. A bias of $b^\mathrm{h}(M) = 1$ means that halos trace the matter distribution perfectly (that is, $\delta_\mathrm{halos}(\vec{x}) = \delta_\mathrm{m}(\vec{x})$), while $b^\mathrm{h}(M) > 1$  or  $b^\mathrm{h}(M) < 1$ indicate that halos are respectively more or less clustered than the total matter.
The halo density profile describes the density of matter as a function of radial distance from the center of a dark matter halo. This profile is crucial for understanding the internal structure of halos and is usually modelled using profiles like the Navarro-Frenk-White (NFW; \citealt{Navarro1996_NFW_profile}):
\begin{equation}
    \rho(r) = \frac{\rho_s}{\frac{r}{R_s} \left(1 + \frac{r}{R_s}\right)^2} \; ,
\end{equation}
with $\rho_s$ and $R_s$ the halo-dependent scaling density and scaling radius respectively, or the Einasto profile.
The auxiliary halo model function $I_\mu^\beta$ allows integrating over all halos to get the total power spectrum (or correlation function), encapsulating the contributions from halos of all masses and sizes.\\

The halo model is at the core of commonly adopted semi-analytical models like \texttt{halofit} \citep{Smith2003_halofit}, which is based on empirical fitting formulas obtained from $N$-body DM simulations. Originally introduced for a CDM universe, it has been updated in \citet{Takada2013} to use better simulation and to account for DE with a constant equation of state (reaching a quoted accuracy of 5\% for $k \leq 1 \, h \, {\rm Mpc}^{-1}$ at $0 \leq z \leq 10$ and 10\% for $1 \, h \, {\rm Mpc}^{-1} \leq k \leq 10 \, h \, {\rm Mpc}^{-1}$ at $0 \leq z \leq 3$) and in \citet{Bird2012_nu} to account for massive neutrinos, which as seen in Sect.~\ref{sec:massive_neutrinos} suppress the nonlinear power spectrum. We will take this (dubbed \texttt{TakaBird}) as the reference model for the nonlinear clustering of matter, and discuss other possible options in Chap.~\ref{chap:scalecuts}.\\
We note that this model does \textit{not} account for baryonic feedback, which has a non-negligible effect on very small scales (up to 15\% suppression of power at $k \sim 10 \, h \, {\rm Mpc}^{-1}$ and $z=0$, \citealt{Chisari2018_baryons}) and is even harder to model.\\

Lastly, we note that the Vlasov-Poisson system of equations can be solved numerically: this is the approach adopted in $N$-body simulations, where the phase-space is volume discretized and its evolution is solved (time-)step by step. DM halos can then be identified and populated with galaxies through, e.g., a halo occupation distribution (HOD) or the abundance matching technique. We will not explore the details of numerical simulations here, but we will rely on some results from the \Euclid \enquote{Flagship} DM simulation \citep{Potter2016} in the next chapters.

\section{Neutrinos}\label{sec:massive_neutrinos}
\ML{Neutrinos are chargeless, weakly interacting leptons the existence of which was 
postulated in 1930 by Wolfgang Pauli to explain the apparent non-conservation of energy in beta decay processes, specifically in the context of neutron decay (see \citealt{Ramond1999_neutrinos} for a brief historical review)}:
\begin{equation}
    \ML{n \rightarrow p + e^- + \bar{\nu}_e \; .}
\end{equation}
They can be categorised according to their type, or \textit{flavor} (electronic, $\nu_e$, muonic, $\nu_\mu$, or tauonic, $\nu_\tau$) or their mass ($\nu_1,\nu_2, \nu_3 $) eigenstates. Since they interact so little with matter, they are called a \textit{free streaming} component, i.e. they can travel for enormous distances without interacting; this makes detecting them a very hard job. The free streaming length can reach cosmological scales, and is sometimes defined as the distance travelled in the Hubble time $t_H=H_0^{-1}$ (while they are relativistic). \\

Neutrinos - and photons - are usually assumed to be the only light (sub-KeV) relic particles still existing after Big Bang Nucleosynthesis (BBN). The study of these particles in Cosmology mainly revolves around the effect they leave on BBN, CMB anisotropies and large-scale clustering. This is because cosmological experiments are sensitive to the effective number of neutrino species $N_{\rm eff}$ - defined as the relativistic energy density beyond that of photons, in units of one neutrino, in the instantaneous decoupling limit \citep{Verde2010} - and the sum of their masses $\Sigma m_\nu$. \\
\ML{The \textit{effective} number of neutrino species is related to the ratio $\frac{\rho_\nu}{\rho_\gamma}$ via the relation}
\begin{equation}
    \frac{\rho_\nu}{\rho_\gamma} = \frac{7}{8}N_{\rm eff}\left(  \frac{4}{11} \right)^{4/3} \; .
\end{equation}
\ML{If the three species decouple instantaneously from the primordial plasma (at a temperature $T\simeq 2$MeV), $N_{\rm eff} $ will indeed be equal to the number of neutrino species $N_{\nu} = 3$; in a more realistic treatment, accounting for non-instantaneity of the decoupling process, we can use the same formula with a slightly different value for $N_{\rm eff}$ (e.g., the Standard Model prediction $N_{\rm eff} = 3.044$, see \citealt{Bennett2021_Neff})}.

Neutrinos are characterised by being a relativistic component at early times, before switching to the non-relativistic regime (\enquote{nr}) at a redshift corresponding to the moment their average momentum equals their mass $m_i$, with $i$ running over the different mass eigenstates:
\begin{equation}
    z^{\rm nr}_i = \frac{m_i}{0.53 \mathrm{meV}}-1 \; .
\end{equation}
While relativistic, they contribute to the radiation component ($\Omega_{\rm r} = \Omega_\gamma + \Omega_\nu$), then they become part of the matter component ($\Omega_{\rm m} = \Omega_{\rm b} + \Omega_{\rm c} + \Omega_\nu$). This transition happens at $z\simeq 110$ for $m_\nu = 0.6 $ eV, thus in the matter-dominated regime. As mentioned, when non-relativistic, neutrinos make a very good Dark Matter candidate, being so weakly interacting; however they have large thermal velocity, given by $v_i/c = 0.53(1+z)\mathrm{meV}/m_i$. Mixed Dark Matter models treat them as a Hot DM fraction.\\

\begin{figure}
\begin{center}
\includegraphics[scale=0.4]{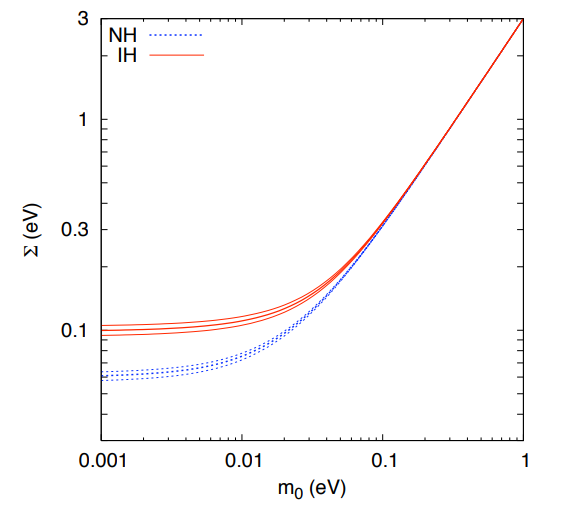}
\caption{Total neutrino mass as a function of the lightest state $m_0$ within the $3\sigma$ regions. Blue dotted (red solid) lines correspond to normal (inverted) hierarchy for neutrino masses, where $m_0 = m_1$ $(m_0 = m_3)$. Source: \citep{Lesgourgues2012_nu}}
\label{fig:neutrino_mass_degeneracy}
\end{center}
\end{figure}

One of the most important properties of these particles is their mass; the phenomenon of neutrino oscillation - the variation of the flavour during their propagation - suggests that at least two mass eigenstates have non-zero mass. In particular, oscillation of solar and atmospheric neutrinos has allowed measuring the differences of squared neutrino masses, while remaining blind to the absolute mass scale. This allows for two possible mass orderings, or \textit{hierarchies} \ML{\citep{Esteban2020, deSalas2021, Capozzi2021}}:
\begin{align}
\Delta m_{21}^2 = m_2^2-m_1^2 = & \, 7.37 \times 10^{-5} \, \text{eV}^2 \\
\Delta m^2 =m_3^2 - (m_1^2+m_2^2)/2 = & \, 2.50 \times 10^{-3} \, \text{eV}^2 \,&(\text{Normal Hierarchy})\\
= & -2.46 \times 10^{-3} \, \text{eV}^2 \,&(\text{Inverted Hierarchy})
\end{align}
Because of the small difference between $m_1$ and $m_2$, we have $\Delta m^2 \simeq |\Delta m_{13}^2| \simeq |\Delta m_{23}^2|$. If the lightest neutrino ($m_1$ or $m_3$) is much heavier than the mass differences we deal with almost degenerate species, since the masses would be very close (see Fig.~\ref{fig:neutrino_mass_degeneracy}). If instead the lightest neutrino mass is close to zero, we can estimate the lower limits of $\Sigma m_\nu = m_1 + m_2 + m_3$ as:
\begin{align} 
    &\text{NH}: m_1 \simeq 0 \rightarrow \Sigma m_\nu \simeq \sqrt{\Delta m_{12}^2} + \sqrt{\Delta m_{23}^2} \simeq 0.06 \, \text{eV}  \\
    &\text{IH}: m_3 \simeq 0 \rightarrow \Sigma m_\nu \simeq \sqrt{\Delta m_{13}^2} + \sqrt{\Delta m_{23}^2} \simeq 2\sqrt{\Delta m_{13}^2} + \sqrt{\Delta m_{12}^2} \simeq 0.1 \, \text{eV} 
\end{align}
Since the neutrino temperature today is greater than $|\Delta m_{31}^2|^{1/2}$ and $|\Delta m_{21}^2|^{1/2}$, at least two mass eigenstates are non-relativistic.
The best cosmology can do in this sense, \ML{ to this day}, is to place an upper bound on the total mass $\Sigma m_\nu = m_1 + m_2 + m_3$. Improving this upper limit is one of the aims of \Euclid. \\

The way to experimentally measure the neutrino mass is through its effect on the matter power spectrum. We can relate the neutrino energy density with their mass sum:
\begin{equation}
\Omega_\nu = \frac{\rho_\nu}{\rho_0} = \frac{\Sigma m_i}{93.14h^2\mathrm{eV}} \; ;
\end{equation}
we can define the neutrino density fraction $f_\nu$ as
\begin{equation}
    f_\nu \equiv \frac{\rho_\nu}{\rho_{c} + \rho_{\rm b} + \rho_\nu} = \frac{\Omega_\nu}{\Omega_{\rm m}} \; .
\end{equation}
Neutrinos do not cluster on scales smaller than their (comoving) free streaming length, which is related to their velocity (assumed to be the thermal velocity) via 
\begin{equation}
    \lambda_{\rm fs} = \frac{2\pi a(t)}{k_{\rm fs}} = 2\pi \sqrt{\frac{2}{3}}\frac{v_{\rm th}(t)}{H(t)} \simeq 126 \, \text{Mpc/h}
\end{equation}
with the last value calculated at the current time. The thermal velocity equals $c$ before the non-relativistic transition, and
\begin{equation}
    v_{\rm th} \simeq 158 (1+z)\left(\frac{1 \mathrm{eV}}{m} \right) \mathrm{km} \, \rm{s}^{-1}
\end{equation}
afterwards. On scales larger than $\lambda_{\rm fs}$, neutrinos cluster like CDM: this means that the power spectrum of matter fluctuations, related to the matter density fluctuations $\delta_{\rm m} = \delta_{\rm b} + \delta_{\rm c} + \delta_\nu$ is damped on small scales, because the velocity of neutrinos is larger than the escape velocity of the gravitational potential wells on those scales; the larger the neutrino mass, the more damping occurs \citep{Audren2013_nu}. The lack of clustering on scales $k>k_{\rm fs}\; (\simeq 0.05 \, h$ Mpc$^{-1}$ at current time for $m_\nu = 0.06$ eV) damps the matter power spectrum by a factor \citep{Bird2012_nu}
\begin{equation}
    \frac{\delta P_{\rm m}(k)}{P_{\rm m}(k)} \simeq -8 f_\nu \qquad \mathrm{for} \qquad f_\nu < 0.07 \; .
\end{equation}
On scales $k<k_{\rm fs}$ neutrinos could not free stream for the whole cosmic history, inducing a smaller suppression. The maximum scale at which neutrinos alter the PS is given by
\begin{equation}
    k_{\rm nr} \simeq 0.018 \sqrt{\Omega_{\rm m}\left( \frac{\Sigma m_\nu}{1 \mathrm{eV}} \right)} \frac{h}{\mathrm{Mpc}},
\end{equation}
for which they free-streamed
only immediately after becoming non-relativistic.
\section{Constraints on cosmological parameters from current experiments}
\FB{As stated above, \Euclid's aim is to place tight constraints on the cosmological parameters, for a variety of different models; to better contextualize this statement, we briefly report the current state of the art in the measurement of such parameters. \\
For CMB observations, the reference results are the ones from the \Planck mission, which constitute our fiducial choice. Some constraints for key parameters from the temperature and polarization anisotropies, as well as their cross-correlation and CMB lensing, are \citep{Planck2014} $\Omega_{{\rm m}, 0} = 0.3156 \pm 0.0091$ (3\%), $\sigma_8 =  0.831 \pm 0.013$ (1\%); for the time-varying DE EoS parameters, \Planck alone has a low constraining power, and is combined with external LSS datasets and observations of Supernovae Ia to get \citep{Planck2018_VI} $w_0 = -0.957 \pm 0.080$ (8\%) and $w_a = -0.29^{+0.32}_{-0.26}$ (100\%).}\\
\FB{Concerning LSS surveys, we quote the recent values from the Dark Energy Survey (DES: \citealt{Crocce2017, Hoyle2017}) Year 3 results \citep{DES_Y3_2022}; we refer to the combination of weak lensing cosmic shear and photometric galaxy clustering, which will be formally introduced in Sect.~\ref{chap:observables}, as well as the combination with datasets from previous LSS experiments, such as the Baryon Oscillation Spectroscopic Survey (BOSS, \citealt{Dawson2016_BOSS}). For the baseline $\Lambda$CDM parameters, DES reports $\Omega_{{\rm m}, 0} = 0.339^{+0.032}_{-0.031}$ (10\%) and $\sigma_8 = 0.733^{+0.039}_{-0.049}$ (6\%) uncertainties; and $w_0 \gtrsim -1.4$, $w_a = -0.9 \pm 1.2$, which improve to $w_0 = -0.95 \pm 0.08$ (8\%), $w_a = -0.4 \pm 0.4$ (100\%) when including other external LSS datasets. \Euclid-only photometric forecast quote in the optimistic case a precision of 0.3\% on $\Omega_{{\rm m}, 0}$ and 0.1\% on $\sigma_8$, and 2\% and 10\% on $w_0$ and $w_a$ respectively, suggesting the magnitude of the increase in precision. The forecast setup leading to these results will be illustrated in detail in the following Chapters.}

\chapter{Statistics}\label{chap:statistics}

One of the peculiarities of Cosmology is the fact that only one realization of the phenomenon to be characterized (the Universe itself) is accessible to the observer. Moreover, Cosmology has an inherently probabilistic nature: most of the processes we observe, especially in large-scale structure (LSS) studies, originate in the inflationary paradigm from quantum random fluctuations of the primordial Universe, which constitute its initial conditions. This makes the Bayesian approach to probability the most suited to analyze and interpret the data in the context of the different models under consideration.\\

This chapter will introduce the statistical framework and specific tools used in this work (and in general in Cosmology) to define meaningful physical observables, to forecast the expected constraints on the model parameters of interest and to perform inference on the actual data. We will mainly rely on Bayes' theorem, and introduce the Fisher Information Matrix (FM) as a powerful forecasting tool. We will then outline the two-point observables used in this work (following the \textit{de facto} standard in cosmological analyses), in particular the power spectrum. These tools constitute the building blocks of modern cosmological analyses. 
\section{Bayes' Theorem}\label{sec:bayes_th}
The pivotal idea in all of Statistics is arguably the concept of probability, around which two main schools of thought exist: the frequentist and the Bayesian. In the frequentist framework, probability is defined as
\begin{equation}
    P(A) = \lim_{{N \to \infty}} \frac{{n(A)}}{{N}}
\end{equation}
where $ P(A) $ is the probability of event $ A $ occurring, and $ n(A) $ is the number of times the event $ A $ occurs in $ N $ equiprobable trials. It is therefore defined as the relative frequency, in the limit of infinite trials.\\
This definition hinges on an asymptotic behaviour, therefore assuming that it is indeed possible to repeat the experiment -- which we can think of as a random \enquote{draw}, or realization, of a physical process -- an infinite (very large) number of times. Cosmology offers a perfect example of a situation in which this is not possible.

The Bayesian approach, by contrast, defines probability as the \textit{degree of belief} in a hypothesis, which can be influenced by some prior knowledge about the said hypothesis. This makes probability not an inherent property of the process, but rather a subjective measure, contingent upon the available information to the observer. For two events, or hypotheses, $A$ and $B$, we have:
\begin{equation}
    P(A|B) = P(B \cap A)P(B)
    \label{eq:conditional_prob}
\end{equation}
where $P(A|B)$ is the \textit{conditional probability}, the probability of the occurrence of $A$ once $B$ has occurred, i.e, conditioned to the occurrence of $B$; $P(A \cap B)$ is the \textit{joint probability}, the probability for both $A$ and $B$ to occur simultaneously. Finally, $P(B)$ is the \textit{marginal probability} for $ B $ to occur, which can be obtained by summing or integrating over all possible outcomes of $ A $ that are compatible with $ B $.
From the commutative property of joint probability $P(A \cap B) = P(B \cap A)$ and Eq.~\eqref{eq:conditional_prob} we easily obtain Bayes' theorem:
\begin{equation}
    P(A|B) = \frac{P(B|A)P(A)}{P(B)} \; ,
\end{equation}
\ML{which holds also in the frequentist paradigm, being it a direct consequence of Eq.~\eqref{eq:conditional_prob}}. This expression can be cast in a form more pertinent to our discussion if we let $A$ and $B$ be respectively the ensemble of model parameters, organised in a vector $\vec{\theta} = (\theta_1, \theta_2 ... \theta_m)$, and the observed (or synthetic) data, $\vec{d}:d_i=(x_i, y_i), \, i=1,2, ..., n$; here, $x_i$ is the independent variable, e.g., the values of the multipoles $\ell$ when measuring the angular power spectra $C(\ell)$ (see Sect.~\ref{sec:angular_ps_theory}). In this way, we have:
\begin{equation} \label{eq:bayes_th}
P(\vec{\theta}|\vec{d}) = \frac{P(\vec{d}|\vec{\theta})P(\vec{\theta})}{P(\vec{d})} \; .
\end{equation}
The meaning of the above probability density functions (PDFs) is as follows:
\begin{itemize}
    \item $P(\vec{\theta}|\vec{d})$ is the \textit{posterior}, the probability of the model parameters $\vec{\theta}$ being true given the measurement of the data $\vec{d}$. It is the quantity we want to compute, from which we can extract the expected values of the parameters, their uncertainties and their covariance. \ML{This interpretation is unique to the Bayesian approach, in which parameters are random variables themselves; the frequentist approach, on the other hand, does not assign probabilities to parameters, which are seen as unknown but fixed quantities inherent to the system under study.}    
    \item $P(\vec{d}|\vec{\theta})$ is the \textit{likelihood}, the probability of the data $\vec{d}$ given a model $\vec{\theta}$. It is often written as $\mathcal{L}(\vec{\theta})$, and treated as a function of the model parameters only \ML{(for a fixed data vector, the observations themselves)}. Parameter estimation aims at finding the particular vector $\vec{\theta}^\star$ which maximizes $\mathcal{L}$ (the \enquote{argmax} of $\mathcal{L}$); in our case of interest, this also means maximizing the posterior (see Eq.~\ref{eq:posterior}).
    \item $P(\vec{\theta})$ is the \textit{prior}, and encodes any knowledge, or \textit{prejudice} we may possess on the model parameters before performing the experiment; these can come from theoretical predictions or from other experiments. Such knowledge is parameterized as a PDF for the different parameters $\theta_i$. Broad, flat prior distributions are often used to let the posterior distribution be influenced almost exclusively by the data at hand, \ML{although flat priors can still carry information (for example, on the credible intervals of a parameter's values) and can be non-flat in a different parameter basis}.
    \item $P(\vec{d})$ is the \textit{evidence}. This is the probability of $\vec{d}$ as determined by summing (or integrating) across all possible values of $\vec{\theta}$, weighted by the probability of $\vec{\theta}$, and it acts as a normalisation factor. \\
    To find the analytical form of $P(\vec{d})$ we can integrate Eq.~\eqref{eq:bayes_th} in $\diff\vec{\theta}$ and solve for the evidence. Since $P(\vec{d})$ does not depend on $\vec{\theta}$ and the posterior is normalised to 1 (that is, $\int P(\vec{\theta}|\vec{d})\diff\vec{\theta} = 1$), we get
    \begin{equation}
    P(\vec{d}) = \int P(\vec{\theta}) \mathcal{L}(\vec{d}|\vec{\theta})\diff \vec{\theta}
    \end{equation}
    This operation is in general numerically expensive because the dimensionality of the problem is set by the number of model parameters (including the nuisance parameters, describing quantities which are not of direct interest, such as systematic effects). Luckily, the overall normalization of the posterior is not important in parameter inference, since we are only interested in finding the argument that maximises the distribution, and not the value of the maximum itself. This is not the case for model selection, which we will not treat here.
\end{itemize}
in the case of flat prior probability distributions, Bayes' theorem simplifies to
\begin{equation}\label{eq:posterior}
    P(\vec{\theta} |\vec{d}) \propto \mathcal{L}(\vec{d} | \vec{\theta}) \; .
\end{equation}
In order to construct the likelihood we need a model, which provides a function $f(\vec{x},\vec{\theta})$ such that $ f(x_i,\vec{\theta})$ approximates $ y_i $ well enough $ \forall i$ \citep{Kerscher2019_model_selection}. \ML{In our case, the function $f$ is the theoretical modelling of the angular PS: $y_i = C(x_i = \ell_i, \vec{\theta})$}. Moreover, an error model for the data is needed. It should be noted that in the context of this work, we will not test different models (say, $f(x,\vec{\theta})$ vs $g(x,\vec{\phi})$); we will focus instead on the standard $\Lambda$CDM concordance model, extending the parameter space when necessary (e.g., adding curvature as a free parameter or using the CPL parameterization for the Dark Energy equation of state as in Eq.~\ref{eq:CPL}). Thus $f$ will be fixed throughout, and when incurring no risk of ambiguity the model parameter vector $\vec{\theta}$ itself will be referred to as the \enquote{model}. \\

If the measurements $y_i$ are drawn from a normal distribution, or if the sample size is sufficiently large (by virtue of the central limit theorem), the likelihood takes the form of a multivariate Gaussian, or
\begin{equation}
\label{eq:likelihood}
    \mathcal{L}(\vec{\theta}) = \prod_{j=1}^n \frac{1}{\sqrt{2\pi} \sigma_j} \mathrm{exp}\left(- \frac{(y_j - f(x_j,\vec{\theta}))^2}{2\sigma_j ^2} \right) \; ,
\end{equation}
where we introduced the variance of the data $\sigma_j^2$; $n$ is the number of data points. Here, we assumed the data to be independent and identically distributed (iid), which means that their covariance matrix is diagonal.
If instead the data are correlated, we define a covariance matrix with elements $\mathrm{Cov}_{ij} = \langle(y_i - f(x_i,\vec{\theta}))(y_j - f(x_j,\vec{\theta}))\rangle$, and write $\mathcal{L}$ as
\begin{equation}
\mathcal{L}(\vec{\theta}) = 
\frac{1}{\left[(2\pi)^{n} \rm{det}(\mathrm{Cov})\right]^{1/2}}
\mathrm{exp} \left( -\frac{1}{2}
\vec{\Delta}^T \mathrm{Cov}^{-1} \vec{\Delta}
 \right) \; ,
\end{equation}
where $ \vec{\Delta} $ is the residual vector between the observed data and the model prediction. The term $ \vec{\Delta}^T \mathrm{Cov}^{-1} \vec{\Delta} $ is commonly referred to as $ \chi^2 $. The $ \chi^2 $ statistic follows a chi-squared distribution with $ n $ degrees of freedom under the null hypothesis that the model is an adequate description of the data. Minimizing the $ \chi^2 $ is equivalent to maximizing the likelihood, thus providing the best-fit model parameters $ \vec{\theta}^\star $. \\

Note that the likelihood can take other functional forms beyond the (multivariate) Gaussian. This, however, has been found to be a good approximation for the \Euclid photometric survey, producing a negligible effect on the best-fit parameters and their uncertainties, as discussed in \cite{Upham2021_gauss_lik}. We also stress the fact that, even if $\mathcal{L}$ is Gaussian in the data, this is not necessarily the case for the model parameters: indeed, this is true only if $f$ is a linear function. The likelihood is in general a complicated function of $\vec{\theta}$, which is one of the reasons why we need to sample (explore) it using numerical techniques. A \enquote{good} experiment (i.e., a constraining dataset), however, will allow exploring the likelihood function sufficiently close to the peak; in this region, the log-likelihood can be approximated well by a second-order expansion, the exponential of which yields precisely a normal distribution. \\

For multiple experiments that produce independent datasets $D_j$ the total likelihood function is the product of the ones for each dataset:
\begin{equation}
    \mathcal{L}_{\rm tot} = \mathcal{L}(D_{\rm tot}|\vec{\theta}) = \prod_{j=1}^n \mathcal{L}_j(D_j|\vec{\theta}) \; .
\end{equation}
This is because, for independent datasets, the joint probability distribution is simply the product of the individual PDFs, which, in the context of parameter estimation, translates to the product of individual likelihood functions. \\

In general, the cross-correlation of data from different cosmological probes offers an extremely powerful tool for improving the constraining power, by leveraging the different nature of the probes and of the different systematics affecting them. An example of this is the cross-correlation of weak lensing and galaxy clustering, which is able to mitigate the systematic uncertainty affecting the latter probe as a tracer of the dark matter distribution, parameterized by the galaxy bias parameters.
These LSS probes, as in many other cases, are not statistically independent. The datasets exhibit covariance, either because they originate from overlapping regions of the sky, share common observational techniques, or are influenced by similar systematic effects. In such cases, the total likelihood function $\mathcal{L}_{\text{tot}}$ becomes

\begin{equation}
    \mathcal{L}_{\text{tot}}(D_{\text{tot}}|\vec{\theta}) = \exp\left[-\frac{1}{2} \vec{\Delta}^T  \mathrm{Cov}^{-1} \vec{\Delta} \right] \; , 
\end{equation}

where $\Delta$ is again the vector of residuals and $\rm{Cov}$ is the total covariance matrix, which encapsulates the covariance of each dataset as well as the \enquote{cross-covariance} between them.
As will be shown in Sect.~\ref{sec:covgauss}, the covariance matrix is often decomposed into signal and noise terms, $\mathrm{Cov} = \mathrm{Cov}_{\text{signal}} + \mathrm{Cov}_{\text{noise}}$, respectively due to the observed signal and observational noise or other systematics. An accurate covariance matrix, which can be derived from analytical models, simulations or bootstrap resampling techniques is crucial for unbiased parameter inference.
\section{Fisher Information Matrix} \label{sec:fisher_theory}
In order to understand if an experiment will be able to improve significantly our knowledge of some physical quantities (i.e., to reduce the uncertainties on some model parameters) \textit{before actually performing it}, we need a way to translate the information on the experimental setup into the predicted parameter uncertainties; in other words, we need a way to propagate the expected uncertainty on the data in the expected uncertainty on the parameters. In this way, we can optimize the survey design to achieve the best possible constraining power.
This is possible through the Fisher Information Matrix \citep{Fisher1935}.\\
The Fisher Matrix describes the curvature of the likelihood function, which in turn encodes the sensitivity of the experiment on a given parameter (i.e., with respect to a given direction in parameter space), and hence the associated uncertainty. If the likelihood falls steeply from the maximum in some direction, the uncertainty on the corresponding parameter will be small, and vice-versa. To construct the FM, we need three ingredients \citep{Dodelson2020}:
\begin{itemize}
    \item A model with its parameters. We choose some particular parameter values, dubbed \textit{fiducial}, which we assume describe the real Universe. The ${\rm argmax}$ of the likelihood is therefore known by construction. The FM can indeed only provide information on the forecasted uncertainties, not on the best-fit values.
    \item The theory (model) prediction as a function of the model parameters
    \item The expected covariance of the data around the fiducial.
\end{itemize}
Note that in doing this we are assuming the true Universe to be described by the chosen fiducial cosmology; we assume the argmax of the likelihood function to be known, and we compute the curvature of the function around the maximum. This means that, in general, we cannot use the FM formalism to quantify the shift (bias) in the best-fit parameters from an incorrect modelling or from neglecting some systematic effects. Appendix A of \cite{Camera2017_FMbias} shows an interesting approach to circumvent this limitation in the case of nested models (two models are \enquote{nested} if the parameter space of one is contained in the parameter space of the other). \\
We start by expanding the logarithmic form of Eq.~\eqref{eq:likelihood} (the \enquote{log-likelihood}) as a function of $\vec{\theta}$, around the fiducial vector $\vec{\theta}_{\rm fid}$ \citep{Verde2010}:
\begin{equation}
    \mathrm{\ln} \mathcal{L}(\vec{\theta}) = \mathrm{ln} \mathcal{L}(\vec{\theta}_{\rm fid}) + \frac{1}{2}\sum_{ij}(\theta_i-\theta_{{\rm fid}, \, i})
    \left.{\frac{\partial^2 \mathrm{ln}\mathcal{L}(\vec{\theta})}{\partial\theta_i \partial \theta_j}}\right|_{\vec{\theta}_{\rm fid}}
    (\theta_j - \theta_{{\rm fid},\, j}) + ... \; .
\end{equation}
The first term vanishes by construction; the term containing the second derivatives,
\begin{equation}
   \mathcal{H}_{ij} =- \left.\frac{\partial^2 \mathrm{ln}\mathcal{L}(\vec{\theta})}{\partial\theta_i \partial \theta_j}\right|_{\vec{\theta}_{\rm fid}} \; ,
\end{equation}
is the Hessian matrix. It expresses the local curvature of a multi-variate function, describing how fast the function decreases from the maximum along different directions in the parameter space, and it is used to test whether a point from the domain corresponds to a local maximum, minimum or saddle. In this context, we use it to obtain the parameter covariance. A non-zero covariance between two parameters indicates that variations in the first result in variations in the other. It is a measure of the degree of the linear joint variability of the two parameters.

We can now define the Fisher Matrix as (minus) the expected value of Hessian of our model’s log-likelihood. By virtue of the Schwarz theorem, it is symmetric, like $\mathcal{H}_{ij}$.
\begin{equation}
\label{fisher}
F_{\alpha\beta} \equiv \left\langle -\frac{\partial ^2 \mathrm{ln} \mathcal{L}}{\partial\theta_\alpha \partial\theta_\beta} \right\rangle \; .
\end{equation}
Again, the brackets indicate an ensemble average, which can be seen as the expected value for a stochastic process. We compute this average over the observed data that we would get if the Universe was indeed described by our fiducial model \citep{Verde2010}.
This matrix encodes the information that the data vector $\vec{d}$ carries about the model parameters $\vec{\theta}$, and can also be seen as the covariance of the score function:
\begin{equation}
    s(\vec{\theta}) = \nabla_{\vec{\theta}} \ln \mathcal{L}
\end{equation}
which is used to evaluate the goodness of our estimated model $\vec{\theta}$. Under the assumption that the log-likelihood can be approximated to second order as a function of the parameters, and therefore that it is Gaussian in $\vec{\theta}$, the FM allows computing (lower limits of the) parameter uncertainties from the knowledge of a fiducial model and the measurement uncertainties. \\
If $\mathcal{L}$ is Gaussian, the log-likelihood is simply proportional to the $\chi^2$ and $F$ depends only on the expected mean and covariance of the data \citetalias{ISTF2020}:
\begin{equation}
\label{eq:FMgauss}
F_{\alpha\beta} =  \frac{1}{2}\mathrm{tr} \left[ \frac{\partial {\rm Cov}}{\partial\theta_\alpha} {\rm Cov}^{-1} \frac{\partial {\rm Cov}}{\partial\theta_\beta} {\rm Cov}^{-1}\right] + \sum_{pq} \frac{\partial \mu_p }{\partial\theta_\alpha}({\rm Cov}^{-1})_{pq}\frac{\partial  \mu_q}{\partial\theta_\beta}
\end{equation}
with $\vec{\mu}$ mean of the data $\vec{\mu} = (\mu_1(\vec{\theta}),\mu_2(\vec{\theta}), ..., \mu_n(\vec{\theta}))$, $\vec{y}=(y_1,y_2, ..., y_n)$ the data and ${\rm Cov} = \langle (\vec{y} - \vec{\mu})(\vec{y} - \vec{\mu})^T\rangle $ is their expected covariance. $p$ and $q$ run over the data vector indices, from 1 to $n$. \\
Since the uncertainties of the actual data will not depend on the cosmological model (we assume the data covariance to be model-independent, a choice commonly adopted in the literature: see e.g. \citet{Joachimi2021_kids1000, Friedrich2021_DES_Y3_cov}), we take the derivatives $\partial {\rm Cov}/\partial \theta_\alpha = 0$, leaving only the second term; this is the final expression we will use in this work.
\begin{equation}
\label{eq:FMgauss_definitive}
F_{\alpha\beta} = \sum_{pq} \frac{\partial \mu_p }{\partial\theta_\alpha}({\rm Cov}^{-1})_{pq}\frac{\partial  \mu_q}{\partial\theta_\beta} \; .
\end{equation}
It can be shown that the inverse FM is simply the covariance matrix of the model parameters, which is exactly what we want to compute: 
\begin{equation} \label{paramcov}
{\rm C}_{\alpha\beta} = (F^{-1})_{\alpha\beta}
\end{equation}
where $F^{-1}$ is the inverse FM and ${\rm C}$ is the parameter covariance matrix ${\rm C}_{\alpha\beta} = \rho_{\alpha\beta} \sigma_{\alpha} \sigma_{\beta}$, with $\rho_{\alpha\beta}$ the correlation coefficient and $\sigma_{\alpha}$ the \textit{marginalised} 1$\sigma$ uncertainty on the parameter $\theta_\alpha$. It is important not to confuse the \textit{data} covariance matrix, ${\rm Cov}_{pq}$, with the \textit{parameter} covariance matrix, ${\mathrm C}_{\alpha \beta} $. \\
The marginalized (1$\sigma$) uncertainty on parameter $\theta_\alpha$ is then trivially obtained as
\begin{equation}
\sigma_{\alpha} = \sqrt{{\rm C}_{\alpha\alpha}} = (F^{-1})_{\alpha\alpha}^{1/2} \; .
\end{equation}
The uncertainties derived in this way are actually lower bounds, or optimistic estimates, of the actual parameter uncertainties, as stated by the Cramér-Rao inequality; indeed, this estimate is exact only in the case of a Gaussian likelihood:
\begin{equation}
\label{eq:cramer_rao}
    \sigma_{\alpha}^{\rm non-Gauss} > (F^{-1})_{\alpha\alpha}^{1/2} \; .
\end{equation}
Marginalised uncertainties on a given parameter are obtained by integrating out the other parameters, i.e. by integrating over their possible values within some intervals set by the chosen priors. To marginalise is then to find the distribution of one or more random variables by exhausting all the cases on the others:
\begin{equation}
    p(x,y|\vec{d}) = \iint p(x,y,z,w|\vec{d})dzdw \; .
\end{equation}
The process is used, for example, to make inference over some physical parameters while correctly taking into account the uncertainty on the \textit{nuisance} parameters (called in this example $z$ and $w$), describing astrophysical or instrumental systematic uncertainties, and thus being able to provide rigorous confidence intervals on the parameters of interest. As seen above, the FM framework makes it easy to compute these uncertainties, avoiding the need to compute expensive multidimensional integrals of the posterior distribution. \\
Conditional uncertainty, on the other hand, is the uncertainty on a parameter conditioned to the specific value of another (see Sect.~\ref{sec:bayes_th}): 
\begin{equation}
    \ML{\sigma_\alpha^{\rm unmarg.} = \sqrt{1/F_{\alpha\alpha}}} \; .
\end{equation}
Computing the conditional uncertainties of the cosmological parameters means assuming perfect knowledge of the nuisance ones, which are fixed at a given value. This is an often unrealistic approach, which is why conditional uncertainties are rarely used, but can be convenient in some situations: for example, we can produce forecasts for flat models by conditioning our uncertainties to $\Omega_{k, 0}=0$, without having to re-run the analysis. \ML{This can easily be achieved through the Fisher Matrix: it is sufficient to cut the parameter covariance matrix ${\rm C}$ by removing the rows and columns corresponding to the parameters we wish to fix to their fiducial value (to keep using our example, the row and column corresponding to $\Omega_{k, 0}$), and then invert it again to obtain the new (reduced) FM.}\\
\begin{figure}[ht]
  \begin{minipage}{.6\textwidth}
    \includegraphics[width=\linewidth]{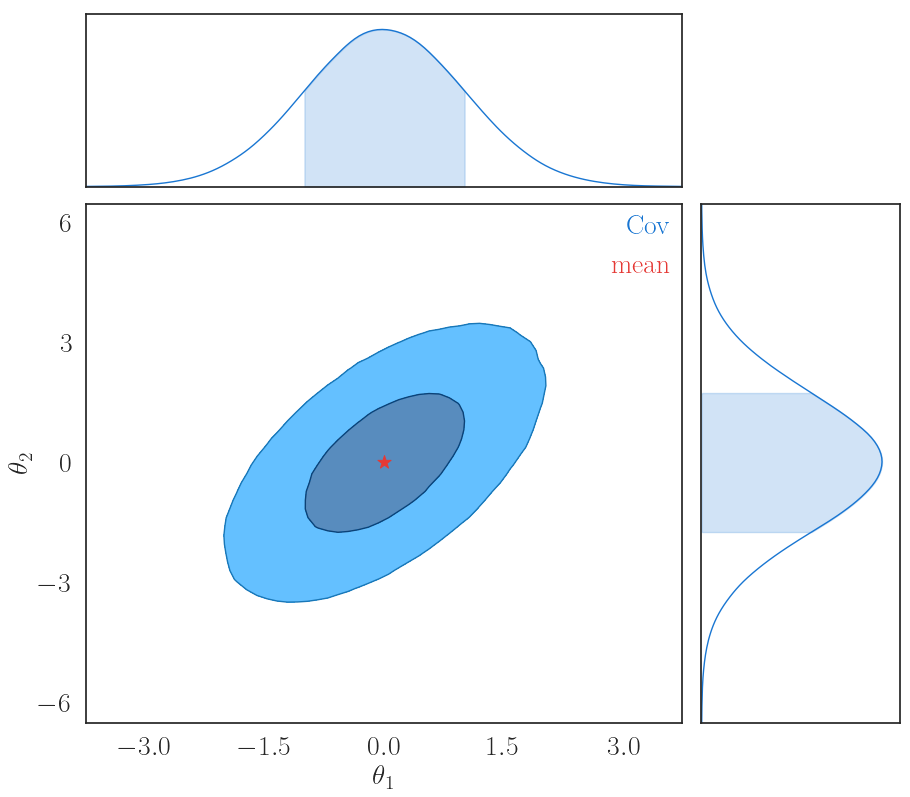}
  \end{minipage}
  \begin{minipage}{.4\textwidth}
    \caption{The 1$\sigma$ and 2$\sigma$ contours for the joint probability distribution for the two parameters $\theta_1$ and $\theta_2$. On the top and right-hand side are the marginal distributions, obtained by integrating over all possible values of the other parameter. The conditional distribution would be instead obtained by slicing the joint PDF at the desired value of the parameter we condition over. Note that the 1$\sigma$ contour of a multivariate 2D Gaussian distribution has a smaller probability content than in the 1D case ($\sim 39\%$ vs $\sim 68\%$): in this kind of contour plots, we will refer to the 68\% intervals as \enquote{1$\sigma$}, with a slight abuse of notation.}
    \label{fig:confidence_ellipse}
  \end{minipage}
\end{figure}

To summarize, the algorithm we used to produce the forecasts is the following:
\begin{enumerate}
\itemsep0pt
    \item Choose a fiducial model $f$, and a fiducial (physical and nuisance) parameter vector $\vec{\theta}$
    \item Compute the data covariance matrix
    \item Compute the derivatives $\partial f/\partial \vec{\theta}$ either numerically or analytically
    \item Compute the FM
\end{enumerate}
The numerical derivatives are computed through the algorithm proposed in Appendix B of \citet{Camera2017_FMbias}, in which the model is evaluated in 15 points in the neighbourhood of the fiducial value for added numerical stability. \\

One of the advantages of FM analysis is that Fisher matrices from independent data sets can be added up, because the log-likelihood of independent datasets is the sum of the individual log-likelihoods and by virtue of the sum rule of derivatives. We can leverage this property as a quick way to include (Gaussian) priors of a given variance in our forecast; to do this we construct a separate FM, with all entries set to 0 except the diagonal ones at the index of the parameter of interest. For example, if we want to add a prior on the first parameter, $\theta_1$, we can use
\begin{equation} 
F^{\rm tot} = F + F^{\rm pr} = F + \begin{pmatrix}
(\sigma_1^{\rm pr})^{-2} & 0 & 0 & 0 \\
0 & 0 & \dots & 0 \\
0 & \vdots & \ddots & \vdots \\
0 & 0 & \dots & 0
\end{pmatrix}
\end{equation}
where $F^{\rm pr}$ is the priors FM and $\sigma_1^{\rm pr}$ is the standard deviation of the Gaussian prior on the parameter $\theta_1$. This method offers a very convenient way to gauge the impact of different priors on the analysis, and will be used in Sect.~\ref{sec:req_on_prior}.\\

\FB{As a final note, we point out that the FM analysis suffers from some limitations, most notably the above-mentioned approximation of Gaussianity of the posterior distribution in the model parameters. This implies that the various 2D projections used to visualize the joint posterior contours over the different couples of parameters are ellipses by construction (see Fig.~\ref{fig:confidence_ellipse}), and can only represent linear degeneracies between parameters. Some of the information present in the true multidimensional posterior volume is therefore lost in the FM representation. This is not always a good approximation, as discussed in \citet{Wolz2012}; however, this depends on the probe considered and the precision of the experiment. One example of this is the forecasted constraint on $n_{\rm s}$ for the weak lensing probe in isolation, which is likely overly optimistic considering that the probe is not well-suited to constraint such a parameter (see Sect.~\ref{sec:shear_ps}). \citet{Casas2023}, however, showed a very good agreement between forecasts obtained for the reference \citetalias{ISTF2020} likelihood in the FM and Monte Carlo Markov Chain approach (see next section) when considering the full photometric survey. This is mainly due to the high precision of \Euclid, which will, as mentioned above, allow the likelihood to be characterized in a tight neighbourhood of its mode.}
\section{Monte Carlo Markov Chain}\label{sec:mcmc}
A different approach to the problem of characterizing the likelihood function is the Monte Carlo Markov Chain (MCMC) technique. Contrary to the FM, this enables us to find the best-fit values of the parameters \ML{-- albeit it is more effective in characterizing the distribution around its maximum, than in finding point estimates such as the mode --} and in this way to also quantify the bias induced by discrepancies between different models. Moreover, this method is well-suited to treat non-Gaussian posterior distributions.\\
The core idea is to sample the target distribution (the posterior) through a random walk in parameter space. A simple method to construct a Markov Chain (a sequence in which each step only depends on the previous one) is the Metropolis-Hastings algorithm \citep{Hastings1970}. The procedure followed to choose the next step, $\vec{\theta}^{i+1}$, given the current position, $\vec{\theta}^{i}$ (with $\vec{\theta}^0$ being the initial point of the chain), is as follows: we define the \textit{acceptance ratio} $r$ as
\begin{equation}
    r = \frac{p(\vec{\theta}^{i+1}) q(\vec{\theta}^{i} | \vec{\theta}^{i+1})}
    {p(\vec{\theta}^{i}) q(\vec{\theta}^{i+1} | \vec{\theta}^{i})} \; ,
\end{equation}
where $p$ is the target distribution (the posterior) and $q$ is the proposal distribution, the conditional probability of $\vec{\theta}^{i+1}$ given $\vec{\theta}^{i}$ (often taken as Gaussian). At each step, we sample $\vec{\theta}^{i+1}$ from $q$ and a random number $x$ from a uniform distribution. If $r > x$ we accept the new step, otherwise we reject it and set $\vec{\theta}^{i+1} = \vec{\theta}^{i}$; we then repeat the procedure. This simple algorithm allows the random walker to also explore areas of lower likelihood, for a complete sampling of the hypersurface. An advantage of this approach is the possibility to run many chains in parallel. \\
Ensuring convergence of the chains, especially in high-dimensional spaces, can be computationally intensive; a commonly adopted way to check this is the Gelman-Rubin convergence diagnostic \citep{Gelman1992}. \ML{After a sufficient number of steps, the different chains should all be close to the stationary distribution. To assess this, it is possible to compare the variance of the chain means, and the mean of chain variances. If all the chains have converged, then the between-chain variation should be close to zero}.\\
Once the chains have been obtained, the posterior distribution is simply given by their histogram: the more often a walker has visited a position in parameter space, the higher the value of the posterior.
\section{Two-point statistics}\label{sec:2_point_stat}
A key step in the process of parameter inference is the comparison (fit) of the theory to the data. The observations produced by the \Euclid mission will allow us to build shear and density maps, characterizing the specific realization of the Universe we live in. It is impossible, and most likely unnecessary, to predict the totality of the features of such a specific realization; we can instead compress the information present in the maps into some summary statistics, for which our current theory is able to produce predictions. The following sections present the summary statistics used in the present work. 

\subsection{Power Spectrum}
The most commonly used, and simple, summary statistic is the two-point correlation function (2PCF), and its Fourier transform, the power spectrum. Given two fields $f(\vec{x})$ and $g(\vec{x}')$, the 2PCF gives their (co)variance as a function of the positions $\vec{x}, \vec{x}'$:
\begin{equation}
   \xi _{fg}(\vec{x},\vec{x}') = \langle f(\vec{x})g(\vec{x}+\vec{x}') \rangle \; ,
\end{equation}
where the brackets indicate the expected value over some ensemble of realizations, or over the volume (see \citealt{Peacock2003}). Under the hypotheses of homogeneity and isotropy, $\xi$ becomes function of only the distance: $\xi_{fg}(\vec{x},\vec{x}') = \xi_{fg}(|\vec{x}'-\vec{x}|) = \xi_{fg}(r)$. More specifically, for a homogeneous Poisson process, the probability (number) of points (e.g., galaxies) in two volumes $V_1$, $V_2$ is given by
\begin{equation}
\diff P_{12} = \bar{n}^2 \diff V_1 \diff V_2 \; ,
\end{equation}
where $\bar{n}$ is the average number density of points. The 2PCF then quantifies the \textit{excess} probability, with respect to the case of uniform random distribution, (which has a null 2PCF), of finding two points at a separation $\vec{r}$: 
\begin{equation}
    \diff P_{12} = \bar{n}^2 \left[1 + \xi_{12}(r)\right] \diff V_1 \diff V_2 \; .
\end{equation}
If $f(\vec{x})$ is an integrable function we can perform the Fourier transform:
\begin{equation}
    \mathcal{F}(f(\vec{x})) = \Tilde{f}(\vec{k}) = \int_{-\infty}^{+\infty} \diff^3 \vec{x} f(\vec{x})e^{-i \vec{k} \cdot \vec{x}} 
\end{equation}
and its inverse:
\begin{equation}
    \mathcal{F}^{-1}(\Tilde{f}(\vec{k})) = f(\vec{x}) = \int_{-\infty}^{+\infty} \frac{\diff^3 \vec{k}}{(2\pi)^3} \Tilde{f}(\vec{k})e^{i \vec{k} \cdot \vec{x}}.
\end{equation}
Sometimes $\Tilde{f}(\vec{k})$ is called the \textit{spectrum} of $f(\vec{x})$. The $\mathcal{F}$ operator allows going from the $\vec{x}$ (in our example the spatial coordinates vectors) to the $\vec{k}$ (wave vectors, conjugated to $\vec{x}$) spaces, allowing choosing the most convenient basis for the problem at hand. \\
We can now consider the 2PCF function of $\Tilde{f}(\vec{k})$  and $\Tilde{g}(\vec{k})$:
\begin{equation}
\langle \Tilde{f}(\vec{k})\Tilde{g^*}(\vec{k}') \rangle = 
\int\diff^3\vec{x} \int \diff^3 \vec{x}' \langle f(\vec{x})f(\vec{x}') \rangle e^{-i\vec{k} \cdot \vec{x}}e^{i\vec{k}' \cdot \vec{x}'} \; ,
\end{equation}
which under homogeneity assumption becomes
\begin{equation}
\langle \Tilde{f}(\vec{k})\Tilde{g^*}(\vec{k}') \rangle = 
\int\diff^3\vec{x} \int \diff^3 \vec{x}' \xi_{fg}(\vec{x}'-\vec{x}) e^{-i\vec{k} \cdot \vec{x}}e^{i\vec{k}' \cdot \vec{x}'} \; ;
\end{equation}
performing the variable change $\vec{x}' \rightarrow \vec{z} = \vec{x}'-\vec{x}$, we have
\begin{equation}
\langle \Tilde{f}(\vec{k})\Tilde{g^*}(\vec{k}') \rangle = 
\int\diff^3 \vec{x}  e^{-i(\vec{k}-\vec{k}') \cdot \vec{x}}
\int \diff^3 \vec{z} \xi_{fg}(\vec{z})e^{i\vec{k}' \cdot \vec{z}} \; ,
\end{equation}
Remembering the definition of the Dirac delta $\delta^3_{\rm D}(\vec{k}) = (2\pi)^{-3}\int\diff^3 \vec{x} e^{i\vec{k} \cdot \vec{x}}$ we finally get
\begin{equation}
\langle \Tilde{f}(\vec{k})\Tilde{g^*}(\vec{k}') \rangle = 
(2\pi)^3 \delta^3_{\rm D}(\vec{k}-\vec{k}')P_{fg}(\vec{k})
\end{equation}
or equivalently
\begin{equation}
\label{eq:pk_vs_2pcf}
\langle \Tilde{f}(\vec{k})\Tilde{g}(\vec{k}') \rangle = (2\pi)^3 \delta_{\rm D} ^3(\vec{k}+\vec{k}')P_{fg}(\vec{k})
\end{equation}
with $P_{fg}(\vec{k})$ the power spectrum (or the \enquote{cross-spectrum}):
\begin{equation}
    P_{fg}(\vec{k}) = \int\diff^3 \vec{x} \xi_{fg}(\vec{x})e^{-i\vec{k} \cdot \vec{x}} = \mathcal{F}(\xi_{fg}(\vec{x})).
\end{equation}
The PS of $fg$ is therefore the Fourier transform of the two-point correlation function $\xi_{fg}$. The Dirac delta $ \delta_{\rm D}^3(\vec{k} \pm \vec{k}') $ ensures the power to be non-zero only when $\vec{k} = -\vec{k}'$. This comes from the isotropy of the Universe; the power spectrum should only depend on the magnitude of the wavevector $\vec{k}$ and not its direction. The factor $(2\pi)^3$ arises from the normalization convention used in the Fourier transform.\\
In the case of spatial isotropy, $\xi_{fg}(r)$ and $P_f(k)$ are connected via the relation
\begin{equation}
P_{fg}(k)=2 \pi \int_0^{\infty} \diff  r r^2 \xi_{fg}(r) \int_{-1}^1 \diff u e^{-i k r u} \; ,
\end{equation}
where $u$ is the cosine of the angle between $\vec{k}$ and $\vec{x}$. Performing the $u$ integration, one finally gets:
\begin{equation}
P_{fg}(k)=4 \pi \int_0^{\infty} \diff r r^2 \xi_{fg}(r) \frac{\sin (k r)}{k r} \; .
\end{equation}
Note that, since $\xi_{fg}(r)$ is dimensionless, $P_{fg}(k)$ has dimension of a volume. The PS is also often expressed in dimensionless form, $\Delta^2_{fg}(k) = k^3 P_{fg}(k)/2\pi^2$\\

We can give a concrete example in the case of matter density fluctuations, introducing the time- and position-dependent density contrast:
\begin{equation} \label{delta}
    \delta_\mathrm{m}(\vec{x},z) = \frac{\rho_{\rm m}(\vec{x},z) - \bar{\rho}_{\rm m}(z)}{\bar{\rho}_{\rm m}(z)}
\end{equation}
which describes the density fluctuations around the mean $\bar{\rho}_{\rm m}(z)$ as a function of position and redshift; we can construct similar fields for the different components, such as galaxies (in which case we will use the galaxy number density $n_{\rm g}(\vec{x}, z)$ instead of $\rho_{\rm m}$). Switching to the Fourier space (i.e., projecting $\delta_\mathrm{m}$ on the plane-wave basis) and taking the expected value (cf. Eq.~\ref{eq:pk_vs_2pcf}), we obtain the \textit{matter} PS, which as we shall see plays a central role in LSS analyses:
\begin{equation}\label{Pm}
\langle \Tilde{\delta}_{\rm m}(\vec{k},z)\Tilde{\delta}_{\rm m}(\vec{k}',z) \rangle = 
(2\pi)^3\delta^3_{\rm D}(\vec{k}+\vec{k}')P_{\rm mm}(\vec{k},z)
\end{equation}
under the homogeneity and isotropy assumptions we have the simplification $P_{\rm mm}(\vec{k},z) \rightarrow P_{\rm mm}(|\vec{k}|,z)=P_{\rm mm}(k,z)$. We will refer to the matter power spectrum as $P_{\rm \delta\delta}$ or $P_{\rm mm}$ interchangeably, since unless specified otherwise $\delta$ will represent the matter density contrast field.

\begin{figure}
    \centering
    \includegraphics[width=0.8\textwidth]{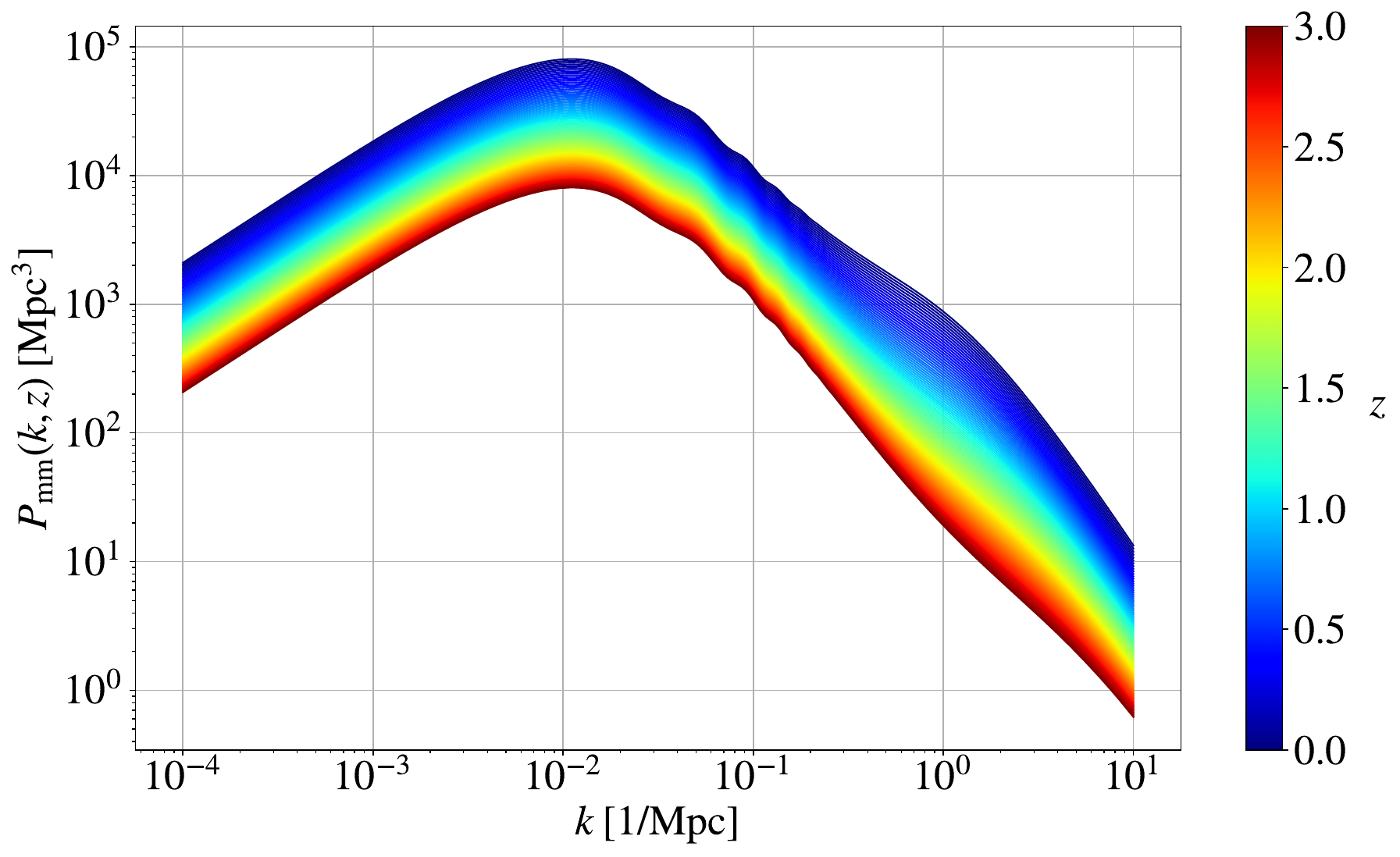}
    \caption{The nonlinear matter PS as a function of scale $k$ and redshift (see color bar). The slope of the PS changes sign after the scale of matter-radiation equality $k_{\rm eq}\sim 10^{-2} {\rm Mpc}^{-1}$: large-scale perturbations enter the horizon after equality, when the Universe is matter-dominated, hence continuing their super-horizon growth. This is why they retain the primordial shape, which follows a power law form (see Eq.~\ref{eq:Harrison_pk}). Smaller scales, on the other hand, enter the horizon earlier -- during radiation domination -- and their growth is suppressed, giving rise to the relative lack of small-scale power at large $k$. Because structures at low $z$ have had more time to grow, the power increases as $z$ decreases.}
    \label{fig:pk_vs_z_rainbow}
\end{figure}

The $\delta(\vec{k},z)$ fields are interpreted as random variables drawn from some probability density functions. In the case of Gaussian PDFs, all the information is contained in the variance (the density contrast fields have zero mean by construction), which is precisely the power spectrum \citep{Piattella2018}. In other words, if the perturbations are Gaussian all predictions
can be derived from the power spectrum, which is therefore a complete statistic. \\
The PS encodes information on the amplitude of the matter fluctuations at a given scale (see Fig.~\ref{fig:pk_vs_z_rainbow}), making it possible to see at a glance the magnitude of the density perturbations (or, in general, of the function we are computing the power spectrum of), and thus to understand the different physical processes at play.
The power spectra of relevance in our analysis are the matter (matter-matter), galaxy-galaxy and matter-galaxy PS. The latter two can be derived from the former simply by defining a relation between the galaxy and matter statistics, the \textit{galaxy bias} (see Sect.~\ref{sec:gal_bias}). The accurate characterization of galaxy bias is one of the most daunting challenges in the use of galaxy statistics for Cosmology and the reason why weak lensing cosmic shear is such a powerful probe, being it sensitive to the total matter distribution instead of just the baryonic part. Moreover, cross-correlating the galaxy and matter fields can greatly help lift the degeneracy between galaxy bias and other cosmological parameters.
\subsection{Angular power spectrum}\label{sec:angular_ps_theory}
The \Euclid satellite is equipped with a near-infrared spectrometer for accurate determination of redshifts of up to $\mathcal{O}(10^7)$ sources \citepalias{ISTF2020}. Such observations will allow constructing the 3-dimensional power spectrum exploiting the high resolution in the radial direction, but are time-consuming. A different approach, widely used nowadays in galaxy surveys (see e.g. \citealt{de_Jong2013_KiDS, DES2005}), is to estimate redshifts through observations in different photometric bands, or filters. This yields a much larger number of redshift estimates ($\mathcal{O}(10^9)$), at the cost of lower precision (see Sect.~\ref{sec:z_distribution}), which translates into a large uncertainty over the radial dimension of the measured field. We then divide these observations into redshift bins, which will likely overlap because of photometric-redshift uncertainties (see Sect.~\ref{sec:tomography}).\\
To exploit this binned data, we need to define the 2-dimensional version of the 3D observables introduced in the present section, that is, the angular power spectra.
To do that, we need a formalism describing the projection of fields on the (celestial) sphere. Since two angles are required to specify each direction in the sky, we can use $(\theta, \phi)$ as our coordinates. As mentioned, for each $z$ bin a different map is constructed, which allows greatly enhancing the statistical power of the survey.\\

In the same way as we defined the two-point correlation function and its Fourier space version, the power spectrum, we can define the \textit{angular} two-point correlation function and its Fourier transform on the sphere, the \textit{angular power spectrum}.\\
To illustrate this, we take as an example the above-mentioned matter density contrast, this time expressed as a function of the unit vector $\hat{\vec{n}} = (\cos \theta \sin \phi, \sin \theta \sin \phi, \cos \phi)$; we drop the subscript ${\rm m}$ to simplify the notation, without loss of generality. Since $ \delta(\hat{\vec{n}})$ is now defined on a spherical (non-flat) surface, the basis functions of the decomposition are no longer sines and cosines (or complex exponentials), but \textit{spherical harmonics} $Y_{\ell m}(\hat{\vec{n}})$:
\begin{equation}
Y_{\ell m}(\hat{\vec{n}}) = Y_{\ell m}(\theta, \phi) =
\sqrt{\frac{2\ell+1}{4\pi}\frac{(\ell-m)!}{(\ell+m)!}}e^{im\phi}\mathcal{P}^m_\ell(\cos\theta) 
\end{equation}
with $\ell$ and $m$ integer numbers satisfying the condition $\ell\geq 0$, $ -\ell \leq m \leq \ell$. Just like $\vec{k}$ was the conjugate of $\vec{x}$ in the Fourier transform, the subscripts $\ell$ and $m$ are the conjugate of $\hat{\vec{n}}$ in the spherical harmonics transform. The \enquote{multipole} $\ell$ is related to the angular scale of the features described by the spherical harmonic decomposition. Specifically, $\ell$ corresponds to the number of oscillations of the function along a meridian, with larger $\ell$ values representing smaller angular scales. The phase and number of oscillations of the function in the azimuthal direction $\phi$ is described instead by $m$. The value of $m$ also dictates the symmetry properties of the spherical harmonic: for $m = 0$, the function is axisymmetric; for $m \neq 0$, the function will have a $m$-fold azimuthal symmetry. Finally, the terms $\mathcal{P}^m_\ell(\cos\theta)$ represent the \textit{Legendre Polynomials}. \\
The $Y_{\ell m}$ functions satisfy the symmetry condition $Y_{\ell m}^* =(-1)^{m}Y_{\ell m}(\theta ,\phi )$ and the orthonormality relation $\int Y_{\ell m} Y^*_{\ell ' m'} \diff ^2 n = \delta^{\rm K}_{\ell \ell '}\delta^{\rm K}_{m m '}$, which make them an appropriate basis choice to represent functions on a sphere and ensure that the $a_{\ell m}$ coefficients can be uniquely determined. In the last equation, $\delta^{\rm K}$ is the Kronecker delta. \\

With these definitions, we can introduce the following  expansion, where the $a_{\ell m }$ (\textit{multipole moments}) terms play the role of expansion coefficients \citep{Dodelson2020}: 
\begin{equation}
\label{eq:spharmonics}
    \delta(\hat{\vec{n}}) = \sum_{\ell = 1}^\infty \sum_{m = -\ell}^\ell  Y_{\ell m}(\hat{\vec{n}}) = \sum_{\ell m}a_{\ell m } Y_{\ell m}(\hat{\vec{n}}).
\end{equation}
which can be inverted to obtain the $a_{\ell m}$s using the orthonormality relation:
\begin{equation}
    a_{\ell m } =  \int \diff^2 \hat{\vec{n}} \delta(\hat{\vec{n}})Y^*_{\ell m} \; .
\end{equation} 
The $a_{\ell m}$ (complex) coefficients provide information about the amplitude of the mode identified by $\ell$ and $m$ (and no longer by the wave vector $\vec{k}$). We can connect the value of $\ell$ to the angular scale through the relation 
\begin{equation}
\alpha \simeq \frac{\pi}{\ell} \; ,
\end{equation}
with $\alpha$ expressed in radiants. If the field $\delta(\hat{\vec{n}})$ follows a Gaussian distribution, so will the $a_{\ell m}$ coefficients, with a null expectation value $\langle a_{\ell m}\rangle = 0$ \citep{Poutanen2004_cl_theory}. We can compute the variance of the multipole moments $\langle a_{\ell m}a^*_{\ell ' m'} \rangle$ -- the second moment of the field distribution around the mean, which again completely defines a Gaussian field --  which under the hypothesis of symmetry under rotations (derived from the isotropy assumption of the Cosmological Principle) becomes 
\begin{equation}
\label{eq:Cl}
    \langle a_{\ell m }a^*_{\ell' m'} \rangle =\delta_{\ell\ell'}^{\rm K}\delta_{mm'}^{\rm K} C_\ell \; ,
\end{equation}
where $C_\ell$ is the angular power spectrum. This harmonic space observable is the one employed in this work and many cosmological analyses. Eq.~\eqref{eq:Cl} tells us that the $a_{\ell m}$ variance does not depend on $m$, which means that for a fixed $\ell$ all the coefficients have the same variance. When we compute the $C_{\ell}$s, then, we estimate the variance from a limited population, with $2\ell + 1$ samples (i.e., different values of $m$) at our disposal. If the observations cover the whole celestial sphere, we can define an unbiased estimator (an estimator the mean of which does not systematically under- or over-estimate the true value of the parameter under consideration) of the true $C_\ell$s as \citep{Montani2011_primordial_cosmology}:
\begin{equation}
    \hat{C}_\ell = \frac{1}{2\ell +1}\sum_m |a_{\ell m}^2| \; ,
\end{equation}
which is an average over $m$ for every $\ell$. There is an intrinsic limitation to the accuracy with which we can measure the $C_\ell$s, which arises precisely from the limited number of large-scale modes. This is called \textit{cosmic variance}. More specifically, the minimum variance of a measured $C_\ell$ is $2C_\ell^2/(2\ell+1)$ so the relative uncertainty related to the cosmic variance is
\begin{equation}
    \frac{\Delta C_\ell}{C_\ell} = \sqrt{\frac{2}{2\ell+1}}C_\ell \; .
\end{equation}
An additional common source of uncertainty in Cosmology is \textit{sample} variance, which arises from the limitedness of the portion of sky observed (most likely even for full-sky observations, because of the necessity to mask out foreground sources). This term goes instead as $\propto 1/f_{\rm sky}$, with $f_{\rm sky} = \Omega_\mathrm{S}/4\pi$ the fraction of the sky observed ($\Omega_\mathrm{S}$ being the survey area in steradiants).\\

Just as the 3D PS is related to the 2PFC, the angular PS can be related to the \textit{angular} two-point correlation function:
\begin{equation}
    w(\vec{\theta}) = \langle \delta(\hat{\vec{n}}) \delta^*(\hat{\vec{n}}') \rangle = \langle \delta(\hat{\vec{n}}) \delta(\hat{\vec{n}}') \rangle \; ,
\end{equation}
since the density contrast function is real-valued and the correlation function only depends on the amplitude of the angular separation, not its direction. Using Eq.~\eqref{eq:spharmonics} we have
\begin{align} 
w(\theta) &= \bigg \langle \sum_{\ell m}a_{\ell m } Y_{\ell m}(\hat{\vec{n}})
\sum_{\ell' m'}a^*_{\ell' m' } Y^*_{\ell' m'}(\hat{\vec{n}}') \bigg\rangle \\
&= \sum_{\ell m} \sum_{\ell' m'} \langle a_{\ell m} a^*_{\ell' m'} \rangle
Y_{\ell m}(\hat{\vec{n}})
Y^*_{\ell' m'}(\hat{\vec{n}}')
\end{align}
which we can connect to the angular PS via \citep{Dodelson2020}:
\begin{equation}
    w(\theta) = \frac{1}{4\pi} \sum_\ell \sum_{m=-\ell}^\ell |a_{\ell m }|^2 \mathcal{P}_\ell(\cos \theta) = \frac{1}{4\pi}\sum_\ell C_\ell(2\ell+1)\mathcal{P}_\ell(\cos\theta)
\end{equation}
having performed the sum over $m$ and having  used Eq.~\eqref{eq:Cl}. \\
%
%
%
\chapter{Galaxy Clustering and Gravitational Lensing}\label{chap:observables}
In the present Chapter, we will discuss the main observables used by the \Euclid mission to explore the fundamental properties of the Universe: the large-scale clustering of galaxies and the weak gravitational lensing of light. In fact, \Euclid's main aim is to construct the most accurate and complete catalogue of galaxy positions and shapes, to \enquote{map the dark Universe}. These two highly complementary probes are amongst the most promising ways to explore the low-redshift Universe, with the potential to increase the precision in the measurement of key cosmological quantities by more than one order of magnitude \citep{laureijs2011euclid, Amendola2018_Euclid}.
The following discussion will introduce the theory behind galaxy number counts and weak lensing cosmic shear, as well as the modelling chosen and systematics included to produce the fiducial spectra used in the subsequent forecasts.
\section{Galaxy Clustering}
The main observable studied in the context of galaxy clustering surveys is the degree of spatial correlation between galaxies' positions. This correlation is the result of the interplay between many different factors, the main one being the \enquote{tug of war} between gravity and the expansion of the Universe, accelerated by Dark Energy (in the concordance cosmological model) at early times. The initial shape and amplitude of the matter power spectrum, the epoch of matter-radiation equality and the contribution of massive neutrinos and baryonic physics further contribute to determining the large-scale distribution of the \textit{tracer} of matter under consideration, galaxies. The uncertainty over the bias affecting this tracer, the \textit{galaxy bias}, further complicates the picture. On the other hand, the dependence on such a rich set of phenomena hints at the large amount of information which can potentially be extracted from this probe. This section will introduce the main theoretical and practical aspects allowing its use to estimate the parameters of the cosmological model. 
\subsection{Spectroscopic Galaxy Clustering}\label{sec:GC_spectro}
As detailed in Sect.~\ref{sec:euclid_instruments}, \Euclid will be equipped with a slitless near-infrared spectrometer for accurate measurement of galaxies' spectra, which allows a very accurate measure of redshift -- the requirement on the redshift uncertainty is, in fact, $\sigma_z^{\rm sp} < 0.001(1 + z)$ \citep{laureijs2011euclid}.
The primary target of the spectroscopic survey will be H$\alpha$ emitters, because of the strength of the spectral line and its being associated with star-forming regions. The resulting galaxy map will have a high radial resolution, allowing the reconstruction of the 3D galaxy power spectrum, which is not possible with the photometric survey. This can then be related to the PS of matter by detailing the above-mentioned relation between galaxies and matter, the galaxy bias. In its simplest form, this relation is simply linear \citep{Peacock1994_gal_bias}; we refer to this modelling as \enquote{linear galaxy bias}:
\begin{equation}\label{eq:gal_bias_def}
    \delta_{\rm g}(k, z) = b(z) \, \delta_{\rm m}(k, z) \implies P_{\rm gg}(k, z) = b^2(z) \, P_{\rm mm}(k, z) \; .
\end{equation}
This of course can be made more complex (and realistic) by allowing $b(z)$ to acquire scale dependence, and/or by including higher-order bias terms. An extensive review of galaxy bias is given in \citet{Desjacques_2018}. \\

\begin{figure}
\begin{center}
\includegraphics[scale=0.4]{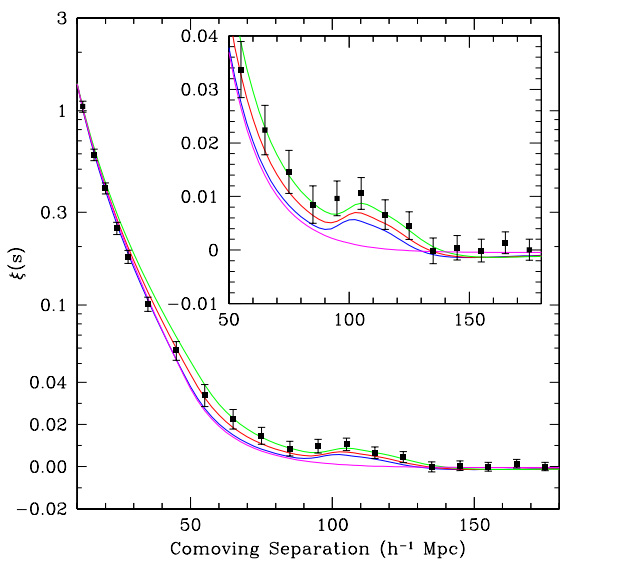}
\caption{Redshift-space correlation function of the
SDSS LRG sample (Sloan Digital Sky Survey - Luminous
Red Galaxies), showing the excess of pairs at separation $s$. The BAO peak is clearly visible. The green, red and blue lines represent $\Omega_m h^2 = (0.12, 0.13, 1.14)$ respectively, while the magenta line is the CDM model with $\Omega_{\rm b} = 0$, highlighting the statistical significance of the peak. Source: \citep{Eisenstein2005_BAO}}\label{fig:BAO}
\end{center}
\end{figure}

Another probe which is possible to investigate thanks to the precision of the spectroscopic survey is the Baryon Acoustic Oscillations (BAO) signal. This is the imprint left by the acoustic density wave oscillations in the photon-baryon plasma -- coupled by Compton scattering -- in the early Universe, created by the interplay of radiation pressure and the DM-sourced gravitational pull. This plasma has a high sound speed ($\sim 170~000 \; {\rm km} \, {\rm s}^{-1}$), since the photons provide the dominant \ML{pressure} contribution. 
At recombination ($\sim$ 400~000 yrs after the Big Bang) the photons become free to travel unbounded to the newly formed hydrogen atoms, thus depriving the baryonic matter of the radiation pressure and leaving it \enquote{frozen} on the shell of the propagation sphere at that time. 
The maximum radius of this sphere, i.e. the maximum distance travelled by the acoustic waves before recombination (the sound horizon), is the BAO scale. The $\sim 127$ kPc (400 000 light years, ly) travelled before recombination corresponds to roughly $\sim 150$ Mpc (500 million ly) today, due to the Universe expansion. This is a powerful standard ruler, measured to high precision in CMB experiments, which we can observe at different redshifts in the angular and radial direction to constrain the expansion history of the Universe. Surveys that aim at detecting this signal must cover an area wider than this typical scale; moreover, since the peak in the two-point correlation function (Fig.~\ref{fig:BAO}) corresponding to this characteristic separation distance is quite small, a large number of objects must be observed to reach the statistical power needed to measure it. The \Euclid survey is optimized to capture the BAO signal, because of the large sample size and wide survey area, which can accommodate the BAO scale. 

Finally, spectroscopic GC allows studying the so-called \textit{redshift space distortions} \citep{Kaiser1987_RSD}. Galaxies move relative to us both because they follow the Hubble flow and because of their peculiar velocities, induced by the local gravitational potential. Consequently, when one maps the galaxy distribution in redshift space, an anisotropy is observed, manifesting as elongations along the line of sight in the galaxy distribution, commonly referred to as the \enquote{Fingers of God}, and squashing in the perpendicular direction. This anisotropy can be measured to constrain the growth rate (Eq.~\ref{eq:growth_rate}).
\subsection{Photometric galaxy clustering}\label{sec:gc_phot_theory}
In addition to the spectroscopic survey, \Euclid will build the largest photometric galaxy catalogue to date, characterizing the position and morphology of billions ($\sim 1.5 \times 10^9$) of sources. This work revolves primarily around the photometric survey.\\

\begin{figure}
    \centering
    \includegraphics[width=\textwidth]{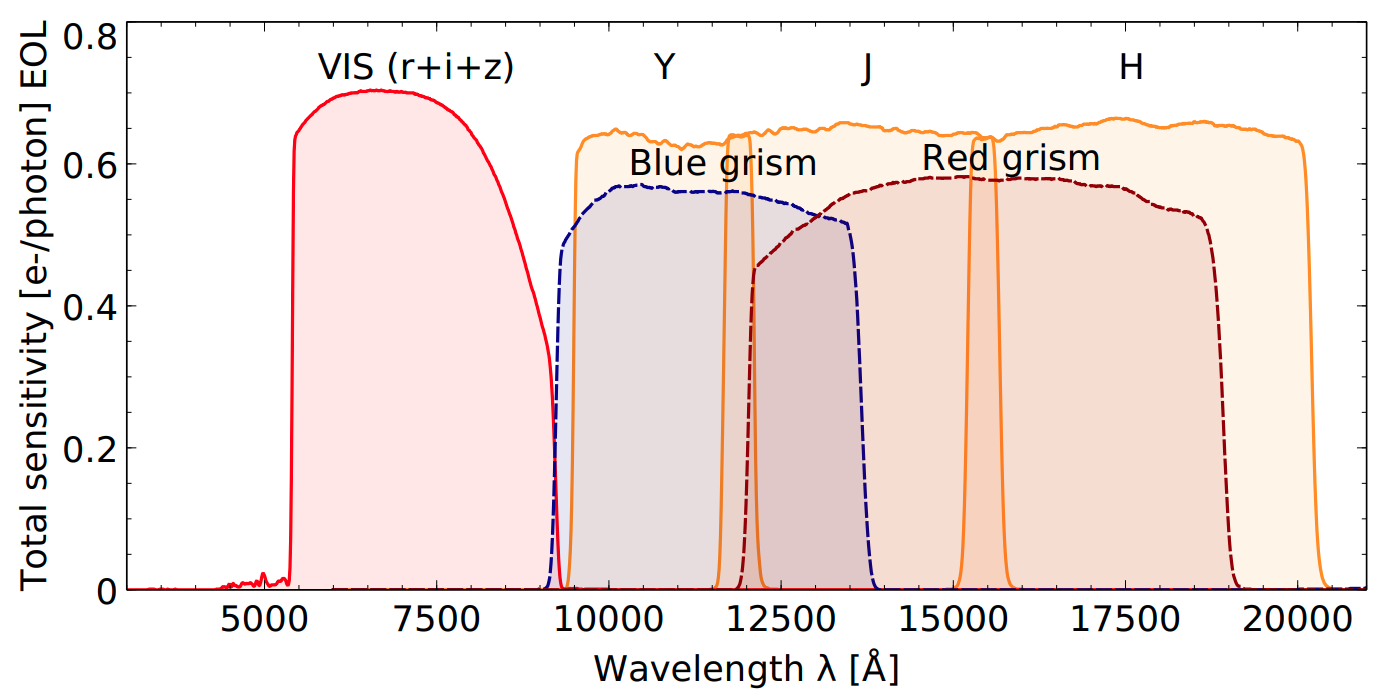}
    \caption{Forecasted sensitivity of the different bands of the VIS and NISP instrument, in units of electron per photon, at the end of the nominal life (6 yrs) of the instruments. The significant width of the photometric filters affects the precision of photometric redshift estimates. One of the advantages of space-based observations is the possibility to cover wavelength ranges poorly transmitted by the atmosphere. Source: \citep{Scaramella2022_survey}}
    \label{fig:NISP_response}
\end{figure}

While with spectroscopy the redshift is determined by measuring the shift of a particular spectral line (such as the H$\alpha$ emission line), or set of lines, between the rest-frame wavelength and the observed one, the photometric survey lacks the spectral resolution to do this. In fact, the light is measured in (i.e., integrated over) different filters, or bands, which cover a broad range of frequencies (Fig.~\ref{fig:NISP_response}). The redshift can then be obtained by color-color diagrams, machine learning approaches or template fitting, comparing the observed, low-resolution spectral energy distribution (SED) to a set of model SED templates at different redshifts, obtained through more accurate techniques \citep{Bolzonella2000_SEDfitting}. A comparison of the performance of different methods for the \Euclid survey is presented in \citet{Desprez2020_photoz}. The redshift estimates obtained in this way suffer significant uncertainty (see Sect.~\ref{sec:z_distribution}), with a requirement of $\sigma_z^{\rm ph} < 0.05(1+z)$ but are much quicker to obtain, hence allowing a larger number of estimates w.r.t. the spectroscopic case.

The consequence of the poor radial resolution of the photometric survey is the need to bin our observations into redshift slices, inside of which we study 2D-projected statistics of the galaxy field instead of the three-dimensional one. Having introduced the formalism used to describe 2D-projected fields on a sphere in Sect.~\ref{sec:angular_ps_theory}, we can derive the expression for the angular photometric galaxy clustering (GCph) PS which will be used in the rest of the work.\\
The uncertain source distances are described for GCph by a distribution $\mathcal{K}(\chi)$, defined as
\begin{equation}\label{eq:W_of_chi_gg}
    \mathcal{K}(\chi) = \frac{1}{N_{\rm g}}\frac{\diff N_{\rm g}}{\diff \chi} \; ,
\end{equation}
with $N_{\rm g}$ the total number of galaxies. This distribution allows projecting the galaxy density contrast field in the radial direction, acting as a \textit{projection kernel}:
\begin{equation}
    \Delta_{\rm g}(\hat{\vec{n}}) = \int_0^\infty\diff \chi \mathcal{K}(\chi) \delta_{\rm g}(\vec{x} = \hat{\vec{n}}\chi, \eta(\chi)) \; ,
\end{equation}
where $\eta(\chi)=\eta_0-\chi$ and $\Delta_{\rm g}$ is the projected density contrast field. Inserting the Fourier transform of $\delta_{\rm g}$ \ML{(indicated without a tilde)} and expanding in spherical harmonics, we get \citep{Dodelson2020}:
\begin{equation}
\begin{aligned}
\Delta_{\rm{g}}(\hat{\vec{n}}) & =\int_0^{\infty} \diff \chi \mathcal{K}(\chi) \int \frac{\diff^3 \vec{k}}{(2 \pi)^3} e^{i \vec{k} \cdot \hat{\vec{n}} \chi} \delta_{\rm g}(\vec{k}, \eta(\chi)) \\
& =4 \pi \int \frac{\diff^3 \vec{k}}{(2 \pi)^3} \sum_{\ell m} i^\ell Y_{\ell m}(\hat{\vec{n}}) Y_{\ell m}^*(\hat{\vec{k}}) \int_0^{\infty} \diff \chi \mathcal{K}(\chi) j_\ell(k \chi) \delta_{\rm g}(\vec{k}, \eta(\chi))
\end{aligned}
\end{equation}
where the coefficients of the expansion are
\begin{equation}
\Delta_{\rm{g}, \ell m}=4 \pi i^\ell \int \frac{\diff^3 \vec{k}}{(2 \pi)^3} Y_{\ell m}^*(\hat{\vec{k}}) \int_0^{\infty} \diff \chi \mathcal{K}(\chi) j_\ell(k \chi) \delta_{\rm g}(\vec{k}, \eta(\chi)) \; .
\end{equation}
As seen in Eq.~\eqref{eq:Cl}, the angular power spectrum is simply given in terms of the expectation value of the product of these coefficients:
\begin{equation}
\begin{aligned}
    \left\langle\Delta_{\rm{g}, \ell m} \Delta^*_{\rm{g}, \ell^{\prime} m^{\prime}}\right\rangle= & (4 \pi)^2 i^{\ell-\ell^{\prime}} \int \frac{\diff^3 \vec{k}}{(2 \pi)^3} \int \frac{\diff^3 \vec{k}^{\prime}}{(2 \pi)^3} Y_{\ell m}^*(\hat{\vec{k}}) Y_{\ell^{\prime} m^{\prime}}\left(\hat{\vec{k}}^{\prime}\right) \int_0^{\infty} \diff \chi \mathcal{K}(\chi) j_\ell(k \chi) \\
    & \times \int_0^{\infty} \diff \chi^{\prime} \mathcal{K}\left(\chi^{\prime}\right) j_{\ell^{\prime}}\left(k^{\prime} \chi^{\prime}\right)\left\langle\delta_{\rm g}(\vec{k}, \eta(\chi)) \delta_{\rm g}^*\left(\vec{k}^{\prime}, \eta\left(\chi^{\prime}\right)\right)\right\rangle .\\
    & = \delta^{\rm K}_{\ell\ell'}\delta^{\rm K}_{m m'}C_{\rm gg}(\ell) \; ,
\end{aligned}
\end{equation}
where the angular power spectrum is defined as
\begin{equation}\label{eq:cl_gg_nolimb}
\begin{aligned}
    C_{\rm{gg}}(\ell)= & \frac{2}{\pi} \int k^2 \diff k \int_0^{\infty} \diff \chi \mathcal{K}(\chi) j_\ell(k \chi) \int_0^{\infty} \diff \chi^{\prime} \mathcal{K}\left(\chi^{\prime}\right) j_\ell\left(k \chi^{\prime}\right) \\
    & \times P_{\rm{gg}}\left(k, \eta(\chi), \eta\left(\chi^{\prime}\right)\right) .
\end{aligned}
\end{equation}
Which involves the unequal-time power spectrum $P_{\rm{gg}}\left(k, \eta(\chi), \eta\left(\chi^{\prime}\right)\right)$, because the radial projection spans different times. Next section will show how, on small scales, the contribution from this PS at $\chi'\neq \chi$ can neglected.
\subsubsection{Limber approximation}\label{sec:limber}
The general form of the angular PS (Eq.~\ref{eq:cl_gg_nolimb}) involves a double integral over the comoving distance and an integral over the scale $k$. This is extremely expensive numerically because the integrand is modulated by the highly oscillatory spherical Bessel functions, and hence has to be evaluated in a large number of points. To reduce the numerical burden to a level appropriate for the many thousands of evaluations required for accurately sampling the likelihood in parameter space, we make use of the Limber approximation \citep{Limber1953, Kaiser1998}.

This approximation is based on the consideration that, on small angular scales ($\ell \gtrsim 20$) the product of spherical Bessel functions $j_\ell(x=k\chi)$ is sharply peaked around $k\chi = \nu$, with $\nu = \ell + 1/2$ \citep{Dodelson2020}. Thus, we can approximate $j_\ell(k\chi)$ with a scaled delta function:
\begin{equation}
    j_\ell(k\chi) \rightarrow \sqrt{\frac{\pi}{2\nu}}\delta_{\rm D}(\nu - k\chi)
\end{equation}
This approximation is valid if the PS varies slowly in the interval $\Delta k \sim 1/(\ell\chi)$ over which the Bessel functions are non-null; in this case, it can be approximated as constant and pulled out of the integral. Under this approximation, then, the expression for the angular PS becomes much simpler:
\begin{equation}\label{eq:cl_gg_limber}
    C_{\rm gg}(\ell) \simeq \int \diff\chi\frac{\mathcal{K}^2(\chi)}{\chi^2}
    P_{\rm gg}\left[ k_\ell = \frac{\ell + 1/2}{\chi}, \eta(\chi) \right]
\end{equation}
where, as mentioned in the last paragraph, we are only left with the equal-time PS and $k_\ell$ is the Limber wavenumber.

The combined effect of the Limber and flat-sky approximations is negligible except for very large angular scales ($\ell \lesssim 10$, see e.g. \citealt{Lemos2017_limber_flatsky}). This approximation works best when the kernels cover much larger support in radial distance $\chi$ than the physical scale being probed, and when they have significant overlap \citep{Leonard2023_N5K}. \ML{This is indeed the case for the cosmic shear signal, for which we will also use the Limber and flat-sky approximations.} 
\subsubsection{Tomography}\label{sec:tomography}
Photometric LSS surveys rely on the large sample size to make up for their low radial resolution. Their statistical power can be largely enhanced by repeating the observations in different redshift slices (bins), and constructing the angular PS for each of these. This is necessary, for example, to achieve high-precision dark energy measurements \citepalias{ISTF2020}. Beyond increasing the number of spectra to the number of bins, this makes it also possible to study the cross-correlation between redshift bins. Furthermore, we can use the tomographic approach both for the auto- and for the cross-spectra, which correlate different fields to break the parameter degeneracies affecting the individual probes and to maximize the constraining power from a multi-probe survey such as \Euclid.\\

To facilitate comparison against other codes (and to follow the recipe and notation outlined in \citetalias{ISTF2020}, the main official reference for the \Euclid forecasts), we choose $z$ as our radial coordinate, so the general form of the tomographic $C_{ij}^{AB}(\ell)$ between observables $A$ and $B$ in redshift bins $i$ and $j$ (both of which can of course coincide), in the context of the Limber approximation, is
\begin{equation}\label{eq:cl_AB_tomo_limb_ISTF} 
    C_{ij}^{AB} (\ell) = \frac{c}{H_0}
    \int \diff z \frac{\mathcal{K}_i^A(z)\mathcal{K}_j^B(z)}
    {E(z)\chi^2(z)}P_{AB}\left[ 
    \frac{\ell + 1/2}{\chi(z)},z
    \right] \; .
\end{equation}
This equation is symmetric under the exchange of redshift bin indices ($i \leftrightarrow j$) only if $A=B$, that is, for the auto-spectra. \ML{Tomography is used for both the photometric probes, as well as their cross-correlation}; the number of spectra for weak lensing ($A=B={\rm L}$) and photometric galaxy clustering ($A=B={\rm G}$) is $N(N+1)/2$, while for the cross-correlation it is $N^2$.\\

The different angular power spectra studied in this work are obtained by plugging the appropriate radial kernel and power spectra in Eq.~\eqref{eq:cl_AB_tomo_limb_ISTF}. For the photometric galaxy clustering case, we express the radial kernel \eqref{eq:W_of_chi_gg} as a function of redshift as
\begin{equation}\label{eq:wigdef}
    {\cal{K}}_{i}^{\rm G}(z) = \frac{H_0}{c} \frac{n_{i}(z)}{\bar{n}} E(z) \; .
\end{equation}
and the galaxy-galaxy power spectrum as indicated in Eq.~\eqref{eq:gal_bias_def}. In the last equation, $n_{i}(z)$ represents the distribution of galaxies in the $i$-th redshift bin and $\bar{n}$ the total number density. \\
In general, the redshift distributions used for galaxy number counts and weak lensing may differ. In this case, with a slightly confusing notation, the literature refers to the redshift distribution entering the galaxy kernel as the \textit{lens} distribution, $n_{i}^{\rm L}(z)$, and to the one entering the lensing kernel as the \textit{source} redshift distribution, $n_{i}^{\rm S}(z)$. \\
The next section will discuss the modelling of these distributions, which are a key ingredient in our analysis.
\subsection{Redshift distribution}\label{sec:z_distribution}
First, we assume that the same galaxy population is used to probe both the weak lensing cosmic shear and the GCph PS. This assumption is likely to be relaxed in case, for example, the data are complemented by ground-based photometry, covering different sky regions (e.g., northern or southern hemispheres): in fact, for \Euclid, a strong interaction with existing and future surveys, such as the Vera C. Rubin Observatory Legacy Survey of Space and Time \citep[LSST,][]{ivezic2018lsst} is planned. We therefore set:
\begin{equation}
n_{i}^{\rm L}(z) = n_{i}^{\rm S}(z) = n_i(z) \; ,
\label{eq:samenz}
\end{equation}
The same equality applies of course for the total source and lens number density, $\bar{n}^{\rm L}$ and $\bar{n}^{\rm S}$.\\

In \citepalias{ISTF2020}, the intrinsic or \enquote{true} number density redshift distribution is modelled analytically as a function of redshift as \citep{Smail1994_nofz}:
\begin{equation} 
n(z) \propto \left( \frac{z}{z_0}\right)^2 e^{-(z/z_0)^{3/2}} \, ,
\label{eq:n_of_z_analytical}
\end{equation}
where $z_{\rm m}$ is the median redshift of the sample and $z_0 = z_{\rm m}/\sqrt{2}$. This distribution then needs to be mapped into the one actually measured in each redshift bin, taking into account the unavoidable uncertainties due to the less precise and less accurate estimate of the redshifts obtained through photometry (compared to spectroscopy). \\
The number density $n_i(z)$ for each redshift bin is then formally defined through a convolution with some function $p_{\rm ph}(z_{\rm p}|z)$, which is the conditional probability distribution function that describes the likelihood of a galaxy with a true redshift $ z $ being observed at a photometric redshift (\enquote{photo-$z$}) $z_{\rm p}$:
\begin{equation} \label{eq:ni_def}
n_i(z) = \frac{\int_{z_i^-}^{z_i^+} \mathrm{d} z_{\rm p} \, n(z) p_{\rm ph}(z_{\rm p}|z)}
{\int_{z_{\text{min}}}^{z_{\text{max}}} \mathrm{d} z \int_{z_i^-}^{z_i^+} \mathrm{d} z_{\rm p} \, n(z) p_{\rm ph}(z_{\rm p}|z)} \; .
\end{equation}
where the interval boundaries $z_i^-$  and $z_i^+$ are the left and right edges of each redshift bin. These boundaries can be selected to yield $\mathcal{N}_\mathrm{b}$ equipopulated or equidistant bins; in the former case, which is the reference choice in \citetalias{ISTF2020}, 10 equipopulated bins with a number density $ \bar{n}/\mathcal{N}_\mathrm{b} = 3 $ gal/arcmin$^2$ and $ \bar{n} = 30$  are chosen.

The photometric redshift PDF $ p_{\rm ph}(z_{\rm p}|z) $ can be parametrized by a bimodal Gaussian model:
\begin{align} \label{eq:pph_def}
p_{\rm ph}(z_{\rm p}|z) = & \; (1-f_{\text{out}})p_{\rm ph}^{\rm in}(z_{\rm p}|z) + (f_{\text{out}})p_{\rm ph}^{\rm out}(z_{\rm p}|z) \nonumber \\
= & \; \frac{1 - f_{\text{out}}}{\sqrt{2\pi}\sigma_{\rm in}(1+z)} \exp \left[ -\frac{1}{2}\left( \frac{z - c_{\rm in} z_{\rm p} - z_{\rm in}}{\sigma_{\rm in}(1+z)} \right)^2 \right] \\
& + \frac{f_{\text{out}}}{\sqrt{2\pi}\sigma_{\rm out}(1+z)} \exp \left[ -\frac{1}{2}\left( \frac{z - c_{\rm out} z_{\rm p} - z_{\rm out}}{\sigma_{\rm out}(1+z)} \right)^2 \right] \, .
\end{align}
Here,  $f_{\text{out}}$ is the fraction of \enquote{catastrophic outliers}, i.e., sources with substantially erroneous redshift measurements. Consequently,  $1 - f_{\text{out}}$ is the fraction of galaxies with accurately measured redshifts. As can be seen in the equation, the standard deviation of the distributions $\tilde{\sigma}$, is an increasing function of $z$: $\tilde{\sigma}_i = \sigma_i(1+z)$, with $i = {\rm in, out}$. The model incorporates a set of tunable parameters  $(c_{\rm in}, z_{\rm in}, \sigma_{\rm in}, c_{\rm out}, z_{\rm out}, \sigma_{\rm out}, f_{\text{out}})$ to investigate various observational scenarios. In \citepalias{ISTF2020}, these are fixed in the analysis to their fiducial values:
\begin{center}
\begin{tabular}{|c|c|c|c|c|c|c|}
\hline
      $ c_{\rm in} $ & $ z_{\rm in} $ & $ \sigma_{\rm in} $ & $ c_{\rm out} $ & $ z_{\rm out} $ & $ \sigma_{\rm out} $ & $ f_{\text{out}} $   \\
      \hline
    \hline
     1.0 & 0.0 & 0.05 & 1.0 & 0.1 & 0.05 & 0.1 \\
     \hline
\end{tabular}
\end{center}
Note that the actual data analysis will allow accounting for uncertainties in the mean of the redshift distribution (i.e., its first moment), by introducing additional nuisance parameters, the shifts in the source redshift distribution $\Delta z_i$ in the $i$-th bin: $n_i(z) \rightarrow n_i(z + \Delta z_i)$. This is in line with the standard approach used in the recent literature -- see e.g. the DES Year 3 results: \citealt{DES_Y3_2022} --, and is motivated by the fact that the uncertainty on mean of the source redshift distribution is the one, amongst the different moments of the distribution, to have the greatest impact on the final constraints (see \citealt{Reischke2023}).\\

\begin{figure}[!ht]
    \centering
    \includegraphics[width=0.6\linewidth]{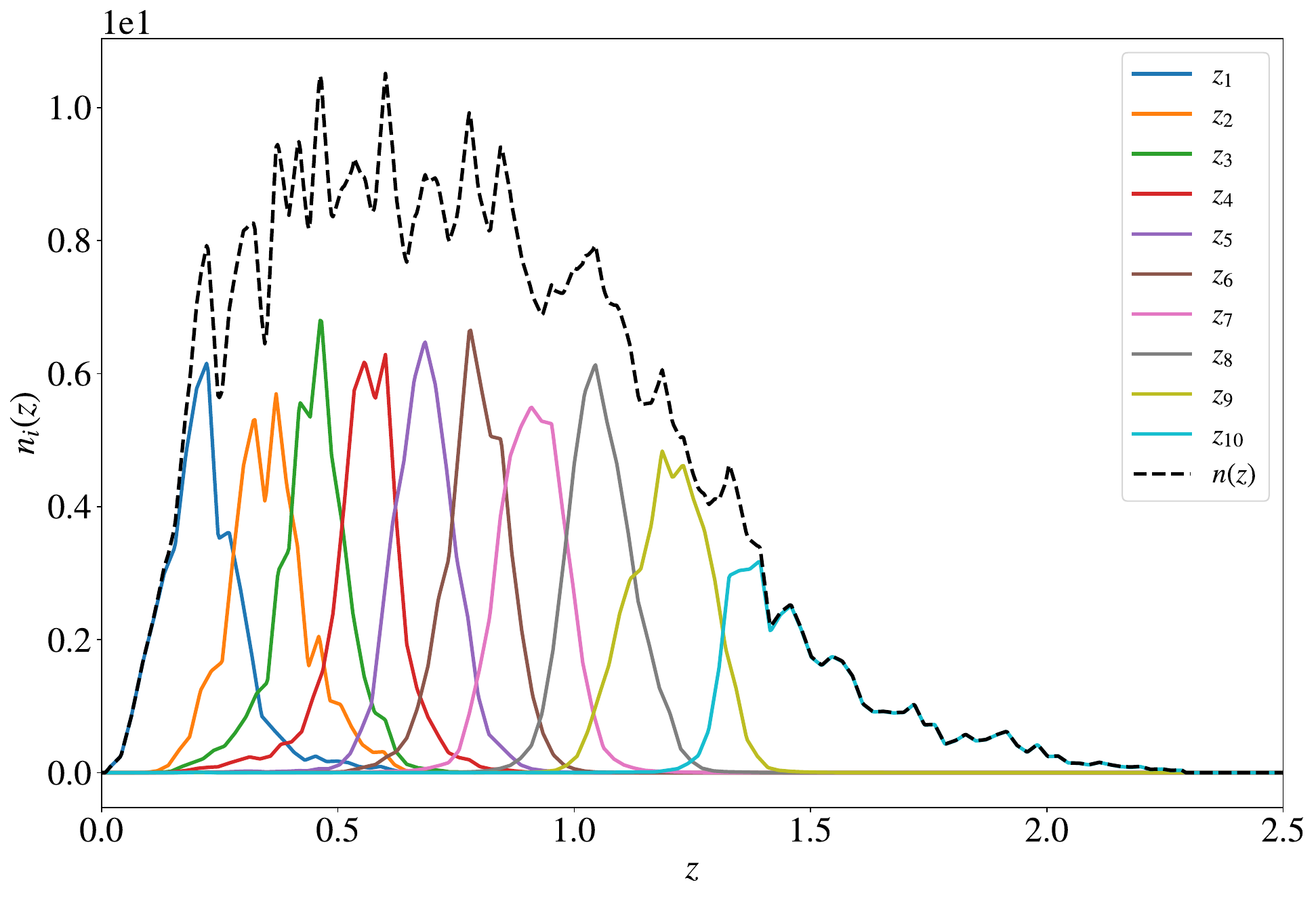}
    \caption{Sources (and lenses) redshift distribution per redshift bin, as well as their sum, obtained from the Flagship 1 simulation as described in Sect.~\ref{sec:z_distribution}}
    \label{fig:niz_FS1_for_PhD_thesis}
\end{figure}

A more realistic galaxy redshift distribution than the analytical one assumed in Eq.~\eqref{eq:n_of_z_analytical} can be obtained from simulations. In our forecasts, we will use the results from \citet{Pocino2021}, in which the $n(z)$ is constructed from photometric redshift estimates in a 400 ${\rm deg}^2$ patch of the Flagship 1 $N$-body dark matter simulation \citep{Potter2016}, using the training-based directional neighbourhood fitting algorithm \citep[DNF,][]{DeVicente2016}. \\
The training set is a random subsample of objects with true (spectroscopic) redshifts known from the Flagship 1 simulation. We choose the fiducial case presented in \citet{Pocino2021}, which takes into account a drop in completeness of the spectroscopic training sample with increasing magnitude. A cut in magnitude $\IE < 24.5$ is applied, isotropic and equal for all photometric bands, corresponding to the optimistic \Euclid setting. The DNF algorithm then produces a first estimate of the photo-$z$, $z_{\rm mean}$, using as a metric the objects' closeness in colour and magnitude space to the training samples. A second estimate of the redshift, $z_{\rm mc}$, is computed from a Monte Carlo draw from the nearest neighbour in the DNF metric. The final distributions for the different redshift bins, $n_i(z)$, are obtained by assigning the sources to the respective bins using their $z_{\rm mean}$, and then taking the histogram of the $z_{\rm mc}$ values in each of the bins -- following what has been done in real surveys such as DES. This redshift distribution is shown in Fig.~\ref{fig:niz_FS1_for_PhD_thesis}. The total galaxy number density measured from the simulation is  $\bar{n} = 28.73 \, {\rm arcmin}^{-2}$.\\

As a reference setting, we choose to bin the galaxy distribution into ${\cal N}_{\rm b} = $ 10 equipopulated redshift bins, with edges
\begin{align}
    z_{\rm edges} = \{ & 0.001, 0.301, 0.471, 0.608, 0.731, 0.851, \nonumber \\ 
    & 0.980, 1.131, 1.335, 1.667,
       2.501 \} \; .
    \label{eq:zbins}
\end{align}
This choice of redshift binning will be discussed and varied in Sect.~\ref{sec:z_bin_variations}.
\subsection{Linear galaxy bias}\label{sec:gal_bias}
As mentioned, one of the main systematic uncertainties affecting GCph measurements is the modelling of galaxy bias. Several different approaches have been presented in the literature to compute it. Simulations are a viable option, but it is much more difficult to simulate a realistic galaxy field than a \ML{dark matter field, being baryonic effects and the physics of galaxy formation and evolution of utmost importance in the former case}.

In this work, we choose to model the galaxy bias as linear and scale-independent (Eq.~\ref{eq:gal_bias_def}). We note that, at the same time, we are using the nonlinear recipe for the matter power spectrum $P_{\rm mm}(k, z)$. This is reminiscent of the hybrid 1-loop perturbation theory (PT) model adopted by, e.g., the DES Collaboration in the analysis of the latest data release \citep{Krause21, Pandey22}, but we drop the higher-order bias terms; it has been referred to as \enquote{minimal bias model} in recent publications \citep[e.g.][]{Sugiyama2023_HSC_minimal_bias_mod}. This simplified model has been chosen in order to be consistent with \citetalias{ISTF2020} forecasts, against which we will validate some of our results. However, we move on from the simple analytical prescription of \citetalias{ISTF2020}:
\begin{equation}
    b(z) = \sqrt{(1+z)}
\end{equation}
and use the fitting function presented in \citet{Pocino2021}, obtained from direct measurements from the \Euclid Flagship galaxy catalogue, based in turn on the Flagship 1 simulation:
\begin{equation}\label{eq:pocinoBias}
    b(z) = \frac{Az^B}{1+z} + C \; ,
\end{equation}
with $(A, B, C) = (0.81, 2.80, 1.02)$. \\

The galaxy bias is modelled to be constant in each bin with the fiducial value obtained by evaluating Eq.~\eqref{eq:pocinoBias} at effective values $z^{\rm eff}_i$ computed as the median of the redshift distribution (cf. Fig.~\ref{fig:niz_FS1_for_PhD_thesis}) considering only the part of the distribution at least larger than 10\% of its maximum. The $z^{\rm eff}_i$ values obtained in this way are
\begin{align}
    z^{\rm eff} = \{ & 0.233, 0.373, 0.455, 0.571, 0.686,
    \nonumber \\ 
    & 0.796, 0.913, 1.070, 1.195, 1.628 \} \; .
    \label{eq:zeff} 
\end{align}
We therefore have ${\cal N}_{\rm b}$ additional nuisance parameters
\begin{equation}
    \vec{\theta}_{\rm gal. \, bias} = \{b_1, b_2, \ldots, b_{{\cal N}_{\rm b}}\} \; ,
\end{equation}
with fiducial values
\begin{align}
    \vec{\theta}^{\rm fid}_{\rm gal. \, bias} = \{ & 1.031, 1.057, 1.081, 1.128, 1.187, \\ \nonumber
    & 1.258, 1.348, 1.493, 1.628,
       2.227 \} \; .
    \label{eq:zeff} 
\end{align}
Although we choose to work with a linear galaxy bias for the power spectra, higher-order terms are relevant in other ingredients of our forecasts; their modelling will be introduced in Sect.~\ref{sec:higher_order_bias}.  
\subsection{Magnification bias}\label{sec:magnification_bias}
Another systematic affecting GCph is \textit{magnification bias}. The effect is caused by foreground lensing magnification (see Sect.~\ref{sec:shear_convergence} for a formal introduction), which modulates the observed number of sources in a magnitude-limited survey. This effect has been found to significantly bias best-fit cosmological parameters for \Euclid in \citealt{Lepori2022_magbias}.\\
Magnification bias can be accounted for as an additional contribution to the galaxy number counts angular power spectrum, in a similar way to the intrinsic alignment term in weak lensing cosmic shear (see Sect.~\ref{sec:IA}):
\begin{equation}
    C^{\mathcal{GG}}_{ij}(\ell) = C^{\rm gg}_{ij}(\ell) + C^{\rm{g}\mu}_{ij}(\ell) + C^{\mu\mu}_{ij}(\ell)
\end{equation}
expressing these contributions in the general form of Eq.~\eqref{eq:cl_AB_tomo_limb_ISTF}, we have:
\begin{gather}
    C_{i j}^{{\rm g g}}(\ell)=\frac{c}{H_0}\int \frac{\mathcal{K}_i^g(z) \mathcal{K}_j^g(z)}{\chi^2(z) E(z)} P_{\rm g g}(k_\ell, z) \diff z \\ 
    C_{i j}^{{\rm g} \mu}(\ell)=\frac{c}{H_0}\int \frac{\mathcal{K}_i^g(z) \mathcal{K}_j^\mu(z)+\mathcal{K}_i^\mu(z) \mathcal{K}_j^g(z)}{\chi^2(z) E(z)} P_{\rm g m}(k_\ell, z) \diff z \\ 
    C_{i j}^{\mu \mu}(\ell)=\frac{c}{H_0}\int \frac{\mathcal{K}_i^\mu(z) \mathcal{K}_j^\mu(z)}{\chi^2(z) E(z)} P_{\rm m m}(k_\ell, z) \diff z \; .
\end{gather}
We used the superscript $\mathcal{GG}$ and gg to indicate respectively the galaxy clustering PS including or neglecting magnification bias, denoted with $\mu$. The magnification bias radial kernel is given by:
\begin{equation}\label{eq:w_mu}
    \mathcal{K}_i^\mu(z)=\frac{3}{2}\frac{H_0}{c} \Omega_{{\rm m}, 0}(1+z) \Tilde{\chi}(z) \int_z^{z_{\rm max}} \diff z' 
    b_{\mu, i}(z')
    \frac{n_{i}(z^\prime)}{\bar{n}}
    \left[ 1- \frac{\Tilde{\chi}(z)}{\Tilde{\chi}(z')}\right]
\end{equation}
$z_{\rm max}$ being the maximum redshift reached by the survey for the source distribution and $\Tilde{\chi}(z) = \chi(z)/(c/H_0)$ the dimensionless comoving distance; $b_{\mu, i}(z)=2\left[\beta_i(z)-1\right]$, with $\beta_i(z)$ the magnification bias in redshift bin $i$, which can be estimated as $\beta_i(z)=(5 / 2) s_i(z)$. Finally, $s_i(z)$ the logarithmic slope\footnote{In the literature, sometimes $s(z)$ itself is referred to as magnification bias.} of the cumulative number density $N$ as a function of the limiting magnitude, which can be measured directly from the luminosity function of the galaxy sample \citep{Lepori2022_magbias}: 
\begin{equation}
    \frac{2}{5} s(z, F_{\rm lim}) \equiv -\frac{\partial \log_{10} N(z, F>F_{\rm lim})}{\partial  \log_{10} F_{\rm lim}} \; ,
\end{equation}
where $F$ is the measured flux and $F_{\rm lim}$ is the survey flux limit.
The magnification kernel (Eq.~\ref{eq:w_mu}) is very similar to the lensing kernel (Eq.~\ref{eq:w_gamma_ISTF}), being magnification a lensing effect. \FB{It should be noted that the $n_i(z)$ and $\bar{n}$ terms entering the expression for the magnification kernel refer to \textit{lens} redshift distribution}. Magnification bias will be included in our forecasts in Chap.~\ref{chap:scalecuts}\ML{; unless otherwise specified, when left free to vary, we will not impose priors either on these or on the galaxy bias parameters}.
\section{Weak Lensing}
\begin{figure}
\begin{center}
 \includegraphics[scale=0.5]{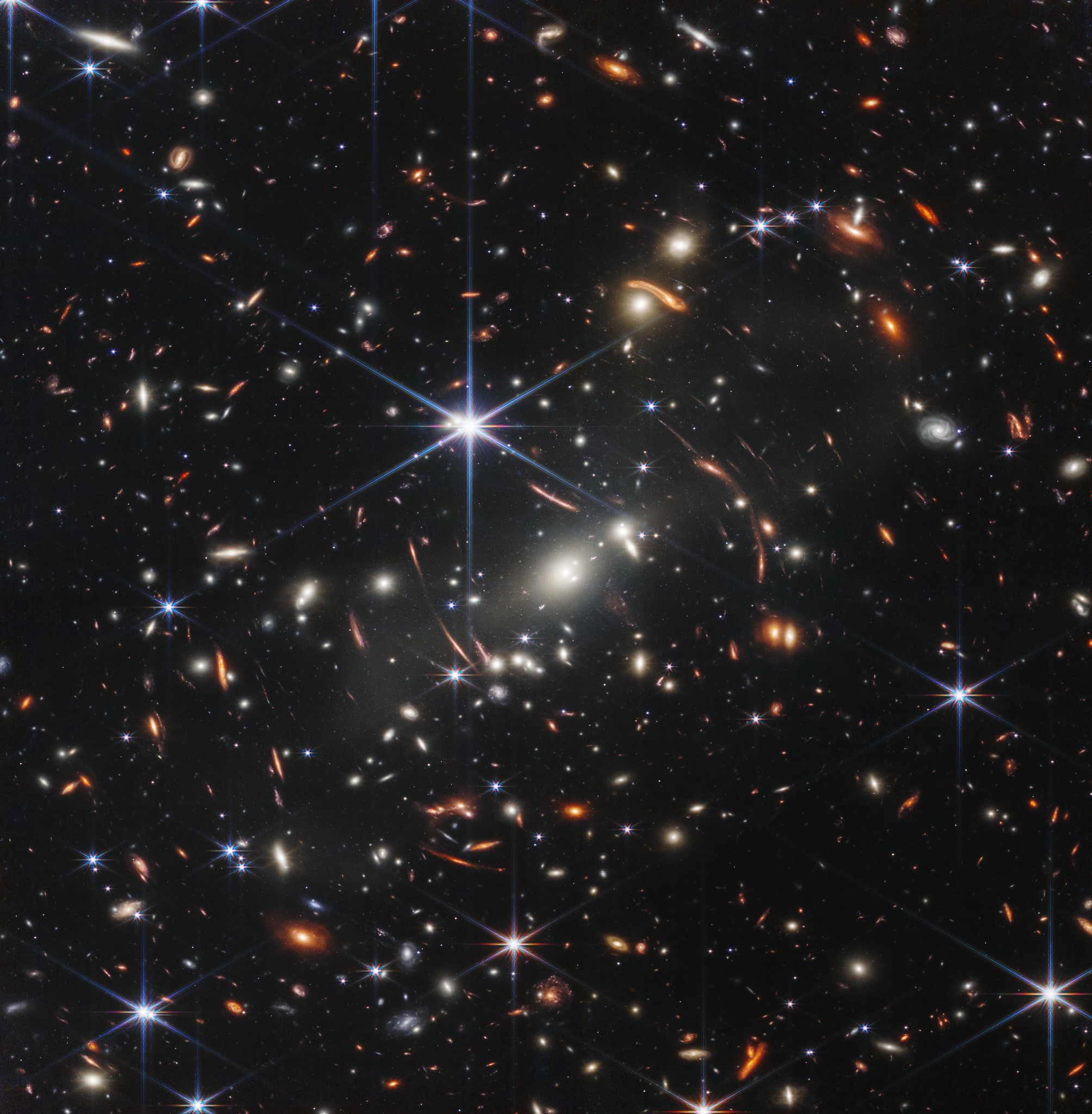}
\caption{An example of \textit{strong} gravitational lensing in one of the first public images by the James Webb Space Telescope: around the centre are visible galaxies belonging to the galaxy cluster SMACS 0723, whose strong gravitational field warps the shape of background galaxies, making them appear in the form of long arcs and multiplying their number. Source: \url{nasa.gov/webbfirstimages/}}
\label{fig:strong_lensing}
\end{center}
\end{figure}

A key insight of General relativity is that mass-energy (and momentum) distort the curvature of the spacetime manifold, hence changing the geodesic path of particles travelling along their worldline. In particular, photons' trajectories are distorted by the presence of matter (\textit{lenses}) between the source and the observer, resulting in the shear, magnification and multiplication of the images of distant objects. Such effect is called \textit{gravitational lensing}, and constitutes one of the main ways in which we are able to infer the global geometry and map the matter distribution of our Universe. \ML{A spectacular example of this phenomenon can be found in Fig.~\ref{fig:strong_lensing}, showcasing the galaxies' shape distortion caused by the gravitational field of a massive cluster in the foreground}. \\
With galaxy clustering we map the 3D distribution of galaxies, which needs galaxy bias as an additional ingredient (suffering significant uncertainty) to allow the reconstruction of the total (baryonic + dark) matter distribution. Being dependent on the overall gravitational potential, on the other hand, weak lensing (WL) gives us the possibility to map the distribution of \textit{all} gravitationally interacting kinds of matter, whether of baryonic nature or not.\\

In the context of the \Euclid survey we are interested in particular in one sub-category of this phenomenon: \textit{weak lensing cosmic shear}, which is the distortion of a galaxy's shape (ellipticity) caused not by one specific massive object, but by the large-scale structures of the Universe. Being this a very small effect (order $1\%$ deviation from the intrinsic ellipticity,  \citealt{Weinberg2013}), requiring high-resolution imaging and exquisite control over the many systematics involved, it is best measured from space (although ground-based observations have made significant progress in recent years, see e.g. the DES, KiDS and HSC surveys: \citealt{DES2005, de_Jong2013_KiDS, Aihara2018_HSC}). The point spread function (PSF) is an example of one such systematic which space-based observations help mitigate. The observed galaxy shape is smeared by the convolution of the real image with the PSF, which is caused by astronomical seeing and instrumental effects. The PSF anisotropy and dependence on the source redshift and brightness complicate the job of removing this effect. Space experiments are unaffected by the atmospheric component of this effect but not by the instrumental ones, which can however be measured in the laboratory or modelled mathematically. \\

Weak gravitational lensing affects the image of the source in two ways: \textit{shear} and \textit{magnification}. Whereas the first results in a distortion of the galaxies' shape, the second affects their observed size. Of the two, the former effect is easier to observe and the most used in cosmology. In fact, although we do not have access to the information about the real image of the source (the image on the \textit{source plane}), the ellipticity cancels out when averaged over a large number of sources (there is no preferred ellipticity direction); this is not true for the galaxies' sizes. From this recognition, statistical observations about the observed shape distortions can be made, such as their degree of correlation as a function of angular scale. In order to increase the sample size (hence statistical power) as much as possible, $\sim 90\%$ of \Euclid's observational time will be dedicated to the wide survey, covering about $14~700$ ${\rm deg}^2$ ($\sim 36 \%$ of the entire sky). 
The main observable studied here is the \textit{cosmic shear angular power spectrum}, which is the Fourier transform on a sphere of the shear angular correlation function (see Sect.~\ref{sec:angular_ps_theory} for a mathematical introduction). This function is related to the projection of the matter power spectrum over the line of sight, and thus it is sensitive to the cosmological parameters (more specifically, as we will see, to $\Omega_{\rm m, 0}$ and $\sigma_8$). To further enhance the survey constraining power it is possible to engage in a tomographic analysis, which as explained in \ref{sec:tomography} introduces the information on the auto- and cross-correlation between the different redshift bins. \\
The following pages will give a brief theoretical description of weak lensing.
\subsection{Theory}
The presence of mass-energy deflects light by virtue of the effect it has on the geometry of spacetime, which is the 4-dimensional space in which the photons' trajectories, the geodesics, are defined. These are the solutions to the \textit{geodesic equation}:
\begin{equation} \label{eq:geodesic}
    \frac{\diff^2 x^\mu}{\diff\lambda^2} + 
    \Gamma^\mu_{\alpha\beta}
    \frac{\diff x^\alpha}{\diff\lambda}
    \frac{\diff x^\beta}{\diff\lambda} = 0
\end{equation}
where $\lambda$ is the affine parameter and
\begin{equation} \label{eq:christoffel}
    \Gamma^\mu_{\alpha\beta} = \frac{1}{2}g^{\mu\gamma}\left( \frac{dg_{\gamma\alpha}}{\diff x^\beta} + 
\frac{\diff g_{\gamma\beta}}{\diff x^\alpha} -
\frac{\diff g_{\alpha\beta}}{\diff x^\gamma}\right)
\end{equation}
are the \textit{Christoffel symbols}, which account for the change in the basis vectors in curved space when taking the $x^\mu$ derivatives, and only depend on the metric and its derivatives. As discussed in the introduction, the Einstein field equations (Eq.~\ref{eq:EFE}) tell us how to build the metric tensor once in possess of the information about the mass-energy density distribution, and Eqs.~\eqref{eq:christoffel} and~\eqref{eq:geodesic} allow computing the resulting particle trajectories. \\

The following analysis of the WL effect is based on the following assumptions \citep{Dodelson2020}:
\begin{itemize}
\item \ML{Background} spacetime is described by the FLRW metric
\item Deflection angles are small
\item The time evolution of the perturbations is slow: $ \diff\Phi/\diff t \ll  \diff\Phi/\diff x_i$
\item Perturbations are small, so that a first-order theory provides a good enough description 
\item The lenses are thin, i.e. with a linear scale much smaller than the curvature scale of the Universe.
\end{itemize}
To make quantitative considerations about light deflection in the weak regime, we can make use of simple but powerful concepts from geometrical optics, such as \textit{Fermat's principle}: the path followed by light is the one which minimizes the travel time
\begin{equation} \label{eq:travel_time}
    t = \int_A^B \frac{\diff l}{v\left[\vec{x}(l)\right]}  =
    \int_A^B \frac{n}{c} \diff l \; ,
\end{equation}
which is the path $\vec{x}(l)$ for which the variation
\begin{equation}
\delta \int_A^B n\left[\vec{x}(l)\right]\diff l 
\end{equation}
is null. Here the curvature acts as a medium with refraction index $n(\vec{x}(l)) = c/v(\vec{x}(l))$ in which the light travels at speed $v(\vec{x}(l))$; $l$ is the geodesic parameter. \\
Calling $\Phi$ the gravitational potential of the massive structure which distorts the light path (the \textit{lens}), we can write the line element as
\begin{equation}
\diff s^2 = \left( 1 + \frac{2\Phi}{c^2} \right) c^2 \diff t^2
- \left( 1 - \frac{2\Phi}{c^2} \right)\diff \vec{x}^2 \; ,
\end{equation}
Using the null geodesic condition $\diff s = 0$ we can then compute the new travel speed $c'$:
\begin{equation}
c' = \frac{|\diff \vec{x}|}{\diff t} = c 
\left(1 + \frac{2\Phi}{c^2}\right)^{1/2}
\left(1 - \frac{2\Phi}{c^2}\right)^{-1/2} \simeq c \left( 1 + \frac{2\Phi}{c^2} \right)
\end{equation}
having used the fact that the weak regime is defined by the condition $\Phi/c^2 \ll 1$. Now we have an expression for the refraction index with which to compute the integral~\eqref{eq:travel_time}: using again the weak potential approximation, we have in fact
\begin{equation}
n \equiv \frac{c}{c'} = \frac{1}{1+\frac{2\Phi}{c^2}}
\simeq 1-\frac{2\Phi}{c^2} \; .
\end{equation}
We can now make use of the Euler-Lagrange equation:
\begin{equation}\label{eq:euler_lagrange}
\frac{\partial}{\partial \lambda}
\frac{\partial L}{\partial \dot{\vec{x}}}
- \frac{\partial L}{\partial \vec{x}} = 0 \qquad \text{with} \qquad
\dot{\vec{x}} \equiv
\frac{\diff \vec{x}}{\diff \lambda} \; ,
\end{equation}
in which the Lagrangian is $L=n(\vec{x}(\lambda))\left| \frac{\diff x}{\diff \lambda} \right|$, 
having used $\lambda$ to re-parametrise the light path: $\diff l = \left| \frac{\diff x}{\diff \lambda} \right| \diff \lambda$. \\
Choosing $\lambda$ such that $|\dot{\vec{x}}| = 1$, way we obtain a set of normal vectors tangent to the curve, $\dot{\vec{x}} = \vec{e}$. With this notation, the Euler-Lagrange equation yields the result \cite{Meneghetti2022}:
\begin{equation}
\frac{\diff}{\diff \lambda}
(n\vec{e}) - \vec{\nabla}n = 0
\quad \rightarrow \quad
n \dot{\vec{e}} = 
\vec{\nabla}n - 
\vec{e}(\vec{\nabla}n \cdot \vec{e}) \; ,
\end{equation}
the right-hand side being the gradient $\vec{\nabla}n$ minus its projection along the direction tangential to the geodesic $\vec{e}$. We can then interpret such difference as the gradient of $n$ in the direction perpendicular to the light path, $\vec{\nabla}_\perp n$:
\begin{equation}
\dot{\vec{e}} = 
\frac{1}{n} \vec{\nabla}_\perp n = 
\vec{\nabla}_\perp \ln n
\end{equation}
and, since $n = 1- 2\Phi/c^2$, $\ln n \simeq -2 \Phi/c^2$, so the last expression becomes
\begin{equation}
\dot{\vec{e}} \simeq -\frac{2}{c^2}\vec{\nabla}_\perp \Phi \; .
\end{equation}
This equation relates the angle of the vector tangent to the geodesic curve to the gravitational potential experienced by the photon along its path. Integrating $-\dot{\vec{e}}$  along the whole path we obtain the total deflection angle $\hat{\vec{\alpha}}$:
\begin{equation}\label{eq:defl_angle_integral}
\hat{\vec{\alpha}} = \frac{2}{c^2} \int_{\lambda_A}^{\lambda_B}
\vec{\nabla}_\perp\Phi\diff\lambda \; .
\end{equation}
Exploiting the fact that this angle is small - that is, the actual path is not very different from the one induced by the gravitational pull of the lens - we can use the \textit{Born approximation}, integrating Eq.~\eqref{eq:defl_angle_integral} over the unperturbed path. 
\begin{figure}
\begin{center}
 \includegraphics[scale=0.35]{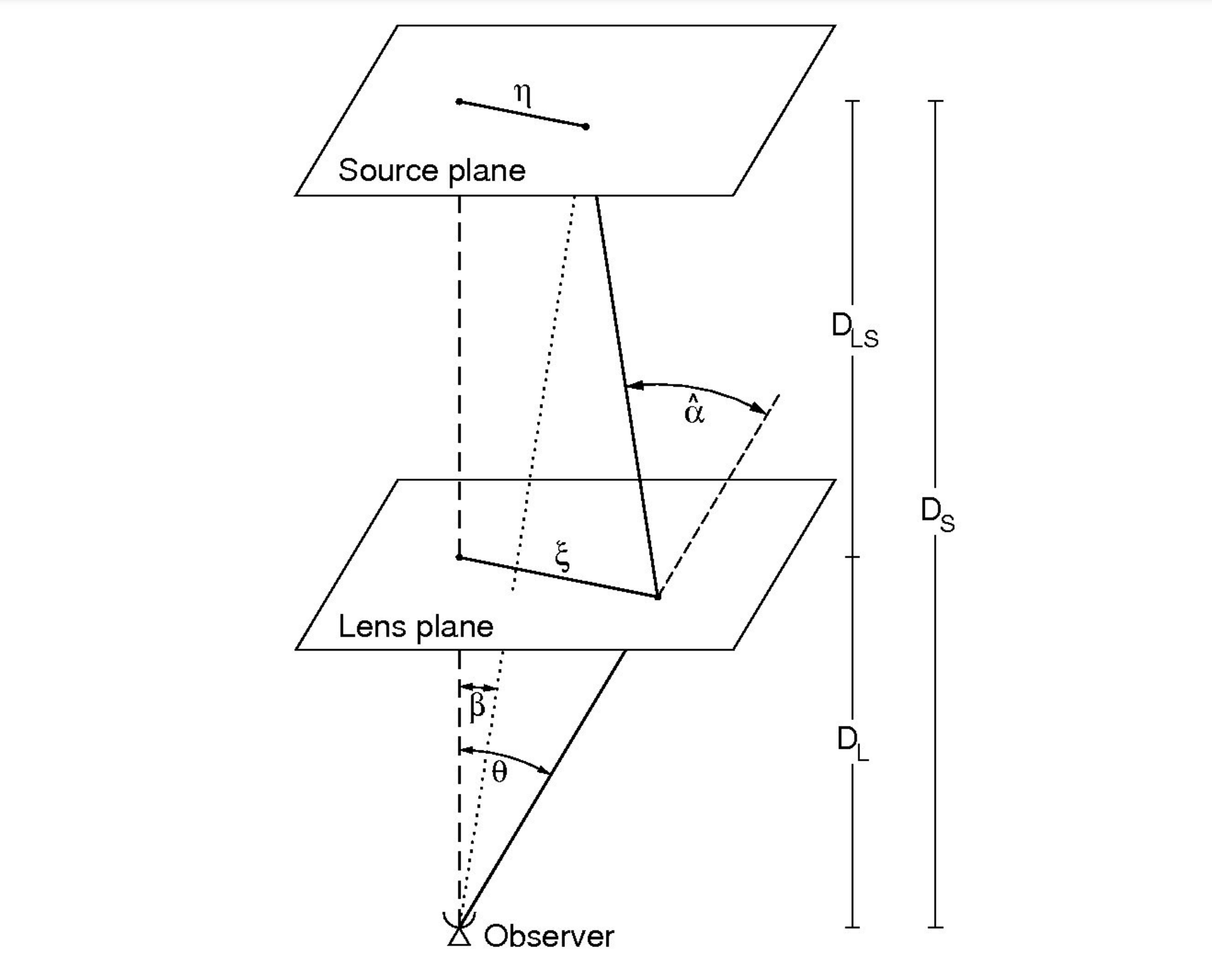}
\caption{The photon path gets distorted by a small angle $\hat{\vec{\alpha}}$ due to the presence of intervening overdensity. The description of the angles and distances is in the text. Adapted in \citet{Meneghetti2022} from \citet{Bartelmann2001_lensing}} \label{fig:lens}
\end{center}
\end{figure}
\subsection{Lens equation}
The main quantities at play are depicted in Fig.~\ref{fig:lens}. $D_{\rm S}$ is the angular diameter distance (ADD, see Sect.~\ref{sec:distances}) of the source, placed at redshift $z_{\rm S}$; $D_{L}$ is the ADD of the lens, placed at redshift $z_{\rm L}$, and $D_{\rm LS}$ is the ADD from the lens to the source. In curved space such distances are not necessarily additive, so the relation $D_{\rm S} = D_{\rm LS} + D_{\rm L}$ is not true in general. $\vec{\eta} = D_{\rm S} \vec{\beta}$ is the actual position of the source on the source plane, $\vec{\xi} = D_{\rm L} \vec{\theta}$ is the impact parameter (the distance of the geodesic from the lens once the photon reaches the lens plane) and $\hat{\vec{\alpha}}$ is the above-mentioned deflection angle. At all times, the photon position can be characterized by a 2D vector specifying its angular distance from the perpendicular to the lens and source plane (the optical axis). The photon starts with angular vector
$\vec{\beta}$, but appears to us to be coming from $\vec{\theta}$. \\
If the angles ($\vec{\hat{\alpha}}, \vec{\beta}, \vec{\theta}$) are small, the distances are related via the \textit{lens equation}:
\begin{equation}
\vec{\theta}D_{\rm S} = \vec{\beta}D_{\rm S} + \hat{\vec{\alpha}}D_{\rm LS}
\end{equation}
which can be rewritten introducing the \textit{reduced deflection angle} $\vec{\alpha}(\vec{\theta})$  as
\begin{equation}
\vec{\theta} = \vec{\alpha} + \vec{\beta} \quad \text{with} \quad \vec{\alpha} = \vec{\alpha}(\vec{\theta}) \equiv
\frac{D_{\rm LS}}{D_{\rm S}}\hat{\vec{\alpha}}(\vec{\theta}) \; .
\end{equation}
As a final remark, we can introduce the quantities $\xi$ and $\eta = \xi D_{\rm S}/D_{\rm L}$ as typical scales of the problem, with the aim of defining the \ML{dimensionless} vectors
\begin{equation}
\vec{x} = \frac{\vec{\xi}}{\xi} \quad \text{and} \quad
\vec{y} = \frac{\vec{\eta}}{\eta}
\end{equation}
and to scale the deflection angle:
\begin{equation}
\vec{\alpha}(\vec{x}) \equiv \frac{D_{\rm L}D_{\rm LS}}{\xi D_{\rm S}}
\hat{\vec{\alpha}}(\xi \vec{x}).
\end{equation}
Thanks to these we can rewrite the lens equation as
\begin{equation} \label{eq:lens_2}
    \vec{y} = \vec{x}-\vec{\alpha}(\vec{x})
\end{equation}
\subsection{Lensing potential, shear and convergence matrices}\label{sec:shear_convergence}
\begin{figure}
\begin{center}
 \includegraphics[scale=0.35]{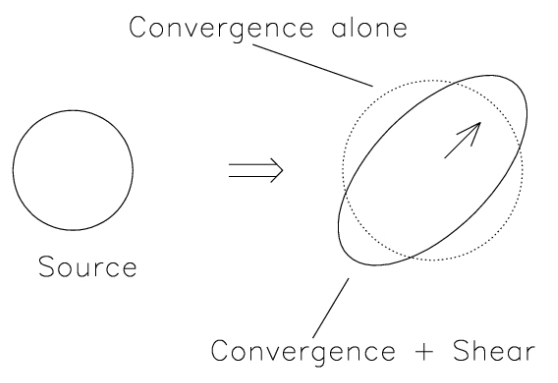}
\caption{The effect of convergence and shear. Source: \citep{Meneghetti2022}}
\label{fig:convergence_shear}
\end{center}
\end{figure}

In many cases of interest - including the one under examination - the lens is an extended matter distribution. In these situations it is useful to define a function of the gravitational potential, the \textit{effective lensing potential}:
\begin{equation} \label{eq:psi}
\hat{\Psi}(\vec{\theta}) \equiv  \frac{D_{\rm LS}}{D_{\rm L} D_{\rm S}}
\frac{2}{c^2} \int \diff z \Phi(D_{\rm L} \vec{\theta}, z) \; ,
\end{equation}
which is the rescaled three-dimensional gravitational potential, projected on the lens plane. Its \ML{dimensionless} version $\Psi = \frac{D_{\rm L}^2}{\xi_0^2}\hat{\Psi}$ has the useful properties:
\begin{enumerate}
    \item $\vec{\nabla}_x \Psi(\vec{x}) = \vec{\alpha}(\vec{x})$: the gradient along the $\vec{x}$ vector (therefore in the direction perpendicular to the geodesic) gives the scaled deflection angle.
    \item $\vec{\Delta}_x \Psi(\vec{x}) = 2\kappa(\vec{x})$: the laplacian along the $\vec{x}$ vector gives twice the \textit{convergence}, which we will define below.
\end{enumerate}
The image distortion caused by weak lensing can be conveniently described through the Jacobian matrix:
\begin{equation}
A_{ij} \equiv \frac{\partial y_i}{\partial x_j} = 
\left( \delta^{\rm K}_{ij} - \frac{\partial \alpha_i(\vec{x})}{\partial x_j} \right) = 
\left( \delta^{\rm K}_{ij} - \frac{\partial^2 \Psi(\vec{x})}{\partial x_i\partial x_j} \right) \; .
\end{equation}
The first equality comes from the lens equation \eqref{eq:lens_2} and the second one from the first of the two points above. Defining the \textit{distortion tensor} $\Psi_{ij} = \partial^2 \Psi / \partial x_i \partial x_j$ and $I$ the identity matrix we can consider the isotropic, traceless part of  $A_{ij}$:
\begin{align}
\left( A - \frac{1}{2}\textbf{tr}A\cdot I
\right)_{ij} &= \delta_{ij} - \Psi_{ij} - \frac{1}{2} (1-\Psi_{11} + 1-\Psi_{22})\delta^{\rm K}_{ij} \\
&= \begin{pmatrix} \label{eq:A_of_Psi}
-\frac{1}{2}(\Psi_{11}-\Psi_{22}) & -\Psi_{12} \\
-\Psi_{12} & \frac{1}{2}(\Psi_{11}-\Psi_{22}) 
\end{pmatrix}.
\end{align}
This matrix is called \textit{shear matrix}. This can be rewritten introducing the shear pseudo-vector $\vec{\gamma} = (\gamma_1, \gamma_2)$:
\begin{align}
\gamma_1(\vec{x}) &= \frac{1}{2}(\Psi_{11}-\Psi_{22})\label{eq:gamma_plus_vs_psi} \\
\gamma_2(\vec{x}) &= \Psi_{12} = \Psi_{21} \label{eq:gamma_x_vs_psi}\; ,
\end{align}
so that Eq.~\eqref{eq:A_of_Psi} can be recast in the form:
\begin{equation}
\left( A - \frac{1}{2}\textbf{tr}A\cdot I
\right)_{ij} =
\begin{pmatrix} \label{shear}
\gamma_1 & \gamma_2 \\
\gamma_2 & -\gamma_1
\end{pmatrix} \; .
\end{equation}
The remaining part of $A_{ij}$ is 
\begin{align}
\frac{1}{2}\textbf{tr}A\cdot I &= \left(
1-\frac{1}{2}(\Psi_{11}+\Psi_{22}) \right) \delta_{ij} \\
&= \left(
1-\frac{1}{2}\Delta_x \Psi \right)\delta_{ij}=
\begin{pmatrix} \label{eq:convergence}
1-\kappa & 0 \\
0 & 1-\kappa
\end{pmatrix} \; ;
\end{align}
this is the \textit{convergence matrix}. Putting together equations \eqref{shear} and \eqref{eq:convergence}, we obtain a new expression for the Jacobian:
\begin{equation} \label{Aij}
A_{ij}=
\begin{pmatrix}
1-\kappa-\gamma_1 & -\gamma_2\\
-\gamma_2 & 1-\kappa+\gamma_1
\end{pmatrix} \; ,
\end{equation}
where two key quantities appear: the \textit{convergence} $\kappa$, quantifying the (isotropic) magnification, and the \textit{shear} $\vec{\gamma}$, whose components govern the image distortion (see Fig.~\ref{fig:convergence_shear}). The former will not be discussed in detail since this work will focus on weak lensing cosmic shear as a cosmological probe. 
\subsection{Ellipticity and shear}\label{sec:ellipticity_shear}
The last step is to relate the shear to some observable quantity. The case of interest is that of weak lensing cosmic shear by large-scale structures. \\
A circular shape is mapped into an elliptical one via the matrix $A_{ij}$, with the semi-major and semi-minor axes $a$ and $b$ functions of the distortion parameters $\gamma$ and $\kappa$:
\begin{equation}
a = \frac{r}{1-\kappa-\gamma} \quad \text{and} \quad
b = \frac{r}{1-\kappa+\gamma}
\end{equation}
where $\pm\gamma = \pm\sqrt{\gamma_1^2 + \gamma_2^2}$ are the eigenvalues of the shear matrix and $r$ is the circular source radius. The \textit{ellipticity} $\epsilon$ is defined as 
\begin{equation}
\epsilon = \frac{a-b}{a+b} = \frac{2\gamma}{2(1-\kappa)} = \frac{\gamma}{1-\kappa} \simeq \gamma
\end{equation}
assuming $\kappa,\gamma\ll 1$. The observed ellipticity is the sum of the source's intrinsic ellipticity and the shear
\begin{equation}
\epsilon_i = \epsilon^{\rm I}_i + \gamma_i \; ;
\end{equation}
if we average over a large number of sources:
\begin{equation}
\langle\epsilon\rangle = \langle\epsilon^{\rm I}\rangle + \langle\gamma\rangle = \langle \gamma \rangle \; .
\end{equation}
\ML{The last equality assumes that} the distribution of galaxy shapes and orientations is completely random - which is not exactly the case, due to, e.g., intrinsic alignment, as we shall discuss. In this case, the average over the intrinsic ellipticity cancels out: this means that in the WL limit the ellipticity is a direct measure of the shear. \FB{The statistical description
of errors associated with the intrinsic variation in galaxy shapes, uncorrelated with the lensing signal, is commonly referred to as
\enquote{shape noise}, and it is by far the largest source of uncertainty in
WL surveys \citep{Gurri2021}}.
\subsection{Shear power spectrum}\label{sec:shear_ps}
From the shear field, two main statistics can be extracted: the two-point shear correlation function and its spherical harmonic transform, the cosmic shear angular power spectrum. In the context of \Euclid, the preferred quantity is the latter. Both choices have advantages and disadvantages; the covariance matrix terms, for example, have a simpler expression in angular space. A large number of additional higher-order statistics can also be constructed (see \citealt{Ajani2023_HOWLS} for a recent application to the \Euclid survey), maximizing the information extracted from the shear field, but these will not be discussed in the present work. \\
To derive the cosmic shear angular PS \ML{in the flat-sky approximation}, we take the Fourier transform \ML{(instead of the spherical harmonics transform)} of the shear field \citep{Weinberg2013}:
\begin{equation}
    \tilde{\gamma}_{1, 2}(\vec{\ell})=\int \diff^2 \,\vec{\theta} \gamma_{1, 2}(\vec{\theta}) e^{-i \vec{\ell} \cdot \vec{\theta}}  \quad \leftrightarrow \quad \gamma_{1, 2}(\theta)=\int \frac{\diff^2 \vec{\ell}}{(2 \pi)^2} \, \tilde{\gamma}_{1, 2}(\vec{\ell}) e^{i \vec{\ell} \cdot \vec{\theta}}  \; .
\end{equation}
We can simplify the problem by rotating the basis of the Fourier space to align them with the direction of the wavevector describing the plane-wave perturbation \FB{\citep{Crittenden2002}}. We call these rotated components $E$ and $B$ modes; they are generated by galaxies which are respectively stretched in the direction of the wave vector and squashed perpendicular to it, or at $45^{\circ}$ angles. 
\begin{align}
\tilde{\gamma}_E(\vec{\ell})& =\cos \left(2 \phi_{\vec{\ell}}\right) \tilde{\gamma}_{1}(\vec{\ell})+\sin \left(2 \phi_{\vec{\ell}}\right) \tilde{\gamma}_{2}(\vec{\ell}) \label{eq:gamma_tilde_E}\\
\tilde{\gamma}_B(\vec{\ell})& =\cos \left(2 \phi_{\vec{\ell}}\right) \tilde{\gamma}_{2}(\vec{\ell})-\sin \left(2 \phi_{\vec{\ell}}\right) \tilde{\gamma}_{1}(\vec{\ell}) \label{eq:gamma_tilde_B}\; ,
\end{align}
where $\tan \phi_1=\ell_2 / \ell_1$, and $\ell_1$ and $\ell_2$ being the components of $\vec{\ell}$ in the old coordinate system. The angular PS are then defined as (see Sect.~\ref{sec:angular_ps_theory}):
\begin{align}
\left\langle\tilde{\gamma}_E^*(\vec{\ell}) \tilde{\gamma}_E\left(\vec{\ell}^{\prime}\right)\right\rangle&=(2 \pi)^2 C_{E E}(\ell) \delta^2_{\rm D}\left(\vec{\ell}-\vec{\ell}^{\prime}\right) \\
\left\langle\tilde{\gamma}_B^*(\vec{\ell}) \tilde{\gamma}_E\left(\vec{\ell}^{\prime}\right)\right\rangle&=(2 \pi)^2 C_{B E}(\ell) \delta^2_{\rm D}\left(\vec{\ell}-\vec{\ell}^{\prime}\right) \\
\left\langle\tilde{\gamma}_B^*(\vec{\ell}) 
\tilde{\gamma}_B\left(\vec{\ell}^{\prime}\right)\right\rangle&=(2 \pi)^2 C_{B B}(\ell) \delta^2_{\rm D}\left(\vec{\ell}-\vec{\ell}^{\prime}\right) \; .
\end{align}
Isotropy makes these PS depend only on the magnitude of $\vec{\ell}$ and not its direction, and reflection symmetry guarantees that $C_{E B}(\ell)=0$. We now need to connect these to the lensing potential $\Psi$; this can be done by leveraging Eqs.~\eqref{eq:gamma_plus_vs_psi} and~\eqref{eq:gamma_x_vs_psi}, which we rewrite in angular coordinates:
\begin{align}
\gamma_{1}(\theta)&=\frac{1}{2}\left(\frac{\partial^2 \Psi}{\partial \theta_1^2}-\frac{\partial^2 \Psi}{\partial \theta_2^2}\right) \\
\gamma_{2}(\theta)&=\frac{\partial^2 \Psi}{\partial \theta_1 \partial \theta_2} .
\end{align}
Using the replacement $\partial / \partial \theta_i \rightarrow i \ell_i$, we find in Fourier space
\begin{align}
\tilde{\gamma}_{1}(\vec{\ell})&=\frac{1}{2}\left(\ell_1^2-\ell_2^2\right) \tilde{\Psi}(\vec{\ell})=\frac{1}{2} \ell^2 \cos \left(2 \phi_1\right) \tilde{\Psi}(\vec{\ell}) \\
\tilde{\gamma}_{2}(\vec{\ell})&=\ell_1 \ell_2 \tilde{\Psi}(\vec{\ell})=\frac{1}{2} \ell^2 \sin \left(2 \phi_{\vec{\ell}}\right) \tilde{\Psi}(\vec{\ell}) .
\end{align}
Substitution back into equations~\eqref{eq:gamma_tilde_E} and \eqref{eq:gamma_tilde_B} we get the important results that cosmic shear possesses only $E$ modes, in other words, it has no curl component:
\begin{align}
\tilde{\gamma}_E(\vec{\ell})&=\frac{\ell^2}{2} \tilde{\Psi}(\vec{\ell}) \rightarrow C_{EE}(\ell) = \frac{\ell^4}{4}C_{\Psi\Psi}(\ell)\label{eq:CEE_vs_Cphiphi}\\
\tilde{\gamma}_B(\vec{\ell})&=0 \; .
\end{align}
\FB{One potential source of spurious $B$ modes is shape noise, which is a priori equally distributed among these and $E$ modes}. Equation~\eqref{eq:CEE_vs_Cphiphi} connects the $E$ mode, shear PS to the PS of the lensing potential $\Psi$, $C_{\Psi\Psi}(\ell)$ This potential is related to the gravitational one by the weighted projection \eqref{eq:psi}, and the gravitational potential is related in turn to the matter overdensity by the Poisson equation (in an expanding background), which in Fourier space is given by:
\begin{equation}
\Phi(k)=\left[\frac{3}{2} \Omega_{{\rm m}, 0} \frac{H_0^2}{c^2}(1+z)\right] k^{-2} \delta_{\rm m}(k) \implies P_{\Phi}(k)=\left[\frac{3}{2} \Omega_{{\rm m}, 0} \frac{H_0^2}{c^2}(1+z)\right]^2 k^{-4} P_{\delta\delta}(k) \; .
\end{equation}
Where $P_{\delta\delta} = P_{\delta_{\rm m}\delta_{\rm m}} = P_{\rm mm}$ is the matter PS; this allows to obtain the the analogous of Eq.~\eqref{eq:cl_gg_limber}:
\begin{equation}\label{eq:C_EE_weinberg}
C_{E E}(\ell)=\int_0^{\chi} \diff \chi'
    \frac{\mathcal{K}^2\left(\chi', \chi\right)}{\chi'^2}
    P_{\delta\delta}\left(k_\ell = \frac{(\ell+1/2)}{ \chi'}\right)  \; ,
\end{equation}
with 
\begin{align}\label{eq:W_weinberg}
\mathcal{K}\left(\chi', \chi\right)=\frac{3}{2} \Omega_{\rm m, 0} \frac{H_0^2}{c^2}\left(1+z'\right) \frac{\chi'\left(\chi-\chi'\right)}{\chi}
\end{align}
the lensing weight function, describing the contributions to lensing of sources at comoving distance $\chi$ from lens structures at distance $\chi'$. Note that this vanishes as the lens approaches the source $\left(\chi' \rightarrow \chi\right)$. This result is \ML{valid} in the flat case and in the Limber, flat-sky approximation used throughout this work. This weight function has to be re-weighted taking into consideration the lens redshift distribution, i.e., the probability distribution of their distance $n(\chi)$:
\begin{equation} 
    \mathcal{K}_{\rm eff} = \int_0^{\chi_{\rm max}} n(\chi)\mathcal{K}(\chi', \chi)\diff\chi
\end{equation}
As can be seen from Eqs.~\eqref{eq:C_EE_weinberg} and \eqref{eq:W_weinberg}, the shear signal will be primarily sensitive to the square of the matter density parameter and the integral of the matter PS, which in the linear regime goes as $\sim\sigma_8^2$ and in the nonlinear one close to $\sim\sigma_8^3$ \citep{Weinberg2013}. It will therefore best constrain the degenerate combination $\Omega_{{\rm m, 0}}^2\sigma_8^3$, with the bulk of the information coming from the small-scale regime. This is why often analyses report constraints on some combination of the two parameters (in particular, $S_8 = \sigma_8 \, \Omega^{0.5}_{{\rm m}, 0}$, see \citealt{Hall2021_S8}). \\

\noindent
Recasting the expression to the general form given in Eq.~\eqref{eq:cl_AB_tomo_limb_ISTF}, we get:
\begin{equation}\label{eq:cl_gammagamma_ISTF} 
    C_{ij}^{\gamma\gamma} (\ell) \simeq \frac{c}{H_0}
    \int \diff z \frac{\mathcal{K}_i^\gamma(z)\mathcal{K}_j^\gamma(z)}
    {E(z)\chi^2(z)}P_{\delta\delta}\left[ 
    \frac{\ell + 1/2}{\chi(z)},z
    \right].
\end{equation}
and\footnote{We remind that Eq.~\eqref{eq:w_gamma_ISTF} assumes a spatially flat Universe. For the general case, one must replace the term in brackets with $S_k(\chi^{\prime} - \chi)/S_k(\chi^{\prime})$, with $S_k(\chi)$ given in Eq.~\eqref{eq:S_k(r)}}
\begin{equation}\label{eq:w_gamma_ISTF}
\mathcal{K}_i^\gamma(z) = \frac{3}{2}\frac{H_0}{c}\Omega_{{\rm m}, 0}
(1+z)\Tilde{\chi}(z)\int_z^{z_{\rm max}} \diff z' \frac{n_{i}(z^\prime)}{\bar{n}}\left[ 1- \frac{\Tilde{\chi}(z)}
{\Tilde{\chi}(z')}
\right] \; .
\end{equation} 
The present discussion has been devoted to the modelling of the cosmic shear signal. Real observations are affected by many different systematic effects, some of which we proceed to describe in the following sections.
\subsection{Intrinsic alignment}\label{sec:IA}
Amongst the most challenging and significant systematic effects for upcoming WL surveys is the intrinsic alignment (IA) of galaxies. 
Intrinsic alignment refers to the tendency of galaxies to have shapes or orientations that are correlated, independently of any lensing effects. These alignments can be broadly classified into two categories: (i) intrinsic-intrinsic alignments (II), where the shapes of two galaxies are directly correlated, and (ii) intrinsic-shear alignments (${\rm I}\gamma$), where the shape of a foreground galaxy is correlated with the background \citep{Troxel2015_IA}. IA mimics the cosmological shear signal, introducing a spurious contribution to the measured power. \\

\begin{figure}
\centering
\includegraphics[width=\linewidth]{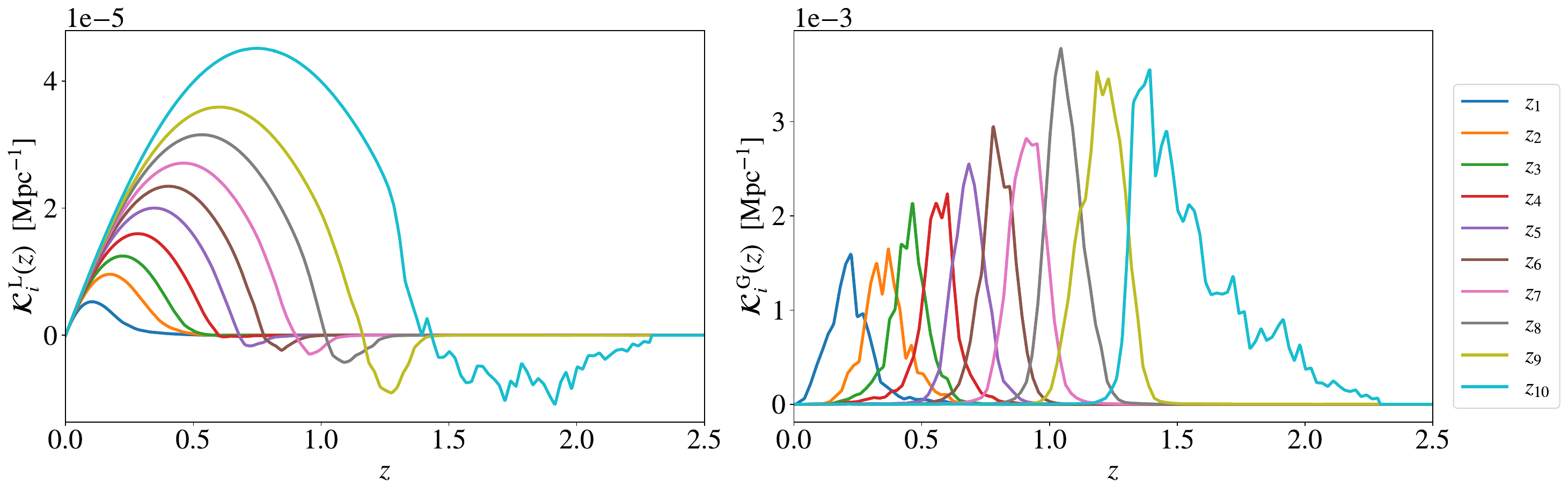}
\caption{Weight functions, or radial kernels $\mathcal{K}_i^A(z)$, for the two photometric probes. The analytic expressions for these are, respectively, Eq.~\eqref{eq:wildef} (left, WL) and Eq.~\eqref{eq:wigdef} (right, GCph). At high redshifts the IA term dominates over the shear term in the lensing kernels, making them negative. Moreover, the wiggly behaviour of the source redshift distribution $n_i(z)$ is smoothed out in the integration (Eq.~\ref{eq:wildef}), but this is only true for the shear kernel $\mathcal{K}_i^{\gamma}(z)$, not for the IA one; this means that, in the $z$ region where latter term dominates over the former, the total lensing kernel ceases to be smooth.}
\label{fig:WF_FS1_for_PhD_thesis}
\end{figure}

As seen in Sect.~\ref{sec:ellipticity_shear}, we can consider the observed ellipticity as the contribution of two terms
\begin{equation} \label{eq:epsilon_WL}
    \epsilon = \epsilon^{\rm I} + \gamma
\end{equation}
where $\epsilon^{\rm I}$ is the intrinsic, or unlensed, ellipticity. Taking the two-point correlation function of Eq.~\eqref{eq:epsilon_WL} we have:
\begin{equation}
    C_{ij}^{\epsilon\epsilon}(\ell) =C_{ij}^{\gamma\gamma}(\ell)  + C_{ij}^{{\rm I}\gamma}(\ell) +C_{ij}^{\gamma {\rm I}}(\ell)+ C_{ij}^{\rm II}(\ell) \; .
\end{equation}
The cross terms $C_{ij}^{{\rm I}\gamma}(\ell)$ and $C_{ij}^{\gamma {\rm I}}(\ell)$  represent respectively the correlation between background shear and foreground intrinsic alignment and the correlation between foreground shear and background ellipticity; $P_{\delta {\rm I}}$ and \ML{$P_{{\rm I} \delta}$} are the corresponding power spectra. Since foreground shear and background ellipticity should not be correlated
the $C_{ij}^{\gamma {\rm I}}(\ell)$ term vanishes.
For the $C_{ij}^{{\rm I}\gamma}(\ell)$ and $C_{ij}^{\rm II}(\ell)$ the \textit{linear alignment} model can be used:
\begin{align}
C_{ij}^{{\rm I}\gamma} (\ell) &=
\frac{c}{H_0}
\int \diff z \frac{\mathcal{K}_i^\gamma(z)\mathcal{K}_j^{\rm IA}(z) +\mathcal{K}_i^{\rm IA}(z)\mathcal{K}_j^\gamma(z)}
{E(z)\chi^2(z)}P_{\delta {\rm I}}(k_\ell,z)
\\
C_{ij}^{\rm II} (\ell) &=
\frac{c}{H_0}
\int \diff z \frac{\mathcal{K}_i^{\rm IA}(z)\mathcal{K}_j^{\rm IA}(z)}
{E(z)\chi^2(z)}P_{\rm II}(k_\ell,z) \; .
\end{align}
The IA weight function is equal to the GCph one, as long as the source and lens redshift distributions coincide:
\begin{equation} \label{eq:w_IA_ISTF}
    {\mathcal{K}}_i^{\rm IA}(z) = {\mathcal{K}}_{i}^{\rm G}(z) = \frac{H_0}{c} 
    \frac{n_{i}(z)}{\bar{n}}E(z).
\end{equation}
$P_{\delta {\rm I}}(k,z)$ and $P_{\rm II}(k,z)$ can be related to the matter power spectrum thanks to models such as the \textit{extended nonlinear alignment} model (eNLA);  with respect to the nonlinear alignment model (NLA, \citealt{Bridle2007_NLA}) this accounts for the luminosity dependence of the IA:
\begin{align}
P_{\delta {\rm I}} (k,z) &= -\mathcal{A}_{\rm IA}\mathcal{C}_{\rm IA}
\Omega_{{\rm m}, 0}\frac{\mathcal{F}_{\rm IA}(z)}{D(z)}P_{\delta\delta} (k,z) \\
P_{\rm I I} (k,z) &= \left[-\mathcal{A}_{\rm IA}\mathcal{C}_{\rm IA}
\Omega_{{\rm m}, 0}\frac{\mathcal{F}_{\rm IA}(z)}{D(z)}\right]^2
P_{\delta\delta} (k,z) \; .
\end{align}
In these equations, the parameters of the eNLA model appear: ${\cal{A}}_{\rm IA}$ is the overall IA amplitude, ${\cal{C}}_{\rm IA}$ a constant (whose value is fixed at $0.0134$, \citetalias{ISTF2020}), $D(z)$ is the linear growth factor, and ${\cal{F}}_{\rm IA}(z)$ a function modulating the dependence on redshift:
\begin{equation}
{\cal{F}}_{\rm IA}(z) = (1 + z)^{\, \eta_{\rm IA}} \left[\langle L \rangle(z)/L_{\star}(z)\right]^{\, \beta_{\rm IA}} \; ,
\label{eq:fiadef}
\end{equation}
where $\langle L \rangle(z)/L_{\star}(z)$ is the redshift-dependent ratio of the mean luminosity $\langle L \rangle(z)$ over the characteristic luminosity of WL sources $L_{\star}(z)$ as estimated from an average luminosity function \citep[see e.g.][and references therein]{Joachimi_2015}. The IA nuisance parameters vector is then: 
\begin{equation}
\vec{\theta}_{\rm IA} = \{{\cal{A}}_{\rm IA}, \eta_{\rm IA}, \beta_{\rm IA}\} \; ,
\end{equation}
with fiducial values set using numerical simulations or fitting of the data \citepalias{ISTF2020}
\begin{equation}
\vec{\theta}_{\rm IA}^{\,\rm fid} =\{1.72, -0.41, 2.17\} \; . 
\end{equation}
All of the IA parameters, except for ${\cal{C}}_{\rm IA}$, will be varied in the analysis.\\
We can finally define a total lensing kernel as \citep[see e.g.][]{Kitching2017, Kilbinger2017, Taylor2018}
\begin{equation}
{\mathcal{K}}_{i}^{\rm L}(z) = {\mathcal{K}}_{i}^{\gamma}(z) - 
\frac{{\cal{A}}_{\rm IA} {\cal{C}}_{\rm IA} \Omega_{{\rm m},0} {\cal{F}}_{\rm IA}(z)}
{D(z)} {\mathcal{K}}_{\rm IA}(z) \; ,
\label{eq:wildef}
\end{equation}
so that the total lensing PS takes the form of Eq.~\eqref{eq:cl_gammagamma_ISTF}:
\begin{equation}\label{eq:cl_LL_ISTF} 
    C_{ij}^{\rm LL} (\ell) \simeq \frac{c}{H_0}
    \int \diff z \frac{\mathcal{K}_i^{\rm L}(z)\mathcal{K}_j^{\rm L}(z)}
    {E(z)\chi^2(z)}P_{\delta\delta}\left[ 
    \frac{\ell + 1/2}{\chi(z)},z \right] \; .
\end{equation}
\ML{The lensing kernel constructed in this way, along with the galaxy one, is shown in Fig.~\ref{fig:WF_FS1_for_PhD_thesis}.}
\subsection{Multiplicative shear bias}\label{sec:mult_shear_bias}
An additional ingredient which we include to mimic as closely as possible the analysis of the actual data is the multiplicative shear bias, $m$, defined as the multiplicative coefficient of the linear bias expansion of the shear field $\vec{\gamma}$, see e.g. \citet{Cragg2023} for a recent study for stage-IV surveys:
\begin{equation}\label{eq:shear_bias_def}
\hat{\vec{\gamma}} = (1+\vec{m}) \, \vec{\gamma} + \vec{c}
\end{equation}
with $\hat{\vec{\gamma}}$ the measured shear field, $\vec{\gamma}$ the true one, $\vec{m}$ the multiplicative and $\vec{c}$ the additive shear bias parameters (which we will fix to 0 in the present analysis). \FB{The latter two are, in principle, complex numbers, albeit they are often assumed to be real.}  The multiplicative shear bias can come from astrophysical or instrumental systematics (such as the effect of the point spread function), which affect the measurement of galaxy shapes.
We take the $m_i$ parameters (one for each redshift bin) as constant and with a fiducial value of 0 in all bins. To include this further nuisance parameter, which is mathematically equivalent to linear galaxy bias insofar it appears as a multiplicative factor to the $C(\ell)$s, one just has to update the shear angular PS as:
\begin{equation}\label{eq:cijbias}
    \displaystyle{C_{ij}^{\rm LL}(\ell) \rightarrow (1 + m_i) (1 + m_j) C_{ij}^{\rm LL}(\ell)}
\end{equation}
where $m_i$ is the $i$-th bin multiplicative bias. We will then have:
\begin{equation}
\vec{\theta}_{\rm shear \, bias} = \{m_1, m_2, \ldots, m_{{\cal N}_{\rm b}}\} \; ,
\end{equation}
with fiducial values
\begin{equation}\label{eq:bzvalue}
\vec{\theta}_{\rm shear \, bias}^{\,\rm fid} = \left \{ 0, 0, \ldots, 0
\right \} \; .
\end{equation}
\subsection{Putting it all together: the 3$\times$2pt analysis}
\begin{figure}
    \centering
    \includegraphics[width=1.1\textwidth]{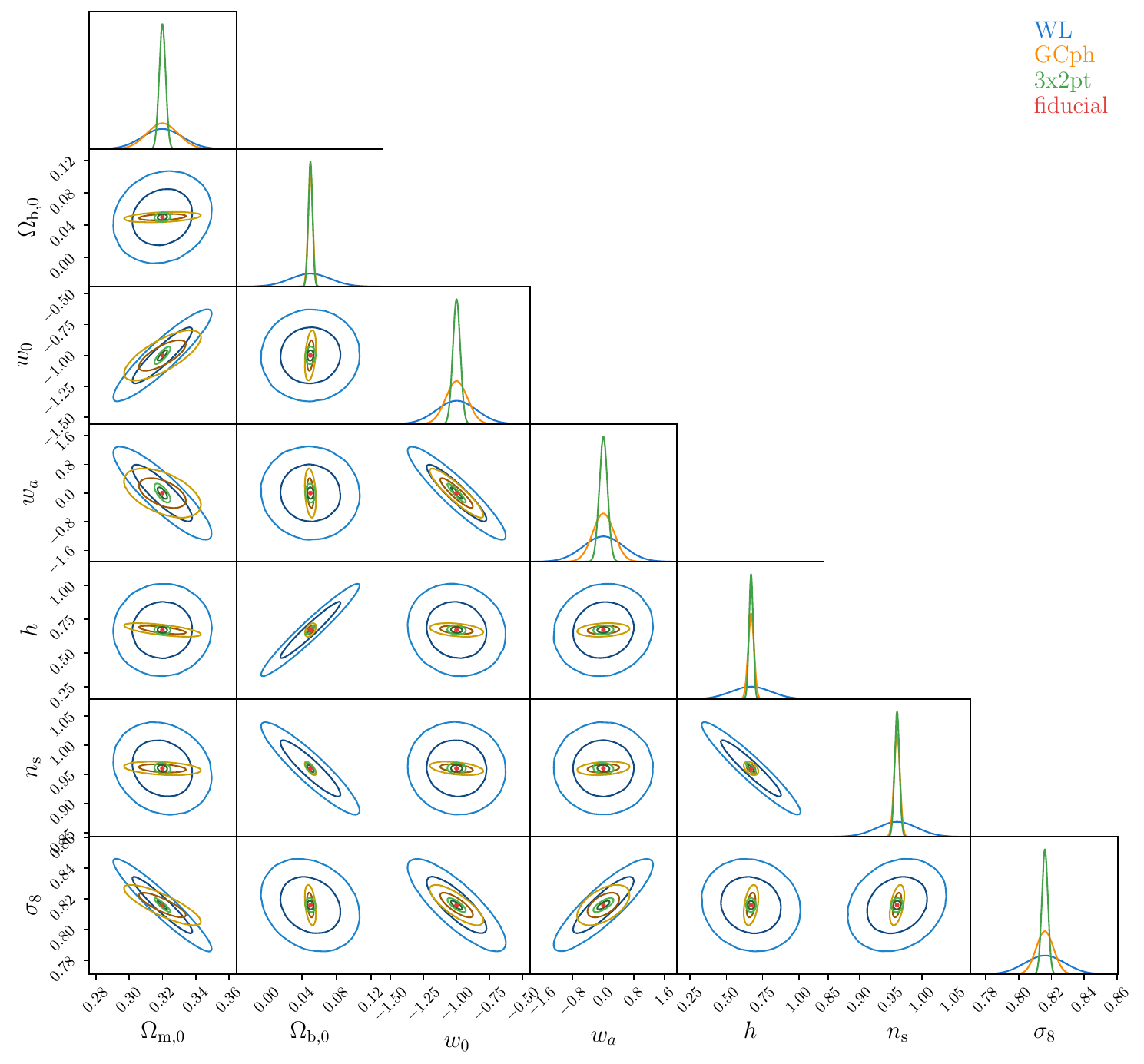}
    \caption{Fisher analysis 68\% and  95\% confidence contours for the photometric probes: WL, GCph and 3$\times$2pt. The fiducial cosmology is indicated in red. As discussed, WL best constrains $\Omega_{{\rm m}, 0}$ and $\sigma_8$ (rather, a degenerate combination of the two), but significantly improves the GCph constraints when added in the 3$\times$2pt analysis, by for example, reducing the correlation between the DE EoS parameters and the standard parameters $n_{\rm s}$, $h$, and $\Omega_{{\rm b}, 0}$, as also found in \citepalias{ISTF2020}. Plot produced with the \texttt{ChainConsumer} package \citep{Hinton2016_chainconsumer}.}
    \label{fig:triangle_ISTF_GO}
\end{figure}

%
Despite the unprecedented accuracy and control over systematics of the WL and GCph surveys, the true power of \Euclid will come from the capability of adding together and cross-correlating these two probes. This approach is nowadays the state of the art in multi-probe cosmological surveys \citep{Joachimi2021_kids1000, DES_Y3_2022} because it combines the strengths of multiple data sets. Different observables have in fact different parameter degeneracies, are affected by different systematics, and constrain different sets of parameters. Just to give an example, the combined use of cosmic shear and galaxy clustering is able to break the degeneracy between galaxy bias and cosmological parameters, which significantly degrades the constraints from GCph when marginalizing over the former set of parameters. The addition of WL and/or the cross-spectra adds information which is unaffected by galaxy bias, improving the constraints on the cosmological parameters while at the same time providing information on $b(z)$ \citep{Tutusaus2020_XC}. Some example confidence contours from a \citetalias{ISTF2020}-like Fisher analysis are shown in Fig.~\ref{fig:triangle_ISTF_GO}. The different degeneracy directions are visible, for example in the $\Omega_{{\rm m}, 0}$-$\sigma_8$ plane.\\

\begin{figure}
    \centering
    \adjustbox{center}{\includegraphics[width=1.1\textwidth]{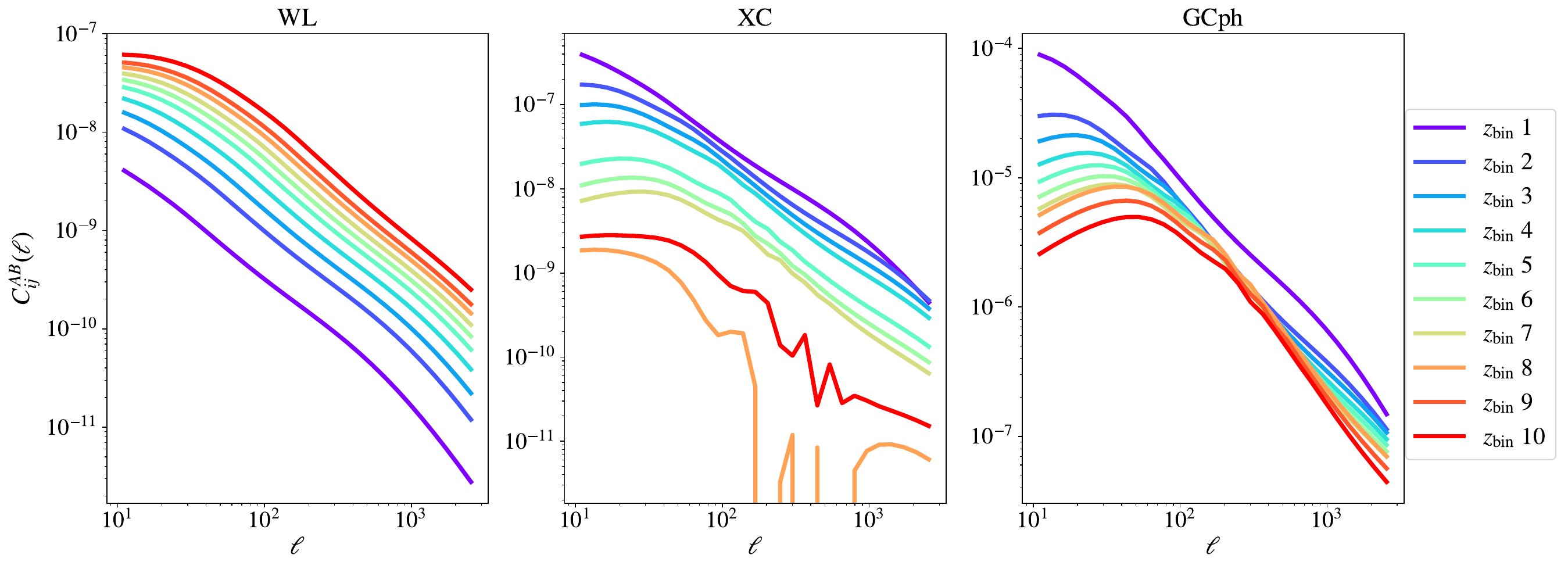}}
    \caption{Angular PS for the photometric probes and their cross-correlation, for the different $z_i, z_j$ redshift bins (with bin $i = j$). The BAO wiggles, visible in the GCph spectra, are smoothed out in the rather featureless (and low-amplitude) lensing signal.}
    \label{fig:cls}
\end{figure}

The cross-correlation of WL and GCph can be referred to as \enquote{XC} (for \enquote{cross-correlation}), \enquote{GGL} (for \enquote{galaxy-galaxy lensing}) or as \enquote{GL}, which will be our preferred choice, following the convention defined in Sect.~\ref{sec:tomography}: G for GCph, L for WL. We consider this notation more convenient, because for two-point statistics it allows writing, e.g., $C^{\rm LL}_{ij}(\ell)$ instead of the more cumbersome $C^{\rm WLWL}_{ij}(\ell)$; this is even more true for four-point functions, such as the covariance matrix, which correlates four tracers.\\
The 3$\times$2pt data vector is then constructed by concatenating the LL, GL and GG angular PS, derived by plugging the appropriate radial kernel (Eq.~\ref{eq:wigdef} for galaxy clustering, Eq.~\ref{eq:wildef} for weak lensing) in Eq.~\eqref{eq:cl_AB_tomo_limb_ISTF}. These angular PS are shown in Fig.~\ref{fig:cls}. As anticipated, the corresponding power spectra are given by the following expressions:
\begin{equation}\label{eq:pk_mm_gm_gg}
P_{AB}(k, z) = \left \{
\begin{array}{ll}
\displaystyle{P_{\rm mm}(k, z)} & \displaystyle{A = B = {\rm L}} \\
 & \\
\displaystyle{b(z) P_{\rm mm}(k, z)} & \displaystyle{(A, B) = {\rm (L, G) \;  or \; (G, L)}} \\
 & \\
\displaystyle{b^2(z) P_{\rm mm}(k, z)} & \displaystyle{A = B = {\rm G},} \\
\end{array}
\right.
\end{equation}
with $P_{\rm mm}(k, z)$ the \textit{nonlinear} matter-matter PS and $b(z)$ the linear, scale-independent and deterministic first-order galaxy bias, described in Sect.~\ref{sec:gal_bias}.\\

We note that the 3$\times$2pt analysis can be further enhanced by combining it with the spectroscopic galaxy clustering (GCsp) probe. As mentioned in Sect.~\ref{sec:fisher_theory}, the log-likelihoods for independent datasets can be simply added up. This is what is often done for surveys which cover different parts of the sky, come from different experiments or study uncorrelated probes. The combination of the photometric and spectroscopic probes is a notable example; recent works \citep{Taylor2022_ph_sp_XC} suggest that these can be treated as independent even in case the observed region overlaps, because the probes correlate significantly only on large scales, where cosmic variance dominates. This greatly simplifies the analysis, since modelling the covariance between 3D and 2D observables is not a trivial task.
\\

\ML{The preceding sections illustrated galaxy clustering and weak lensing cosmic shear as main probes for \Euclid. Having established the theoretical framework for these observables, we now transition to detailing the mission in charge of their measurement.}
\section{\Euclid}
\Euclid is a medium-class ESA mission, born from the union of the DUNE and SPACE proposals in the ESA Cosmic Vision 2015-2025 program. The satellite was launched on the 1st of July, 2023 on a SpaceX Falcon 9 rocket from Cape Canaveral, Florida, USA; after about one month of travel, it reached the second Sun-Earth Lagrangian point (L2), at a distance of approximately $1.5 \times 10^{6}$ km. From there, during the course of the designed six years of nominal survey operations, it will produce the most advanced catalogue of galaxies' shapes and morphology to date, advancing our understanding of Cosmology and producing a wealth of data for legacy science studies. This section will introduce the main technical aspects of the survey.
\subsection{Satellite}\label{sec:euclid_instruments}
\begin{figure}
    \centering
    \begin{subfigure}{0.8\textwidth}
        \centering
        \includegraphics[width=\linewidth]{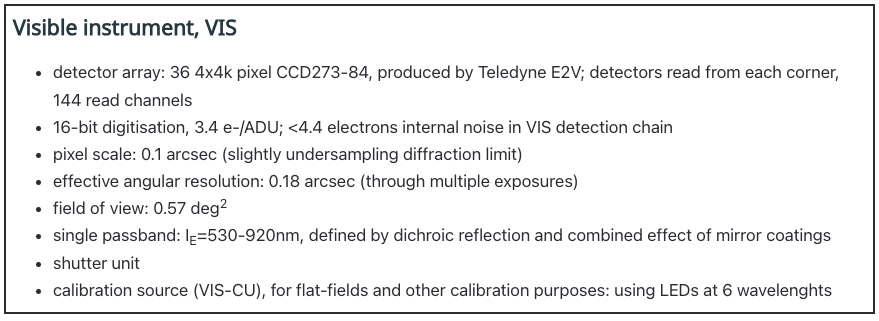}
    \end{subfigure}
    \begin{subfigure}{0.8\textwidth}
        \centering
        \includegraphics[width=\linewidth]{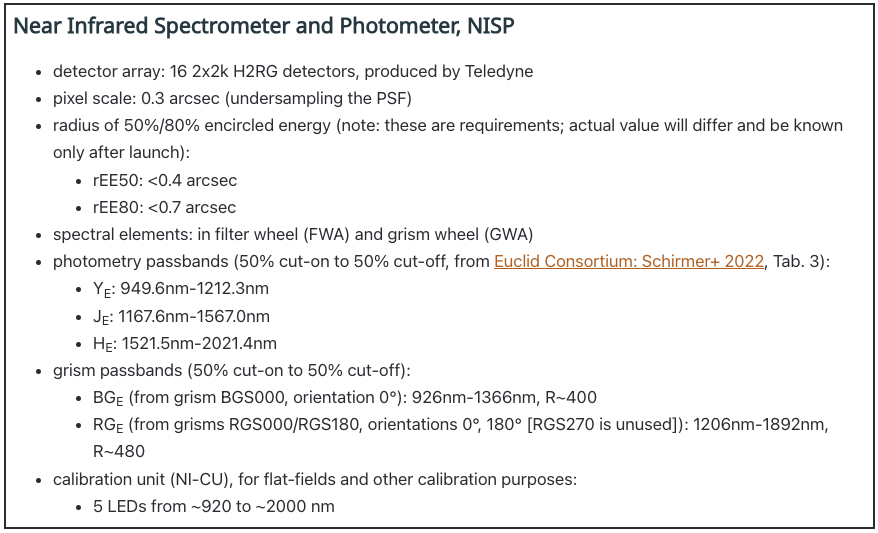}
    \end{subfigure}
    \caption{VIS and NISP specifics, from the \href{https://shorturl.at/xDISX}{Euclid Consortium website}}
    \label{fig:vis_nisp_sheets}
\end{figure}

\Euclid is placed in a Lissajous-type orbit around the L2 point, where the gravitational pull of the Sun and the Earth balances the centrifugal force experienced by the satellite. The halo orbit is quite large ($10^6$ km), and the satellite does not experience Earth eclipses, which would affect the thermal stability of the system.\\

The satellite is equipped with an axial Korsh telescope measuring 1.2 m in diameter, composed of three curved silicon carbide mirrors operating below 130 K. The light is directed through the exit pupil and split with the use of a dichroic plate; the transmitted fraction is sent to the Near Infrared Spectrometer and Photometer (NISP), and the reflected fraction to the visible (VIS) instrument. \ML{The telescope and instruments are part of the Payload Module (PLM)}; the Service Module (SVM), on the other hand, houses the spacecraft's essential subsystems, including power distribution, propulsion, and communication systems. The whole satellite has a size of $4,5 \times 3 \times 3$ m and a mass of 2200 kg. VIS and NISP are its two main scientific instruments; their characteristics are summarized in Fig.~\ref{fig:vis_nisp_sheets}.\\

The main aim of the VIS instrument is to provide high-resolution images to characterize the morphology of the sources to the accuracy needed for the study of cosmic shear. The focal plane is composed of $6\times6$ $4096\times4132$ CCDs with 12-micron pixels (609 million pixels in total) and a spectral range from 550 to 950 nm \citep{Cropper2016_VIS}, covering a wide instantaneous field of view of 0.57 deg$^2$, which is almost three times the solid angle of the full Moon. This high-quality panoramic imager reaches a resolution of 0.18 arcsec, achieving a signal-to-noise ratio of 10 or more for 1.5 billion galaxies down to magnitude $\IE = 24.5$ in 4200 seconds exposures\footnote{\url{https://www.euclid-ec.org}}. Data from the Gaia mission \citep{Gaia2016} will be used to remove bright stars contaminating the field of view. \\

The main aim of the NISP instrument is to provide galaxies' redshift, either through photometry or spectroscopy. The focal plane is composed of $4\times4$, $2040\times2040$ 18 micron pixels (64 million pixels in total), covering the  950 to 2020 nm range \citep{Schirmer2022_NISP}, and can operate as a photometer (NISP-P) and a spectrometer (NISP-S) -- although not simultaneously. It shares the same 0.57 deg$^2$ FoV of the VIS instrument. The incoming light is sent through two wheels, one containing the three photometry filters (Y, J, and H; see Fig.~\ref{fig:NISP_response}), and the other four spectrometry grism, which splits light from every star and galaxy by wavelength directly on the image (a technique called \enquote{slitless spectroscopy}). This allows measuring all the spectra in each exposure, at the cost of a more laborious analysis to disentangle them, for which the different orientations of the grisms are essential.\\

\begin{figure}
    \centering
    \begin{subfigure}{0.30\textwidth}
        \includegraphics[width=\textwidth]{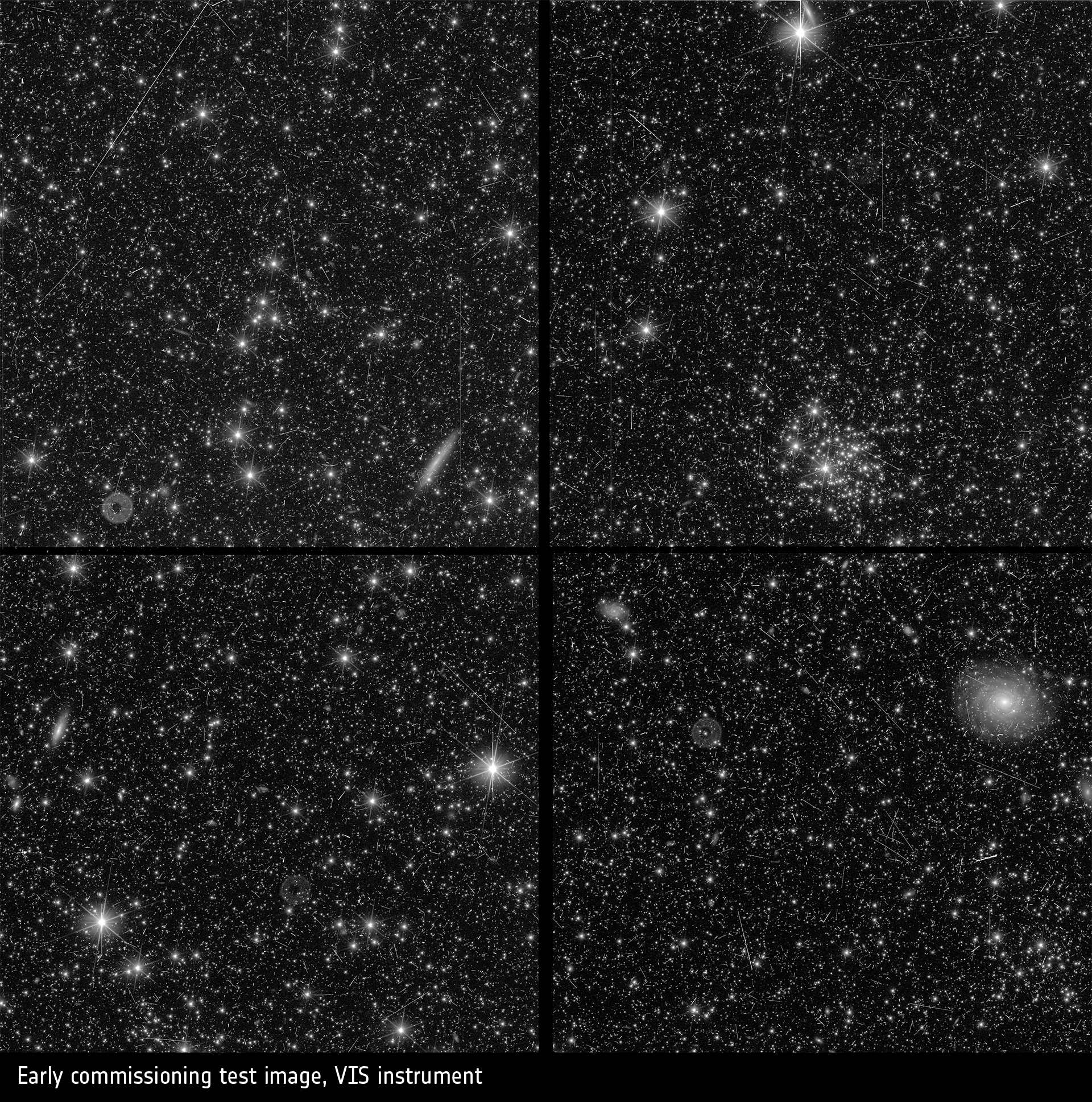}
    \end{subfigure}\hspace{0.025\textwidth}%
    \begin{subfigure}{0.30\textwidth}
        \includegraphics[width=\textwidth]{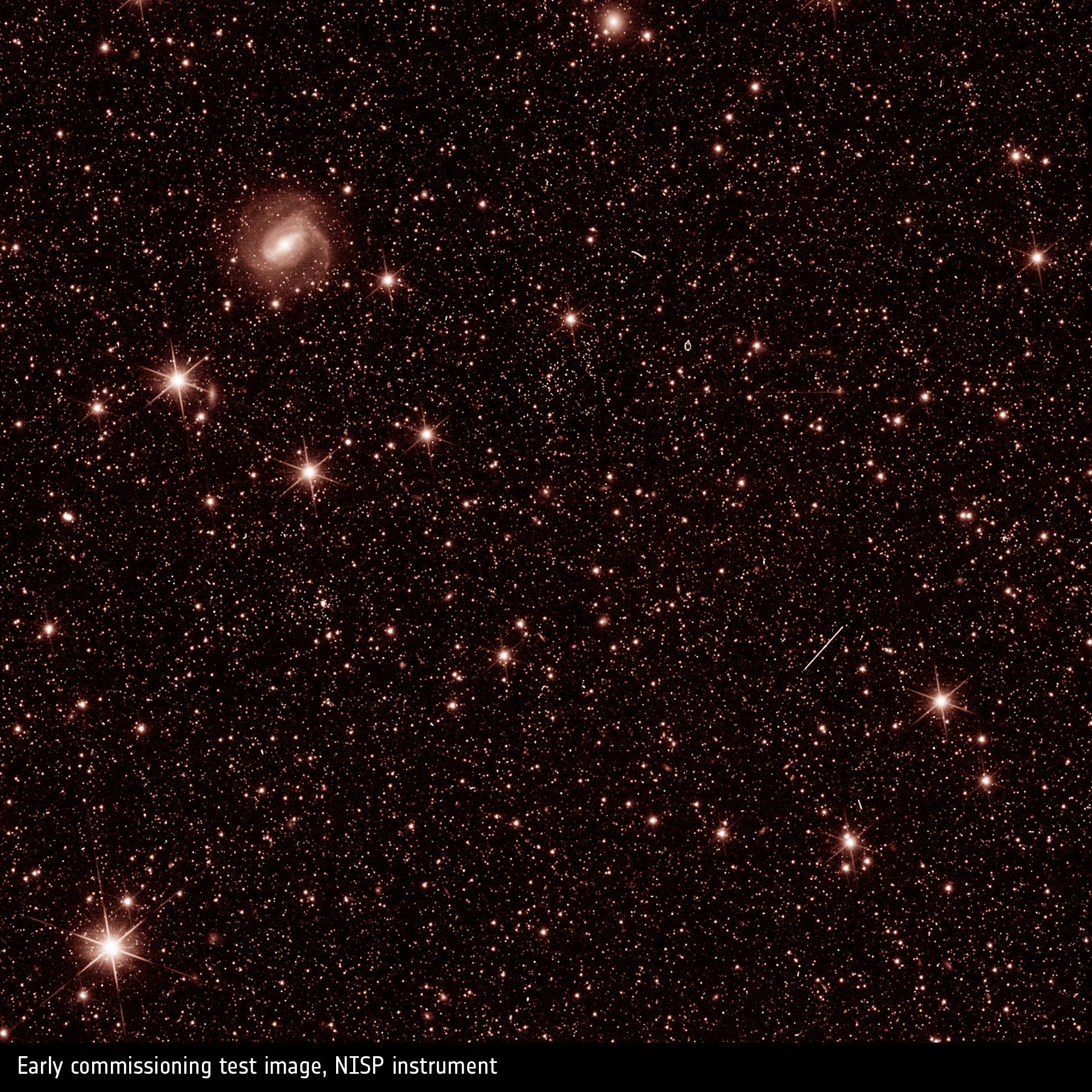}
    \end{subfigure}\hspace{0.025\textwidth}%
    \begin{subfigure}{0.30\textwidth}
        \includegraphics[width=\textwidth]{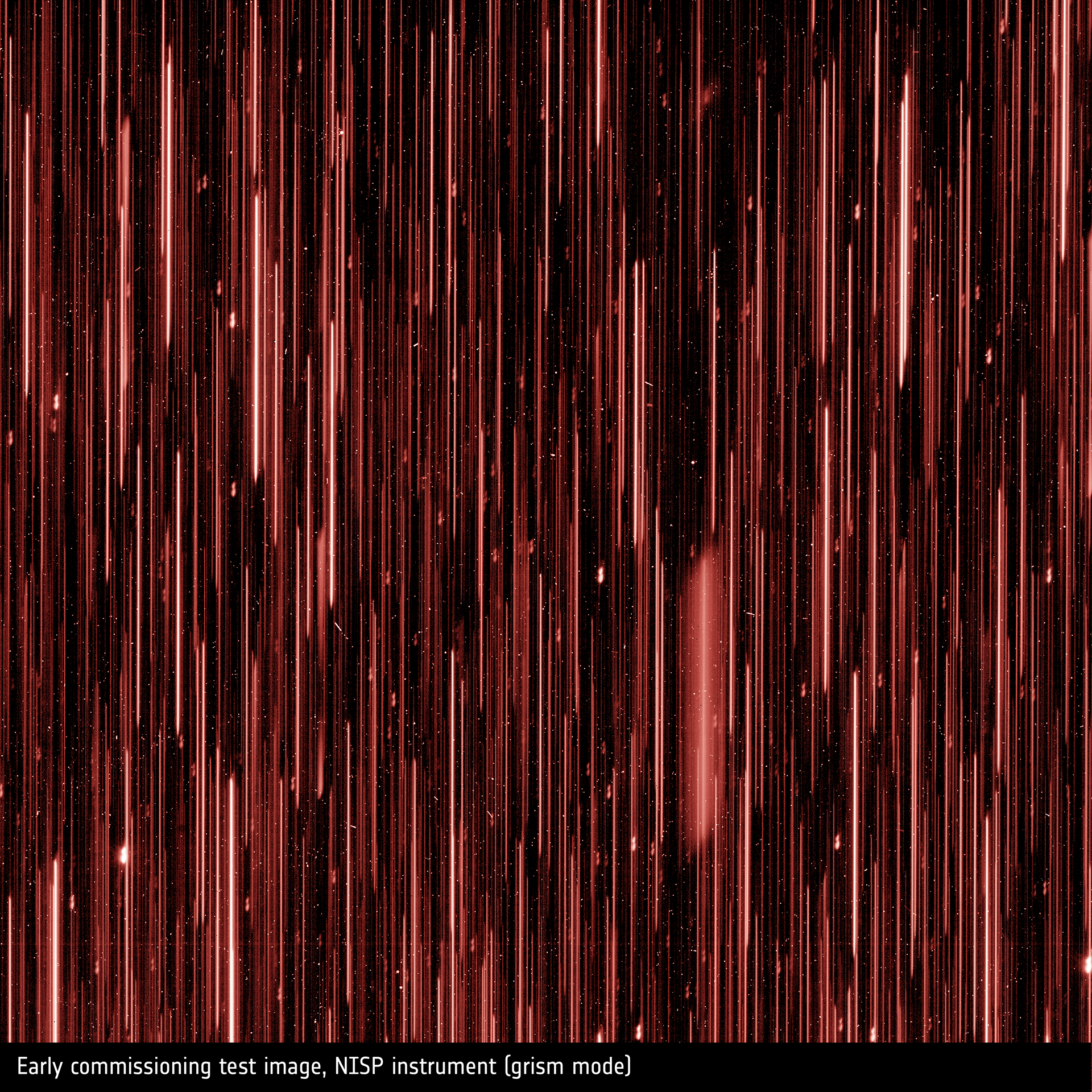}
    \end{subfigure}
    \caption{Early commissioning test images, the first publicly released by the Euclid Consortium. Left: 566 s exposure from the VIS instrument, showing 4 out of the 36 available CCDs. Some cosmic rays and optical artefacts are present in the largely unprocessed image. Center: NISP-P instrument, 100 s exposure, Y band, zoomed in from one of the 16 detectors. Right: NISP-S instrument. The light from stars and galaxies is spread by the grism into its different wavelengths, allowing measurement of the spectra of all the different sources in the image.
    Source: \href{https://shorturl.at/xDISX}{Euclid Consortium website}.}\label{fig:early_commissioning_imgs}
\end{figure}

Some early commissioning test images, amongst the first ever captured by the satellite, are shown in Fig.~\ref{fig:early_commissioning_imgs} 
\subsection{Survey strategy and data releases}
The observational strategy will consist of both wide and deep surveys. The wide survey \citep{Scaramella2022_survey} will explore the darkest sky (free from the contaminating light of the sun and the galaxy) over an area of $14~700$ deg$^2$, which corresponds to more than one-third of the whole celestial sphere. This is the main part of the survey operation, to which \Euclid will devote the most time (about $\sim 90\%$ of the total observational time). It will cover patches of 400 deg$^2$ per month, and the orientation of the telescope will be changed every six months to explore the other hemisphere.\\

The wide survey will be complemented by three deep field surveys covering 40 deg$^2$, reaching 2 magnitudes deeper than the wide survey in order to calibrate the instruments and provide data for the secondary scientific objectives (the so-called \enquote{legacy science}, such as the detection of faint high-redshift galaxies, AGNs and supernovae). Furthermore, the deep surveys will be used to assess the purity and completeness of the wide surveys. Purity is a measure of how well a sample isolates a particular type of object or phenomenon of interest: in this case, a high-purity sample would predominantly contain galaxies that genuinely belong to the category under investigation (H$\alpha$ emitters), with minimal contamination from interlopers or artefacts. Completeness, conversely, quantifies the fraction of the total objects of interest that are successfully identified and included in the sample, i.e., it measures how exhaustive the survey is in capturing the galaxies that meet the selection criteria.\\

\begin{figure}
    \centering
    \includegraphics[width=0.8\textwidth]{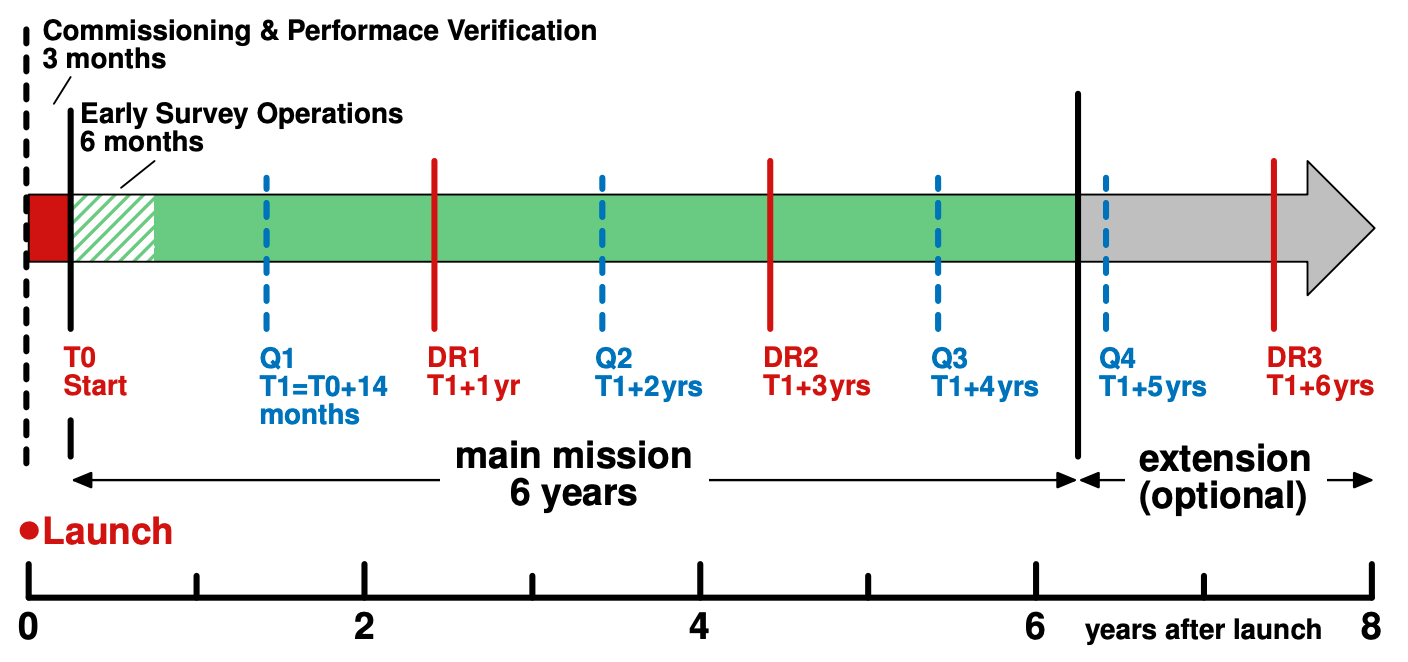}
    \caption{The expected survey timeline for the \Euclid mission. Source: \citep{Scaramella2022_survey}.}
    \label{fig:euclid_timeline}
\end{figure}

The timeline for the survey operations is sketched in Fig.~\ref{fig:euclid_timeline}. The main Data Releases (DR) will cover respectively 2500, 7500 and 14700 deg$^2$. In addition, four Quick Data Releases (Q) are foreseen, each of $\sim 50$ deg$^2$ \citep{Scaramella2022_survey}. The Euclid Consortium will publicly release the data after a proprietary period of about one year \citep{laureijs2011euclid}.

\subsection{Data processing}
Over the mission's six-year span, \Euclid will collect over 500,000 visible and near-infrared images, which will be relayed to Earth on a daily cadence.\\
The analysis and processing of this massive volume of data, expected to reach tens of petabytes, are under the responsibility of the Science Ground Segment (SGS). The SGS is divided into two core sections: the ESA Science Operations Centre (SOC), and the Euclid Consortium Science Ground Segment (EC SGS). Among the SOC's responsibilities are survey planning, managing the downlinked data and daily quality reports, whereas the EC SGS is in charge of the reduction of the data from (real or simulated) images to observables (such as the $C(\ell)$ described in Sect.~\ref{sec:angular_ps_theory}), as well as the integration of photometric data from other ground-based surveys, which will provide data for the missing g, r, i and z bands.
\chapter{Super-sample covariance}\label{chap:SSC}
As mentioned in the previous chapters, we are entering the era of precision cosmology, characterized by a large amount of high-quality data. This wealth of information will allow us to constrain the cosmological parameters with unprecedented accuracy and precision, thus making it possible to select between competing cosmological models and advance in this way our understanding of the dark and visible Universe. However, these new and exciting possibilities come with their own unique set of challenges, mainly the need to tame the systematic effect that could bias the parameters' estimates at this level of precision. Indeed, upcoming Stage IV surveys such as the Vera C. Rubin Observatory Legacy Survey of Space and Time \citep[LSST,][]{ivezic2018lsst}, the Nancy Grace Roman Space Telescope \citep{spergel2015widefield}, and the \Euclid mission \citep{laureijs2011euclid} will have stringent requirements on the accuracy of the modelling of both the theory and the covariance of the observables under study. The covariance matrix is in fact one of the crucial ingredients in a likelihood analysis, as the measurement of a physical quantity is meaningless without an estimate of the associated uncertainty. The following two chapters focus on the description of super-sample covariance, an important contribution to sample variance for upcoming high-resolution LSS surveys, and the computation of its impact on the \Euclid photometric survey, which is the aim of one of the Key Project papers \citep{Sciotti2023} of the Photometric Galaxy Clustering Work Package within the GC SWG of the Euclid Collaboration.
\section{Introduction}\label{sec:ssc_introduction}
As illustrated in Chap.~\ref{chap:statistics}, the covariance matrix is a measure of the joint variability of two quantities, allowing to quantify their correlation or anti-correlation; its diagonal elements describe the covariance of the variable with itself, which is the variance, i.e., the square of its uncertainty. The likelihood analysis allows translating data into (cosmological) parameters, and the covariance of the data into the covariance of the parameters. This can be accomplished through a FM or MCMC approach.\\

The covariance matrix of the two-point harmonic space observables gets contributions from the measurement noise (e.g., the shot and shape noise), and the sample variance due to the incomplete sampling of Fourier modes caused by the finiteness of the survey volume (in the radial and angular direction). In harmonic space, it is composed of three terms: the Gaussian (G), super-sample (SSC) and connected non-Gaussian (cNG) covariance. The first represents the covariance of the observables if the statistical distribution of the corresponding underlying field (e.g., the galaxy number density field $\delta_{\rm g}(\vec{x}, a)$ in the case of GCph) was perfectly Gaussian. The second and third arise because of the non-Gaussian coupling of the different Fourier modes of the fields.\\

As seen in Sect.~\ref{sec:lin_struct_formation}, the time evolution of the density contrast field $\delta(\vec{x}, a)$ at linear order, encapsulated in Eq.~\eqref{eq:delta_m_linear}, is scale-independent, which means that all Fourier modes evolve at the same rate. Linear theory holds for small density perturbations around the mean, i.e., $\rho(\vec{x}) \sim \bar{\rho} \rightarrow \delta(\vec{x}) \sim 0$. However, contrary to the temperature perturbations in the CMB, which show a level of anisotropy of order $10^{-4}$, the density contrast field $\delta$ can take much more extreme values: from -1 for the emptiest regions, to $\sim 200$ for the dark matter halos. Moreover, nonlinear clustering becomes increasingly more important at smaller scales, which contain most of the cosmological information (especially for WL). This makes the linear treatment inadequate for the investigation of the small scales of LSS. A more appropriate description involves the inclusion (at least) of the second-order terms when solving the differential equations describing the time evolution of the density and velocity fields as a function of time (Eqs.~\ref{eq:ddelta_deta} and~\ref{eq:dtheta_deta}), which involve mode-coupling terms.\\

This coupling of modes is captured in the covariance matrix by the cNG and SSC terms. The first of these, also referred to as the \enquote{non-Gaussian covariance in the absence of survey window effects} \cite{Krause2017}, describes mode coupling within the survey volume. Recent works have shown this term to be subdominant with respect to the SSC for upcoming Stage IV surveys (the parameter constraints obtained with a G + SSC + cNG covariance are $\lesssim 5\%$ larger than the ones obtained with a G + SSC covariance for a \Euclid-like cosmic shear survey, see \citealt{Barreira2018response_approach}), and will not be treated in the present work.\\

\begin{figure}[h!]
    \centering
    \includegraphics[width=\hsize]{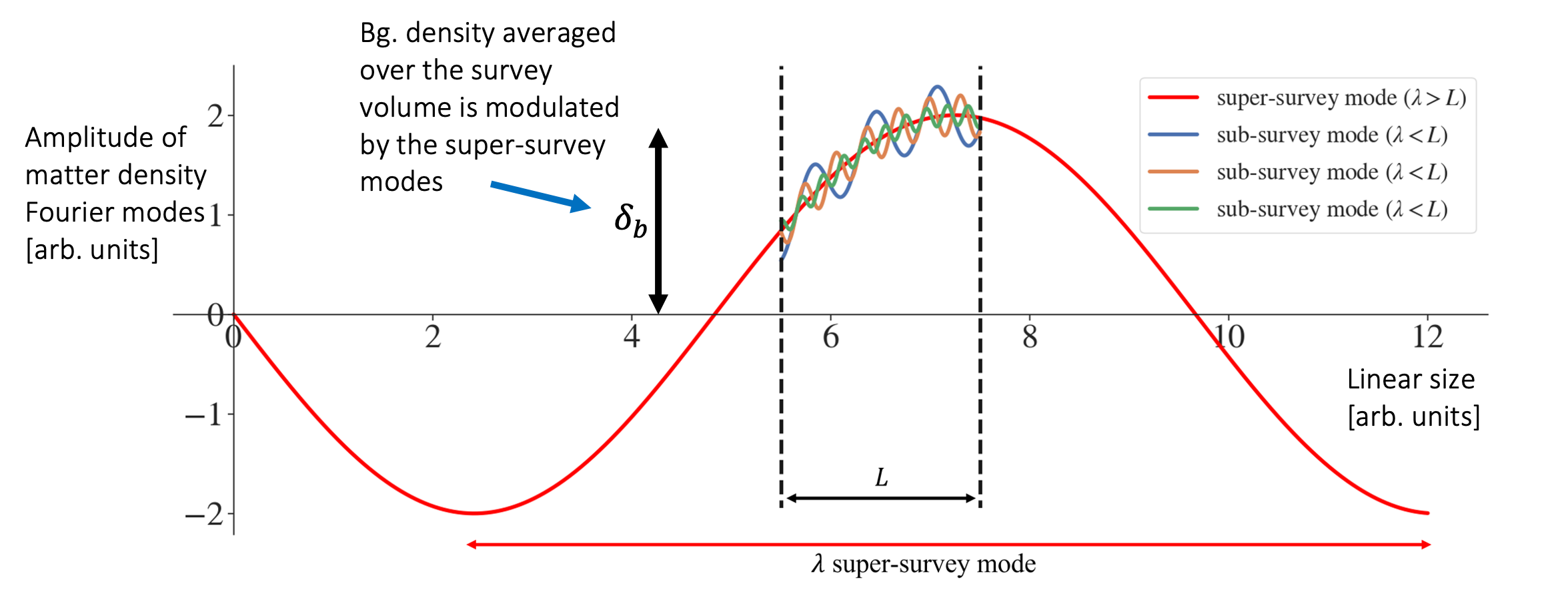}
    \caption{A sketch of the mode modulation induced by a long-wavelength background density perturbation $\delta_\mathrm{b}$, or super-survey mode, in red. The blue, orange and green curves represent the sub-survey (or hard) modes, which have $\lambda < L$ and can thus be sampled. The net effect is a shift of the observed modes, which biases the measurement of the density perturbations.}
    \label{fig:ssc_sketch}
\end{figure}

The second term, instead, describes the effects of the coupling of modes respectively larger and smaller than the survey typical linear size $L = V_W^{1/3}$ (or the smallest survey dimension, $V_W$ being the survey volume). SSC has been first introduced for cluster counts in \citet{Hu2003} -- sometimes being referred to as `beat coupling', see \citet{Rimes2006, Hamilton2006} -- and has gardened a lot of attention in recent years \citep{Takada2013, Li2014, Barreira2018response_approach, Digman2019, Bayer2022, Yao2023}; see also \citet{Linke2023} for an insightful discussion on SSC in real space. \\
The above-mentioned length $L$ sets a natural scale for the maximum wavelength $\lambda$ (minimum wavenumber $k$) which can be sampled by any real survey. In fact, Fourier modes with $k < 1/L$ simply cannot be accommodated by the survey volume; moreover, the nonlinear mode coupling described above intertwines the evolution of these \enquote{super-survey}, or \enquote{soft} modes with the evolution of \enquote{sub-survey}, or \enquote{hard} modes. The net result of these two factors is a modulation of the observables within the survey volume by an unobservable background perturbation, called $\delta_\mathrm{b}$, which biases our measurement. This is the fundamental quantity characterizing the SSC effect, and is defined by 
\begin{equation}
    \delta_{\mathrm{b}} \equiv \frac{{\bar \rho}_{\mathrm{m} W}}{{\bar \rho}_{\mathrm{m}}} - 1 \; ,
    \label{eq:delta_b_def}
\end{equation}
where ${\bar \rho}_{\mathrm{m} W}$ is the matter density averaged \textit{within the survey volume} and ${\bar \rho}_{\mathrm{m}}$ is the matter density averaged over the whole Universe \citep{Lima2018}. In other words, $\delta_\mathrm{b}$ is the spatial average (denoted by the brackets) of the density contrast $\delta_\mathrm{m}(\vec{x}, z)$, or mean density fluctuation, over the survey volume (window):
\begin{align}
 &\langle\delta_\mathrm{m}(\vec{x}, z)\rangle_{\text{universe}} = 0 \; , \\
 &\langle\delta_\mathrm{m}(\vec{x}, z)\rangle_{W} = \delta_\mathrm{b}(z)\; .
\end{align}
The modulation induced by the super-survey modes is therefore equivalent to a change in the background density of the observed region, which affects and correlates all LSS probes -- see Fig.~\ref{fig:ssc_sketch}.
This is accounted for as an additional term in the data covariance matrix, which becomes non-diagonal in $(\ell_1, \ell_2)$ -- or in $(k_1, k_2)$ -- again because the different modes do not evolve independently. Being the most affected by nonlinear dynamics, the smaller scales are heavily impacted by SSC, where the effect is expected to be the dominant source of statistical uncertainty for the two-point statistics of WL: it has in fact been found to increase unmarginalised uncertainties on cosmological parameters up to a factor of about 2
\citep[for a \Euclid-like survey, see][]{Barreira2018cosmic_shear, beauchamps2021}. In the case of photometric galaxy clustering (again, for a \Euclid-like survey), \citet{Lacasa_2019} -- hereafter \citetalias{Lacasa_2019} -- found the cumulative signal-to-noise to be decreased by a factor of  $\sim 5.7$ at $\ell_\mathrm{max} = 2000$.
These works, however, either do not take into account marginalised uncertainties or the variability of the probe responses, do not include cross-correlations between probes, or do not follow the full specifics of the \Euclid photometric survey described below.\\

In the following two Chapters, we detail the theory, approximations, survey settings, assumptions and numerical pipelines underlying the present treatment of SSC for the WL, GCph and 3$\times$2pt probes in \Euclid. We first validate the forecast constraints on the cosmological parameters both including and neglecting the SSC term; these are produced using two independent codes, whose only shared feature is their use of the public \texttt{Python} module \texttt{PySSC}\footnote{\texttt{\url{https://github.com/fabienlacasa/PySSC}}}\footnote{\texttt{\url{https://pyssc.readthedocs.io/en/latest/index.html}}} (\citetalias{Lacasa_2019}) to compute the fundamental elements needed to build the SSC matrix. Secondly, we investigate the impact of SSC on the marginalised uncertainties and on the dark energy Figure of Merit (FoM), both obtained through a Fisher forecast of the constraining power of \Euclid's photometric observables.\\

These Chapters are organized as follows: Sect.~\ref{sec:ssc_theory} presents a general overview of the theoretical framework underpinning SSC, with the application to the projected observables relevant for this work discussed in Sect.~\ref{sec:ssc_projected} and the approximations used in presented in Sect.~\ref{sec:ssc_approx}. In  Sect.~\ref{sec:specifics} we outline the theoretical model and specifics used to produce the forecasts, while Sect.~\ref{sec:validation} provides technical details on the codes' implementation and validation. Then, we study in Sect.~\ref{sec:impact} the impact of SSC on \Euclid constraints, for different binning schemes and choices of systematic errors and priors. Finally, we present our conclusions in Sect.~\ref{sec:conclu}.
\section{Theory and approximations}\label{sec:ssc_theory}
The starting point to obtain an analytical model for SSC is, due to the nature of the effect, a formalism describing the observables in the presence of a finite window function, which as we shall see is sufficient in itself to introduce mode couplings. Furthermore, the reaction, or response, of the observables to a shift in background density must be quantified, using semi-analytical models or simulations. The present section will introduce these. 
\subsection{Power spectrum estimator and covariance in the presence of a survey window}\label{sec:pk_cov_window}

We begin by introducing the different terms in the covariance matrix of the matter power spectrum in the presence of a finite window function $W(\vec{x})$ describing the survey volume (e.g., with a value of 1 for $\vec{x}$ in the observed region and 0 outside), mainly following the discussion presented in \cite{Scoccimarro1999, Takada2013, Li2014}. In the following, we assume the mean density of the field to be known, which is appropriate for WL (in this case, we probe the matter density directly, whose background value is fixed by the cosmological parameters); we will then generalize the result to the case of a density contrast field measured w.r.t. the mean of the survey region, as in the case of GCph.\\

The observed field in the presence of a window function is, in real space, $\delta_W(\vec{x}) = \delta(\vec{x})W(\vec{x})$. This product becomes a convolution in Fourier space:
\begin{equation}
    \tilde{\delta}_W(\vec{k})=\int \frac{\diff ^3 \vec{q}}{(2 \pi)^3} \tilde{W}(\vec{q}) \tilde{\delta}(\vec{k}-\vec{q}) \; ,
    \label{eq:delta_W_tide}
\end{equation}
where the tilde denotes Fourier-transformed quantities. We define a binned estimator (or \enquote{band-power} estimator) $\hat{P}\left(k_i\right)$ of the true underlying power spectrum $P(k)$, which as seen in Sect.~\ref{sec:2_point_stat} is defined by the relation
\begin{equation}\label{eq:PS_def}
    \left\langle\tilde{\delta}(\vec{k}) \tilde{\delta}\left(\vec{k}'\right)\right\rangle=(2 \pi)^3 \delta_{\rm D}^3\left(\vec{k}+\vec{k}'\right) P(k) \; ,
\end{equation}
as
\begin{equation}
    \hat{P}\left(k_i\right) \equiv \frac{1}{V_W} \int_{|\vec{k}| \in k_i} \frac{\diff ^3 \vec{k}}{V_{k_i}} \tilde{\delta}_W(\vec{k}) \tilde{\delta}_W(-\vec{k})
\label{eq:Pk_W_estimator}
\end{equation}
where $V_W = \int \diff ^ 3 \vec{x} W(\vec{x})$ is the survey volume and the integral is computed over a thin spherical shell of width $\Delta k$, with volume $V_{k_i} \approx 4 \pi k_i^2 \Delta k$ for $ \Delta k \ll k_i$. Its ensemble average is 
\begin{equation}\label{eq:PS_estimator_ensemble}
    \left\langle\hat{P}\left(k_i\right)\right\rangle=\frac{1}{V_W} \int_{|\vec{k}| \in k_i} \frac{\diff ^3 \vec{k}}{V_{k_i}} \int \frac{\diff ^3 \vec{q}}{(2 \pi)^3}|\tilde{W}(\vec{q})|^2 P(\vec{k}-\vec{q}) \; ,
\end{equation}
having used Eqs.~\eqref{eq:delta_W_tide},~\eqref{eq:PS_def}, and~\eqref{eq:Pk_W_estimator}. The observed power spectrum is then a convolution of the underlying power spectrum with the Fourier transform of the window function. Such convolution has the effect of combining (or \enquote{smearing}) the density modes that are separated by less than the Fourier width of the window, which in this case is $1/L$. The Fourier transform of the window function acts in fact as a filter in Fourier space, with its width determining the range of modes that are combined. This is a general property of convolutions: they combine the information from the input function (in this case, the true power spectrum) over the range determined by the other function (in this case, the Fourier transform of the window function).\\
The estimator $\hat{P}(k_i)$ is biased low for modes outside the window function support, i.e., for $k < 1/L$, whereas the bias becomes progressively smaller for shorter-wavelength modes within the survey volume, $k \gg 1/L$. In this limit, we have in fact an unbiased estimator:
\begin{align}
    \left\langle\hat{P}\left(k_i\right)\right\rangle & \simeq \frac{1}{V_W} \int_{|\vec{k}| \in k_i} \frac{\diff ^3 \vec{k}}{V_{k_i}} P(k) \int \frac{\diff ^3 \vec{q}}{(2 \pi)^3}\left|\tilde{W}\left(\vec{q}\right)\right|^2 \\ \nonumber
    & \simeq P\left(k_i\right) \frac{1}{V_W} \int \frac{\diff ^3 \vec{q}}{(2 \pi)^3}\left|\tilde{W}\left(\vec{q}\right)\right|^2=P\left(k_i\right) .
\end{align}
Having used $P(|\vec{k}-\vec{q}|) \simeq P(k)$ over the integration range of $\diff ^3 \vec{q}$ which the window function supports (first line); indeed, for $q \gg 1/L$ the above integral is suppressed by the window function $|{\tilde W}(\vec{q})|$, which means that the result
is only non-negligible if $k \gg q$ \citep{Barreira2018response_approach}.  
We have also assumed that $P(k)$ varies slowly within the $k$ bin (second line); the last equality relies on the general identity of the window function:
\begin{equation}
    V_W=\int \diff ^3 \vec{x} W^n(\vec{x})=\int\left[ \, \prod_{a=1}^n \frac{\diff ^3 \vec{q}_a}{(2 \pi)^3} \tilde{W}\left(\vec{q}_a\right)\right](2 \pi)^3 \delta_{\rm D}^3\left(\vec{q}_{1 \ldots n}\right),
    \label{eq:V_W_identity}
\end{equation}
having adopted the notation $\vec{q}_{1 \ldots n}=\vec{q}_1+\ldots +\vec{q}_n$.\\

Having constructed our unbiased power spectrum estimator, we can now proceed to study its statistical properties, namely its covariance matrix. In the following discussion, we will neglect measurement noise contributions such as shape and shot noise (that is, we will consider only sample variance). The covariance matrix is then defined as:
\begin{align}
    \mathrm{Cov}_{i j} \equiv & \; \mathrm{Cov}\left[P(k_i), P(k_j)\right]=\left\langle\hat{P}(k_i) \hat{P}(k_j)\right\rangle-\left\langle\hat{P}(k_i)\right\rangle\left\langle\hat{P}(k_j)\right\rangle \\ 
    \simeq & \; \frac{1}{V_W}\left[\frac{(2 \pi)^3}{V_{k_i}} 2 P(k_i)^2 \delta_{ij}^{\rm K}+\bar{T}^W(k_i, k_j)\right]
    \label{eq:cov_terms}
\end{align}
where the third equality holds in the same $k \gg 1 / L$ limit. The Kronecker delta function enforces the first term of the covariance, the Gaussian (or \enquote{disconnected}) term, to be diagonal in $k_i, k_j$ (but not necessarily in other quantities, such as the different redshift bins for a tomographic analysis). This is because, in the Gaussian limit, each Fourier mode is an independent Gaussian random variable. Hence, the Gaussian covariance is simply proportional to the number of independent $k$ modes (Gaussian random variables) in the shell $i$:
\begin{equation}
    \mathrm{Cov}_{ij}^{\mathrm{G}}=\frac{1}{N_\mathrm{modes}(k_i)} P^2\left(k_i\right) \delta_{ij}^{\rm K} \, ,
    \label{eq:cov_gauss_kikj}
\end{equation}
with
\begin{equation}
    N_\mathrm{modes}\left(k_i\right)=\frac{V_{k_i} V_W}{2(2 \pi)^3} \simeq \frac{2 \pi k_i^2 \Delta k V_W}{(2 \pi)^3}
    \label{eq:N_modes_def}
\end{equation}
The factor 2 in the denominator of Eq.~\eqref{eq:N_modes_def} arises from the reality condition of the density field $\tilde{\delta}_W^*(\vec{k})=\tilde{\delta}_W(-\vec{k})$. \\

The second term is the non-Gaussian contribution and is described by the trispectrum, the Fourier transform of the connected 4-point correlation function:
\begin{equation}
    \left\langle \tilde{\delta}(\vec{k}_1) \tilde{\delta}(\vec{k}_2) \tilde{\delta}(\vec{k}_3) \tilde{\delta}(\vec{k}_4) \right\rangle _c =(2 \pi)^3 \delta_{\rm D}^3(\vec{k}_{1234}) T(\vec{k}_1, \vec{k}_2, \vec{k}_3, \vec{k}_4)
\end{equation}
where the Dirac delta once again enforces momentum conservation: $\vec{k}_1 + \vec{k}_2 + \vec{k}_3 + \vec{k}_4 = 0$. The 4-point function appears naturally from the definition of covariance itself, which involves the (expectation value of the) product of two two-point functions. We note that both terms in Eq.~\eqref{eq:cov_terms} scale as the inverse of the survey volume, but the non-Gaussian term does not scale as $1/V_{k_i}$, the volume of the shell. This means that, contrary to the Gaussian contribution, we cannot average over more Fourier modes $k_i$ (i.e., use finer band-powers) to decrease the non-Gaussian covariance; the only way to reduce this term is to increase the survey volume.\\
The trispectrum itself is then convolved with the window function, analogously to Eq.~\eqref{eq:PS_estimator_ensemble}:
\begin{align}
    \bar{T}^W(k_i, k_j) = & \;
    \frac{1}{V_W} \int_{|\vec{k}| \in k_i} \frac{\diff ^ 3 \vec{k}}{V_{k_i}} \int_{|\vec{k}|^{\prime} \in k_j} \frac{\diff ^ 3 \vec{k}'}{V_{k_j}} \int\left[ \, \prod_{a=1}^4 \frac{\diff ^ 3 \vec{q}_a}{(2 \pi)^3} \tilde{W}(\vec{q}_a)\right](2 \pi)^3 \delta_{\rm D}^3(\vec{q}_{1234}) \\ \nonumber
    & \times  T(\vec{k}+\vec{q}_1,-\vec{k}+\vec{q}_2, \vec{k}'+\vec{q}_3,-\vec{k}'+\vec{q}_4) \; .
    \label{eq:T_bar_W}
\end{align}
This convolution with the window function means that different 4-point configurations separated by less than the
Fourier width of the window function $1/L$ and involving contributions from super-survey modes 
contribute to the covariance. \\

If, as assumed for the PS, the trispectrum is slowly varying as a function of scale, it can be pulled out of the innermost integral; moreover, we can again use Eq.~\eqref{eq:V_W_identity} to eliminate one $1/V_W$ factor, leaving us with 
\begin{align}
    \mathrm{Cov}_{ij} \approx & \; \mathrm{Cov}_{i j}^\mathrm{G} + \mathrm{Cov}_{ij}^{T 0} \\
    \approx & \; \frac{1}{V_W} \frac{(2 \pi)^3}{V_{k_i}} 2 P^2\left(k_i\right) \delta_{ij}^{\rm K}+\mathrm{Cov}_{ij}^{T 0} \, ,
\end{align}
being
\begin{equation}
    \label{eq:T0}
    \mathrm{Cov}_{i j}^{T 0}=\frac{1}{V_W} \int_{|\vec{k}| \in k_i} \frac{\diff^3 \vec{k}}{V_{k_i}} \int_{\left|\vec{k}'\right| \in k_j} \frac{\diff^3 \vec{k}'}{V_{k_j}} T(\vec{k},-\vec{k}, \vec{k}',-\vec{k}') \; .
\end{equation}
This equation implies that only parallelogram configurations of the trispectrum contribute to the non-Gaussian covariance of the band-power estimates. \\

\begin{figure}
    \centering
    \includegraphics[width=0.5\hsize]{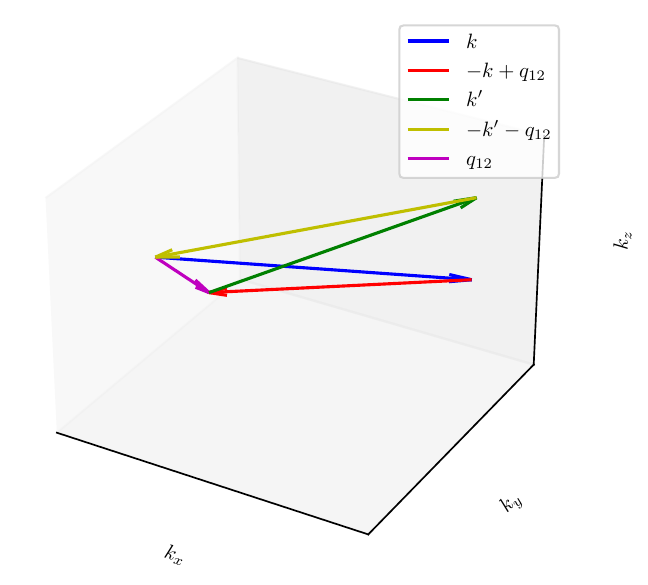}
    \caption{A sketch of a possible squeezed quadrilateral configuration. The magnitude of the long-wavelength mode $q_{12}$ has been exaggerated for clarity. The units are arbitrary.}
\label{fig:squeezed_quadrilateral_plot}
\end{figure}

If, on the other hand, the small-scale ($k \gg 1/L$) trispectrum exhibits significant variations or features within a range of wavenumbers $\Delta k \lesssim 1/L$ (i.e., comparable to the inverse of the survey size), additional effects arise in the non-Gaussian term of the covariance matrix. These are described in the pioneering work of \cite{Takada2013} by the so-called \textit{squeezed quadrilaterals} configurations of the trispectrum (see Fig.~\ref{fig:squeezed_quadrilateral_plot}), which have pairs of sides nearly equal and opposite. Reminding the condition $\vec{q}_{1234}=\vec{0}$, we can perform a change of variables to rewrite the trispectrum term of Eq.~\eqref{eq:T_bar_W}:
\begin{align}
\vec{k} + \vec{q_1} & \, \ML{\rightarrow} \, \vec{k} \\
\vec{k}' + \vec{q_3} & \, \ML{\rightarrow} \, \vec{k'} \\
\vec{q_1} + \vec{q_2} & \, \ML{\rightarrow} \, \vec{q}_{12} \; .
\end{align}
which gives $T(\vec{k}+\vec{q}_1,-\vec{k}+\vec{q}_2, \vec{k}'+\vec{q}_3,-\vec{k}'+\vec{q}_4) \leftrightarrow T\left(\vec{k},-\vec{k}+\vec{q}_{12}, \vec{k}',-\vec{k}'-\vec{q}_{12}\right)$. The configurations of interest can then be expressed as
\begin{equation}
\lim_{q_{12} \rightarrow 0} T\left(\vec{k},-\vec{k}+\vec{q}_{12}, \vec{k}',-\vec{k}'-\vec{q}_{12}\right) \; ,
\end{equation}
always under the approximation $q_{12} \ll k, k'$. Again, $k, k'$ are hard modes and $q_{12}$ is a soft mode that approaches zero in the squeezed limit; this limit therefore captures the correlation between the PS at $k$ and $ k'$ that is induced by a shared long- (infinite-) wavelength mode $ q_{12}$.
\\
The key ansatz introduced in the above-mentioned paper is that these particular configurations in the trispectrum are determined by the response of the PS to a change in the background density $\delta_\mathrm{b}$, given by the infinite-wavelength mode $q_{12}$. This is called the \textit{trispectrum consistency relation}:
\begin{equation}
\bar{T}\left(\vec{k},-\vec{k}+\vec{q}_{12}, \vec{k}',-\vec{k}'-\vec{q}_{12}\right) \approx T\left(\vec{k},-\vec{k}, \vec{k}',-\vec{k}'\right)+\frac{\partial P(k)}{\partial \delta_\mathrm{b}} \frac{\partial P(k')}{\partial \delta_\mathrm{b}} P_{\mathrm{L}}\left(q_{12}\right) \, ,
\label{eq:trisp_consistency_relation}
\end{equation}
where the overbar indicates an angle average over $\vec{q}_{12}$, because a constant mode $\delta_\mathrm{b}$ cannot quantify a directional dependence. In the last equation the \textit{linear} matter power spectrum appears: the perturbation $\delta_\mathrm{b}$ must be a mode in the linear regime.\\
The complete covariance matrix becomes therefore :
\begin{align}
    \mathrm{Cov}_{ij} & =\mathrm{Cov}_{ij}^\mathrm{G}+\mathrm{Cov}_{ij}^{T 0}+\sigma^2 \frac{\partial P\left(k_i\right)}{\partial \delta_\mathrm{b}} \frac{\partial P\left(k_j\right)}{\partial \delta_\mathrm{b}} \\
    & =\mathrm{Cov}_{ij}^\mathrm{G}+\mathrm{Cov}_{ij}^\mathrm{cNG} + \mathrm{Cov}_{ij}^\mathrm{SSC} \; .
\end{align}
\ML{The second term, $\mathrm{Cov}^{T0}_{ij}$, is given again in Eq.~\eqref{eq:T0} and can be interpreted as the nonlinear correlations which are not mediated by a long-wavelength (super-survey), constant mode. This is the connected non-Gaussian term, although the term \enquote{connected} is slightly deceptive, being the SSC also related to connected configurations of the trispectrum.}\\
The last term, the SSC, quantifies the correlation between the PS in bins $k_i$ and $k_j$ induced by the shared background fluctuation $\delta_\mathrm{b}$, whose variance is given by 
\begin{equation}
    \sigma^2 \equiv \frac{1}{V_W^2} \int \frac{\diff^3 \vec{q}}{(2 \pi)^3}|\tilde{W}(\vec{q})|^2 P_\mathrm{L}(q) \; ;
    \label{eq:sigma2_def_kspace}
\end{equation}
Combining the two expressions above, we can give the explicit form of the SSC:
\begin{equation}\label{eq:cov_ssc_realspace}
    \mathrm{Cov}^\mathrm{SSC}(k_i, k_j) = \frac{1}{V_W^2} \frac{\partial P\left(k_i\right)}{\partial \delta_\mathrm{b}} \frac{\partial P\left(k_j\right)}{\partial \delta_\mathrm{b}} \int \frac{\diff^3 \vec{q}}{(2 \pi)^3}|\tilde{W}(\vec{q})|^2 P_\mathrm{L}(q) \; .
\end{equation}
This term is composed of two main ingredients: the (co)variance of the background density mode $\delta_\mathrm{b}$, given by Eq.~\eqref{eq:sigma2_def_kspace}, and the \textit{response functions} $\partial P\left(k_i\right)/\partial \delta_\mathrm{b}$, which quantify the reaction of a given statistics (in this case, the matter power spectrum) to a coherent change in background density. Of these two terms, the latter is less straightforward to compute; different methods will be presented in the following section.

We conclude this section by noting that, contrary to the Gaussian and connected non-Gaussian terms, the SSC dependence on the survey volume is non-trivial, and encoded within the $\sigma^2$ term: $\mathrm{Cov}^\mathrm{SSC}_{ij} \propto \sigma^2(V_W)$, while $\mathrm{Cov}^\mathrm{G, cNG}_{ij} \propto 1/V_W$.
\subsection{Response functions}
As stated in the introduction and shown formally in the previous section, super-sample covariance parameterizes our ignorance on the effect of large-scale density modes on the observables of interest. This effect arises precisely because of the sensitivity of these to a coherent change in background density, which is a common characteristic of all LSS probes. Quantifying this sensitivity is the aim of the present section. \\
In the literature, two main approaches are used to solve this problem. The first is to compute the responses using the halo model \citep{Takada2013, Krause2017, Rizzato2019, Lacasa_2019}; this semi-analytical treatment offers the advantage of a more profound physical intuition of the mechanisms at play and the different contributions to the responses but tends to fail on small, highly nonlinear scales \citep{Wagner2014}. The second possible approach is the so-called \textit{separate universe} technique \citep{Wagner2014, Wagner2015, Li2015, Barreira2019}, which is based on the ansatz that a constant background density perturbation is akin to a separate universe with different cosmological parameters. The next two sections will explain these two different methods.
\subsubsection{Halo model response}\label{sec:hm_responses}

The trispectrum consistency relation (Eq.~\ref{eq:trisp_consistency_relation}) tells us that it is possible to compute the responses from the squeezed configurations of the matter trispectrum. In the halo model approach, the trispectrum can be described in a similar way to the PS, but including contributions from up to four halos: $T_{n \mathrm{h}}$ are the $n$-halo terms, describing the correlations between \textit{four} points distributed in $n$ halos (1 for the intra-halo term and 2, 3 or 4 for the inter-halo terms).
\begin{equation}
T=T_\mathrm{1 h}+\left(T_\mathrm{2 h}^{22}+T_\mathrm{2 h}^{13}\right)+T_\mathrm{3 h}+T_\mathrm{4 h} \; .
\label{eq:trispectrum_halomodel_sum}
\end{equation}
The superscripts in the 2-halo terms indicate the number of points in the first and second halos: $T_\mathrm{2 h}^{22}$ will have two points in the first and two points in the second, while $T_\mathrm{2 h}^{13}$ will have one point in the first and three in the second.\\
These terms are in turn given by the halo model auxiliary functions~\eqref{eq:halomodel_auxiliary} and perturbation theory expressions:
\begin{equation}
\begin{aligned}
    & T_{1 h}(\vec{k}_1, \vec{k}_2, \vec{k}_3, \vec{k}_4)=I_4^0\left(k_1, k_2, k_3, k_4\right), \\
    & T_{2 h}^{22}(\vec{k}_1, \vec{k}_2, \vec{k}_3, \vec{k}_4)=P_{\mathrm{L}}\left(k_{12}\right) I_2^1\left(k_1, k_2\right) I_2^1\left(k_3, k_4\right)+2 \; \text{perm.} \\
    & T_{2 h}^{13}(\vec{k}_1, \vec{k}_2, \vec{k}_3, \vec{k}_4)=P_{\mathrm{L}}\left(k_1\right) I_1^1\left(k_1\right) I_3^1\left(k_2, k_3, k_4\right)+3 \; \text{perm.} \\
    & T_\mathrm{3 h}(\vec{k}_1, \vec{k}_2, \vec{k}_3, \vec{k}_4)=B_{\mathrm{PT}}\left(\vec{k}_1, \vec{k}_2, \vec{k}_{34}\right) I_1^1\left(k_1\right) I_1^1\left(k_2\right) I_2^1\left(k_3, k_4\right)+5 \text { perm., } \\
    & T_\mathrm{4 h}(\vec{k}_1, \vec{k}_2, \vec{k}_3, \vec{k}_4)=T_{\mathrm{PT}}\left(\vec{k}_1, \vec{k}_2, \vec{k}_3, \vec{k}_4\right) I_1^1\left(k_1\right) I_1^1\left(k_2\right) I_1^1\left(k_3\right) I_1^1\left(k_4\right),
    \label{eq:trispectrum_halomodel_terms}
\end{aligned}
\end{equation}
where \enquote{perm.} refers to cyclic permutations, obtained by rotating the $k_i$ indices: $k_1 \rightarrow k_2$, $k_2 \rightarrow k_3$, $k_3 \rightarrow k_4$ and $k_4 \rightarrow k_1$. $B_\mathrm{PT}$ and $T_\mathrm{PT}$ are the bispectrum and trispectrum computed from perturbation theory, defined as
\begin{align}
B_\mathrm{PT}\left(\vec{k}_1, \vec{k}_2, \vec{k}_3\right) & = 2 F_2\left(\vec{k}_1, \vec{k}_2\right) P_\mathrm{L}\left(k_1\right) P_\mathrm{L}\left(k_2\right)+2 \; \text{perm.} \\
T_\mathrm{PT}\left(\vec{k}_1, \vec{k}_2, \vec{k}_3, \vec{k}_4\right) & =  4 [ F_2\left(\vec{k}_{13},-\vec{k}_1\right) F_2\left(\vec{k}_{13}, \vec{k}_2\right) P_\mathrm{L}\left(k_{13}\right) P_\mathrm{L}\left(k_1\right) P_\mathrm{L}\left(k_2\right) \nonumber \\
&\quad +11 \; \text{perm.} ] \nonumber \\
&\quad +6\left[F_3\left(\vec{k}_1, \vec{k}_2, \vec{k}_3\right) P_\mathrm{L}\left(k_1\right) P_\mathrm{L}\left(k_2\right) P_\mathrm{L}\left(k_3\right)+3 \; \text{perm.}\right] \; ,
\end{align}
where the mode-coupling kernels $F_2$, $G_2$ and $F_3$ are given respectively in Eqs.~\eqref{eq:F2_kernel}, \eqref{eq:G2_kernel} and~\eqref{eq:F3_kernel}. We remind that these are only exact for Einstein-de Sitter cosmologies, but are approximately valid for other cosmological models by virtue of their weak dependence on $\Omega_\mathrm{m,0}$.\\

To obtain an expression for the response functions, we compare the SSC term from the consistency relation (Eq.~\ref{eq:trisp_consistency_relation}):
\begin{align}
    \delta T & \simeq T\left(\vec{k},-\vec{k}+\vec{q}_{12}, \vec{k}^{\prime},-\vec{k}^{\prime}-\vec{q}_{12}\right) - T\left(\vec{k},-\vec{k}, \vec{k}^{\prime},-\vec{k}^{\prime}\right) \nonumber \\
    & \simeq \frac{\partial P(k)}{\partial \delta_\mathrm{b}} \frac{\partial P(k')}{\partial \delta_\mathrm{b}} P_\mathrm{L}(q_{12})
    \label{eq:delta_T_trisp_consistency}
\end{align}
with the expression of $\delta T$ from the halo model; to leading order in $q_{12}/k$, the variation in the trispectrum induced by the long-wavelength mode $\vec{q}_{12}$ can be expressed by plugging the different terms of Eq.~\eqref{eq:trispectrum_halomodel_terms} for the connected configurations of interest in Eq.~\eqref{eq:trispectrum_halomodel_sum}:
\begin{equation}
\begin{aligned}
& \delta T_\mathrm{1 h} \approx 0 \\
& \delta T_\mathrm{2 h}^{22} \approx P_{\mathrm{L}}\left(q_{12}\right) I_2^1(k, k) I_2^1\left(k^{\prime}, k^{\prime}\right) \\
& \delta T_\mathrm{2 h}^{13} \approx 0 \\
& \delta T_\mathrm{3 h} \approx 2 P_{\mathrm{L}}\left(q_{12}\right) I_2^1\left(k^{\prime}, k^{\prime}\right) \mathcal{F}\left(\vec{k}, \vec{q}_{12}\right) +2 P_{\mathrm{L}}\left(q_{12}\right) I_2^1(k, k) \mathcal{F}\left(\vec{k}^{\prime},-\vec{q}_{12}\right) \\
& \delta T_\mathrm{4 h} \approx 4 P_{\mathrm{L}}\left(q_{12}\right) \mathcal{F}\left(\vec{k}, \vec{q}_{12}\right) \mathcal{F}\left(\vec{k}^{\prime},-\vec{q}_{12}\right) 
\label{eq:delta_T_for_SSC}
\end{aligned}
\end{equation}
having again used the approximation $|\vec{k} + \vec{q}_1| \sim k$, and in which we defined 
\begin{equation}
\mathcal{F}(\vec{k}, \vec{q}) \equiv {\left[P_{\mathrm{L}}(k) F_2(\vec{q},-\vec{k})+P_{\mathrm{L}}(|\vec{k}-\vec{q}|) F_2(\vec{q}, \vec{k}-\vec{q})\right] } I_1^1(k) I_1^1(|\vec{k}-\vec{q}|) \; ,
\end{equation}
We can factor out the terms in $P_\mathrm{L}(q_{12})$ and write $\delta T$ as 
\begin{equation}
\begin{aligned}
    \delta T & \approx P_{\mathrm{L}}\left(q_{12}\right) \Bigl[ I_2^1(k, k) I_2^1\left(k^{\prime}, k^{\prime}\right) + 2 I_2^1\left(k^{\prime}, k^{\prime}\right) \mathcal{F}\left(\vec{k}, \vec{q}_{12}\right) \\
    & \quad + 2 I_2^1(k, k) \mathcal{F}\left(\vec{k}', -\vec{q}_{12}\right) + 4 \mathcal{F}\left(\vec{k}, \vec{q}_{12}\right) \mathcal{F}\left(\vec{k}' -\vec{q}_{12}\right) \Bigr] \\
    & = P_{\mathrm{L}}\left(q_{12}\right) \Bigl[ \left(I_2^1(k, k) + 2\mathcal{F}\left(\vec{k}, \vec{q}_{12}\right)\right)
    \times \left(I_2^1(k', k') + 2\mathcal{F}\left(\vec{k}', -\vec{q}_{12}\right)\right)\Bigr]
    \label{eq:delta_T_refactor}
\end{aligned}
\end{equation}
We now integrate $\mathcal{F}$ over the cosine $\mu_k$ of the angle $\theta$ between $\vec{k}$ and $\vec{q}$ to average over all possible directions, or orientations of $\vec{k}$ with respect to $\vec{q}$:
\begin{equation}
\int_{-1}^1 \frac{\diff \mu_k}{2} \mathcal{F}(\vec{k}, \vec{q}) \approx \frac{1}{2}\left(\frac{68}{21}-\frac{1}{3} \frac{\diff \ln k^3 P_{2 h}}{\diff \ln k}\right) P_{2 h} \quad \quad 
\mu_k = \cos(\theta) = \frac{\vec{k} \cdot \vec{q}}{|\vec{k}| \, |\vec{q}|} \; ,
\end{equation}
1/2 being a normalization factor.\\
In Eqs.~\eqref{eq:delta_T_for_SSC} recognise how only the trispectrum terms involving $P_\mathrm{L}(k)$ contribute to the SSC. Comparing the last line of Eq.~\eqref{eq:delta_T_refactor} with Eq.~\eqref{eq:delta_T_trisp_consistency}, it's immediately apparent that the matter power spectrum responses take the form
\begin{align}
    \frac{\partial P(k)}{\partial \delta_\mathrm{b}} & \approx \left(\frac{68}{21}-\frac{1}{3} \frac{\diff \ln k^3 P_\mathrm{2 h}(k)}{\diff \ln k}\right) P_\mathrm{2 h}(k)+I_2^1(k, k) \\
    & = \left(\frac{68}{21}-\frac{1}{3} \frac{\diff \ln k^3 \left[ I^1_1(k)\right]^2
    P_\mathrm{L}(k)}{\diff \ln k}\right) \left[ I^1_1(k)\right]^2
    P_\mathrm{L}(k)+I_2^1(k, k) \; .
\end{align}
This is the final expression for the SSC responses in the halo model formalism. It allows isolating the different terms to the covariance and to understand their physical meaning \citep{Li2014}:
\begin{enumerate}
    \item The first addendum is called \enquote{beat coupling}, and represents the intuitive fact that the growth of a sub-survey, short-wavelength perturbation is enhanced in a large-scale overdensity.
    \item The second is the \enquote{linear dilation} effect; it arises because the long-wavelength perturbation changes the scale factor $a$ locally, and therefore the physical size of small-scale modes. This corresponds to a rescaling of the argument of the matter power spectrum.
    \item The last term is the \enquote{halo sample variance}, which accounts for the increase in the \textit{halo} number density caused by the large-scale overdensity.
\end{enumerate}

We conclude this section by noting that this treatment assumes that the density contrast used to compute the matter power spectrum is defined with respect to the \textit{global} mean density, as opposed to the \textit{in-survey} mean density. This is appropriate for statistics such as weak lensing, which probes the matter density field directly; in the case of galaxy number counts, on the other hand, the mean (reference) density is computed from the survey and hence is itself affected by the background perturbation $\delta_\mathrm{b}$. In this case, we need to rescale the power spectrum as:
\begin{equation}
P_W(k)=P(k) /\left(1+\delta_\mathrm{b}\right)^2 .
\end{equation}
so the response rescales as
\begin{equation}
\frac{\partial P(k)}{\partial \delta_\mathrm{b}} \rightarrow \frac{\partial P_W(k)}{\partial \delta_\mathrm{b}} \approx \frac{\partial P(k)}{\partial \delta_\mathrm{b}}-2 P(k) \; .
\label{eq:pk_response_rescaling}
\end{equation}
\subsubsection{Separate universe response}
A different method to compute the probe response is through the so-called \enquote{separate universe} technique, which takes advantage of the above-mentioned correspondence between a local constant background density perturbation and a separate universe with different background density. This method is better suited to characterize the response functions in the nonlinear regime, where the validity of the halo model and perturbation theory breaks down \citep{Wagner2015}. \\

The key idea of this approach can be expressed by the definition of the background density perturbation itself (see Eq.~\ref{eq:delta_b_def}, but with the subscript $W$ now indicating the quantities of the separate universe), relating the mean densities averaged over the universe and over the survey (which is now the separate universe):
\begin{equation}
\bar{\rho}_\mathrm{m}\left(1+\delta_\mathrm{b}\right)=\bar{\rho}_{\mathrm{m} W} \; .
\end{equation}
from which follows the relation between the density parameters in the two universes:
\begin{equation}
\frac{\Omega_\mathrm{m} h^2}{a^3}\left(1+\delta_\mathrm{b}\right)=\frac{\Omega_{{\rm m} W} h_W^2}{a_W^3} \; .
\end{equation}
We also have
\begin{equation}
\lim _{a \rightarrow 0} \delta_\mathrm{b}(a)=0 \quad \rightarrow \quad
\lim _{a \rightarrow 0} a_W\left(a, \delta_\mathrm{b}\right)=a
\end{equation}
from the fact that there is no SSC effect at early times. This means that the scale factors of the two different cosmologies agree at high redshift ($a \rightarrow 0$). From this, it follows that the \textit{physical} densities of the two universes match:
\begin{equation}
\Omega_{{\rm m} W} h_W^2=\Omega_\mathrm{m} h^2 \; .
\label{eq:physical_densities}
\end{equation}
from which we obtain a mapping for the scale factors:
\begin{equation}
    a_W=\frac{a}{\left(1+\delta_\mathrm{b}\right)^{1 / 3}} \approx a\left(1-\frac{\delta_\mathrm{b}}{3}\right) \; .
\end{equation}
or equivalently, defining $\delta_a(t)$ such that $a_W(t) = \left[ 1+\delta_a(t)\right] a(t)$:
\begin{equation}
    1 + \delta_\mathrm{b}(t) = \left[ 1+\delta_a(t)\right]^{-3} \; .
    \label{eq:delta_b_vs_delta_a}
\end{equation}
These equations allow writing the Friedmann equations in the two cosmologies to obtain the different expansion rates:
\begin{equation}
\delta H^2=H_W^2-H^2 \approx-\frac{2}{3} H \dot{\delta}_b \; ;
\label{eq:delta_H_square}
\end{equation}
with the growth rate given by 
\begin{equation}
H \dot{\delta}_b=\frac{\Omega_{\rm m} H_0^2}{2 a^2}\left[\frac{5}{D}-\frac{3}{a}-\frac{2 \Omega_k}{\Omega_{\rm m}}\right] \delta_\mathrm{b} \; ,
\end{equation}
using the normalization for the linear growth function $\delta_\mathrm{b}=\left(D / D_0\right) \delta_{b 0}$:
\begin{equation}
\lim _{a \rightarrow 0} D=a \; .
\end{equation}
we can then expand $H_W^2$ on the l.h.s., using the Friedmann equation for the separate universe:
\begin{equation}
\begin{aligned}
H_W^2 & =H_{0 W}^2\left[\frac{\Omega_{{\rm m} W}}{a_W^3}+\Omega_{\Lambda W}+\frac{\Omega_{k W}}{a_W^2}\right] \\
& \approx H^2+\frac{H_{0 W}^2-H_0^2}{a^2}+H_0^2 \delta_\mathrm{b}\left[\frac{\Omega_{\rm m}}{a^3}+\frac{2}{3} \frac{\Omega_k}{a^2}\right] \; .
\end{aligned}
\end{equation}
We can plug this term in Eq.~\eqref{eq:delta_H_square} to get the change in the Hubble constant:
\begin{equation}
\frac{\delta h}{h} \equiv \frac{H_{0 W}-H_0}{H_0} \approx-\frac{5 \Omega_{\rm m}}{6} \frac{\delta_\mathrm{b}}{D} \; .
\label{eq:h_tilde_vs_h}
\end{equation}
The other density parameters are then given by
\begin{equation}
\frac{\delta \Omega_{\rm m}}{\Omega_{\rm m}}=\frac{\delta \Omega_{\Lambda}}{\Omega_{\Lambda}} \approx-2 \frac{\delta h}{h} .
\label{eq:Omega_tilde_vs_Omega}
\end{equation}
having used Eq.~\eqref{eq:physical_densities}; the separate universe will also have a different curvature (e.g., finally, a closed universe for $\delta_\mathrm{b}>0$):
\begin{equation}
\begin{aligned}
\Omega_{k W} & =1-\Omega_{{\rm m} W}-\Omega_{\Lambda W} \\
& =1-\left(\Omega_{\rm m}+\Omega_{\Lambda}\right)\left(1+\frac{5 \Omega_{\rm m}}{3} \frac{\delta_\mathrm{b}}{D}\right)\; .
\label{eq:Omegak_tilde_vs_Omegak}
\end{aligned}
\end{equation}
Having expressed the separate universe parameters as a function of the reference ones, it is possible to set up a simulation for different (small) variations of $\delta_\mathrm{b}$ around 0 to numerically measure the power spectrum response.\\

In particular, we use the results presented in \cite{Wagner2015}, which use the separate universe technique to measure a specific term amongst the three different contributions to the responses presented in the previous section. In this work, the $n$-th order response functions are defined as the coefficients of the expansion of the
power spectrum in the linearly extrapolated (Lagrangian) overdensity $\delta_{L}$: 
\begin{equation}
    P\left(k, t \mid \delta_{L 0}\right)=\sum_{n=0}^{\infty} \frac{1}{n !} R_n(k, t)\left[\delta_{L}\right]^n P(k, t) \; ,
\end{equation}
with $\delta_{L} = \delta_{L 0} \hat{D} = D(t)/D(t_0)\delta_{L0}$ and $\delta_{L0} = \delta_{L}(t=0)$ is the initial overdensity. Equivalently, the response coefficients $R_n(k, t)$ can be defined as the $n$-th derivative
of the power spectrum with respect to $\delta_L$ (computed in $\delta_{L} = 0$), normalized by the power spectrum:
\begin{equation}
    R_n(k, t)=\left.\frac{1}{P(k)} \frac{\diff^n P\left(k, t \mid \delta_L\right)}{\diff\left(\delta_L(t)\right)^n}\right|_{\delta_L=0} \; .
    \label{eq:R_n_def}
\end{equation}
For the first-order response coefficient $R_1(k, t)$, which we are interested in in the context of this work, the response to $\delta_L$ coincides with the response to the evolved nonlinear (Eulerian) overdensity $\delta_\mathrm{b}$. Note that, where no confusion arises, we will refer to $R_n(k, t)$ interchangeably as probe response coefficient or probe response (the only difference between the two being the normalization by the PS, as shown in the equation above).\\
As mentioned, $R_n(k, t)$ is composed of different terms, the first being the \enquote{reference density} effect (analogous to the halo sample variance), which rescales the power spectrum as:
\begin{equation}
P(k) = \left[1+\delta_\mathrm{b}\right]^2 {P_W}(k),
\end{equation}
The second effect (the linear dilation) comes from a change in the comoving coordinates $\vec{x}$ caused by the rescaling of the scale factor:
\begin{equation}
    \vec{x} = \frac{a_W(t)}{a(t)}\vec{x}_W = [1+\delta_a(t)]\vec{x}_W \quad \rightarrow \quad k_W=\left(1+\delta_a\right) k \; ,
\end{equation}
and changes the power spectrum as
\begin{equation}
    P(k) = \left[1+\delta_a\right]^3 P_W\left(\left[1+\delta_a\right] k\right) \; .
\end{equation}
Putting the two together and using Eq.~\eqref{eq:delta_b_vs_delta_a} then gives
\begin{equation}
P(k)=\left[1+\delta_\mathrm{b}\right] P_W\left(\left[1+\delta_a\right] k\right) \; .
\end{equation}
The $\delta_a$ and $\delta_\mathrm{b}$ terms can be expanded in series as a function of $\delta_{L0}$. The coefficients of these expansions can be derived analytically for an Einstein-de Sitter fiducial cosmology, which approximates very well the $\Lambda$CDM case (see Appendix B.1 of \citealt{Wagner2015}):
\begin{equation}
\begin{aligned}
& \delta_a(t)=\sum_{n=1}^{\infty} e_n\left[\delta_{L 0} \hat{D}(t)\right]^n \\
& \delta_\mathrm{b}(t)=\sum_{n=1}^{\infty} f_n\left[\delta_{L 0} \hat{D}(t)\right]^n \; .
\end{aligned}
\end{equation}
The final contribution, which we referred to as \enquote{beat coupling}, is related to the modification in the growth of small-scale perturbations embedded in the local, large-scale overdensity (or underdensity). This is captured by the \textit{growth-only} responses:
\begin{equation}
    G_n(k) = \frac{1}{P(k)}\frac{\diff^n P_W(k)}{\diff \delta_L^n}\bigg|_{\delta_{L} = 0} \; .
    \label{eq:growth_only_def}
\end{equation}
We finally have, from the above expressions:
\begin{equation}
P\left(k \mid \delta_L\right)=\left[1+\delta_\rho\right]\left[\left(1+\sum_{n=1}^{\infty} \frac{1}{n !} G_n(k_W) \delta_L^n\right) P(k_W)\right]_{k_W=\left[1+\delta_a\right] k}
\end{equation}
and, for the first-order response $R_1(k, t)$:
\begin{equation}
R_1(k, t)=f_1+e_1 \frac{\diff \ln P(k, t)}{\diff \ln k}+G_1(k, t) = 1 -\frac{1}{3} \frac{\diff \ln P(k, t)}{\diff \ln k}+G_1(k, t) \; .
\label{eq:R_1_mm_expansion}
\end{equation}
The $G_1$ term can be computed analytically in the linear regime, where the growth is scale-independent: 
\begin{equation}
    G_1^{\text {lin}}=\left.\frac{1}{D^2} \frac{\diff\left(D_W^2\right)}{\diff\delta_L}\right|_{\delta_L=0}
\end{equation}
where $D_W$ is the growth factor in the modified cosmology, which can again be expressed as a series expansion:
\begin{equation}
    D_W(t)=D(t)\left\{1+\sum_{n=1}^{\infty} g_n\left[\delta_{L 0} \hat{D}(t)\right]^n\right\} \; .
\end{equation}
For an Einstein-de Sitter cosmology, the first coefficients are
\begin{equation}
    g_{1, 2} = \left\{\frac{13}{21}, \frac{71}{189} \right\}\; ,
\end{equation}
giving finally 
\begin{equation}
    G_1^{\text {lin}}=\frac{26}{21} \; .
\end{equation}
We note that $R_1(k, t)$ is the response function of isotropic large-scale \textit{density} perturbations; we neglect the contribution from the anisotropic \textit{tidal-field} perturbations to the total response of the power spectrum (and consequently to the SSC), which has been shown in \citet{Barreira2018response_approach} to be subdominant for WL with respect to the first contribution (about 5\% of the total covariance matrix at $\ell \gtrsim 300$).  \\

\begin{figure}
    \centering
    \includegraphics[width=0.7\linewidth]{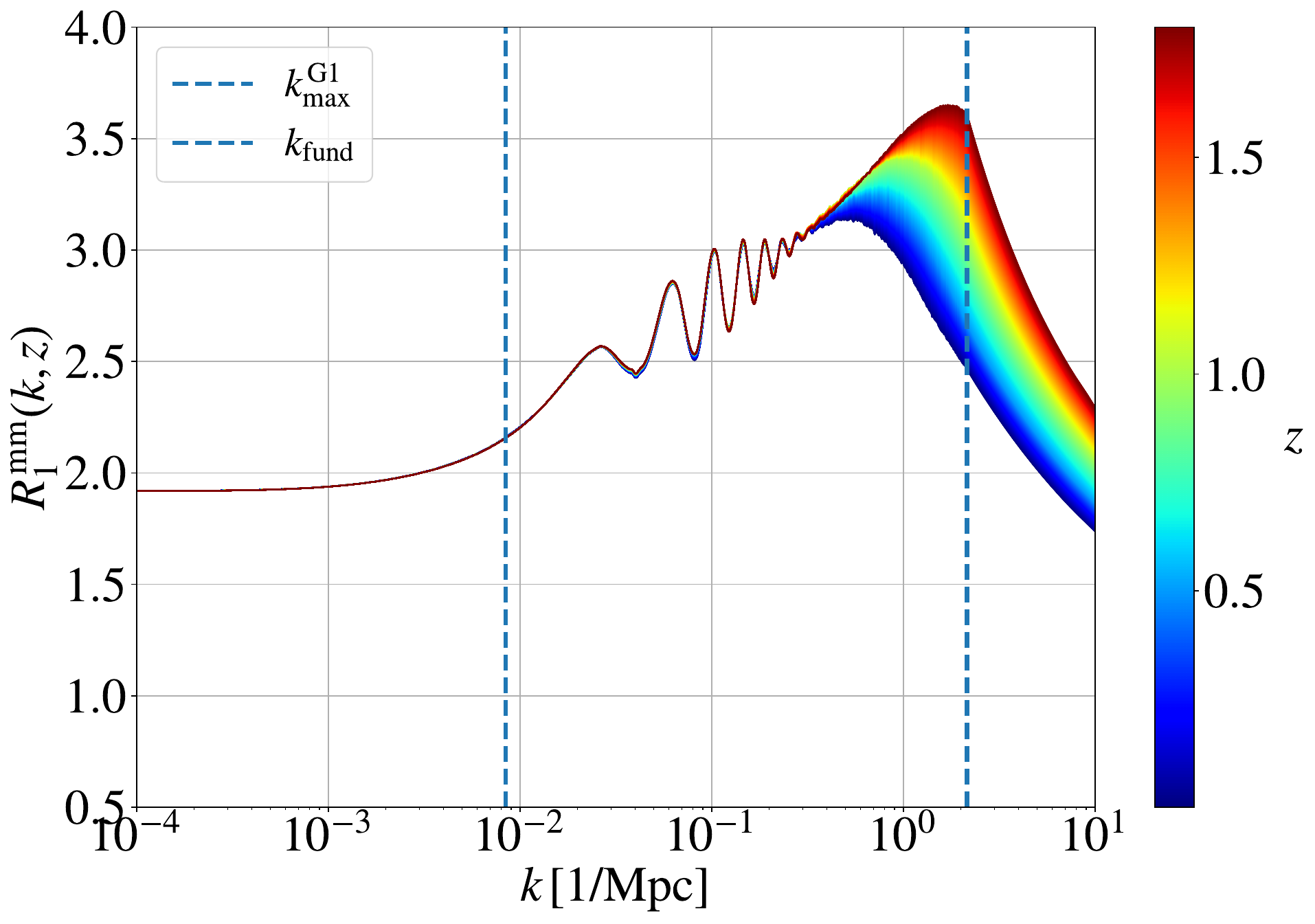}
    \caption{Scale and redshift dependence of the 3D first-order matter (\enquote{mm}) response coefficient. The minimum and maximum wavenumbers ($k_\mathrm{fund}$ and $k^\mathrm{G1}_\mathrm{max}$, respectively the left and right vertical dashed lines) used to compute the growth-only response from the \citealt{Wagner2015} simulations are also shown. An analogous plot is shown in Fig.~1 of \citealt{Barreira2018response_approach}, which uses the same $G_1^\mathrm{mm}$.}
    \label{fig:r1mm_rainbow}
\end{figure}

The $G_1(k)$ term is measured in the nonlinear regime, using the definition (Eq.~\ref{eq:growth_only_def}). The separate universe simulations have linearly evolved, present-day overdensity of
\begin{equation}
    \delta_{L0} = \left\{0, \pm0.01, \pm0.02,
\pm0.05, \pm0.07, \pm0.1, \pm0.2, \pm0.5, \pm0.7, \pm 1\right\} \; .
\end{equation}
The cosmological parameters of the separate universe are mapped from the ones of the reference cosmology using Eqs.~(\ref{eq:h_tilde_vs_h}, \ref{eq:Omega_tilde_vs_Omega}, \ref{eq:Omegak_tilde_vs_Omegak}).The simulations used in this work, presented in \citep{Wagner2014}, have been run with \texttt{Gadget-2} \citep{Springel2005_gadget2} and have the following setup: simulation box with size $L_{\rm box}=500 \,h^{-1} \,\mathrm{Mpc}$ and $N_p=512^3$ particles; starting redshift $z=49$ and a flat, $\Lambda$CDM fiducial cosmology with parameters 
$\Omega_{\rm m} = 0.27$, 
$h = 0.7$, 
$\Omega_\mathrm{b} h^2 = 0.023$, 
$n_\mathrm{s} = 0.95$, 
$\sigma_8 = 0.8$. 
This setup allows measuring the growth-only response in a limited range $k \in\left[k_{\text {fund }}, k_{\text {max }}\right]$, 
where $k_{\text {fund }}=2 \pi / L_{\rm box} \approx 0.012 \,h\, \mathrm{Mpc}^{-1}$ and $k_{\max }=k_{\mathrm{Nyquist}}=N_p^{1 / 3} \pi / L_{\mathrm{box}} \approx 3.2 \, h \, \mathrm{Mpc}^{-1}$. Outside of this range, we take, following \citep{Barreira2018response_approach}:
\begin{align}
& k < k_{\text{{fund}}}: G_1(k) = \frac{{26}}{{21}} \\
& k > k_{\text{{max}}}: G_1(k) = B_1 + \left[G_1\left(k=k_{\text{{max}}}\right)-B_1\right]\left(\frac{k}{k_{\text{{max}}}}\right)^{-1 / 2} \; .
\end{align}
In the last equation above we set $B_1 \approx 0.75$, which gives $R_1(k \rightarrow \infty)=1$: on small scales, the (isotropic) density perturbations response contributes only via the density effect, while the other two vanish. The specific exponent (in this case, $-1/2$) does not have a significant impact on the final results and has been chosen to have a continuous and smooth transition between the different regimes. The response functions obtained in this way are shown in Fig.~\ref{fig:r1mm_rainbow}.\\

Having introduced the different ingredients entering the SSC term and their origin, we move on to the specific treatment for 2D-projected observables relevant to the \Euclid photometric survey.
\section{SSC for projected observables}\label{sec:ssc_projected}
To apply the SSC theory introduced in this chapter to our case scenario, we need to translate our formalism from the three-dimensional $k$ space to the two-dimensional harmonic, or $\ell$, space. This mapping is introduced again in the seminal paper of~\cite{Takada2013} and adapted to the multi-probe, tomographic case in e.g. \cite{Krause2017, LacasaRosenfeld2016, Barreira2018cosmic_shear, upham2021} (the latter for pseudo-$C(\ell)$).\\

We begin by considering the general expression for 2D-projected observables; considering a general field $\delta(\chi\vec{\theta}, \chi)$, we can project it on the sphere by means of a simple radial integration weighted by some radial kernel $\mathcal{K}(\chi)$, as seen in Sect.~\ref{sec:gc_phot_theory}: \\
\begin{equation}
    \Sigma(\vec{\theta})=\int \diff \chi \mathcal{K}(\chi) \delta(\chi \vec{\theta}, \chi) \; .
\end{equation}
We can equivalently use the redshift as our radial coordinate, hence integrating in $\diff z$, by leveraging the distance-redshift relation:
\begin{equation}
    \frac{\diff \chi(z)}{\diff z} = \frac{c}{H(z)} = \frac{c}{H_0 E(z)} \quad \rightarrow \quad \diff \chi =  \frac{c}{H_0 E(z)} \diff z
\end{equation}
which is the choice commonly adopted in \Euclid publications (and in the official \Euclid code, \texttt{CLOE} (\enquote{Cosmology Likelihood for Observables in Euclid}).\\

\noindent
Just as $\delta_W(\vec{x}) = \delta(\vec{x})W(\vec{x})$, the masked projected field $\Sigma_{\mathcal W}(\vec{\theta})$ is given by a product in real (angular) space:
\begin{equation}
\Sigma_{\mathcal W}(\vec{\theta})=\Sigma(\vec{\theta}) \mathcal{W}(\vec{\theta}) \; ,
\end{equation}
with $\mathcal{W}(\vec{\theta})$ the \textit{angular} window function and $\vec{\theta}$ the angular coordinate. As seen in Eq.~\eqref{eq:delta_W_tide}, this product becomes a convolution in harmonic space; using the flat-sky approximation, we can then express the Fourier transform of the projected field as a function of the multipole $\vec{\ell}$ (see Sect.~\ref{sec:angular_ps_theory}) as:
\begin{equation}
    \tilde{\Sigma}_{\mathcal W}(\vec{\ell})=\int \frac{\diff^2 \vec{\ell}^{\prime}}{(2 \pi)^2} \tilde{{\cal W}}\left(\vec{\ell}-\vec{\ell}^{\prime}\right) \tilde{\Sigma}\left(\vec{\ell}^{\prime}\right)  \; ,
\end{equation}
and its binned estimator, in analogy as Eq.~\eqref{eq:Pk_W_estimator}, with 
\begin{equation}
\hat{C}\left(\ell_i\right) \equiv \frac{1}{\Omega_{\mathcal W}} \int_{| \vec{\ell} | \in \ell_i} \frac{\diff^2 \vec{\ell}}{\Omega_{\ell_i}} \tilde{\Sigma}_{\mathcal W}(\vec{\ell}) \tilde{\Sigma}_W(-\vec{\ell}),
\end{equation}
where we have introduced the \textit{angular} survey area $\Omega_{\mathcal W} \equiv \int \diff^2 \vec{\theta} \mathcal{W}(\vec{\theta})$ and
$\Omega_{\ell_i}=\int_{| \vec{\ell} | \in \ell_i} \diff ^2 \vec{\ell} \simeq 2 \pi \ell_i \Delta \ell$ when $\Delta \ell / \ell_i \ll 1$. In the same way as $\hat{P}(k_i)$ was an unbiased estimator for $k \gg 1/L$, for angular modes much larger than the angular width of the window function we have $\left\langle\hat{C}\left(\ell_i\right)\right\rangle=C\left(\ell_i\right)$. The covariance of this estimator is then defined similarly to what was done above:
\begin{equation}
\begin{aligned}
\mathrm{Cov}_{i j} & \equiv\left\langle\hat{C}\left(\ell_i\right) \hat{C}\left(\ell_j\right)\right\rangle-C\left(\ell_i\right) C\left(\ell_j\right) \\
& =\frac{1}{\Omega_{\mathcal W}}\left[\frac{(2 \pi)^2}{\Omega_{\ell_i}} C\left(\ell_i\right) \delta_{ij}^{\rm K}+\bar{\mathcal{T}}^{\mathcal W}\left(\ell_i, \ell_j\right)\right],
\end{aligned}
\end{equation}
with the trispectrum $\bar{\mathcal{T}}^{\mathcal W}$ now given by
\begin{align}
    \bar{\mathcal{T}}^{\mathcal W}\left(\ell_i, \ell_j\right)& =\frac{1}{\Omega_{\mathcal{W}}} \int_{\mid \vec{\ell} \mid \in \ell_i} \frac{\diff^2 \vec{\ell}}{\Omega_{\ell_i}} \int_{ \mid \vec{\ell}^{\prime} \mid \in \ell_j} \frac{\diff^2 \vec{\ell}^{\prime}}{\Omega_{\ell_j}} \int\left[\prod_{a=1}^4 \frac{\diff^2 \vec{q}_a}{(2 \pi)^2} \tilde{W}\left(\vec{q}_a\right)\right](2 \pi)^2 \delta_{\rm D}^2\left(\vec{q}_{1234}\right) \nonumber \\
    & \quad \times \mathcal{T}\left(\vec{\ell}+\vec{q}_1,-\vec{\ell}+\vec{q}_2, \vec{\ell}^{\prime}+\vec{q}_3,-\vec{\ell}^{\prime}+\vec{q}_4\right) \; .
\end{align}
In the Limber approximation (Sect.~\ref{sec:limber}), the angular power spectra and the trispectrum can be related to their 3D counterparts in terms of the Limber wavenumber $\vec{k}_\ell$: this gives us, for the projected power spectrum and trispectrum, 
\begin{align}
    C(\ell) & \approx \int \diff \chi \frac{\mathcal{K}^2(\chi)}{\chi^{2}}  P(\vec{k}_\ell ; \chi), \label{eq:cl_limber_takada} \\
    \mathcal{T}\left(\vec{\ell}_1, \vec{\ell}_2, \vec{\ell}_3, \vec{\ell}_4\right) & \approx \int \diff \chi \; \frac{\mathcal{K}^4(\chi)}{\chi^{6}} T\left(\vec{k}_{\ell_1}, \vec{k}_{\ell_2}, \vec{k}_{\ell_3}, \vec{k}_{\ell_4} ; \chi\right),
\end{align}
so we finally get, using again the trispectrum consistency relation:
\begin{align}
    \mathrm{Cov}_{i j}^{\mathrm{G}} & =\frac{1}{\Omega_{\mathcal{W}}} \frac{(2 \pi)^2}{\Omega_{\ell_i}} C\left(\ell_i\right) \delta_{ij}^{\rm K}, \\
    \mathrm{Cov}_{i j}^{\mathrm{T} 0} & =\frac{1}{\Omega_{\mathcal{W}}} \int_{|\vec{\ell}| \in \ell_i} \frac{\diff^2 \vec{\ell}}{\Omega_{\ell_i}} \int_{\left|\vec{\ell}^{\prime}\right| \in \ell_j} \frac{\diff^2 \vec{\ell}^{\prime}}{\Omega_{\ell_j}} \mathcal{T}\left(\vec{\ell},-\vec{\ell}, \vec{\ell}^{\prime},-\vec{\ell}^{\prime}\right), \\
    \mathrm{Cov}_{i j}^{\mathrm{SSC}} & =\frac{1}{\Omega_{\mathcal{W}}^2} \int \diff \chi \frac{\mathcal{K}^4(\chi)}{\chi^{6}} \frac{\partial P(k_{\ell_i} ; \chi)}{\partial \delta_\mathrm{b}} \frac{\partial P(k_{\ell_j} ; \chi)}{\partial \delta_\mathrm{b}} \int \frac{\diff^2 \vec{\ell}}{(2 \pi)^2} P_{\mathrm{L}}(k_\ell ; \chi)|\tilde{W}(\vec{\ell})|^2 \; . \label{eq:covSSC_angular_takada}
\end{align}
Let us now manipulate the third expression (which is essentially a \enquote{projected version} of Eq.~\ref{eq:cov_ssc_realspace}) further to arrive at the final form used in our work. 
\subsection{Multi-probe, tomographic case}
We now proceed to generalize Eq.~\eqref{eq:covSSC_angular_takada}, which allows computing the angular PS covariance between different $\ell$ bins (or, in its 3D version, between different $k$-bins) to take into account:
\begin{itemize}[noitemsep]
    \item multiple probes,
    \item multiple redshift bins,
    \item the covariance between $\delta_\mathrm{b}$ at different redshifts.
\end{itemize}
This will indeed be necessary for the full 3$\times$2pt tomographic \Euclid analysis.\\

We begin with the $\sigma^2$ term, generalizing the definition~\eqref{eq:sigma2_def_kspace} by taking into account the covariance of the background density at different redshifts $z_1, z_2$ \citep{LacasaRosenfeld2016}:
\begin{equation}
    \sigma^2(z_1, z_2) \equiv \langle \delta_\mathrm{b}(z_1) \delta_\mathrm{b}(z_2) \rangle = \frac{1}{\Omega_{\mathcal W}^2}
    \int \frac{\diff^3 \vec{k}}{(2 \pi)^3}\tilde{W}(\vec{k}, z_1)\tilde{W}(\vec{k}, z_2) P_\mathrm{L}(k, z_{12}) \; .
    \label{eq:sigma2_general}
\end{equation}
Note that we are now expressing the radial dependence in terms of $z$ instead of the comoving distance $\chi$. \FB{This function can also take into account disjoint patches, as long as they are captured by the survey window function and we want to account for the (super-sample) covariance between them}. We also introduced the linear matter cross-spectrum between $z_1$ and $z_2$, $P_\mathrm{L}(k, z_{12}) \equiv P_\mathrm{L}(k, z=0)D(z_1)D(z_2)$. In the full-sky case ($\Omega_{\mathcal W} = 4\pi$), the harmonic space transform of $W$ becomes a Kronecker delta $\delta^{\rm K}_{\ell0}$, reflecting the fact that only the monopole ($\ell = 0$) term contributes, and the projected background covariance, again in the Limber approximation, becomes:
\begin{align}
    \sigma^2(z_1, z_2) & = \frac{1}{(4\pi)^2}
    \int \frac{\diff^3 \vec{k}}{(2 \pi)^3}\mathrm{j}_0(k \chi_1)\mathrm{j}_0(k \chi_2) P_\mathrm{L}(k, z_{12}) \\ 
    & = \frac{1}{(4\pi)^2}\frac{1}{2 \pi^{2}} \int \diff k  k^{2} \mathrm{j}_{0}\left(k \chi_1\right) \mathrm{j}_{0}\left(k \chi_2\right)
    P_\mathrm{L}(k, z_{12}) \; , \label{eq:sigma2_pyssc_4pi2}
\end{align}
indicating with $\mathrm{j}_0(k \chi) = \mathrm{j}_{\ell=0}(k \chi)$ the 0-th order Bessel function and $\chi_i = \chi(z_i)$, and having integrated over the angular $\vec{k}$ coordinates.\\

\begin{figure}[ht]
    \centering
    \begin{minipage}[t]{0.49\textwidth}
        \centering
        \includegraphics[width=\linewidth]{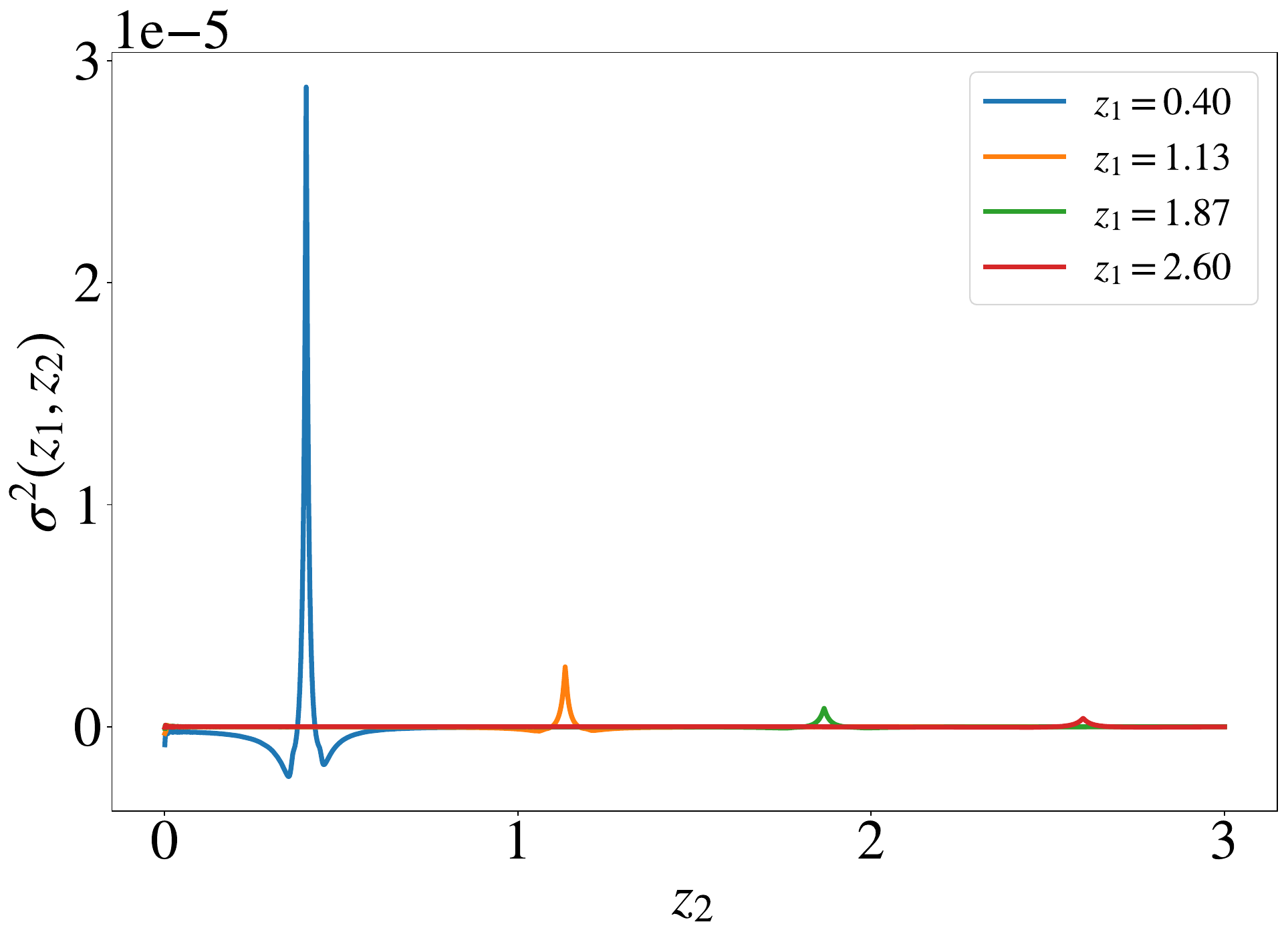}
        \caption{The dimensionless $\sigma^2(z_1, z_2)$, for four different values of $z_1$. The $(4\pi)^{-2}$ factor has not been included, since it is absorbed in the volume element definition as described in the text.}
        \label{fig:sigma2_spikes}
    \end{minipage}\hfill
    \begin{minipage}[t]{0.49\textwidth}
        \centering
        \includegraphics[width=\linewidth]{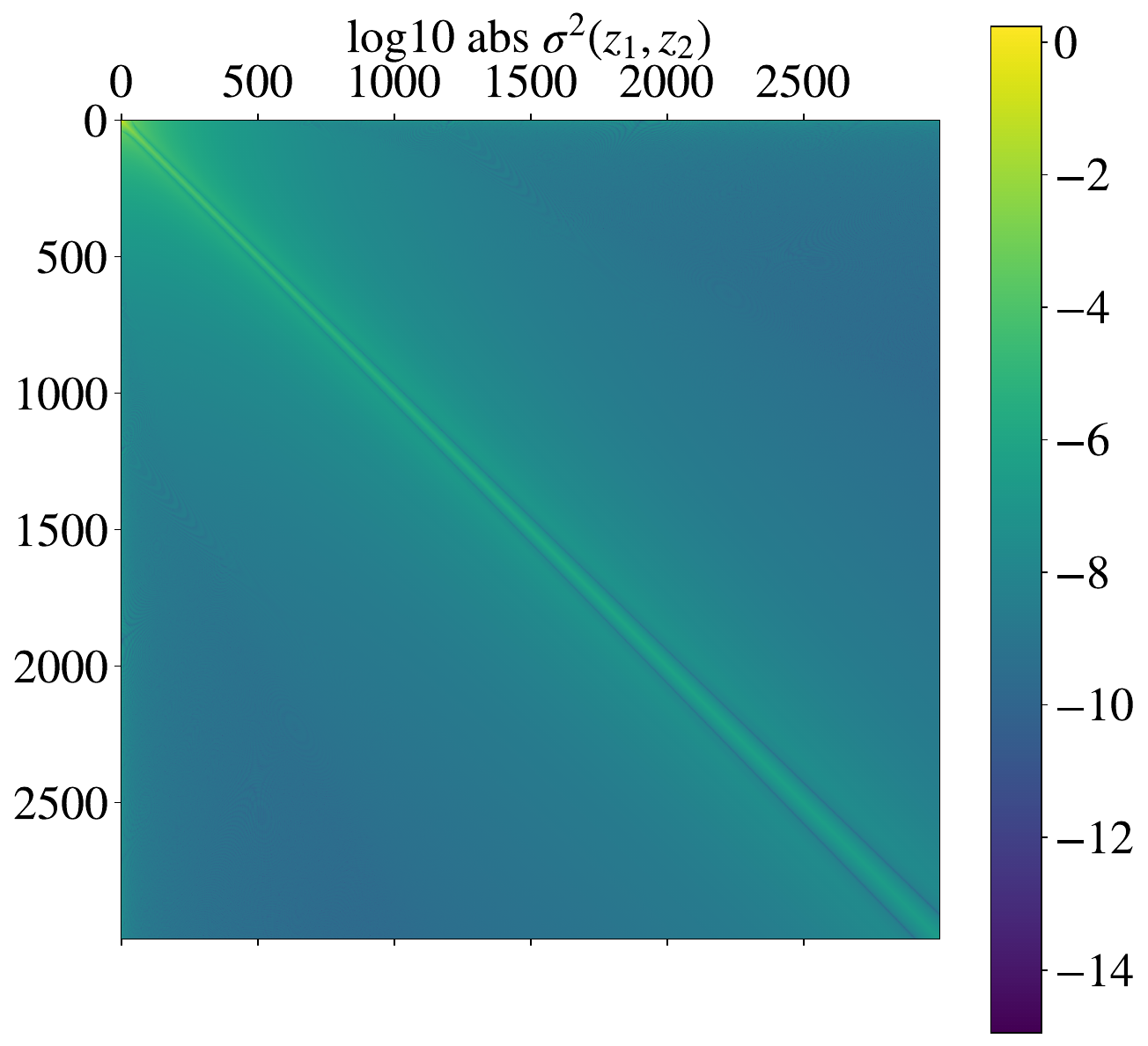}
        \caption{The same $\sigma^2(z_1, z_2)$, but displayed in matrix form. We take the absolute value to be able to plot in logarithmic scale also the negative values around the spikes, caused by the oscillatory nature of the Bessel functions in the integral. The larger values at $z_1 \sim z_2$ are clearly visible, and so is the increase of the function at lower $z$ (indicating larger covariance).}
        \label{fig:sigma2_matshow}
    \end{minipage}
\end{figure}

As for the remainder of the SSC expression, first of all, we can rewrite Eq.~\eqref{eq:cl_limber_takada} as 
\begin{equation}
    \frac{\partial C(\ell)}{\partial \chi} = \frac{\mathcal{K}^2(\chi)}{\chi^2} P(\vec{k}_\ell ; \chi) \; ;
\end{equation}
and, assuming that the radial kernels $\mathcal{K}(\chi)$ vary slowly in $\delta_\mathrm{b}$, we can plug this expression in Eq.~\eqref{eq:covSSC_angular_takada} to get
\begin{equation}\label{eq:cov_ssc_ellspace}
    \mathrm{Cov}_{i j}^{\mathrm{SSC}} =\frac{1}{\Omega_{\mathcal{W}}^2} \int \frac{\diff \chi}{\chi^2} \frac{\partial^2 C(\ell_i)}{\partial \delta_\mathrm{b}\partial \chi} 
    \frac{\partial^2 C(\ell_j)}{\partial \delta_\mathrm{b}\partial \chi}  
    \int \frac{\diff^2 \vec{\ell}}{(2 \pi)^2} P_{\mathrm{L}}(k ; \chi)|\tilde{W}(\vec{\ell})|^2 \; .
\end{equation}
We then use the expression for the volume element per steradian $\diff V = \chi^2 \diff \chi$ and generalize Eq.~\eqref{eq:covSSC_angular_takada} following \citealt{LacasaRosenfeld2016}, i.e., considering the redshift width (and asymmetry) of the $\sigma^2$ term (see Fig.~\ref{fig:sigma2_spikes}) and not treating it as a Dirac delta function of redshifts. In this way the SSC is expressed by a \textit{double} integral over redshift (or distance, or volume), and the general expression becomes:
\begin{equation}
    \mathrm{Cov}_{i j}^{\mathrm{SSC}} = \iint \diff V_1 \diff V_2  
    \frac{\partial^2 C(\ell_i)}{\partial \delta_\mathrm{b}\partial V_1} 
    \frac{\partial^2 C(\ell_j)}{\partial \delta_\mathrm{b}\partial V_2}  
    \sigma^2(z_1, z_2) \; .
\end{equation}
Note the slight abuse of notation, having absorbed the $\Omega_\mathcal{W}^{-2}$ factors in $\sigma^2(z_1, z_2)$ (cf. Eq.~\ref{eq:sigma2_general}) in the volume element per steradian $\diff V$. This expression can be further generalized for the multi-probe case as \citep{LacasaRosenfeld2016}:
\begin{equation}
    \mathrm{Cov}^{\mathrm{SSC}}(\mathcal{O}_1, \mathcal{O}_2) = 
    \iint \diff V_1 \diff V_2  
    \frac{\partial o_1}{\partial \delta_\mathrm{b}}(z_1)
    \frac{\partial o_2}{\partial \delta_\mathrm{b}}(z_2) 
    \sigma^2(z_1, z_2) \; ,
    \label{eq:ssc_lacasa_rosenfeld}
\end{equation}
where $o_i, \; i = 1, 2$ is the \textit{comoving density} of a generic LSS observable $\mathcal{O}_i$ defined as $\mathcal{O}_i \equiv \int \diff V  o_i$. \\
Taking into account the correlations of different probes in different redshift bins necessary for a full 3$\times$2pt tomographic analysis, we can rewrite the Limber angular PS \eqref{eq:cl_limber_takada} as
\begin{align}
    C^{AB}_{ij}(\ell) & = \int \diff \chi  \; \frac{\mathcal{K}^A_i(z) \mathcal{K}^B_j(z)}{\chi^{2}} P_{AB}(k_\ell , z) \label{eq:cijdef_Euclid_Takada} \\
    & = \int \diff V \; K^A_i(z) K^B_j(z) P_{AB}(k_\ell , z)\label{eq:cijdef_pyssc} \; ,
\end{align}
with $K^A_i(z) = \mathcal{K}^A_i(z) \chi^{-2}$, making it apparent that, for the case under study, $o_i = K^A_i(z) K^B_j(z) P_{AB}(k_\ell, z) = \partial C_{ij}^{AB}(\ell)/\partial V$.

The general for SSC~\eqref{eq:ssc_lacasa_rosenfeld} is expensive to compute, involving a double integral over the volume and an integral of two Bessel functions over $k$. Moreover, the $\sigma^{2}(z_1, z_2)$ function is sharply peaked around $z_1 = z_2$, making it necessary to compute it for a large number of values even in the full-sky case. Having derived the general expression for the SSC between two 2D-projected observables in different redshift bins, we can introduce the approximations used to speed up its computation, following the work of \cite{Lacasa_2019}.
\section{SSC approximations}\label{sec:ssc_approx}
As mentioned in the introduction of this Chapter, the code used to compute the SSC term is \texttt{PySSC}, a public \texttt{Python} module written by F. Lacasa et al., and presented alongside the paper \citealt{Lacasa_2019}. To speed up the expensive computation of the SSC integrals, we start from the general form of Eq.~\eqref{eq:ssc_lacasa_rosenfeld} which, for the case of interest, becomes
\begin{align}
    \mathrm{Cov}^{\mathrm{SSC}}\left[C^{AB}_{ij}(\ell), C^{CD}_{kl}(\ell ') \right] & = 
    \iint \diff V_1 \diff V_2  
    K^A_i(z_1)K^B_j(z_1)
    K^C_k(z_2)K^D_l(z_2) \nonumber \\
    & \quad \times 
    \frac{\partial P_{AB}(k_{\ell}, z_1)}{\partial \delta_\mathrm{b}}
    \frac{\partial P_{CD}(k_{\ell '}, z_2)}{\partial \delta_\mathrm{b}} 
    \sigma^2(z_1, z_2) \; ,
\end{align}
and assume \textit{the responses to vary slowly in redshift with respect to $\sigma^2(z_1, z_2)$}. This is the main approximation adopted, tested in the above-mentioned paper. This approximation hinges on the sharp redshift variability of $\sigma^2(z_1, z_2)$, which can be seen from Fig.~\ref{fig:sigma2_spikes}. We can then approximate the responses with their weighted average over the $K^A_i(z)$ kernels, indicated with an overbar:
\begin{align}
    \frac{\partial \bar{P}^{AB}(k_{\ell}, z)}{\partial \delta_{\rm b}}
    & = \frac{ \int \diff V \; K_{i}^{A}(z) K_{j}^{B}(z)\, \partial P_{AB}(k_{\ell}, z) / \partial \delta_{\rm b}}{\int \diff V \; K^A_i(z) K^B_j(z)} \nonumber \\
    & =  \frac{1}{I^{AB}_{ij}} \int \diff V \; K_{i}^{A}(z) K_{j}^{B}(z)\, \frac{\partial P_{AB}(k_{\ell}, z)}{\partial \delta_{\rm b}} \; ;
\label{eq:avg_response}
\end{align}
the denominator on the right-hand side of the first equation acts as a normalization term, which we call $I^{AB}_{ij}$. We then pull the average responses out of the $\diff V$ integral:
\begin{align}
    \mathrm{Cov}^{\mathrm{SSC}}\left[C^{AB}_{ij}(\ell), C^{CD}_{kl}(\ell ') \right] & \simeq 
    \frac{\partial \bar{P}^{AB}(k_{\ell}, z_1)}{\partial \delta_\mathrm{b}}
    \frac{\partial \bar{P}^{CD}(k_{\ell '}, z_2)}{\partial \delta_\mathrm{b}} \nonumber \\
    & \quad \times \iint \diff V_1 \diff V_2  
    K^A_i(z_1)K^B_j(z_1)
    K^C_k(z_2)K^D_l(z_2)
    \sigma^2(z_1, z_2) \; .
\end{align}
Additionally, we can further manipulate Eq.~\eqref{eq:avg_response} by factorising the probe response as
\begin{equation}
    \frac{\partial P_{AB}(k_\ell , z)}{\partial \delta_{\rm b}} = 
    R^{AB}(k_\ell, z) P_{AB}(k_\ell , z) \; ,
    \label{eq:3dresponses}
\end{equation}
where $R^{AB}(k_\ell, z)$ is the first-order response coefficient defined in Eq.~\eqref{eq:R_n_def}, which is given, for the matter-matter case, by Eq.~\eqref{eq:R_1_mm_expansion}. We can introduce the probe response coefficients of the \textit{angular} power spectra $\mathcal{R}^{AB}_{ij}(\ell)$ in a similar way, using Eq.~\eqref{eq:cijdef_pyssc}:
\begin{align}
     \frac{\partial}{\partial \delta_{\rm b}} C^{AB}_{ij}(\ell) & = \frac{\partial}{\partial \delta_{\rm b}} \int \diff V \; K_{i}^{A}(z) K_{j}^{B}(z)\,P_{AB} (k_{\ell}, z) \nonumber \\
     & = \int \diff V \; K_{i}^{A}(z) K_{j}^{B}(z)\, \frac{\partial P_{AB} (k_{\ell}, z)}{ \partial \delta_{\rm b}} \nonumber \\
    & = \int \diff V \; K_{i}^{A}(z) K_{j}^{B}(z)\, R^{AB}(k_{\ell}, z) P_{AB} (k_{\ell}, z) \nonumber \\
     &\equiv \mathcal{R}^{AB}_{ij}(\ell) C^{AB}_{ij}(\ell) \; .
     \label{eq:2dresponses}
\end{align}
We can now substitute Eq.~\eqref{eq:2dresponses} into the r.h.s. of Eq.~\eqref{eq:avg_response} and adopt the full-sky approximation by dividing the SSC by $f_{\rm sky}$, to get the expression which will be used throughout this work:
\begin{equation} 
    {\rm Cov^{SSC}}\left[C^{AB}_{ij}(\ell),C^{CD}_{kl}(\ell')\right] \simeq \\
    \frac{\mathcal{R}^{AB}_{ij}(\ell)C^{AB}_{ij}(\ell)\   \mathcal{R}^{CD}_{kl}(\ell')C^{CD}_{kl}(\ell')\ S^{A,B;C,D}_{i,j;k,l}}{f_\mathrm{sky}}  \, .
    \label{eq:covssc_sijkl}
\end{equation}
In the above equation, we have defined
\begin{equation}\label{eq:sijkl}
        S^{A,B;C,D}_{i,j;k,l} \equiv \int \diff V_1 \diff V_2 \; \frac{K^A_i(z_1)K^B_j(z_1)}{I^{AB}_{ij}} \, 
        \frac{K^C_k(z_2)K^D_l(z_2)}{I^{CD}_{kl}}\, \sigma^2(z_1, z_2) \, .
\end{equation}
The $S^{A,B;C,D}_{i,j;k,l}$ matrix (referred to as $S_{ijkl}$ from here on) is the volume average of $\sigma^2(z_1, z_2)$, and is a dimensionless quantity; it is the actual output of \texttt{PySSC} -- we therefore need the additional ingredients appearing in Eq.~\eqref{eq:covssc_sijkl} to compute the full SSC term. A description of the way this code has been used, and some comments on the inputs to provide and the outputs it produces, can be found in Sect.~\ref{sec:validation}. Once the $S_{ijkl}$ matrix has been computed, then, Eq.~\eqref{eq:covssc_sijkl} makes the computation of the SSC have the same numerical cost as the computation of the angular PS themselves.\\

The validity of Eq.~\eqref{eq:covssc_sijkl}, commented in Appendix~\ref{sec:Sijkl_approximation}, has been tested in \citetalias{Lacasa_2019} in the case of GCph and found to reproduce the Fisher matrix \citep{tegmark1997} elements and signal-to-noise ratio from the original expression (Eq.~\ref{eq:ssc_lacasa_rosenfeld}):
\begin{itemize}
    \item within 10\% discrepancy up to $\ell \simeq 1000$ for $R^{AB}_{ij}(k_\ell, z) = {\rm const}$; 
    \item within 5\% discrepancy up to $\ell \simeq 2000$ when using the linear approximation in scale for $R^{AB}(k_\ell, z)$ provided in Appendix C of the same work.
\end{itemize}
The necessity to push the analysis to smaller scales, as well as to investigate the SSC impact not only for GCph but also for WL and their cross-correlation, has motivated a more exhaustive characterization of the probe response functions, which will be detailed in the next section.\\

Lastly, we note that in Eq.~\eqref{eq:covssc_sijkl} we account for the sky coverage of the survey through the full-sky approximation: 
$\mathrm{Cov}^\mathrm{SSC}_\mathrm{part. \; sky} \simeq 
\mathrm{Cov}^\mathrm{SSC}_\mathrm{full \; sky}/f_\mathrm{sky}$. The validity of this approximation has been discussed in \citet{beauchamps2021}, and found to agree at the percent level on the marginalized parameter constraints with the more rigorous treatment accounting for the exact survey geometry when considering large survey areas. For this test, the authors considered an area of $15\,000 \ \rm{deg}^2$ and a survey geometry very close to what \Euclid will have, i.e. the full sky with the ecliptic and galactic plane removed. Intuitively, the severity of the SSC decays as $f_{\rm sky}^{-1}$ because larger survey volumes are able to accommodate more Fourier modes, although as mentioned in Sect.~\ref{sec:pk_cov_window} the SSC term has a non-trivial scaling with the survey volume.\\

Having presented the theoretical framework and approximations used in our study, we can proceed to quantify the impact of this additional systematic effect on the \Euclid photometric survey. To do this, we will introduce the additional ingredients and methodology used in the forecast analysis. 
\chapter{\ML{Super-sample covariance} for the \Euclid photometric survey}\label{chap:SSC_for_Euclid}
This chapter will delve into the specifics and results obtained for the computation of the SSC term, computed with the formalism and approximation introduced in the last chapter, to the 3$\times$2pt survey for the \Euclid mission. 
\section{Forecast specifics}
\label{sec:specifics}
The forecast analysis presented below has been performed with the goal of improving the work of the inter-science taskforce for forecasting (\enquote{IST:F}: \citetalias{ISTF2020}, based in turn on the \Euclid Red Book \citealt{laureijs2011euclid}), mainly by considering SSC as an additional source of uncertainty beyond the already considered sample variance, shot and shape noise. We will take into account some updates to the \citetalias{ISTF2020} prescriptions based on more recent results by the Euclid Collaboration. In particular, the updates concern the fiducial value of the linear bias, redshift distribution $n(z)$ and multipole binning, as detailed below.

\subsection{Cosmological model and matter power spectrum}
The baseline fiducial cosmological model adopted in this work is the flat $ w_0w_a$CDM model, i.e., we model the dark energy equation of state with a CPL parameterization (Eq.~\ref{eq:CPL}). We also include the contribution from massive neutrinos with total mass set to the \Planck preferred value \ML{(consistent with oscillation experiments)} $\sum{m_{\nu}} = 0.06 \ {\rm eV}$, which we keep fixed in the FM analysis. The cosmological parameters' vector is then 
\begin{equation}\label{param_vector}
\vec{\theta}_{\rm cosmo} = \left\{\Omega_{{\rm m},0}, \Omega_{{\rm b},0}, w_0, w_a, h, n_{\rm s}, \sigma_8\right\} \; ;
\end{equation}
we will expand this parameter basis in the variations explored later on, e.g. considering non-flat models. The parameters' fiducial values are set to the latest \Planck results (\citealt{Planck2020_params}, \citetalias{ISTF2020}); \ML{for $w_0$ and $w_a$, we choose $(-1, 0)$ respectively, which describe a time-independent equation of state}: 
\begin{equation}\label{theta_fid}
\vec{\theta}_{\rm cosmo}^{\, \rm fid} = \left\{0.32, 0.05, -1.0, 0.0, 0.67, 0.96, 0.816\right\} \; . 
\end{equation}
These are used as input for the evaluation of the fiducial linear and nonlinear matter PS, which is obtained using the \texttt{TakaBird} recipe. This is implemented in both \texttt{CAMB} and \texttt{CLASS}. 
\subsection{Radial kernels}
Fig.~\ref{fig:WF_FS1_for_PhD_thesis} shows the redshift dependence of the radial kernels defined in Eqs.~\eqref{eq:wildef} and \eqref{eq:wigdef}, for the different redshift bins. The two sets of kernels have very different shapes, because of the different nature of the probes they represent. In particular, the left panel shows the very large support in $z$ of the lensing kernels, due to the integrated nature of the effect: the shear signal originating from the sources in the furthest bins is created by the lenses spanning the \textit{whole} radial distance from said sources to the observer. This means that the sources from all bins receive a non-negligible contribution from the low-$z$ region, a fact of great importance in the impact of SSC, as explained in Sect.~\ref{sec:ssc_ref_results}. This also means that, contrary to GCph, the off-diagonal tomographic correlations $i \neq j$, quantified by the angular PS $C_{ij}^{AB}(\ell)$, will never be null, because the overlap between the kernels is non-zero even for distant redshift bins.\\
The large support in $z$ also means that, when performing the integral of Eq.~\eqref{eq:cijdef_Euclid} (or equivalent forms), the signal at a single angular scale $\ell$ will receive contributions from the PS computed in a large set of $k$ values: $k_\ell = \left. (\ell + 1/2)\chi^{-1}(z) \right|_{z \in z_\mathrm{int}}$ with $z_\mathrm{int}$ the $z$ range of the $C_{ij}^{AB}(\ell)$ integral. This implies a non-univocal correspondence between angular and three-dimensional scales, which in turn complicates the process of imposing scale cuts. A potential solution to this problem will be discussed in Chap.~\ref{chap:scalecuts}.\\

For GCph, we choose to include the galaxy bias term $b_i(z)$ in the PS (see Eq.~\ref{eq:pk_mm_gm_gg}) rather than in the galaxy kernel, as opposed to what has been done in \citetalias{ISTF2020}. This is done to compute the galaxy response as described in Sect.~\ref{sec:multi_probe_response}. Since the galaxy bias is assumed constant in each bin, however, in this case the question is of no practical relevance when computing the $S_{ijkl}$ matrix, since the constant bias cancels out.

We remind the reader once again that the above definitions of the lensing and galaxy kernels ($\mathcal{K}_i^A(z), \ A = {\rm L, G}$) differ from the ones used in \citetalias{Lacasa_2019} -- for which we used the notation $K_i^A(z)$. This is simply because of a different definition of the $C_{ij}^{AB}(\ell)$ Limber integral, which is performed in $\diff V$ in \citetalias{Lacasa_2019} (i.e., using Eq.~\ref{eq:cijdef_pyssc}) and in $\diff z$ in \citetalias{ISTF2020}. The mapping between the two conventions is simply given by the expression for the volume element:
\begin{equation}
    \diff V = \chi^2(z)\frac{\diff \chi}{\diff z}\diff z = \frac{c}{H_0} \frac{\chi^2(z)}{E(z)}\diff z \; ,
\end{equation}
and 
\begin{equation}\label{eq:wfmatch}
    K^A_i(z) = \mathcal{K}^A_i(z)/\chi^2(z) \; , 
\end{equation}
so that the \Euclid form of the $C^{AB}_{ij}(\ell)$ integral is simply
\begin{equation}
    C^{AB}_{ij}(\ell) = \frac{c}{H_0}
    \int \diff z \frac{\mathcal{K}^A_i(z) \mathcal{K}^B_j(z)}{E(z)\chi^2(z)} \;  P_{AB}(k_\ell , z) \label{eq:cijdef_Euclid} 
\end{equation}

with $A = {\rm L, G}$. In Fig.~\ref{fig:WF_FS1_for_PhD_thesis} we plot the values of $\mathcal{K}^A_i(z)$ to facilitate the comparison with \citetalias{ISTF2020}. As outlined in Appendix~\ref{sec:validation_appendix}, when computing the $S_{ijkl}$ matrix through \texttt{PySSC}, the user can either pass the kernels in the form used in \citetalias{Lacasa_2019} ($K_i^A$), or the one used in \citetalias{ISTF2020} (${\mathcal K}_i^A$) while specifying a non-default \texttt{convention} parameter.

\subsection{Multi-probe response}\label{sec:multi_probe_response}
\begin{figure}
    \centering
    \includegraphics[width=0.7\linewidth]{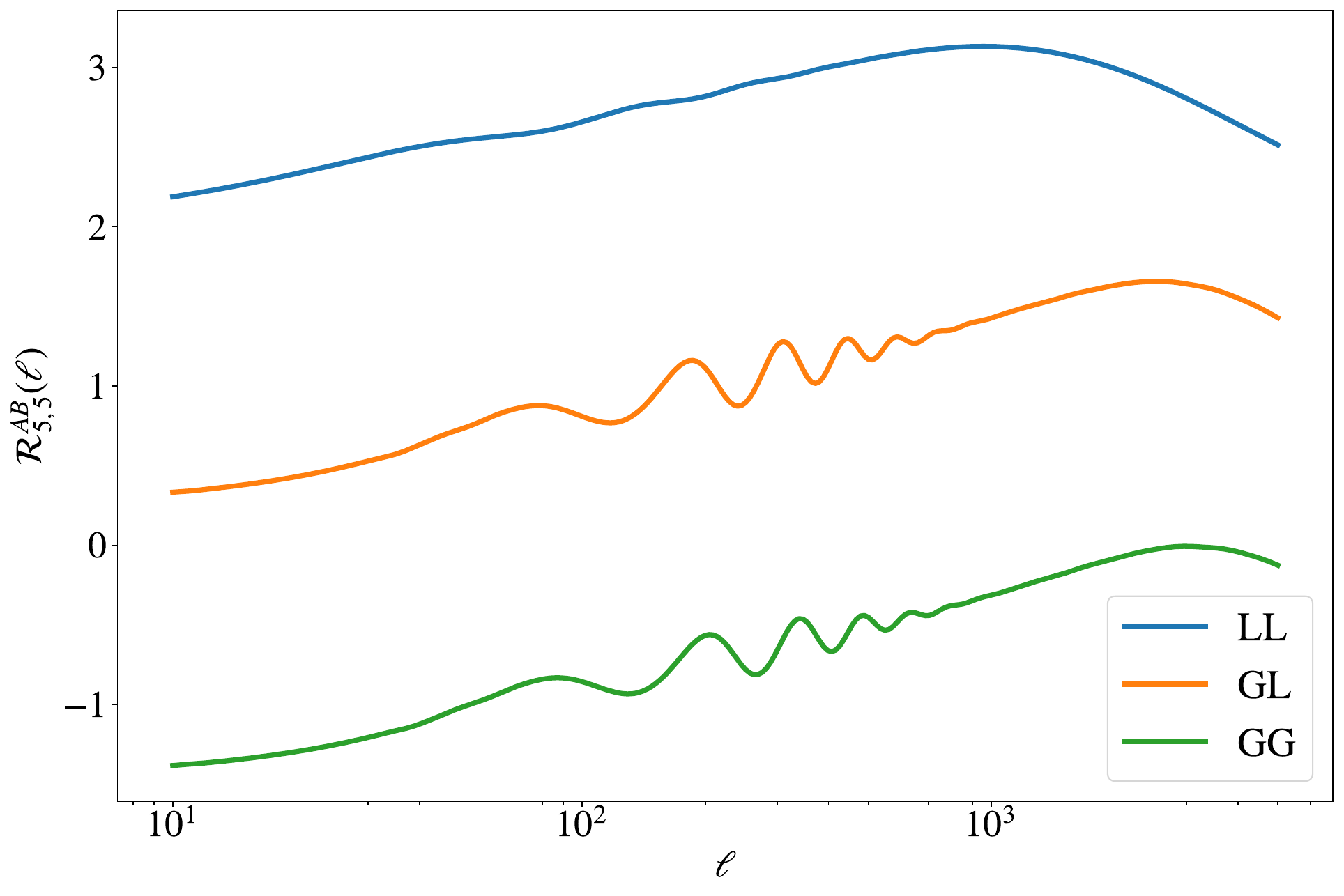}
    \caption{Projected response coefficients for the WL and GCph probes and their cross-correlation, for the central redshift bin considered ($0.8 \lesssim z \lesssim 0.9$; see Sect.~\ref{sec:z_distribution})  -- the shape and amplitude of the functions for different redshift pairs are analogous. For WL, the baryon acoustic oscillation wiggles are smoothed out by the projection, due to the kernels being larger than the GCph ones. The different amplitude of the response is one of the main factors governing the severity of SSC.}
    \label{fig:responses_vinc_allprobes}
\end{figure}

The responses for the different probes can be obtained in terms\footnote{Since we are using the nonlinear matter power spectrum $P_{\rm mm}(k, z)$, we do not force $R^{\rm mm}(k, z)$ to reduce to its linear expression, that is to say, we do not set $G_{1}^{\rm mm} = 26/21$ in Eq.~\eqref{eq:R_1_mm_expansion}.} of $R^{\rm mm}(k, z)$ by plugging the relations between matter and galaxy PS (Eq.~\ref{eq:pk_mm_gm_gg}) into Eq.~\eqref{eq:3dresponses} -- see also \citet{Krause2017} for a similar derivation of the multi-probe case, using the halo model responses\footnote{Note that, as of the time of writing, there is a typo in Eq.~(A11) of the aforementioned paper: the factor in front of the logarithmic derivative of the PS should be -1/3 instead of -1/2.}.\\
We note that the linear modelling for the galaxy bias is not rigorous at high $\ell$ and scale cuts should be performed in order to avoid biasing the constraints, but in the present context, we are more interested in the relative impact of SSC on the constraints than the constraints themselves. Any systematic error due to this approximate modelling should roughly cancel out in the ratio we will compute later on.\\
We also note that we choose to include a perfectly Poissonian shot noise term in the covariance matrix, rather than in the galaxy PS of Eq.~\eqref{eq:pk_mm_gm_gg}, as can be seen in Eq.~\eqref{eq:noiseps}. Again, we restrict our study to the first-order isotropic response $R_1(k, z)$, so we drop the subscript \enquote{1} from hereon:
\begin{equation}\label{eq:Rgg}
    R^{\rm gg}(k, z) = \frac{\partial \ln P_{\rm gg}(k,z)}{\partial \delta_{\rm b}} = R^{\rm mm}(k, z) + 2b_{(1)}^{-1}(z) \left[ b_{(2)}(z) - b_{(1)}^2(z) \right],
\end{equation}
and similarly for $R^{\rm gm}$:
\begin{equation}\label{eq:Rgm}
    R^{\rm gm}(k, z) = \frac{\partial \ln P_{\rm gm}(k,z)}{\partial \delta_{\rm b}} = R^{\rm mm}(k, z) + b_{(1)}^{-1}(z)\left[ b_{(2)}(z) - b_{(1)}^2(z) \right].
\end{equation}
Having used the definitions of the first and second-order galaxy bias, i.e., $b_{(1)}(z) = (\partial n_{\rm g} / \partial \delta_{\rm b})/n_{\rm g}$ and $b_{(2)}(z) = (\partial^2 n_{\rm g} / \partial \delta_{\rm b}^2)/n_{\rm g}$, with $n_{\rm g}$ the total angular galaxy number density, in ${\rm arcmin}^{-2}$. In the following, where there is no risk of ambiguity, we will also drop the subscript in parenthesis when referring to the first-order galaxy bias -- i.e., $b(z) = b_{(1)}(z)$ -- to shorten the notation, and we will indicate the value of the first-order galaxy bias in the $i$-th redshift bin with $b_i(z)$.
Equations~\eqref{eq:Rgg}--\eqref{eq:Rgm} are obtained by differentiating a PS model for a galaxy density contrast defined with respect to the \textit{observed} galaxy number density, and so they already account for the fact that the latter also ``responds'' to the large scale perturbation $\delta_{\rm b}$ -- hence no rescaling of the response, shown in Eq.~\eqref{eq:pk_response_rescaling} is required. This is also the reason why $R^{\rm GG}_{ij}(\ell)$ can have negative values: for galaxy clustering, the (number) density contrast $\delta_{\rm gal}$ is measured w.r.t. the observed, local number density ${\bar n}_{\rm gal}$: $\delta_{\rm gal} = n_{\rm gal}/{\bar n}_{\rm gal} - 1$. The latter also responds to a background density perturbation $\delta_{\rm b}$, and it can indeed happen that ${\bar n}_{\rm gal}$ grows with $\delta_{\rm b}$ faster than $n_{\rm gal}$, which leads to $\delta_{\rm gal}$ decreasing with increasing $\delta_{\rm b}$ (which also implies $\partial C^{\rm GG}_{ij}(\ell) / \partial \delta_{\rm b} < 0$).
We also stress the fact that the second-order galaxy bias appearing in the galaxy-galaxy and galaxy-lensing response coefficients is not included in the signal, following \citetalias{ISTF2020}. Once computed in this way, the response coefficients can be projected in harmonic space using Eq.~\eqref{eq:2dresponses}, and inserted in Eq.~\eqref{eq:covssc_sijkl} to compute the SSC in the \citetalias{Lacasa_2019} approximation. The projected ${\mathcal R}^{AB}_{ij}(\ell)$ functions are shown in Fig.~\ref{fig:responses_vinc_allprobes} for all the probes combinations considered.
\subsection{Higher-order bias}\label{sec:higher_order_bias}
To compute the galaxy-galaxy and galaxy-galaxy lensing probe response terms (Eqs.~\ref{eq:Rgg} and~\ref{eq:Rgm}) we need the second-order galaxy bias $b_{(2)}(z)$. This can be estimated through the halo model\footnote{We neglect here the response of $\langle N\vert M \rangle$ to a perturbation $\delta_{\rm b}$ in the background density.} as (\citealt{VB21, Alex2021}; see also Appendix C of \citetalias{Lacasa_2019}):
\begin{equation}
b_{(i)}(z) = \int{\diff M \; \Phi_{\rm MF}(M, z) b_{(i)}^{\rm h}(M, z) \langle N \vert M \rangle/n_{\rm g}(z)} \; ,
\label{eq:bicalc}
\end{equation}
with 
\begin{equation}
n_{\rm g}(z) = \int{\diff M \; \Phi_{\rm MF}(M, z) \langle N|M \rangle } \; .
\label{eq:ngalvszed}
\end{equation}
We remind the meaning of the different terms entering the last two equations, some of which have been introduced in Sect.~\ref{sec:pk_nonlin_halomodel}; $n_{\rm g}$ is the galaxy number density, $\Phi_{\rm MF}(M, z)$ the halo mass function (HMF), $b_{(i)}^{\rm h}(M, z)$ the $i$-th order \textit{halo} bias, and $\langle N|M \rangle$ the average number of galaxies hosted by a halo of mass $M$ at redshift $z$ (given by the HOD). These are integrated over the mass range $\log_{10} M \in [9,16]$, with the mass expressed in units of solar masses. The expression for the $i$-th order galaxy bias (Eq.~\ref{eq:bicalc}) is the same as Eq.~(C.2) of \citetalias{Lacasa_2019}, but here we are neglecting the scale dependence of the bias evaluating it at $k = 0$ so that $\Tilde{u}(k \, | \, M = 0, z) = 1$, $\Tilde{u}(k \, | \, M, z)$ being the Fourier Transform of the halo profile. Strictly speaking, this gives us the large-scale bias, but it is easy to check that the dependence on $k$ is negligible over the range of interest.

Although Eq.~\eqref{eq:bicalc} allows the computation of both the first and second-order galaxy bias, we prefer to use the values of $b_{(1)}(z)$ measured from the Flagship simulation for the selected galaxy sample; this is to maintain consistency with the choices presented in Sect.~\ref{sec:gal_bias}. For each redshift bin, we vary (some of) the HOD parameters to fit the measured $b_{(1)}(z)$, thus getting a model for $b_{(1)}^{\rm h}(z)$. We then compute $b_{(2)}^{\rm h}(z)$ using as an additional ingredient the following relation between the first and second-order halo bias, which approximates the results from separate universe simulations \citep{Lazeyras_2016} within the fitting range $1 \lesssim b_{(1)}^{\rm h} \lesssim 10$: 
\begin{align} \label{eq:b2vsvb1}
b_{(2)}^{\rm h}(M, z) & = 0.412 - 2.143 \, b_{(1)}^{\rm h}(M, z) \nonumber \\
& + 0.929 \, \left[b_{(1)}^{\rm h}(M, z)\right]^2 
+ 0.008 \, \left[b_{(1)}^{\rm h}(M, z)\right]^3 \; . 
\end{align}
Finally, we plug the $b_{(2)}^{\rm h}$ values obtained in this way back into Eq.~\eqref{eq:bicalc} to get the second-order galaxy bias. The details of the HMF and HOD used and of the fitting procedure are given in Appendix \ref{sec:halomodel_appendix}.
\subsection{Gaussian covariance}\label{sec:covgauss}
\begin{figure}
    \centering
    \includegraphics[width=\textwidth]{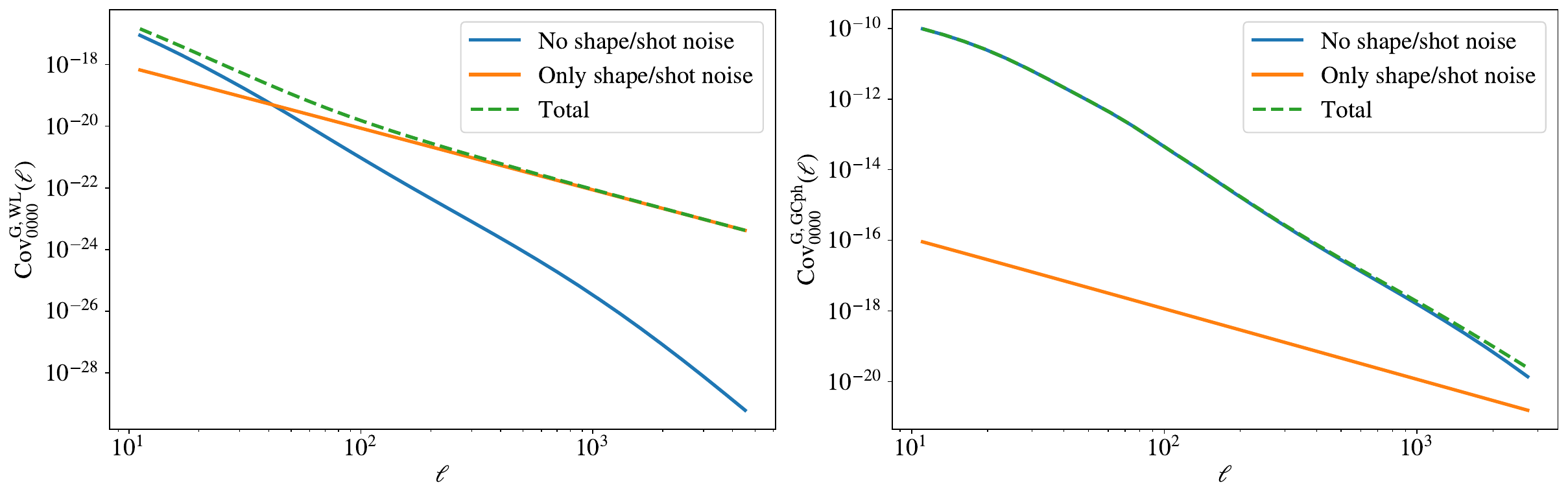}
    \caption{Gaussian covariance with and without the noise terms $N^{AB}_{ij}(\ell)$  (cf. Eq.~\ref{eq:covgauss}), for the first redshift bin: $z_i, z_j, z_k, z_l = 0, 0, 0, 0$. The shape noise is more important for WL than the shot noise for GCph, becoming dominant before $\ell \sim 100$.}
    \label{fig:sample_noise_terms_covg}
\end{figure}

As mentioned in Sect.~\ref{sec:ssc_introduction}, the Gaussian part of the covariance matrix encapsulates the sample variance (and, from hereon, the measurement noise) for perfectly Gaussian-distributed fields. For a multi-probe tomographic harmonic analysis, Eq.~\eqref{eq:cov_gauss_kikj} generalizes to the following expression \citepalias{ISTF2020}:
\begin{align}\label{eq:covgauss}
{\rm Cov^{\rm G}} & \left[C_{ij}^{AB}(\ell), C_{k l}^{C D}(\ell^{\prime})\right]=
\left[(2\,\ell + 1)\,f_{\rm sky} \, \Delta \ell \right]^{-1}
\,\delta_{\ell \ell^{\prime}}^{\rm K}  \nonumber \\
 \times \, \Bigg\{  & \left[C_{ik}^{AC}(\ell) + {N}_{ik}^{AC}(\ell)\right]
\left[C_{jl}^{BD}(\ell') + N_{jl}^{BD}(\ell') \right]  \nonumber \\
+ & \left[C_{il}^{AD}(\ell) + N_{il}^{AD}(\ell) \right]
\left[C_{jk}^{BC}(\ell') + N_{jk}^{BC}(\ell') \right] 
\Bigg\}  \; ,
\end{align}
where the noise PS $ N_{ij}^{AB}(\ell)$, which were lacking in Eq.~\eqref{eq:cov_gauss_kikj}, are, for the different probe combinations:
\begin{equation}
N_{ij}^{AB}(\ell) = \left \{
\begin{array}{ll}
\displaystyle{(\sigma_{\epsilon}^2/\bar{n}_{i}^{\rm G}) \, \delta_{ij}^{\rm K}} & \displaystyle{A = B = {\rm L} \;\; ({\rm WL})} \\
 & \\
\displaystyle{0} & \displaystyle{A \neq B} \\
 & \\
\displaystyle{(1/\bar{n}_{i}^{\rm L}) \, \delta_{ij}^{\rm K}} & \displaystyle{A = B = {\rm G} \;\; ({\rm GCph}}) \; . \\
\end{array}
\right . 
\label{eq:noiseps}
\end{equation}
In the above equations $\sigma_{\epsilon}^2$ is the variance of the total intrinsic ellipticity dispersion of WL sources  -- where $\sigma_{\epsilon} = \sqrt{2}\sigma_\epsilon^{(i)}$, $\sigma_\epsilon^{(i)}$ being the ellipticity dispersion per component of the galaxy ellipse. We assume no correlation between shot and shape noise, resulting in $N_{ij}^{\rm GL}(\ell) = N_{ij}^{\rm LG}(\ell) = 0$, and no inter-bin noise, as enforced by the Kronecker delta $\delta_{ij}^{\rm K}$. The noise spectra are scale-independent, which makes them the dominant contribution at large $\ell$, where the sample variance term goes to zero because of the larger number of modes available (see Eq.~\ref{eq:cov_gauss_kikj} and Fig.~\ref{fig:sample_noise_terms_covg}).
We note that the average densities used in Eq.~\eqref{eq:noiseps} are not the total number densities, but rather those in the $i$-th redshift bin. In the case of ${\cal N}_{\rm b}$ equipopulated redshift bins, they can be simply written as $\bar{n}_{i}^{A} = \bar{n}^{A}/{\cal N}_{\rm b}$ for both $A = ({\rm L, G})$. Finally, we recall that $f_{\rm sky}$ is the fraction of the total sky area covered by the survey, while $\Delta \ell$ is the width of the multipole bin centered on a given $\ell$. From Sect.~\ref{sec:z_distribution} we have that $\bar{n} = 28.73 \, {\rm arcmin}^{-2}$, while we set $\sigma_\epsilon = 0.37$ \citep[from the value $\sigma_\epsilon^{(i)}=0.26$ reported in][]{Martinet_2019} and $f_{\rm sky} = 0.356$ (corresponding to $\Omega_\mathrm{S} = 14~700$ deg$^2$). \\
To make the computation of this covariance term, key to any likelihood analysis, quick and straightforward, I have prepared an optimized \texttt{Python} module, \texttt{Spaceborne\_covg}\footnote{\texttt{\url{https://github.com/davidesciotti/Spaceborne\_covg}}}, available as a public repository. It only requires as inputs the signal, the noise, and the convention of choice for the ordering of the resulting two-dimensional covariance matrix, and it is currently being used by other groups within the Euclid Collaboration (for example, in the validation of the \texttt{CLOE} code and in the Key Paper \enquote{Cosmological constraints on non-standard
cosmologies from simulated \Euclid probes} (D'Amico et al, in prep)).\\

In the context of the present work, we do not consider the other non-Gaussian contribution to the total covariance matrix, the so-called connected non-Gaussian (cNG) term. This additional non-Gaussian term has been shown to be sub-dominant with respect to the Gaussian and SSC terms for WL both in \citet{Barreira2018cosmic_shear} and in \citet{upham2021}. For what concerns galaxy clustering, \citet{wadekar_20} showed that the cNG term was subdominant, but this was for a spectroscopic sample so (i) they had a much larger contribution from shot-noise-related terms compared to what is considered here for the \Euclid photometric sample, and (ii) they considered larger and more linear scales than in the present study. \citet{Lacasa2020_braiding} showed that the cNG term in the covariance matrix of GCph only impacts the spectral index $n_\mathrm{s}$ and HOD parameters, but there are a few differences between that analysis and the present work, such as the modelling of galaxy bias. Thus it is still unclear whether the cNG term has a strong impact on cosmological constraints obtained with GCph. Quantifying the impact of this term for the 3$\times$2pt analysis with \Euclid settings is left for future work.
\subsection{Data vectors and Fisher matrix}
\label{sec:datavectors}
We can now proceed to set some data-related specifications.\\
First, we assume to measure $C_{ij}^{AB}(\ell)$ in 10 equipopulated redshift bins over the redshift range $(0.001, 2.5)$. When integrating Eq.~\eqref{eq:cijdef_Euclid} in $\diff z$, $z_{\rm max}$ must be larger than the upper limit of the last redshift bin to account for the broadening of the bin redshift distribution due to photo-$z$ uncertainties. We have found that the $C_{ij}^{AB}(\ell)$ stop varying for $z_{\rm max} \gtrsim 4$, which is what we take as the upper limit in the integrals over $z$.
This also means that we need to extrapolate the bias beyond the upper limit of the last redshift bin; we then take its value as constant and equal to the one in the last redshift bin, that is, $b(z > 2.501) = b_{10}$.

Second, we assume the same multipole limits as in \citetalias{ISTF2020}, hence examining two cases, namely
\begin{itemize}
\item[-]{{\it pessimistic:}
\begin{displaymath}
(\ell_{\rm min}, \ell_{\rm max}) = \left \{
\begin{array}{ll}
(10, 1500) & {\rm for \ WL} \\
 & \\
(10, 750) & {\rm for \ GCph \ and \ XC}
\end{array}
\right . \; ,
\end{displaymath}}
\item[-]{{\it optimistic:}
\begin{displaymath}
(\ell_{\rm min}, \ell_{\rm max}) = \left \{
\begin{array}{ll}
(10, 5000) & {\rm for \ WL} \\
 & \\
(10, 3000) & {\rm for \ GCph \ and \ XC}
\end{array}
\right . \; .
\end{displaymath}}
\end{itemize}
\ML{These limits were set in \citetalias{ISTF2020} by a preliminary study of the impact of non-Gaussian contributions to the covariance matrix, whose characterization is precisely the aim of the present Chapter. For WL, a $30\%$ signal-to-noise decrease was found when including non-Gaussian covariance, corresponding to an effective $\ell$ cut at $\ell_{\rm max} = 1420$, leading to the pessimistic choice of $\ell_{\rm max}^{\rm WL} = 1500$. Similarly, for GCph, the signal-to-noise saturation above $\ell_{\rm max} \gtrsim 500$ -- a similar value being found also in \citetalias{Lacasa_2019} -- is quoted as the main reason for the pessimistic cut at $\ell_{\rm max}^{\rm GCph} = 750$; however, as acknowledged by the authors and as shown in Sect.~\ref{sec:impact}, the signal-to-noise is not necessarily a faithful indicator of the impact of the additional covariance on parameter constraints. In the rest of the Chapter we keep using these as upper limits on the multipole, to facilitate comparison with the original results of \citetalias{ISTF2020}; these scale cuts will have to be updated by more nuanced analyses, accounting for, amongst other things, the SSC impact on marginalized uncertainties and the additional considerations introduced in Chapter~\ref{chap:scalecuts}.}\\
As for the multipole binning, instead of dividing these ranges into ${\cal N}_{\ell}$ (logarithmically equispaced) bins in all cases as is done in \citetalias{ISTF2020}, we follow the most recent prescriptions of the EC and proceed as follows:
\begin{itemize}
    \item we fix the centers and edges of 32 bins (as opposed to 30) in the $\ell$ range $[10, 5000]$ following the procedure described in Appendix~\ref{sec:appendix_binning}. This will be the $\ell$ configuration of the optimistic WL case.
    \item The bins for the cases with $\ell_{\rm max} < 5000$, i.e., WL pessimistic, GCph and XC, are obtained by cutting the bins of the optimistic WL case with $\ell_{\rm center} > \ell_{\rm max}$. This means that instead of fixing the number of bins and having different bins' centers and edges as done in \citetalias{ISTF2020}, we fix the bins' centers and edges and use a different number of bins, resulting in, e.g., ${\cal N}_\ell^{\, \rm WL, \, opt} > {\cal N}_\ell^{\, \rm GCph, \, opt}$.
\end{itemize}
The number of multipole bins is then ${\cal N}_\ell^{\, \rm WL} = 26$ and ${\cal N}_\ell^{\, \rm GCph} = {\cal N}_\ell^{\, \rm XC} = 22$ in the pessimistic case and ${\cal N}_\ell^{\, \rm WL} = 32$ and ${\cal N}_\ell^{\, \rm GCph} = {\cal N}_\ell^{\, \rm XC} = 29$ in the optimistic case. In all these cases, the angular PS are computed at the center of the $\ell$ bin.
We will both consider the different probes in isolation and combine them in the 3$\times$2pt analysis, for which the $\ell$ binning will be the same as for the GCph one.

The one-dimensional data vector $\vec{C}$ is constructed by simply compressing the redshift and multipole indices (and, in the 3$\times$2pt case, the probe indices) into a single one, which we call $p$ (or $q$).
For Gaussian-distributed data with a parameter-independent covariance, the FM is given by Eq.~\eqref{eq:FMgauss_definitive}:
\begin{equation}\label{eq:fishmat}
    F_{\alpha \beta} =     
    \frac{\partial \vec{C}}{\partial \theta_{\alpha}}
    \, {\rm Cov}^{-1} \,
    \frac{\partial \vec{C}}{\partial \theta_{\beta}}
    =  
    \sum_{pq}
    \frac{\partial C_p}{\partial \theta_{\alpha}}
    \, {\rm Cov}^{-1}_{pq} \,
    \frac{\partial C_q}{\partial \theta_{\beta}} \; .
\end{equation}

We note that the size of the 3$\times$2pt covariance matrix quickly becomes large. For a standard setting with ${\cal N}_{\rm b} = 10$ redshift bins there are respectively (55, 100, 55) independent redshift bin pairs for (WL, XC, GCph), to be multiplied by the different ${\cal N}_{\ell}$. In general, the covariance will be a ${\cal N}_C \times {\cal N}_C$ matrix with 
\begin{align}
{\cal N}_C & = \bigg[
{\cal N}_{\rm b} ({\cal N}_{\rm b} + 1)/2 
\bigg]\bigg[
{\cal N}_\ell^{\, \rm WL} + {\cal N}_\ell^{\, \rm GCph}
\bigg] 
+ {\cal N}_{\rm b}^2  {\cal N}_\ell^{\, \rm XC} \nonumber \\
& = \bigg[{\cal N}_{\rm b} ({\cal N}_{\rm b} + 1) + {\cal N}_{\rm b}^2 \bigg]  {\cal N}_\ell^{\rm 3{\times}2pt},
\label{eq:sizecov}
\end{align}
for the 3$\times$2pt -- where the second line represents the case with the same number of $\ell$ bins for all probes, which is the one under study -- and 
\begin{equation}
{\cal N}_C = \big[{\cal N}_{\rm b} ({\cal N}_{\rm b} + 1)/2 \big] \, {\cal N}_\ell^{\, \rm WL/GCph} \; .
\end{equation}
for the WL and GCph cases. As an example, we will have ${\cal N}_C^{\rm 3{\times}2pt, \, opt} = 6090$.

\begin{figure}
    \centering
    \includegraphics[width=1\textwidth]{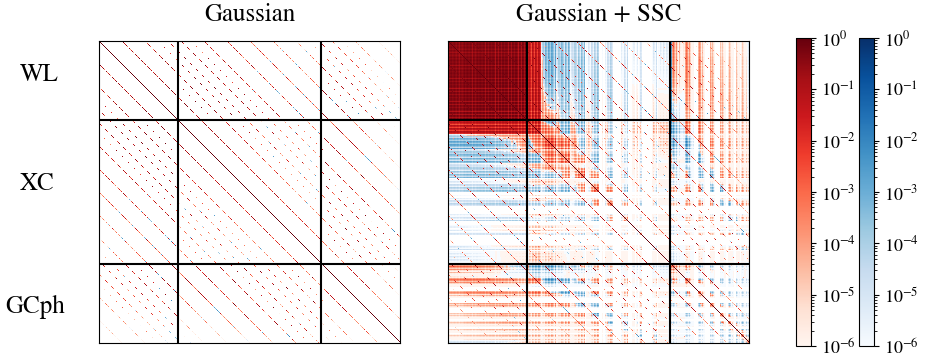}
    \caption{Correlation matrix in log scale for all the statistics of the 3$\times$2pt data-vector in the G and GS cases. The positive and negative elements are shown in red and blue, respectively. The Gaussian covariance is block diagonal (i.e., it is diagonal in the multipole indices, but not in the redshift ones; the different diagonals appearing in the plot correspond to the different redshift pair indices, for $\ell_1 = \ell_2$). The overlap in the WL kernels makes the WL block in the Gaussian + SSC covariance matrix much more dense than the GCph one.}
    \label{fig:Cor_GOvsGS}
\end{figure}

Being diagonal in $\ell$, most elements of this matrix will be null in the Gaussian case (see Fig.~\ref{fig:Cor_GOvsGS}, displaying the G and GS correlation matrices\footnote{Where the correlation matrix is defined as ${\rm Corr}_{ij} = {\rm Cov}_{ij}/\sqrt{{\rm Cov}_{ii}{\rm Cov}_{jj}}$}). This is no longer true with the inclusion of the SSC contribution, which makes the matrix computation -- and the subsequent inversion, to implement Eq.~\eqref{eq:fishmat} -- much more resource-intensive.\\
Given the highly non-diagonal nature of the GS covariance, we can wonder whether the inversion of this matrix needed to obtain the FM is stable. To investigate this, we compute the condition number of the covariance, which is defined as the ratio between its largest and smallest eigenvalues and in this case of order $10^{13}$. This condition number, multiplied by the standard \texttt{numpy float64} resolution ($2.22\times10^{-16}$), gives us the minimum precision that we have on the inversion of the matrix, of about $10^{-3}$. This means that numerical noise in the matrix inversion can cause, at most, errors of order $10^{-3}$ on the inverse matrix. Hence, we consider the inversion to be stable for the purpose of this work.
\section{Forecast code validation}
\label{sec:validation}
In order to validate the SSC computation with \texttt{PySSC}, we compare the $1\sigma$ forecast uncertainties (which correspond to a 68.3\% probability, due to the assumptions of the FM analysis) obtained using two different codes independently developed by two groups, which we call A and B. To produce the FM and the elements needed for its computation (the observables, their derivatives and the covariance matrix), group A uses a private\footnote{Available upon request to the author, Davide Sciotti} code fully written in \texttt{Python} and group B uses $\texttt{CosmoSIS}$\footnote{\texttt{\url{https://bitbucket.org/joezuntz/cosmosis/wiki/Home}}} \citep{Zuntz_2015}.  The only shared feature of the two pipelines is the use of \texttt{PySSC} (to compute the $S_{ijkl}$ matrix). For this reason, and because the SSC is not considered in isolation but added to the Gaussian covariance, we compare the forecast results of the two groups both in the Gaussian and Gaussian + SSC cases.

Following \citetalias{ISTF2020}, we consider the results to be in agreement if the discrepancy of each group's results with respect to the median -- which in our case equals the mean --  is smaller than 10\%. This simply means that the A and B pipelines' outputs are considered validated against each other if
\begin{equation} \label{eq:sigmas}
    \abs{\frac{\sigma_\alpha^i}{\sigma^m_\alpha}-1} < 0.1 \quad {\rm for} \quad i = {\rm A,B}; \quad \sigma^m_\alpha = \frac{\sigma_\alpha^A+\sigma_\alpha^B}{2} \; ,
\end{equation}
with $\sigma_\alpha^A$ the $1\sigma$ uncertainty on the parameter $\alpha$ for group A. The above discrepancies are equal and opposite in sign for A and B.

The \textit{marginalised} uncertainties are extracted from the FM $F_{\alpha\beta}$, which is the inverse of the covariance matrix ${\rm C}_{\alpha\beta}$ of the parameters, as described in Sect.~\ref{sec:fisher_theory}: $(F^{-1})_{\alpha\beta} = {\rm C}_{\alpha\beta}$. We do not consider unmarginalised, or conditional, uncertainties in this work. We then have
\begin{equation} \label{eq:sigma_marg}
    \sigma_\alpha = \sigma_\alpha^{\rm marg.} =  (F^{-1})_{\alpha\alpha}^{1/2} \; .
\end{equation}
We remind that the uncertainties found in the FM formalism constitute lower bounds, or optimistic estimates, on the actual parameters' uncertainties, as stated by the Cramér-Rao inequality.

\begin{figure}[ht]
  \begin{minipage}{.65\textwidth}
    \includegraphics[width=\linewidth]{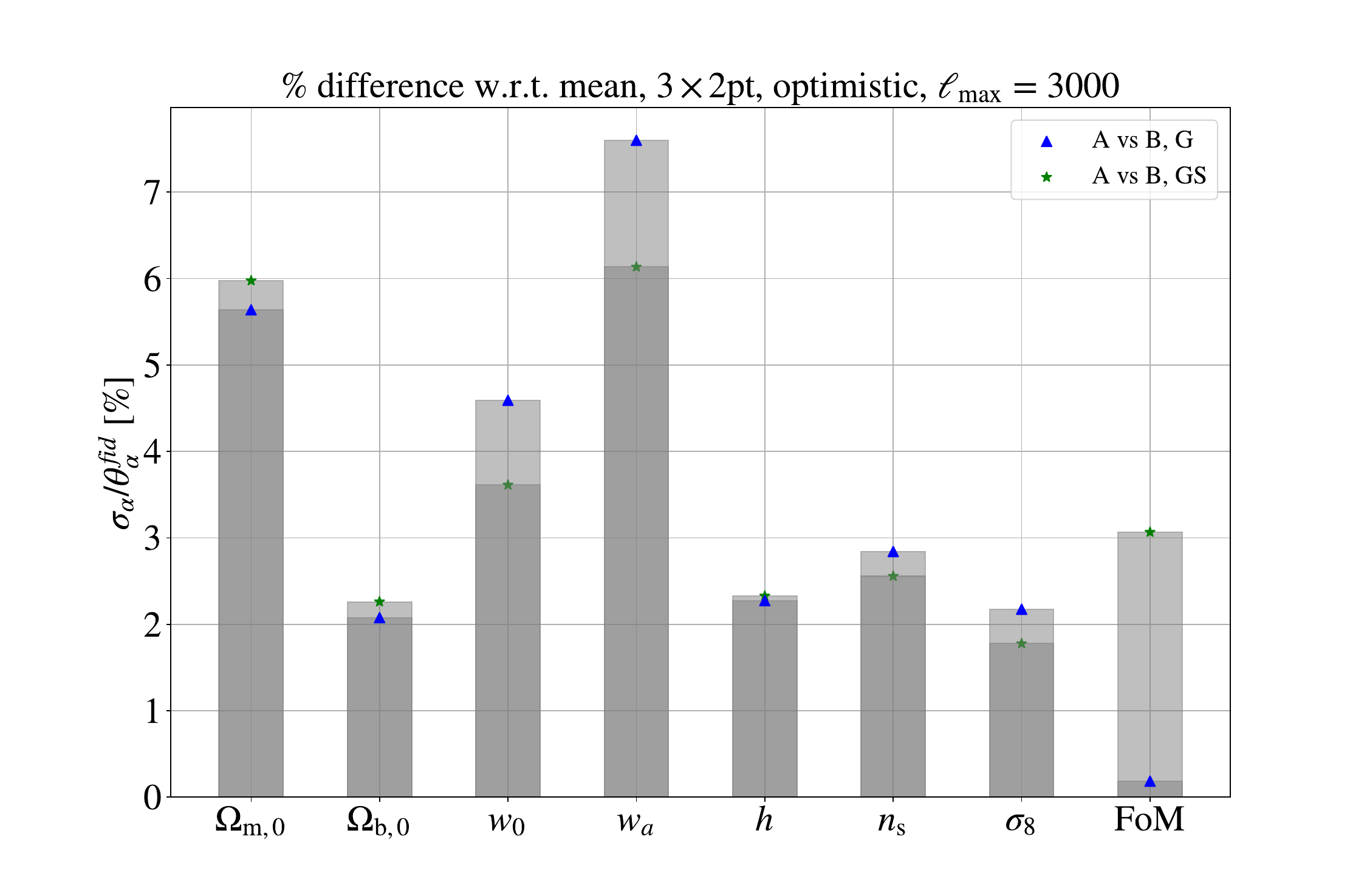}
  \end{minipage}%
  \begin{minipage}{.35\textwidth}
    \caption{Percent discrepancy of the normalized 1$\sigma$ uncertainties with respect to the mean for the WL probe, both in the G and GS cases (optimistic settings). The index $i = {\rm A, B}$ indicates the two pipelines, whilst $\alpha$ indexes the cosmological parameter. The desired agreement level is reached in all cases (WL, GCph probes and pessimistic case not shown).}
    \label{fig:dav_vs_sylv}
  \end{minipage}
\end{figure}

In the following, we normalize $\sigma_\alpha$ by the fiducial value of the parameter $\theta_\alpha$, in order to work with relative uncertainties: $\bar{\sigma}^i_\alpha = \sigma_\alpha^i/\theta_\alpha^{\,\rm fid}; \ \bar{\sigma}^m_\alpha = \sigma^m_\alpha/\theta_\alpha^{\,\rm fid}$, again with $i = {\rm A, B}$. If a given parameter has a fiducial value of 0, such as $w_a$, we simply take the absolute uncertainty. The different cases under examination are dubbed `G', or `Gaussian', and `GS', or `Gaussian + SSC'. The computation of the parameters constraints differs between these two cases only by the covariance matrix used in Eq.~\eqref{eq:fishmat} to compute the FM
\begin{equation}
 \rm{Cov} = 
\begin{cases}
\rm Cov_{\rm G} & \rm{Gaussian}\\
\rm Cov_{\rm GS} = Cov_{\rm G} +  Cov_{SSC} & \rm{Gaussian+SSC} \; .
\end{cases}
\end{equation}
For the reader wishing to validate their own code, we describe the validation process in Appendix~\ref{sec:validation_appendix}. Here we sketch the results of the code validation: in Fig.~\ref{fig:dav_vs_sylv}, we show the percent discrepancy as defined in Eq.~\eqref{eq:sigmas} for the 3$\times$2pt case. Similar results have been obtained for the GCph and WL cases, both for the optimistic and pessimistic settings specified in Sect.~\ref{sec:datavectors}. The constraints are all found to satisfy the required agreement level (less than $10\%$ discrepancy with respect to the mean). In light of these results, we consider the two forecasting pipelines validated against each other.
\section{SSC impact on forecasts}
\label{sec:impact}
We can now finally investigate how the inclusion of SSC degrades the constraints with respect to the Gaussian case. To this end, we will look in the following at the quantity
\begin{equation}
\mathcal{R}(\theta) = \sigma_{\rm GS}(\theta)/\sigma_{\rm G}(\theta) \; ,
\label{eq:ratiodef}
\end{equation}
where $\sigma_{\rm G}(\theta)$ and $\sigma_{\rm GS}(\theta)$ are the usual  marginalised uncertainties on the parameter $\theta$ computed, as detailed above, with Gaussian or Gaussian\,+\,SSC covariance matrix. We run $\theta$ over the set of cosmological parameters listed in Eq.~\eqref{param_vector}, i.e., $\theta \in \{\Omega_{{\rm m},0}, \Omega_{{\rm b},0}, w_0, w_a, h, n_{\rm s}, \sigma_8\}$.

In addition, we examine the Figure of Merit (FoM) as defined in \citet{albrecht2006}, a useful way to quantify the joint uncertainty on several parameters. We parameterize the FoM following \citetalias{ISTF2020} to focus on the joint uncertainty on the dark energy equation of state parameters $w_0$ and $w_a$, such that  
\begin{equation}
    {\rm FoM} = \sqrt{ {\rm  det}(\tilde{F}_{w_0 w_a})} \; .
    \label{eq:fom}
\end{equation}
This quantity is inversely proportional to the area of the 2$\sigma$ confidence ellipse in the plane spanned by the parameters $(w_0, w_a)$. $\tilde{F}_{w_0 w_a}$ is the Fisher sub-matrix obtained by marginalising over all the parameters but $w_0$ and $w_a$, and is computed by inverting $F_{\alpha\beta}$ (that is, taking the parameters' covariance matrix), removing all the rows and columns but the ones corresponding to $w_0$ and $w_a$ and re-inverting the resulting $2 \times 2$ matrix.

We will also use the notation ${\cal{R}}({\rm FoM})$ as a shorthand for ${\rm FoM}_{\rm GS}/{\rm FoM}_{\rm G}$. We note that, since we expect the uncertainties to be larger for the GS case, we will have ${\cal{R}}(\theta) > 1$, and the FoM being inversely proportional to the area of the uncertainty ellipse, ${\cal{R}}({\rm FoM}) < 1$. 
\subsection{Reference scenario}\label{sec:ssc_ref_results}
Let us start by considering the case with ${\cal N}_{\rm b} = 10$ equipopulated redshift bins, which we will take in the following as a reference. Table~\ref{tab:ratio_ref} gives the values of the ${\cal{R}}$ ratios for the different parameters and the FoM in both the pessimistic and optimistic scenarios, for the single or combined probes.

In accordance with previous results in the literature \citep[see e.g.][]{Barreira2018cosmic_shear, upham2021}, we find that the WL constraints are dramatically affected by the inclusion of SSC. The impact is so severe that the FoM is reduced by a factor of about $2$ in both the pessimistic and optimistic scenarios. The marginalised uncertainties worsen by a large factor for those parameters which correlate the most with the amplitude of the signal: indeed, the largest ${\cal{R}}(\theta)$ values are obtained for $(\Omega_{{\rm m,0}}, \sigma_8)$, while ${\cal{R}}(\theta)$ does not meaningfully deviate from unity for $\theta = (w_a, h, n_{\rm s})$, and $w_0$ sits in between the two extreme cases. This is because the SSC effect is essentially an unknown shift, or perturbation, in the background density, with which $\Omega_{{\rm m,0}}$ and $\sigma_8$ are highly degenerate.\\

\begin{table*}
\centering
\caption{Ratio between the GS and G constraints for all cosmological parameters and the FoM in the reference scenario, for both pessimistic (P) and optimistic (O) assumptions. We remind the reader that in the reference case, we marginalize over the galaxy bias nuisance parameters while holding the multiplicative shear bias ones fixed.}
\begin{tabular}{l | c c c c c c c c c | l}
\hline
\multicolumn{1}{l |}{$\mathcal{R}(x)$} & $\Omega_{{\rm m},0}$ & $\Omega_{{\rm b},0}$ & $w_0$ & $w_a$ & $h$ & $n_{\rm s}$ & $\sigma_8$ & \multicolumn{1}{| l }{FoM}\\
\hline
\hline
\multicolumn{1}{l |}{WL, P} & 
1.998 & 1.001 & 1.471 & 1.069 & 1.052 & 1.003 & 1.610 & \multicolumn{1}{| l }{0.475} \\
\multicolumn{1}{l |}{WL, O} & 
1.574 & 1.013 & 1.242 & 1.035 & 1.064 & 1.001 & 1.280 & \multicolumn{1}{| l }{0.451} \\
\hline
\hline
\multicolumn{1}{l |}{GCph, P} & 
1.002 & 1.002 & 1.003 & 1.003 & 1.001 & 1.001 & 1.001 & \multicolumn{1}{| l }{0.996} \\
\multicolumn{1}{l |}{GCph, O} & 
1.069 & 1.016 & 1.147 & 1.096 & 1.004 & 1.028 & 1.226 & \multicolumn{1}{| l }{0.833} \\
\hline      
\hline
\multicolumn{1}{l |}{3$\times$2pt, P} & 
1.442 & 1.034 & 1.378 & 1.207 & 1.028 & 1.009 & 1.273 & \multicolumn{1}{| l }{0.599} \\
\multicolumn{1}{l |}{3$\times$2pt, O} & 
1.369 & 1.004 & 1.226 & 1.205 & 1.018 & 1.030 & 1.242 & \multicolumn{1}{| l }{0.622} \\
\hline
\end{tabular}
\label{tab:ratio_ref}
\end{table*}

The results in Table~\ref{tab:ratio_ref} also show that GCph is not as strongly affected by SSC. This is an expected result, being the GCph probe response coefficients lower (in absolute value) than the WL ones, as can be seen in Fig.~\ref{fig:responses_vinc_allprobes}. This is due to the additional terms that account for the response of the galaxy number density $n_{\rm g}$ (see Eq.~\ref{eq:Rgg}), which is itself affected by the super-survey modes.
Moreover, the constraints from GCph alone are obtained by marginalising over a larger number of nuisance parameters than WL -- the galaxy bias parameters, which are strongly degenerate with the amplitude of the signal. This works as a sort of effective systematic covariance which makes the SSC less dominant than in the WL case. Lastly, as can be seen from Fig.~\ref{fig:WF_FS1_for_PhD_thesis}, all WL kernels have non-zero values for $z \rightarrow 0$, contrary to the GCph ones. In this limit, the effective volume probed by the survey tends to 0, hence making the variance of the background modes $\sigma^2$ tend to infinity. We thus have a larger $S_{ijkl}$ matrix, which is one of the main factors driving the amplitude of the SSC. We nevertheless note, for GCph, a $17\%$ decrease of the FoM in the optimistic case, which is related to the inclusion of nonlinear modes that are more sensitive to the SSC, as we discuss later. \\

The full 3$\times$2pt case sits in between the two extremes as a consequence of the data vector containing the strongly affected WL probe, and the less affected GCph one. The contribution from the XC probe is again an intermediate case because of its lower response coefficient, so the final impact on the FM elements will be intermediate between the WL and GCph cases, as the ${\cal{R}}(\theta)$ values in Table~\ref{tab:ratio_ref} indeed show.
\begin{figure}
 \centering
    \includegraphics[width=0.9\textwidth]{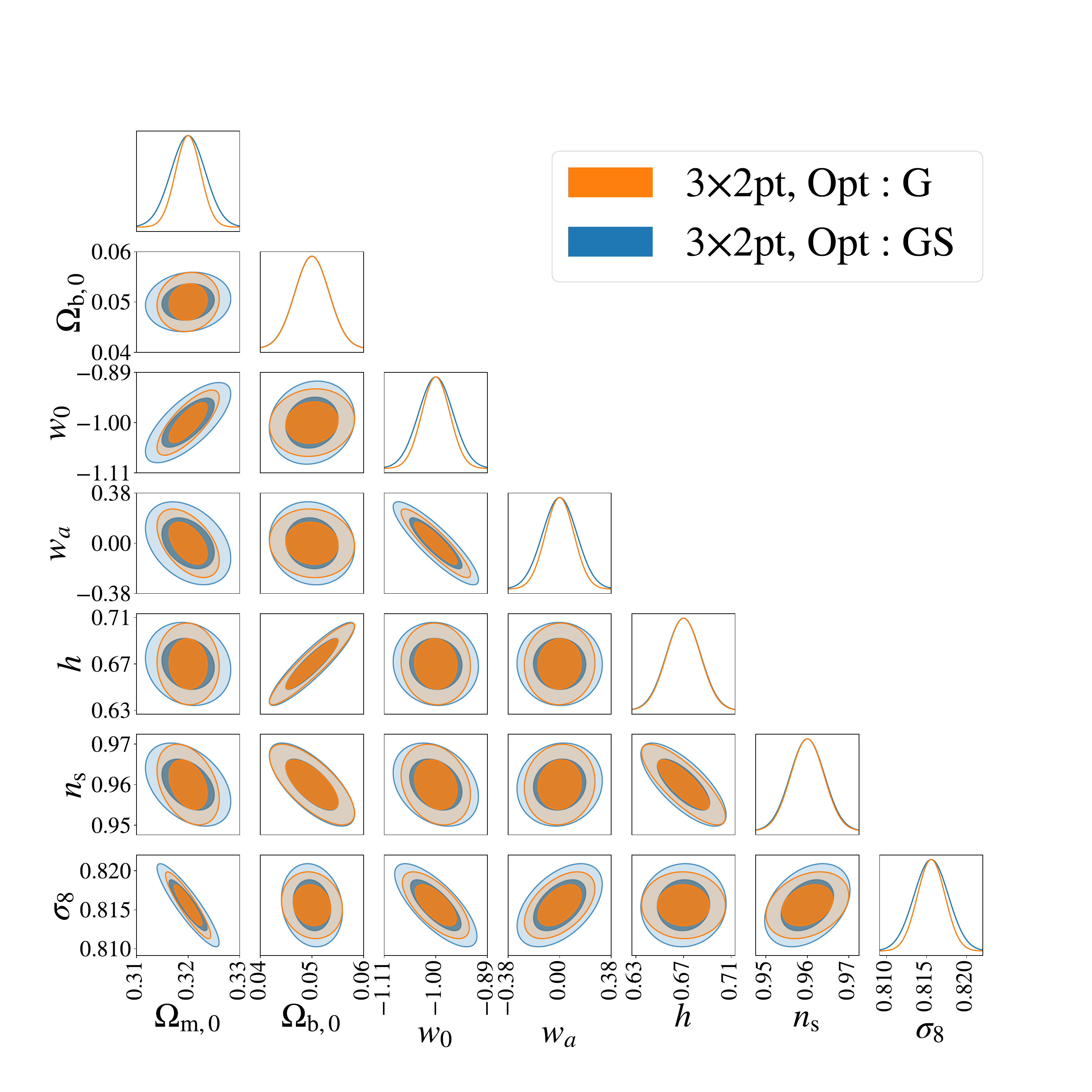}
    \caption{Contour plot for the G and GS constraints, considering the full 3$\times$2pt analysis in the optimistic case, in the reference scenario. The addition of SSC does not seem to substantially modify the correlations between cosmological parameters. For clarity, the nuisance parameters are shown separately in Fig.~\ref{fig:nuisance}.}
    \label{fig:triangle_plot}
\end{figure}
Comparing the optimistic and the pessimistic cases for the two individual probes, we can see that there is a different behaviour of the SSC as a function of the maximum multipole. Indeed, for WL the $\mathcal{R}(\theta)$ ratio for the most affected\footnote{This is not the case for the unconstrained parameters, but the small difference is likely related to numerical artifacts.} parameters is larger in the pessimistic than in the optimistic case. This is consistent with the results of \citet{upham2021} showing that the diagonal elements of the WL total covariance matrix are more and more dominated by the Gaussian term as we move to higher $\ell$. This is because of the presence of the scale-independent shape noise in the Gaussian covariance (see Eq.~\ref{eq:covgauss} for $A = B = {\rm L}$), which largely dominates over the SSC on small scales. As such, the relative importance of off-diagonal correlations decreases at large $\ell$ which is precisely what happens when moving from the pessimistic to the optimistic case. This causes the SSC impact to be smaller in the optimistic case, although we note that the ${\cal{R}}(\theta)$ are still remarkably large. Indeed, the  ${\cal{R}}$ values for the FoM are roughly the same, pointing to the importance of SSC in both scenarios.

As also seen in \citet{Lacasa2020_braiding}, we observe the opposite behaviour for the GCph probe, which is more impacted by the SSC in the optimistic case. This is because the impact of the shot noise at these scales is lower than the shape noise for WL, so the SSC still dominates in that multipole range.

In Fig.~\ref{fig:triangle_plot} we show the comparison of the 2D contours for all cosmological parameters between G and GS in the case of the 3$\times$2pt analysis, in the optimistic case. Again, we can clearly see that the most impacted parameters are $\theta = (\Omega_{{\rm m},0}, w_0, \sigma_8)$. In addition, this shows that SSC does not seem to strongly affect the degeneracy between cosmological parameters; however, we note that, in general, the Gaussian contours are more elliptical (more degenerate) than the GS ones. This is because, as noted in \citealt{Lacasa2020_braiding}, in the former case we attribute artificially small uncertainties to the small scales (because of the very small cosmic variance), which consequently hold the majority of the constraining power, leading to parameter degeneracies. In the GS case, on the other hand, the constraining power is more evenly distributed between the different scales.\\

\begin{figure*}
\centering
    \includegraphics[width=\textwidth]{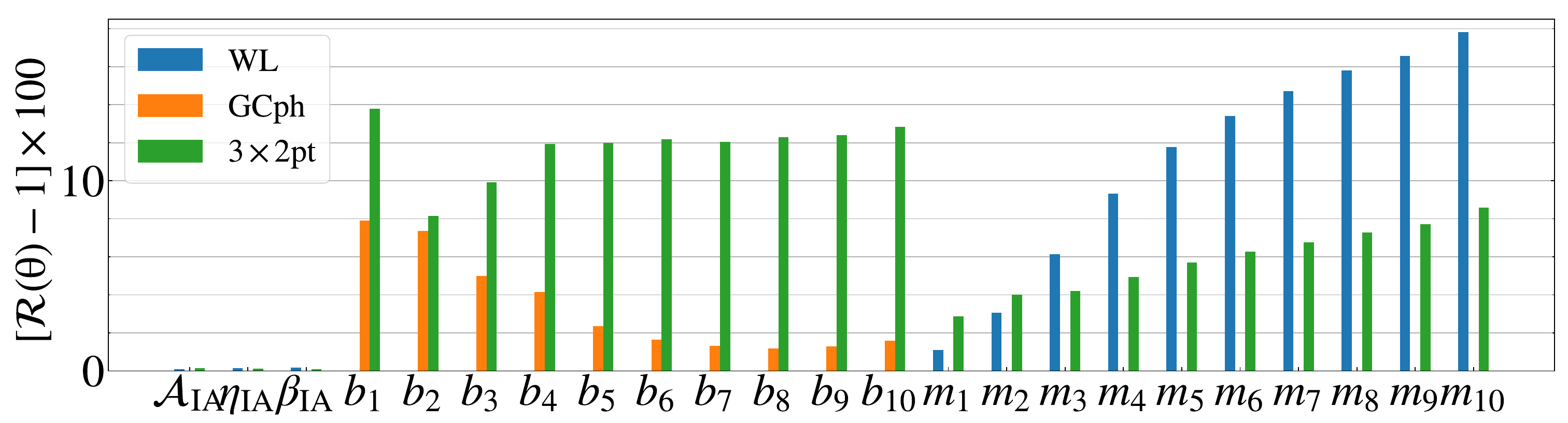}
    \caption{Percent increase of the marginalised $1\sigma$ uncertainty of the nuisance parameters, for all probe choices, in the optimistic case and for the reference scenario.}
    \label{fig:nuisance}
\end{figure*}

To conclude this section, it is also worth looking at the impact of SSC on the astrophysical nuisance parameters. Indeed, although an issue to be marginalised over when looking at cosmological ones, the IA and the galaxy bias parameters are of astrophysical interest. We show the impact of SSC on the constraints on these quantities in Fig.~\ref{fig:nuisance}, and, as an anticipation of the next section, we also show the constraints for other WL-related nuisance parameters, the multiplicative shear bias parameters $m_i$. 
For IA-related nuisance parameters, the uncertainty increase due to SSC is lower than 0.5\%. The uncertainty on $b_i$ and $m_i$ in each of the ten redshift bins is however significantly affected by SSC, showing an increase between 1 and 14\% for $b_i$ and between 1 and 18\% for $m_i$, depending on the probe combination choice. This is because both of these nuisance parameters simply act as a multiplicative factor on the power spectrum and are thus highly degenerated with the effect of SSC. Again, this is because the first-order effect of SSC is to modulate the overall clustering amplitude because of a shift in the background density  $\delta_{\rm b}$. As mentioned, this cross-talk between SSC and linear galaxy bias could also explain why the GCph probe seems less affected by SSC: some of the difference between G and GS is absorbed by the $b_i$ in the marginalisation. This will be also confirmed for WL in the next section, showing a reduced relative impact of SSC in the presence of multiplicative shear bias. Note that going beyond the linear approximation for the modelling of the galaxy bias will add more nuisance parameters, thus degrading the overall constraints on cosmological parameters and further reducing the relative degradation of constraints due to SSC.

Finally, comparing how uncertainties on $b_i$ and $m_i$ react to the addition of SSC, we can see that surprisingly the $b_i$ are more affected in the 3$\times$2pt case than in the GCph case, while it is the contrary for $m_i$, the uncertainty increase is larger for WL than for 3$\times$2pt. This difference in the behaviour of the uncertainty increase might come from the numerous degeneracies existing between these nuisance parameters and the most constrained cosmological parameters in each case. Though it is not easy to exactly understand this behaviour, we note that in all cases the $\mathcal{R}(\theta)$ for these parameters are of the same order of magnitude and are never completely negligible.
\begin{table*}
\centering
\caption{Same as Table~\ref{tab:ratio_ref} but removing the flatness prior.}
\begin{tabular}{l | c c c c c c c c c c | l}
\hline
\multicolumn{1}{l |}{$\mathcal{R}(x)$} & $\Omega_{{\rm m},0}$ & $\Omega_{{\rm DE, 0}}$ & $\Omega_{{\rm b},0}$ & $w_0$ & $w_a$ & $h$ & $n_{\rm s}$ & $\sigma_8$ & \multicolumn{1}{| l }{FoM}\\
\hline
\hline
\multicolumn{1}{l |}{WL, P} & 
2.561 & 1.358 & 1.013 & 1.940 & 1.422 & 1.064 & 1.021 & 1.433 & \multicolumn{1}{| l }{0.514} \\
\multicolumn{1}{l |}{WL, O} & 
2.113 & 1.362 & 1.004 & 1.583 & 1.299 & 1.109 & 1.038 & 1.559 & \multicolumn{1}{| l }{0.631} \\
\hline
\hline
\multicolumn{1}{l |}{GCph, P} & 
1.002 & 1.001 & 1.002 & 1.002 & 1.003 & 1.001 & 1.000 & 1.001 & \multicolumn{1}{| l }{0.996} \\
\multicolumn{1}{l |}{GCph, O} & 
1.013 & 1.020 & 1.006 & 1.153 & 1.089 & 1.004 & 1.039 & 1.063 & \multicolumn{1}{| l }{0.831} \\
\hline
\hline
\multicolumn{1}{l |}{3$\times$2pt, P} & 
1.360 & 1.087 & 1.043 & 1.408 & 1.179 & 1.021 & 1.009 & 1.040 & \multicolumn{1}{| l }{0.677} \\
\multicolumn{1}{l |}{3$\times$2pt, O} & 
1.572 & 1.206 & 1.013 & 1.282 & 1.191 & 1.013 & 1.008 & 1.156 & \multicolumn{1}{| l }{0.756} \\
\hline
\end{tabular}
\label{tab:ratio_nonflat}
\end{table*}
\subsection{Non-flat cosmologies}

In the previous section, we investigated the SSC on the cosmological parameters under the assumption of a flat model. Actually, the requirement on the FoM assessed in the \Euclid Red Book refers to the case with the curvature as an additional free parameter to be constrained, i.e., the non-flat $w_0w_a$CDM model. This is why in \citetalias{ISTF2020} are also reported the marginalised uncertainties for the parameter $\Omega_{{\rm DE,0}}$, with a fiducial value $\Omega_{{\rm DE,0}}^{\rm fid} = 1 - \Omega^{\rm fid}_{{\rm m,0}}$ to be consistent with a flat Universe. It is worth wondering what the impact of SSC is in this case too. This is summarised in Table~\ref{tab:ratio_nonflat}, where we now also include the impact on $\Omega_{{\rm DE,0}}$. 

A comparison with the results in Table~\ref{tab:ratio_ref} is quite hard if we look at the single parameters. Indeed, opening up the parameter space by removing the flatness assumption introduces additional degeneracy among the parameters controlling the background expansion, which are thus less constrained whether SSC is included or not. We can nevertheless note again that WL is still the most impacted probe, while GCph is less affected, and the 3$\times$2pt sits in between. The difference between pessimistic and optimistic scenarios is now less evident with ${\cal{R}}(\theta)$ increasing or decreasing depending on the parameter and the probe. 

Once more, the most affected parameters for WL are $(\Omega_{{\rm m,0}}, \sigma_8)$, the uncertainties on which are now further degraded by the fact that they correlate with the parameter $\Omega_{{\rm DE,0}}$ which is also affected. Although $(w_0, w_a)$ are also degraded by the SSC, a sort of compensation is at work, so that the overall decrease in the FoM is similar to the case with the flatness prior. The motivations that make GCph much less affected still hold when dropping the flatness prior, explaining the corresponding $\mathcal{R}(\theta)$ values.  

We also note an increase of ${\cal{R}}({\rm FoM})$ in the 3$\times$2pt case, meaning a smaller degradation of the FoM due to SSC. The FoM indeed degrades by $24\%$ $(32\%)$ in the non-flat case vs. $38\%$ $(40\%)$ for the flat case in the optimistic (pessimistic) scenario. This can be qualitatively explained by noting that the decrease of both FoM(G) and FoM(GS) is related to a geometrical degeneracy which is the same on all scales, whether or not they are affected by the increase in uncertainty due to the SSC inclusion.
\begin{table*}
\centering
\caption{Same as Table~\ref{tab:ratio_nonflat} but adding multiplicative shear bias nuisance parameters.}
\begin{tabular}{l | c c c c c c c c c c | l}
\hline
\multicolumn{1}{l |}{$\mathcal{R}(x)$} & $\Omega_{{\rm m},0}$ & $\Omega_{{\rm DE, 0}}$ & $\Omega_{{\rm b},0}$ & $w_0$ & $w_a$ & $h$ & $n_{\rm s}$ & $\sigma_8$ & \multicolumn{1}{| l }{FoM}\\
\hline
\hline
\multicolumn{1}{l |}{WL, P} & 
1.082 & 1.049 & 1.000 & 1.057 & 1.084 & 1.034 & 1.025 & 1.003 & \multicolumn{1}{| l }{0.917} \\
\multicolumn{1}{l |}{WL, O} & 
1.110 & 1.002 & 1.026 & 1.022 & 1.023 & 1.175 & 1.129 & 1.009 & \multicolumn{1}{| l }{0.976} \\
\hline
\hline
\multicolumn{1}{l |}{3$\times$2pt, P} & 
1.297 & 1.087 & 1.060 & 1.418 & 1.196 & 1.021 & 1.030 & 1.035 & \multicolumn{1}{| l }{0.674} \\
\multicolumn{1}{l |}{3$\times$2pt, O} & 
1.222 & 1.136 & 1.010 & 1.300 & 1.206 & 1.013 & 1.009 & 1.164 & \multicolumn{1}{| l }{0.745} \\
\hline
\end{tabular}
\label{tab:ratio_mult}
\end{table*}
\subsection{Role of nuisance parameters}

\begin{figure}
    \centering
    \includegraphics[width=0.47\textwidth]{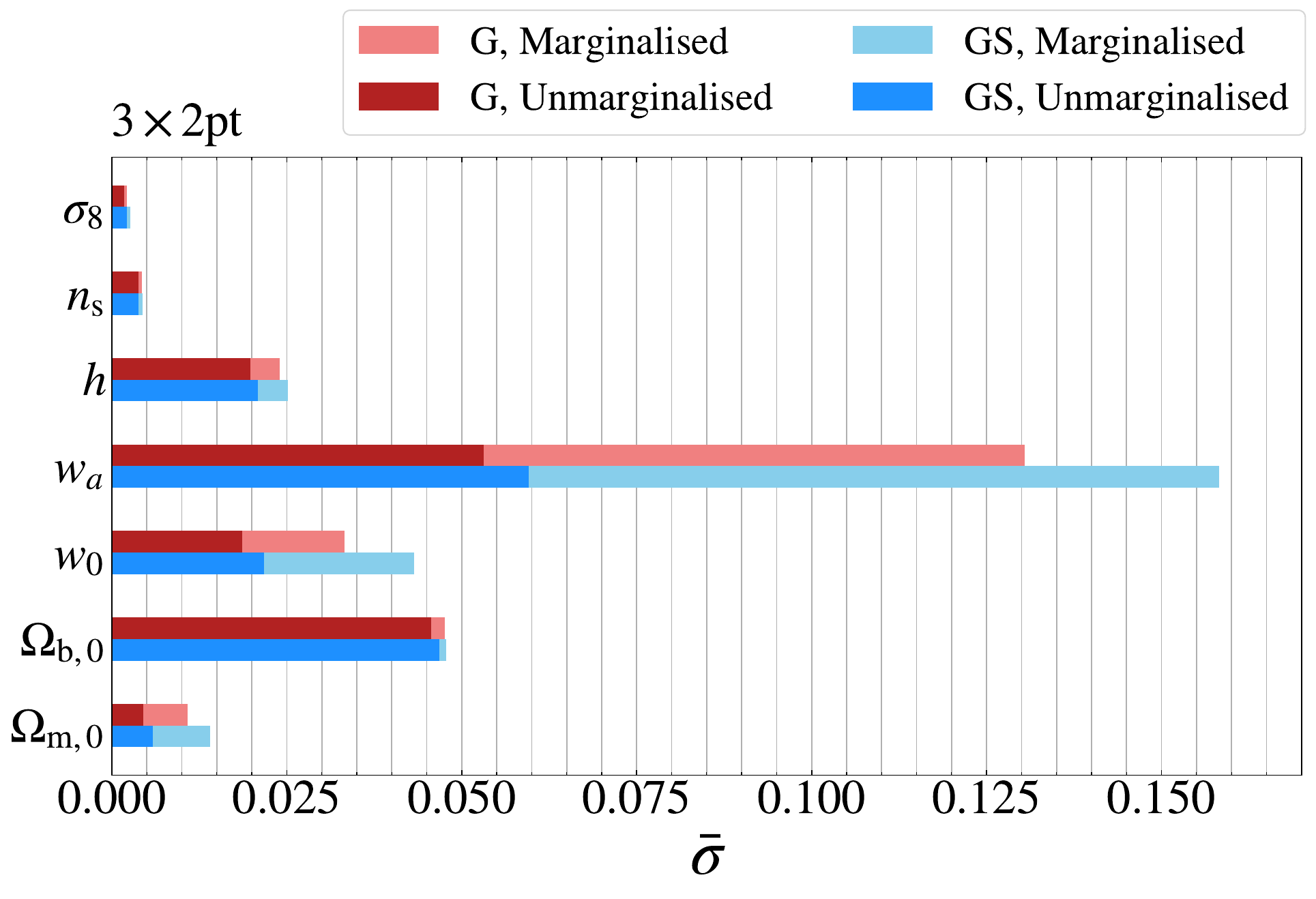}
    \includegraphics[width=0.47\textwidth]{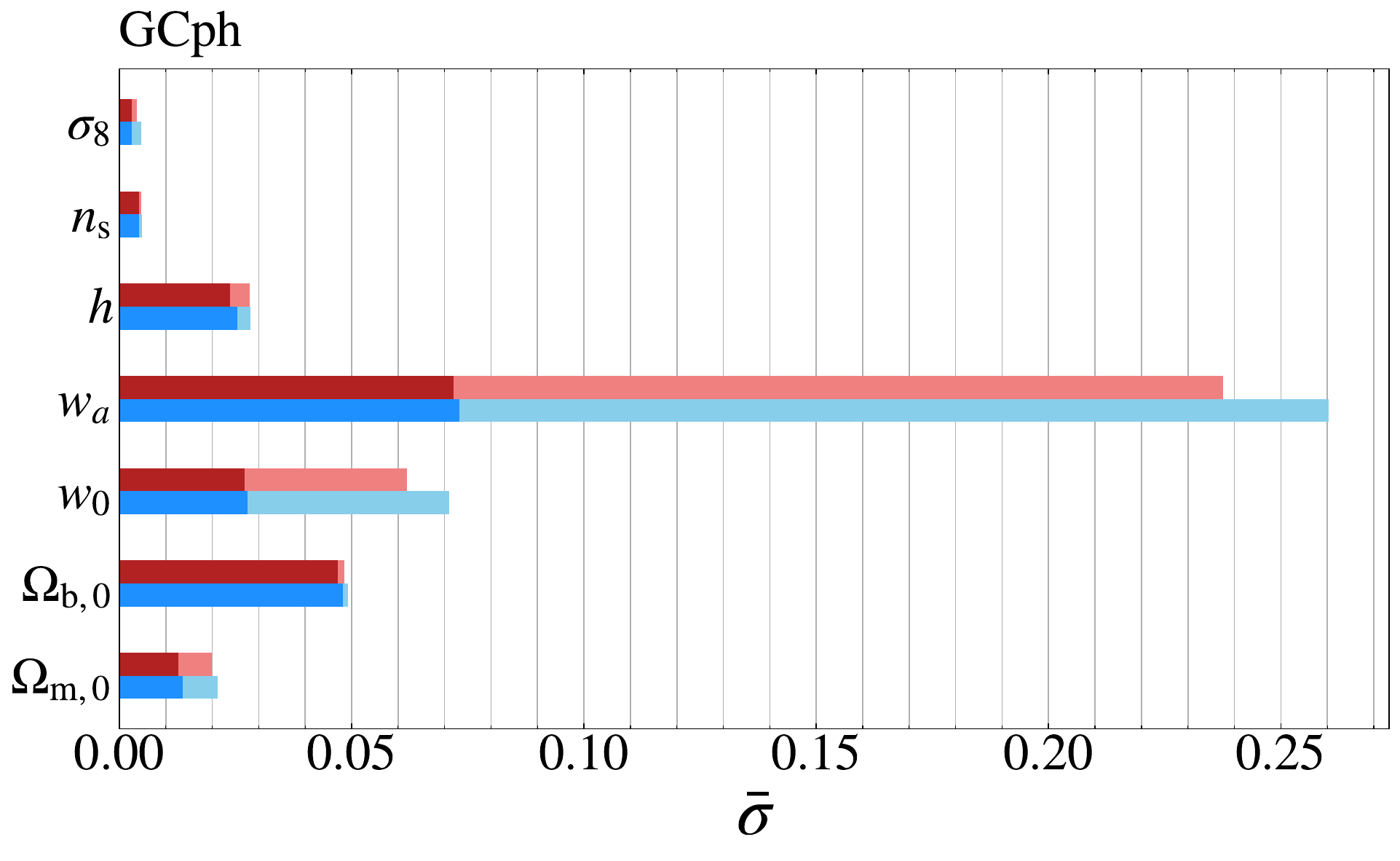}
    \includegraphics[width=0.47\textwidth]{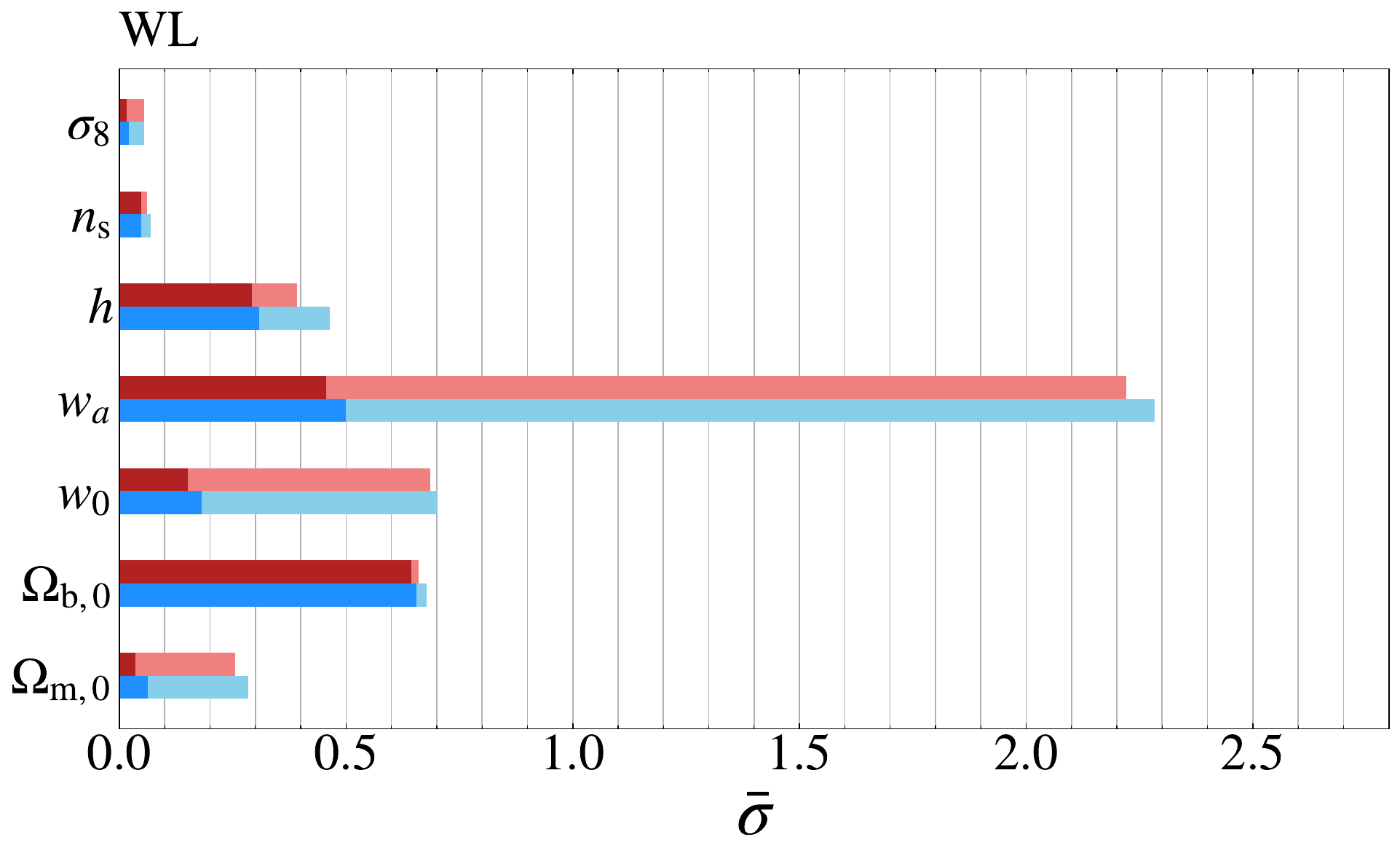}
    \includegraphics[width=0.47\textwidth]{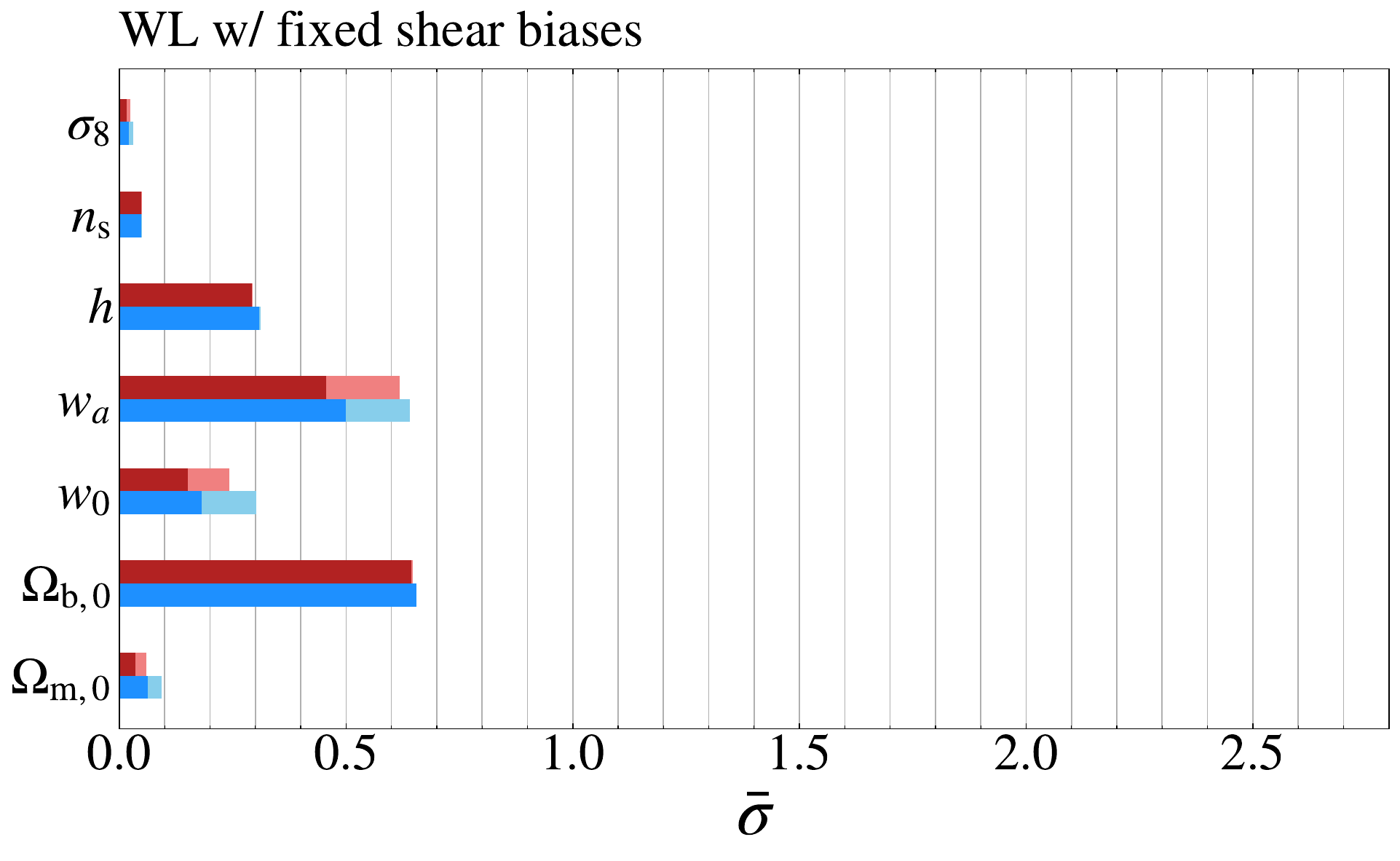}
    \caption{Marginalised and unmarginalised 1$\sigma$ uncertainties on the cosmological parameters, relative to their corresponding fiducial values, in both the G and GS cases for 3$\times$2pt, GCph and WL. For WL, we show the results in the case where the multiplicative shear biases are either varied or fixed, in other words, whether we marginalise over all nuisance parameters or only over the IA ones.}
    \label{fig:barplot}
\end{figure}

\begin{figure}
    \centering
    \includegraphics[width=0.7\textwidth]{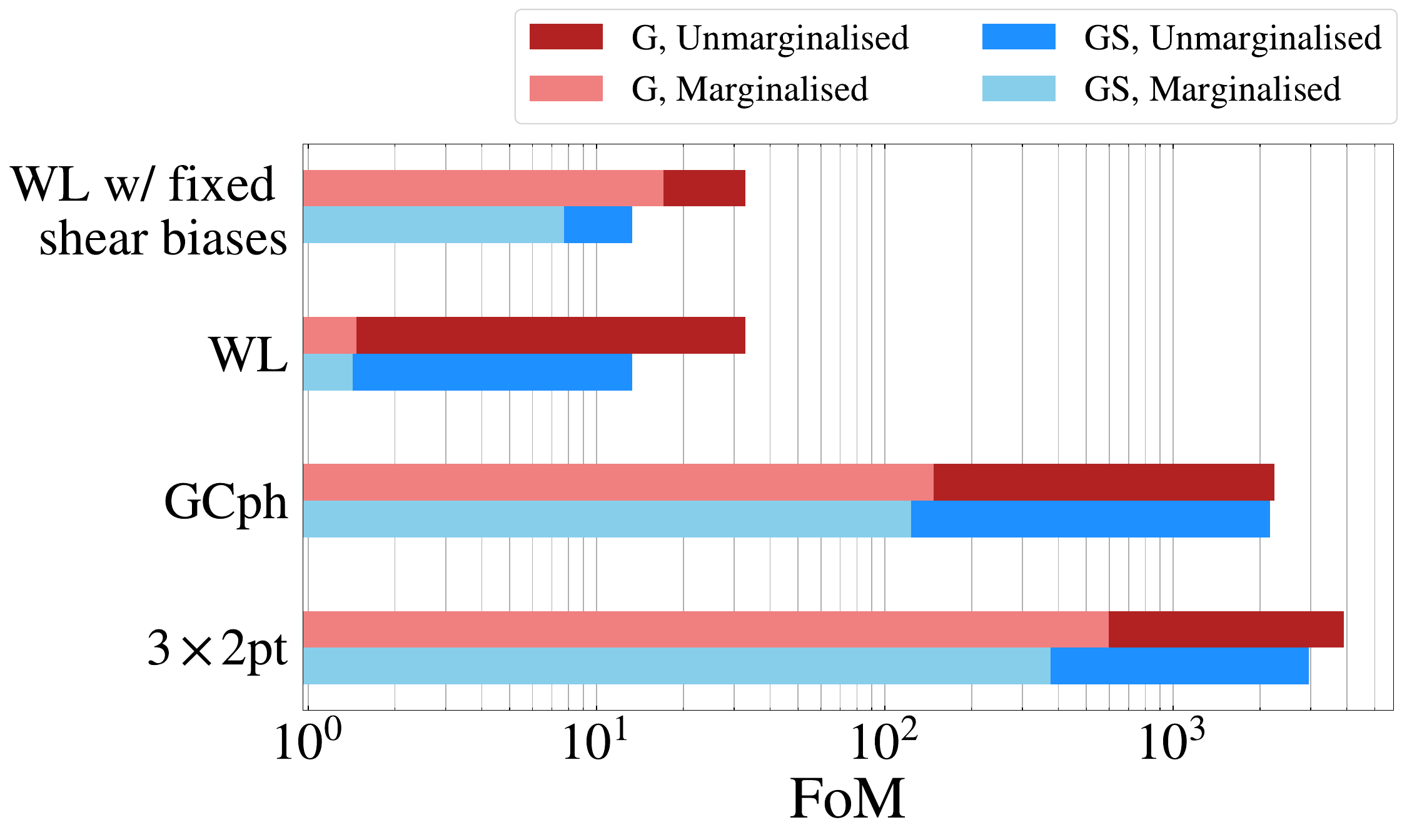}
    \caption{Dark energy FoM for marginalised and unmarginalised constraints in both the G and GS cases, 3$\times$2pt, GCph, WL, and WL with fixed multiplicative shear biases.}
    \label{fig:barplot_fom}
\end{figure}

We can now open up the parameter space by letting the shear bias parameters introduced in Sect.~\ref{sec:mult_shear_bias} free to vary. 
We expand the FM by adding these additional parameters and recompute the ratios of uncertainties with and without SSC obtaining the results shown\footnote{We do not report here the results for GCph since they are the same as the ones shown in Table \ref{tab:ratio_ref}, given that $C_{ij}^{\rm GG}(\ell)$ is unaffected by multiplicative shear bias.} in Table~\ref{tab:ratio_mult}. We remind the reader that the number of nuisance parameters depends on which probe (WL or 3$\times$2pt) one is considering. For the WL case, the ${\cal N}_{\rm b}$ multiplicative shear bias parameters add up to the 3 IA ones leading to the result that the SSC has a very minor impact on the constraints and on the FoM. The values in Table~\ref{tab:ratio_mult} are actually easily explained. We recall that ${\cal{R}}(\theta)$ is a ratio between the constraints with and without the SSC. Adding $m_i$ to the cosmological parameters introduces a degeneracy between $m_i$ itself and the parameters $(\Omega_{\rm m,0}, \sigma_8)$ which set the overall amplitude of $C_{ij}^{\rm LL}(\ell)$. Such a degeneracy is a mathematical one present on the whole $\ell$ range, similar to the galaxy bias parameters for GCph. As a consequence, the constraints on all the parameters and the FoM are strongly degraded in a way that is independent of the presence of SSC. This is shown in Fig.~\ref{fig:barplot} and \ref{fig:barplot_fom}, which exhibits the relative uncertainty $\bar{\sigma}$ and the dark energy FoMs in the G and GS cases for each parameter, if we marginalise or not on over nuisance parameters. Letting the nuisance parameters free to vary, i.e. marginalising over them, tends to increase the uncertainty on cosmological parameters way more than including SSC and this is even more true when these nuisance parameters are simply multiplicative such as $b_i$ and $m_i$.
This is why the ${\cal{R}}$ values drop down to values close to unity in contrast to what we have found up to now. Moreover, introducing more nuisance parameters degenerate with the amplitude of the signal dilutes the SSC effect in a larger error budget; because of this, it is the relative rather than the absolute impact of SSC that decreases. Indeed, marginalising over nuisance parameters is formally equivalent to having additional covariance.

Note that this does not mean that adding nuisance parameters improves the constraints. Indeed, the marginalised uncertainties on all parameters increase (hence the FoM decreases) with respect to the case when the multiplicative shear bias is fixed. The degradation is, however, the same with and without SSC so the ${\cal{R}}(\theta)$ values stay close to unity.

On the contrary, the results for the 3$\times$2pt case show that the SSC still matters. The additional information carried by the GCph and XC data allows the partial breaking of the mathematical degeneracy among $(m_i, \Omega_{\rm m,0}, \sigma_8)$ hence making again the scale-dependent increase of the uncertainties due to the inclusion of SSC important. However, the larger number of nuisance parameters (from 13 to 23) still introduces additional degeneracies with the cosmological ones hence alleviating the impact of SSC. The overall effect is, however, small with the ${\cal{R}}$ values being close to the ones in Table~\ref{tab:ratio_nonflat}. In particular, the FoM degradation is essentially the same in both the pessimistic and optimistic cases.
%
Overall, these results suggest a dependence of the SSC significance on both the number and type of parameters to be constrained. Qualitatively, we can argue that SSC is more or less important depending on whether the additional parameters (with respect to the reference case of a flat model with fixed shear bias) introduce degeneracies which are or not scale-dependent and how strong is the degeneracy between these parameters and the amplitude of the power spectrum. In future works lens magnification effects should be included in the analysis as it was shown to have a significant impact on cosmological constraints \citep{Unruh_19}. However, from our results, we can anticipate that the inclusion of magnification-related nuisance parameters will further dilute the impact of SSC.

\subsection{Dependence on redshift binning}
\label{sec:z_bin_variations}

The results summarised in Tables\,\ref{tab:ratio_ref}--\ref{tab:ratio_mult} have been obtained for a fixed choice of number and type of redshift bins. We investigate here how they depend on these settings given that we expect both the G and GS constraints to change as we vary the number and type of bins. We will consider the case of non-flat models, fixing the multiplicative shear bias parameters in order to better highlight the impact of SSC. For this same reason, we will only consider the WL and 3$\times$2pt cases, since SSC has always a modest impact on GCph.
\begin{figure*}[!ht]
\centering
\includegraphics[width=\textwidth]{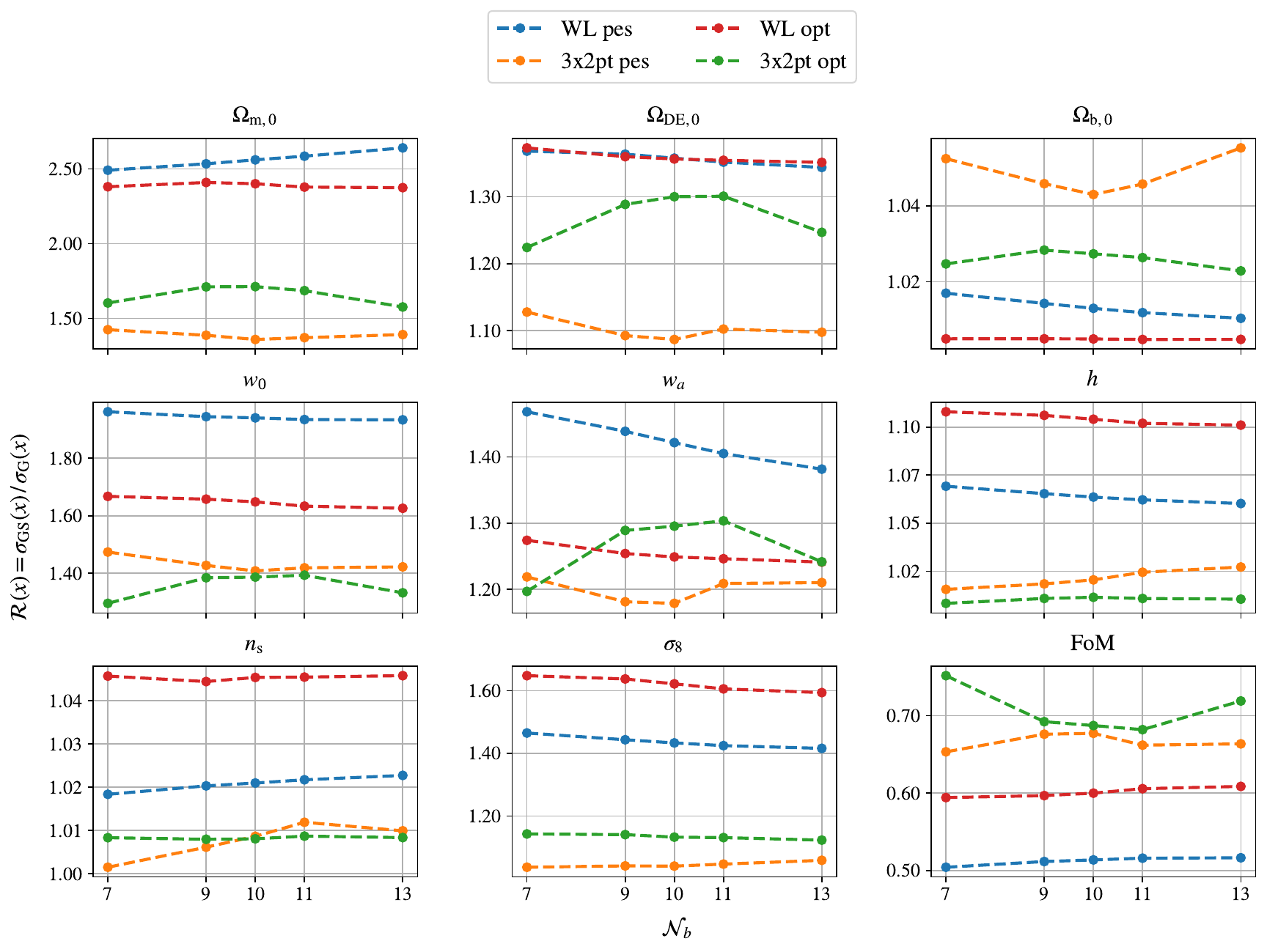}
\caption{Ratio between WL and 3$\times$2pt marginalised uncertainties computed by including or neglecting the SSC contribution, as a function of the number of redshift bins, for the pessimistic and optimistic cases.}
\label{fig:ratiovsnbin}
\end{figure*}
\begin{figure}
     \centering
     \begin{subfigure}[b]{0.49\textwidth}
         \centering
         \includegraphics[width=\textwidth]{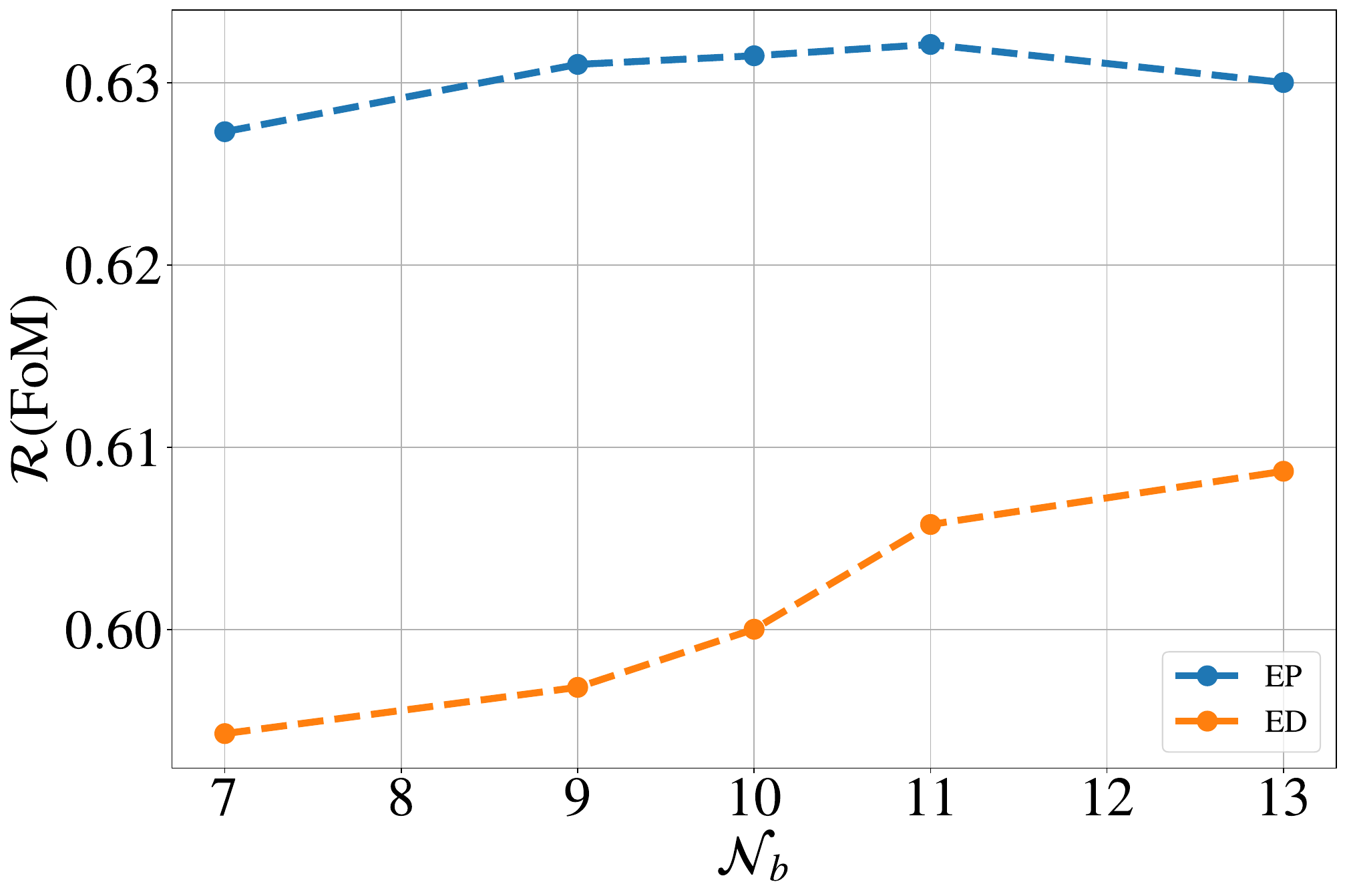}
     \end{subfigure}
     \hfill
     \begin{subfigure}[b]{0.49\textwidth}
         \centering
         \includegraphics[width=\textwidth]{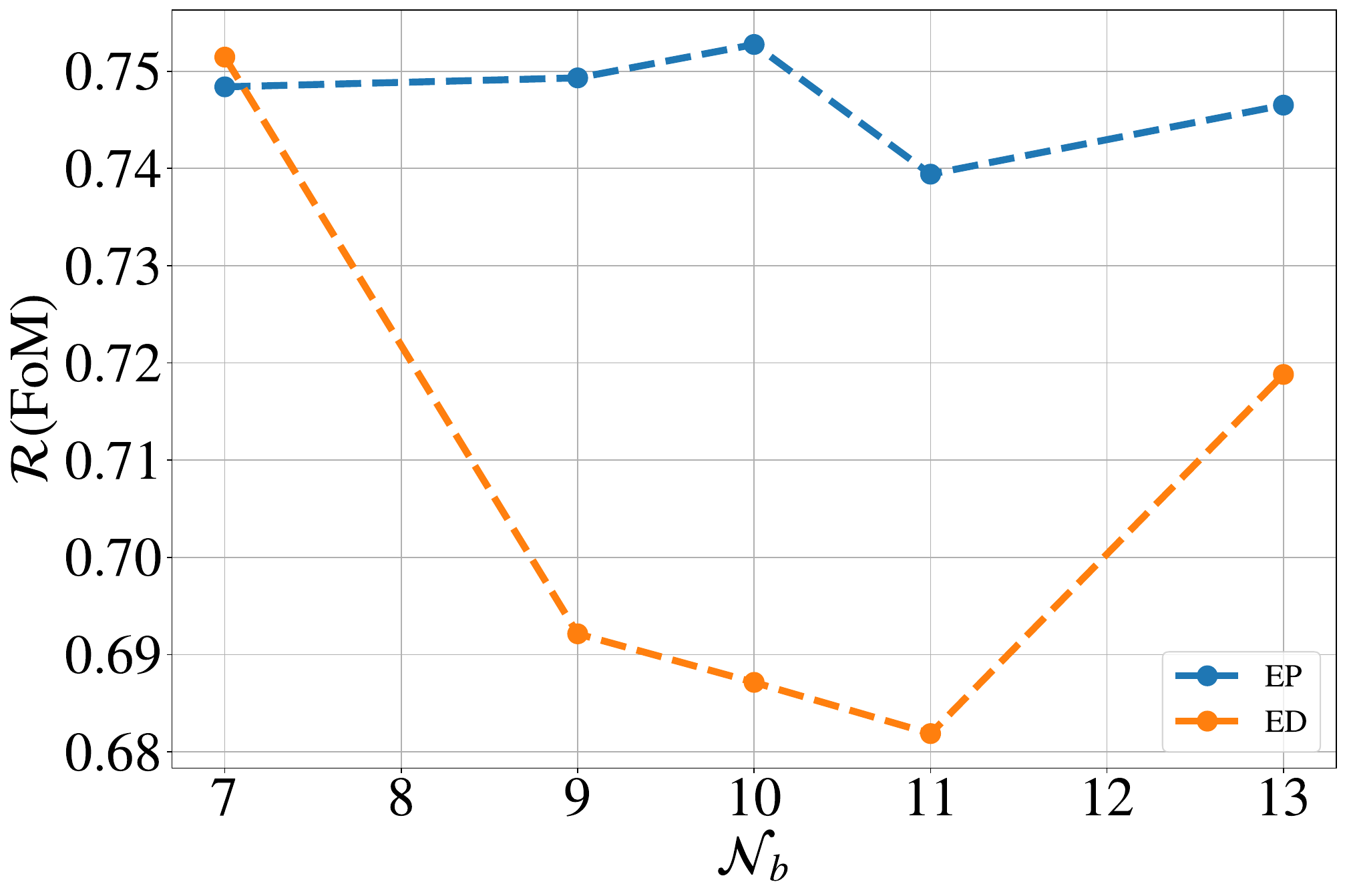}
     \end{subfigure}
        \caption{FoM ratio vs the number of EP and ED redshift bins for WL (left) and 3$\times$2pt (right) in the optimistic scenario.}
    \label{fig:ratioeped}
\end{figure}
Let us first consider changing the number of redshift bins ${\cal N}_{\rm b}$. We show the scaling of ${\cal{R}}(\theta)$ as a function of ${\cal N}_{\rm b}$ for the WL and 3$\times$2pt probes, respectively, in Fig.~\ref{fig:ratiovsnbin} -- for both the pessimistic and optimistic assumptions. The most remarkable result is the weak dependence of ${\cal{R}}({\rm FoM})$ on ${\cal N}_{\rm b}$ as can be inferred from the small range spanned by the curves in the bottom right panel. The scaling of ${\cal{R}}(\theta)$ with ${\cal N}_{\rm b}$ depends, instead, on the parameter and the probe one is looking at. It is quite difficult to explain the observed trends due to the interplay of different contrasting effects. For instance, a larger number of bins implies a smaller number density in each bin, and hence a larger shot noise. As a consequence, the SSC contribution to the total covariance for the diagonal elements will likely be more and more dominated by the Gaussian component because of the larger shot and shape noise terms. However, this effect also depends on the scale so that, should the SSC be the dominant component on the scales to which a parameter is most sensitive, the impact should still be important. On the other hand, a larger number of bins also comes with a larger number of nuisance parameters, which, as shown above, leads to a reduction of the SSC impact. Quantifying which actor plays the major role is hard, which explains the variety of trends in the different panels.

As a further modification to the reference settings, we can change how the redshift bins are defined. Up to now, we have considered equipopulated (EP) bins so that the central bins cover a smaller range in $z$, because of the larger source number density. As an alternative, we divide the full redshift range into ${\cal N}_{\rm b}$ bins with equal length (ED) and recompute the FM forecasts with and without SSC. We show the FoM ratio as a function of the number of bins for EP and ED bins considering WL (left) and 3$\times$2pt (right) probes in the optimistic scenario in Fig.~\ref{fig:ratioeped}. Note that finding the exact number and type of redshift bins used to maximize the constraining power of \Euclid is outside the scope of this paper; this effort is indeed brought forward in the context of the SPV exercise.

In order to qualitatively explain these results, let us first consider the WL case. Given that the bins are no longer equipopulated, the number density of galaxies will typically be larger in the lower redshift bins than in the higher ones. As a consequence, the larger the number of bins, the higher the shape noise in the higher redshift bins, so that the SSC will be subdominant in a larger number of bins, which explains why its impact decreases (i.e. ${\cal{R}}({\rm FoM})$ increases) with ${\cal N}_{\rm b}$. Nevertheless, the impact of SSC will be larger than in the EP case since SSC will dominate in the low redshift bins which are the ones with the largest S/N. However, this effect is less important, so although ${\cal{R}}({\rm FoM})$ is smaller for ED bins than for EP bins, the difference is not greater than 3--5\%.

When adding GCph and XC into the game, the impact of SSC is determined by a combination of contrasting effects. On the one hand, we can repeat the same qualitative argument made for WL also for GCph, thus pointing to ${\cal{R}}({\rm FoM})$ increasing with ${\cal N}_{\rm b}$. No shape or shot noise is included in the XC Gaussian covariance, which is then only determined by how much shear and position are correlated. The larger the number of bins, the narrower they are, and the smaller the cross-correlation between them hence the smaller the Gaussian covariance. This in turn increases the number of elements in the data vector whose uncertainty is dominated by the SSC. Should this effect dominate, we would observe a decrease of ${\cal{R}}({\rm FoM})$ with ${\cal N}_{\rm b}$ with the opposite trend if it is the variation of the shape and the noise of the shot that matters the most. This qualitative argument allows us then to roughly explain the non-monotonic behaviour of ${\cal{R}}({\rm FoM})$ we see in the right panel of Fig.~\ref{fig:ratioeped}. It is worth remarking, however, that the overall change of ${\cal{R}}({\rm FoM})$ for ED bins over the range in ${\cal N}_{\rm b}$ is smaller than $\sim 12\%$, which is also the typical value of the difference between ${\cal{R}}({\rm FoM})$ values for EP and ED bins once ${\cal N}_{\rm b}$ is fixed. 

The analysis in this section, therefore, motivates us to argue that the constraints and FoM degradation due to SSC are quite weakly dependent on the redshift binning.
\subsection{Requirements on prior information}\label{sec:req_on_prior}
The results in the previous paragraph show that the SSC may dramatically impact the constraints on the cosmological parameters. As a consequence, the 3$\times$2pt FoM is reduced by up to $\sim 24\%$ with respect to the case when only the Gaussian term is included in the total covariance. This decrease in the FoM should not be interpreted as a loss of information due to the addition of the SSC. On the contrary, one can qualitatively say that removing SSC from the error budget is the same as adding information that is not actually there. 
It is nevertheless interesting to ask which additional information must be added to recover the Gaussian FoM, which is usually taken as a reference for gauging the potential of a survey. This information can come from priors on the nuisance (or cosmological) parameters. 
In the following section, we will investigate the former option by adding Gaussian priors on the galaxy and multiplicative shear bias parameters. This is easily done in the FM formalism, as explained in Sect.~\ref{sec:fisher_theory}, by adding $(\sigma_\alpha^{\rm p})^{-2}$ to the appropriate diagonal elements of the G and GS FMs ($\sigma_\alpha^{\rm p}$ being the value of the prior uncertainty on parameter $\alpha$).

\begin{figure}
     \centering
     \begin{subfigure}[b]{0.49\textwidth}
         \centering
         \includegraphics[width=\textwidth]{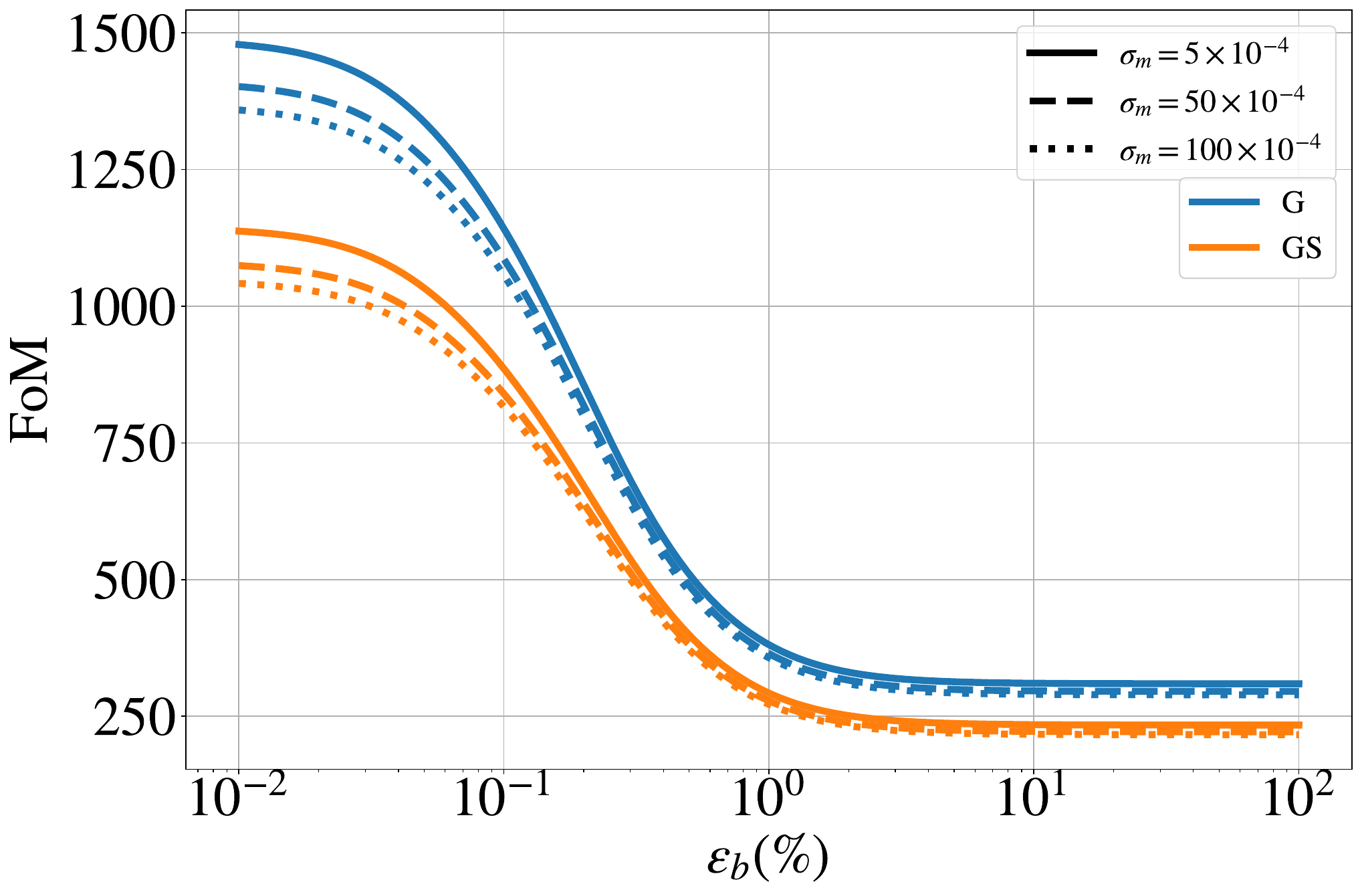}
     \end{subfigure}
     \hfill
     \begin{subfigure}[b]{0.49\textwidth}
         \centering
         \includegraphics[width=\textwidth]{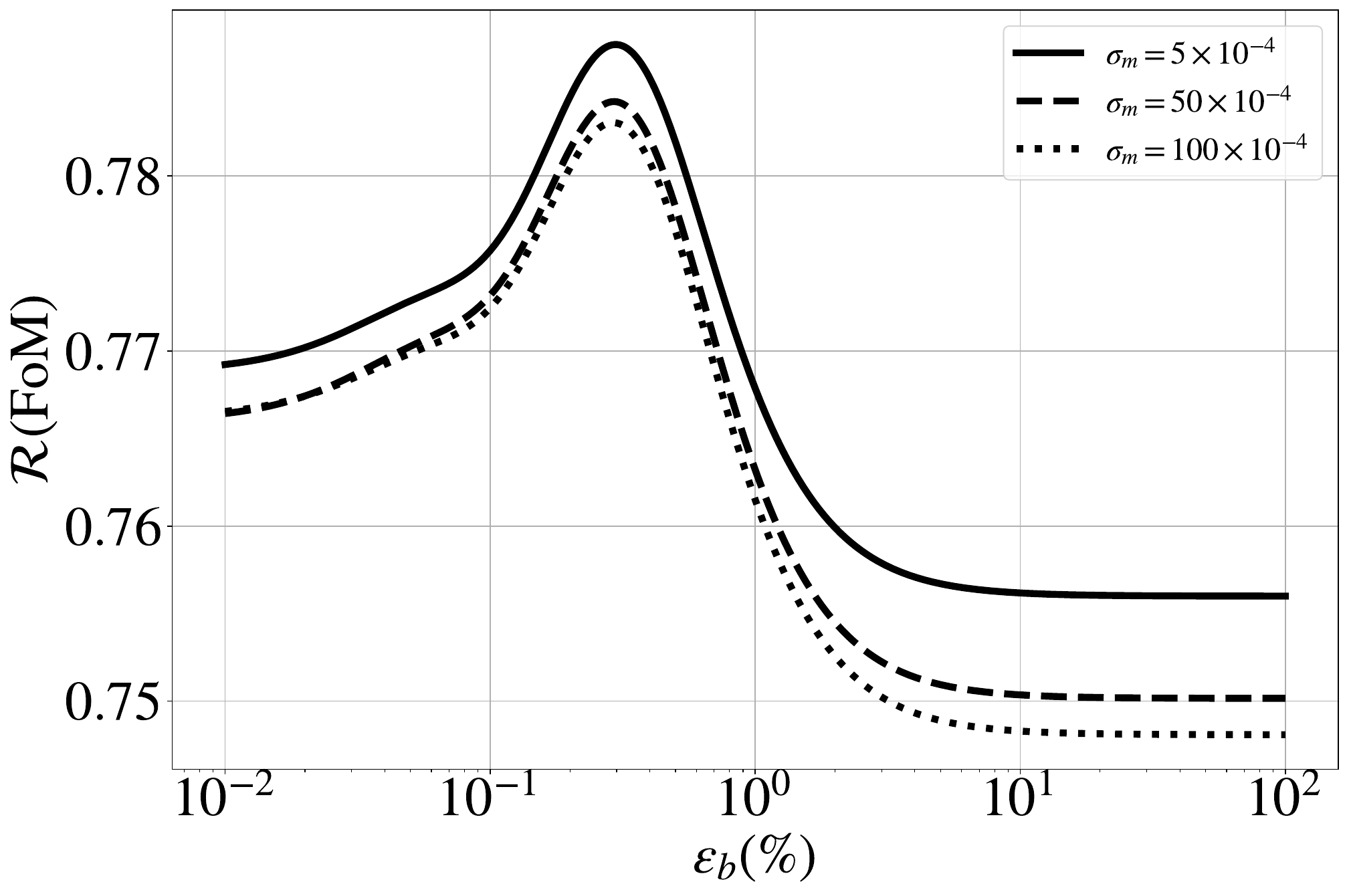}
     \end{subfigure}
        \caption{{\it Left.} 3$\times$2pt FoM in the optimistic scenario with and without SSC as a function of the percentage prior $\varepsilon_b$ on the galaxy bias parameters for $\sigma_m = (5, 50, 100) \times 10^{-4}$ (solid, dashed, dotted lines). {\it Right.} FoM ratio as function of $\varepsilon_b$ for the three $\sigma_m$ values in the left panel.}
    \label{fig:ratiofomvsprior}
\end{figure}

To this end, we consider the realistic case of a non-flat model plus the galaxy bias and multiplicative shear bias as nuisance parameters. As a simplifying assumption, we will assume that all the ${\cal N}_{\rm b}$ bias values $b_i$ are known with the same percentage uncertainty $\varepsilon_b = \sigma_b/b_{\rm fid}$, while we put a prior $\sigma_m$ on all the $m_i$ parameters (having set the fiducial value $m_{\rm fid}$ to 0). We then compute the FoM with and without SSC for the 3$\times$2pt probe in the optimistic scenario and investigate how the ratio ${\cal{R}}({\rm FoM})$ scales with $(\varepsilon_b, \sigma_m)$ obtaining the results shown in Fig.~\ref{fig:ratiofomvsprior}. 

A prior on the nuisance parameters increases both the Gaussian and Gaussian\,+\,SSC FoM so that one could expect their ratio to be independent of the prior itself. This is not exactly the case since the correlation between different multipoles introduced by SSC alters the way the prior changes the FM elements. As a result, we find a non-flat scaling of ${\cal{R}}({\rm FoM})$ as can be seen from the right panel of Fig.~\ref{fig:ratiofomvsprior}. When a strong prior is set on the galaxy bias (i.e., $\varepsilon_b \ll 1$), there is not much gain in improving the knowledge of the multiplicative shear bias so that the solid, dashed, and dotted lines (corresponding to three $\sigma_m$ values) are quite close to each other. This is no longer the case for larger $\varepsilon_b$ values (i.e., weak or no prior on the bias): lowering $\sigma_m$ has now a larger impact on ${\cal{R}}({\rm FoM})$.
The non-monotonic behaviour of ${\cal{R}}({\rm FoM})$ with $\varepsilon_b$ tells us that ${\rm FoM}_{\rm GS}$ increases with decreasing $\varepsilon_b$ faster (slower) than ${\rm FoM}_{\rm G}$ when the galaxy bias is known with an uncertainty smaller (higher) than the sub-percent level. Another way to interpret it is that the information gained in the FoM saturates faster when SSC is included: better constraints on $\varepsilon_b$ do not bring more information as the SSC now dominates the error budget. However, it is worth stressing that, even for a strong prior on the multiplicative shear bias, the FoM ratio can actually be improved by less than a few percent under the (likely unrealistic) assumption of a sub-percent prior on the galaxy bias.

\begin{figure}[ht]
\centering
\includegraphics[width=0.7\hsize]{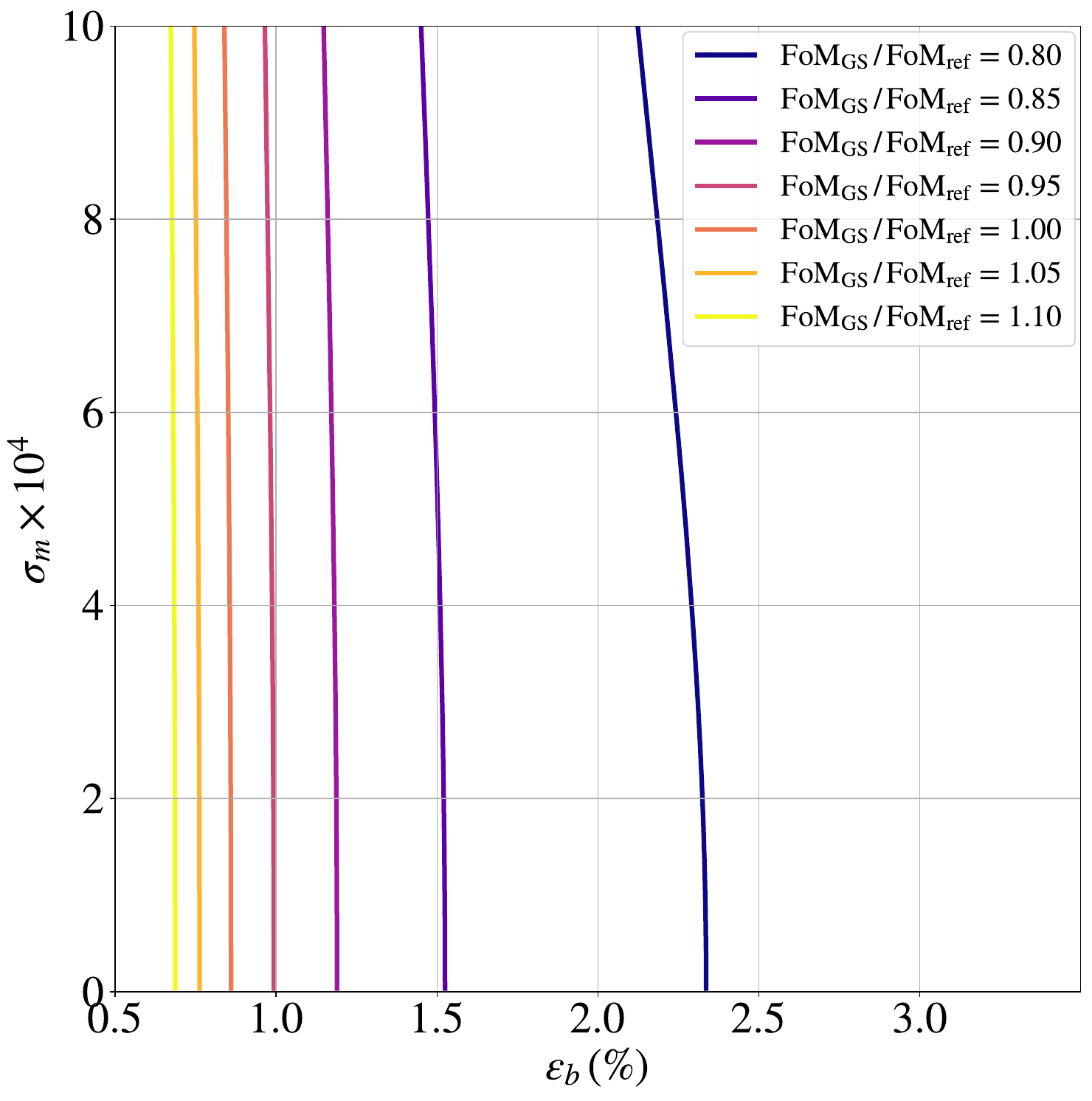}
\caption{${\rm FoM_{GS}}$ contours in the $(\varepsilon_b, \sigma_m)$ plane for ${\rm FoM_{GS}/FoM_{\rm ref}}$ going from 0.8 to 1.1 in steps of 0.05 (from right to left).} 
\label{fig:gsrefvsprior}
\end{figure}

The need for such strong priors comes from the attempt to retrieve the same FoM as a Gaussian case. Alternatively, one can also wonder which additional information must be added through priors to retrieve the idealised FoM value obtained in forecasts that neglect the SSC. In other words, we look for the requirements that must be put on the priors $(\varepsilon_b, \sigma_m)$ in order to make ${\rm FoM}_{\rm GS}/{\rm FoM}_{\rm ref} = 1$, where ${\rm FoM}_{\rm ref} = 295$ is the FoM computed for a non-flat reference case without SSC and with no priors on galaxy bias, but a fiducial prior $\sigma_m = 5 \times 10^{-4}$ on the shear bias. The answer to this question is shown in Fig.~\ref{fig:gsrefvsprior} for the optimistic scenario and 10 equipopulated redshift bins. Some numbers help to better understand how priors can indeed supply the additional information to retrieve the FoM one would obtain in an ideal case where SSC is absent. Solving 
%
\begin{displaymath}
{\rm FoM}_{\rm GS}(\varepsilon_b, \sigma_m) = f \, {\rm FoM}_{\rm ref}
\end{displaymath}
with respect to $\varepsilon_b$, we get 
    \begin{displaymath}
    \varepsilon_b = \left \{
    \begin{array}{ll}
    \displaystyle{(2.34, 1.19, 0.86) \, \%} & \displaystyle{{\rm for} \; \sigma_m = 0.5 \times 10^{-4}} \\
    & \\
    \displaystyle{(2.27, 1.18, 0.85) \, \%} & \displaystyle{{\rm for} \; \sigma_m = 5 \times 10^{-4}} \\
    & \\
    \displaystyle{(1.40, 0.93, 0.72) \, \%} & \displaystyle{{\rm for} \; \sigma_m = 100 \times 10^{-4}} \; , \\
    \end{array}
    \right . 
    \end{displaymath}
%
%
where the three values refer to $f = (0.8, 0.9, 1.0)$. These numbers (and the contours in Fig.~\ref{fig:gsrefvsprior}) show that it is indeed possible to compensate for the degradation due to SSC by adding strong priors on the galaxy bias, which have a much larger impact on the (G and GS) FoM than strong priors on the multiplicative shear bias. However, it is worth noticing that it is actually easier to obtain priors on the multiplicative shear bias provided a sufficient number of realistic image simulations are produced and fed to the shear measurement code to test its performance. It is therefore worth wondering how much the FoM is restored by improving the prior on $m$ for a fixed one on the bias. We find
%
%
\begin{displaymath}
    \frac{{\rm FoM_{GS}}}{{\rm FoM_{\rm ref}}} = \left \{
    \begin{array}{ll}
    \displaystyle{(2.87, 2.86, 2.64)} & \displaystyle{{\rm for} \; \varepsilon_b = 0.1\%} \\
     & \\
    \displaystyle{(0.95, 0.95, 0.88)} & \displaystyle{{\rm for} \; \varepsilon_b = 1\%} \\
     & \\
    \displaystyle{(0.76, 0.76, 0.70)} & \displaystyle{{\rm for} \; \varepsilon_b = 10\%} \; , \\
    \end{array}
    \right .
    \end{displaymath}
with the three values referring to $\sigma_m = (0.5, 5.0, 100) \times 10^{-4}$. As expected, improving the prior on the multiplicative bias with respect to the fiducial one (which, we remind, is included in ${\rm FoM_{\rm ref}}$) does not help a lot in recovering the constraining power. However, a $1\%$ prior on the galaxy bias can almost fully recover the reference FoM thanks to the additional information compensating for the presence of SSC. 

Investigating whether the priors proposed here can be achieved in practice (e.g., through theoretical bias models tailored to galaxy clustering data or $N$-body hydrodynamic simulations) is outside the aim of this work. We refer the interested reader to, e.g., \citet{Alex2021} and \citet{Zen22} for some preliminary results.

\section{\ML{Summary}}\label{sec:conclu}

Precision cosmology asks for precision computation too: previously neglected theoretical contributions must therefore now be taken into account. Motivated by this consideration, we have here computed and studied the impact of SSC on the \Euclid photometric survey, exploring how the different probes and their combination are affected by this additional, non-Gaussian term in the covariance matrix. The analysis of the impact of SSC on the spectroscopic survey, which has been shown to be small in \cite{wadekar_20} for the BOSS  data, is left for future work.  We employed a FM analysis, producing forecasts of the $1\sigma$ marginalised uncertainties on the measurement of the cosmological parameters of the flat and non-flat $w_0w_a$CDM cosmological models. We validated two different forecast pipelines against the results of \citetalias{ISTF2020}, taking as reference survey the one specified therein, and then updated the galaxy bias and the source redshift distributions according to the most recent versions presented in \citet{Pocino2021}. The SSC was computed relying on the analytical approximations and numerical routines presented in \citetalias{Lacasa_2019}, interfacing the public code \texttt{PySSC} with two distinct forecast pipelines to validate the constraints. As a further step forward, we build upon the work of \citetalias{Lacasa_2019} by computing the scale and redshift dependence of the response functions of the different probes, starting from the results of \citet{Wagner2015} and \citet{Barreira2018response_approach}.

We find the severity of the impact, quantified by the ratio $\sigma_{\rm GS}/\sigma_{\rm G}$ between the marginalised uncertainties with and without SSC, to vary substantially between different parameters and probes. For both WL and GCph, the most affected parameters are $(\Omega_{{\rm m,0}}, w_0, \sigma_8)$, while the constraints on $(\Omega_{{\rm b,0}}, h, n_{\rm s})$ are only weakly degraded by SSC. However, there is a great difference between the two probes in how much the constraints worsen because of SSC. In agreement with previous results \citep{upham2021, Barreira2018cosmic_shear}, we found the WL case to be dramatically impacted by SSC so that the corresponding FoM is reduced by as much as $55\%$, while GCph is less affected with the FoM decrease being about $17\%$. The 3$\times$2pt case sits in between these two since it receives contributions from both extreme cases. These results are the consequence of a complicated interplay among three factors. First, SSC originates from the uncertainty in the determination of the background mean density when measuring it over a finite region. This prevents determining the overall amplitude of the matter power spectrum hence increasing the uncertainty on those parameters that concur in setting its amplitude, mainly $\Omega_{{\rm m,0}}$ and $\sigma_8$. Secondly, the elements of the SSC matrix depend on the amplitude of the response functions. Thirdly, the impact depends on how large a contribution the signal receives from the low-$z$ region, where the effective volume probed is smaller, making the variance of the background modes larger. Both the last two factors are more severe for WL than for GCph, hence causing the former probe to be more affected than the latter. \\
Finally, the deviation of a given element of the GS FM from the Gaussian one depends also on its correlations: in other words, the degradation of the constraints on a given parameter can be large if this is strongly correlated with a parameter severely degraded by SSC. Quantifying the impact of SSC on a single parameter is therefore quite hard in general, and must be investigated on a case-by-case basis taking care of the details of the probe and the way it depends on the parameter of interest.

Nuisance parameters to be marginalised over act as a sort of additional contribution to the covariance. As such, the importance of both the Gaussian and SSC contribution to the overall effective covariance becomes less important when the number of nuisance parameters increases. In order to consider cases that mimic the most future \Euclid data, we have opened up the parameter space by adding $\Omega_{{\rm DE,0}}$ (i.e., removing the flatness prior), and the multiplicative shear bias. It turns out that, as long as the additional parameters have a scale-independent degeneracy with the most impacted ones, the relative impact of SSC decreases. We stress, however, that this reduction in the SSC impact does not come for free. On the contrary, the marginalised uncertainties on the parameters are definitely worsened, but the degradation is roughly the same whether the SSC is included or not, hence making the ratio $\sigma_{\rm GS}/\sigma_{\rm G}$ closer to unity for all parameters and probes. This result can be taken as a warning against investing too much effort in refining the estimate of the computationally expensive SSC when no approximations are done. For a \Euclid-like survey, the main concern would indeed be the number of nuisance parameters, which makes less relevant the impact of the SSC itself.

We furthermore note that, in light of the recent theoretical developments presented in \citet{Lacasa2022}, it appears feasible to include the effect of SSC in the form of nuisance parameters, which would be the value of the density background $\delta_\mathrm{b}$ in each redshift bin. This approach is interesting as it would reduce the complexity of the data covariance matrix and would allow for a simpler interpretation of the effect of SSC and how it is correlated to the other cosmological and nuisance parameters. 

Variations in the $z$ binning strategy have contrasting effects: a larger number of bins means a larger number of nuisance parameters (either galaxy bias or multiplicative shear bias for each bin), which leads to a loss of constraining power. Moreover, the larger the number of bins, the larger the Gaussian contribution to the covariance, causing the shot and shape noise to dominate over the SSC for diagonal elements. On the downside, a larger number of bins leads to larger data vectors, thus adding information that can partially compensate for the increase in covariance. The contrasting effects at play conspire in such a way that the degradation of the FoM due to SSC ends up being approximately independent of the number of redshift bins (cf. Fig.~\ref{fig:ratioeped}). 

An interesting development in this sense is to leverage the SSC dependence on the low-$z$ contribution to investigate whether its impact could be mitigated by the use of the BNT transform \citep{Bernardeau_2014}, which transforms redshift bins in such a way as to increase the separation between the WL kernels (see Sect.~\ref{sec:BNT}).

An alternative strategy is to increase the constraining power by adding information through informative priors, hence recovering the FoM when SSC is incorrectly neglected. We investigate this possibility by quantifying the requirements on the prior information needed to recover the Gaussian FoM. Our results show that the main role is played here by the priors on galaxy bias parameters, while the FoM recovery quickly saturates with the prior on the multiplicative shear bias. However, the galaxy bias must be known to sub-percent level in order to recover $\sim 90\%$ of the Gaussian FoM. Investigating whether this is possible is beyond the scope of this work. We nevertheless note that including such remarkable prior information is the same as stating we are able to model the evolution of the bias with redshift. This is actually quite difficult based on the current knowledge of galaxy formation processes. Alternatively, one could investigate whether an empirical fitting formula can be found as a compromise between the need for strong priors on bias and the number of nuisance parameters.

Although some more work is needed to make the results more robust, we can safely conclude that the effect of including the SSC term in the total covariance matrix of \Euclid photometric observables is definitely non-negligible, especially for WL and 3$\times$2pt. However, the degradation of the constraints on cosmological parameters depends on the particular probe and the number and kind of parameters to constrain. The FoM is nevertheless reduced by $32\%$ ($25\%$) for the 3$\times$2pt probe in the pessimistic (optimistic) scenario in the case where all cosmological (including $\Omega_{\rm DE,0}$) and nuisance (multiplicative shear bias) parameters are left free to vary.
\chapter{Scale Cuts}\label{chap:scalecuts}
The treatment of SSC for \Euclid is just a part of the broader effort to characterize the constraining power of the survey. This is an inter-group endeavour, framed in the context of the \enquote{Science Performance Verification}. The aim of this exercise is twofold:
\begin{itemize}
    \item Verifying that the expected performance of \Euclid is still in line with the core science objectives of the nominal mission.
    \item Developing and maintaining tools and codes used to make quick assessments of the performances of the mission for different scenarios.
\end{itemize}
These aims are achieved through the collaboration between teams focusing on virtually all of the scientific aspects of the mission: from the survey strategy to the calibration, to the modelling of the PSF, the choice of the nonlinear model and the binning strategy, just to name a few.

Our work in this activity mainly concerned the optimization of the survey, that is, the search for a reference choice for some of the many ingredients involved in the computation of the final forecasts. In particular, we studied the effect of different prescriptions and codes for the computation of the nonlinear matter PS, the inclusion of systematics such as multiplicative shear bias, the shift in the mean of the redshift distribution, magnification bias, the optimal type and number of redshift bins, and the impact of different magnitude and scale cuts, using as main metric the 3$\times$2pt FoM. 

This Chapter illustrates in detail the last of these. Section~\ref{sec:BNT} presents the BNT transform, an essential ingredient to include scale cuts in the harmonic space photometric analysis; Sect.~\ref{sec:angular_scale_cuts} shows the procedure used, on the basis of BNT transform, to obtain redshift-dependent \textit{angular} scale cuts, and Sect.~\ref{sec:k_cut_forecasts} illustrates the forecast setup and the results obtained. We present our conclusions in Sect.~\ref{sec:bnt_conclusions}.

These results, along with the other themes explored in the SPV, will be published once the exercise is concluded.
\section{Scale cuts and the BNT transform}\label{sec:BNT}
As stated earlier in this work, the current decades have seen Cosmology rise from the status of \enquote{data-starved} \citep{tegmark1997} to \enquote{data-rich} science. This will be increasingly the case now that state-of-the-art Stage IV LSS surveys are beginning to come online. In such a scenario, it is actually the theoretical side to have become the limiting factor in some (important) aspects of the analysis. In particular, \Euclid will be able to probe scales that, at this point in time, are poorly understood. Our analytical models, some of which have been introduced in Sect.~\ref{sec:nonlinear_scales}, are unable to capture either DM structure formation deeply in the nonlinear regime, or the impact of baryonic (collisional) matter, which introduces a whole new layer of complexity. 

A viable option, widely used in the literature, is to resort to magnetohydrodynamic (MHD) numerical simulations (see \citealt{Chisari2019_baryon_sim_comparison} for a review of the most recent simulations, as well as the approximate methods), which however often give different results based on the modelling choices adopted and the numerical implementation. Another possibility is to try to capture baryonic feedback through parameters which are marginalized over or calibrated against simulations.

An alternative, more conservative approach is simply to remove (\enquote{cut}) the elements of the data vector corresponding to the poorly understood scales. This approach has already been adopted, for example, in \citetalias{ISTF2020}, which as illustrated in Sect.~\ref{sec:datavectors} defined \enquote{pessimistic} and \enquote{optimistic} scenarios, depending on our ability to exploit the high-$\ell$ measurements. The cuts hereby defined have two main problems.

Firstly, they are redshift-independent. This means that we assume to be able to model the same maximum angular scale, $\ell_{\rm max}$, without taking into account whether we are considering a close source (which will then correspond to a smaller physical scale, or larger $k_{\rm max}$) or a faraway source (for which the opposite will be true).

Secondly, the correspondence between physical scale $k$ and angular scale $\ell$ is not, in general, well-defined. We have seen that the angular PS is obtained from a radial projection of the three-dimensional PS, weighted by some probe-specific radial kernel: this means that a whole \textit{range} of scales $k$ will contribute to a given $\ell$ mode. The extension of this range is modulated by the redshift width of the kernels, so that narrow kernels will result in a better consistency between $\ell$ and $k$. \\
More specifically (rewriting here Eq.~\ref{eq:cl_AB_tomo_limb_ISTF} for convenience), this correspondence is given by the Limber relation, which tells us how to compute, for a given $\ell$ value, the $k_\ell$ entering the $C(\ell)$ integral as a function of redshift: $k_\ell = \frac{\ell + 1/2}{\chi(z)}$.
\begin{equation}\label{eq:cl_AB_tomo_limb_ISTF_spvchap} 
    C_{ij}^{AB} (\ell) = \frac{c}{H_0}
    \int_0^{z_{\rm max}} \diff z \frac{\mathcal{K}_i^A(z)\mathcal{K}_j^B(z)}
    {E(z)\chi^2(z)}P_{AB}\left[ k_{\ell} = 
    \frac{\ell + 1/2}{\chi(z)},z
    \right] \; .
\end{equation}
Wide kernels will result in a large support for the $z$ integral and a large range of scales contributing to the overall power at the same multipole. This issue is relevant for GCph, because of the broadening of the kernels due to photo-$z$ uncertainties, but much more so for WL, because as discussed the kernel is necessarily very broad, spanning the whole distance between source and observer.\\

A potential solution to this problem has been proposed in \citet{Bernardeau_2014}. The main idea outlined in this work is to transform the lensing kernels so as to make them separable in $z$, by weighting galaxies according to their photometric redshifts. In practice, this means reorganizing the data so that the lensing signal in a given redshift bin becomes only sensitive to a narrow distribution of sources. This makes the $k-\ell$ relation more precise, and largely reduces the overlap between the kernels.\\

To illustrate this transformation, we sketch the steps illustrated in the original paper. We begin by writing the projected convergence (but the transform can be applied also to the shear angular PS, the two differing only by an $\ell$-dependent prefactor) as:
\begin{equation}\label{eq:convergence_field}
    \kappa=\frac{3\Omega_{{\rm m}, 0}H_0^2}{2c^2}\sum_{i} p_{i}\,\int_{0}^{\chi_{i}}\diff \chi\ \frac{S_k(\chi_{i}-\chi)S_k(\chi)}{S_k(\chi_{i})}
    \frac{\delta_{\rm m}(\chi)}{a(\chi)} \; ,
\end{equation}
where $S_k(\chi)$ has been defined in Eq.~\eqref{eq:S_k(r)} and $p_{i}$ are dimensionless weight coefficients. We now take three discrete source planes, $i, i-1, i-2$ at comoving distance $\chi_i > \chi_{i-1} > \chi_{i-2}$: the main idea is to find a set of weights $p_i$ such that the contribution to the projected convergence for $\chi < \chi_{i-2}$ vanishes. Being this already true for $\chi > \chi_i$, because background lenses do not contribute to the signal, this would make the signal only sensitive to lenses in the narrow range $[\chi_{i-2}, \chi_i]$. \\
To do this, we can rewrite Eq.~\eqref{eq:convergence_field} in the form
\begin{equation}
\kappa=\frac{3\Omega_{{\rm m}, 0}H_0^2}{2c^2}\int_{0}^{\chi_{\rm H}}\diff\chi\ \frac{\delta(\chi)}{a(\chi)}\,w(\chi),
\end{equation}
$\chi_{\rm H}$ being the comoving distance to the horizon and having defined the weight function
\begin{align}
    w(\chi) & =\sum_{i,\,\chi_{i}>\chi} p_{i}\,\frac{S_k(\chi_{i}-\chi)S_k(\chi)}{S_k(\chi_{i})}
    \label{eq:bnt_weights} \\
    &=S_k^2(\chi)\left[\frac{1}{g_{k}(\chi)}\sum_{i, \chi_{i}>\chi}p_{i}-\sum_{i, \chi_{i}>\chi}\frac{p_{i}}{g_{k}(\chi_{i})}\right] \label{w3} \; ,
\end{align}
with 
\begin{eqnarray}
\gK(\chi) \equiv\left\{
\begin{array}{cc}
\displaystyle\frac{\tan(\sqrt{K}\chi)}{\sqrt{K}} & \mathrm{for}\,\,K>0,\\
\displaystyle\chi & \mathrm{for}\,\,K=0,\\
\displaystyle\frac{\tanh(\sqrt{-K}\chi)}{\sqrt{-K}} & \mathrm{for}\,\,K<0\ ; .
\end{array}
\right.
\end{eqnarray}
The condition $w(\chi) = 0$ for $\chi \notin [\chi_{i-2}, \chi_i]$ then translates into the system of equations
\begin{equation}
    \left\{
    \begin{aligned}
        \sum_{i=1}^{3}p_{i}&\,=0 \\
        \sum_{i=1}^{3}\frac{p_{i}}{g_k(\chi_{i})}&\,=0 \; ,
    \end{aligned}
    \right.
\end{equation}
which has the unique solution 
\begin{equation}\label{eq:three_plane_solution}
    \frac{p_2}{p_1} =\frac{c(2,3,1)}{c(1,2,3)},\quad \frac{p_3}{p_1} =\frac{c(3,1,2)}{c(1,2,3)} \; ,
\end{equation}
having defined
\begin{equation}
    c(i,j,k)=g_{k}(\chi_{i})\left[g_{k}(\chi_{j})-g_{k}(\chi_{k})\right] \; .
\end{equation}
The generalization to the case of an arbitrary number $n$ of discrete planes $\kappa_j$ requires taking linear combinations of the solutions~\eqref{eq:three_plane_solution}:
\begin{equation}
    \Tilde{\kappa}_i = \mathcal{M}_{ij}\kappa_j
\end{equation}
where the tilde denotes BNT-transformed quantities and the matrix $\mathcal{M}$, which we refer to as the BNT matrix (of size $\mathcal{N}_{\rm b} \times \mathcal{N}_{\rm b}$), is defined by:
\begin{align}
\begin{aligned}
    \mathcal{M}_{ii}= &\; 1,  \\
    \mathcal{M}_{i, i-1}= &\; c(i-1,i-2,i)/c(i-2,i,i-1), \\
    \mathcal{M}_{i, i-2}= &\; c(i-2,i-1,i)/c(i-2,i,i-1) \; .
\end{aligned}
\end{align}
This gives the solution for $n-2$ of the new maps. To make the transformation invertible, we set 
\begin{equation}\label{eq:bnt_matrix_conditions}
    \mathcal{M}_{11}=1,\quad \mathcal{M}_{21}=-1,\quad \mathcal{M}_{22}=1 \; ,
\end{equation}
so that the first two maps, $\Tilde{\kappa}_1$ and $\Tilde{\kappa}_2$, are left unchanged. The normalization is set to have the diagonal elements equal to 1.\\

\begin{figure}
    \centering
    \includegraphics[width=0.7\linewidth]{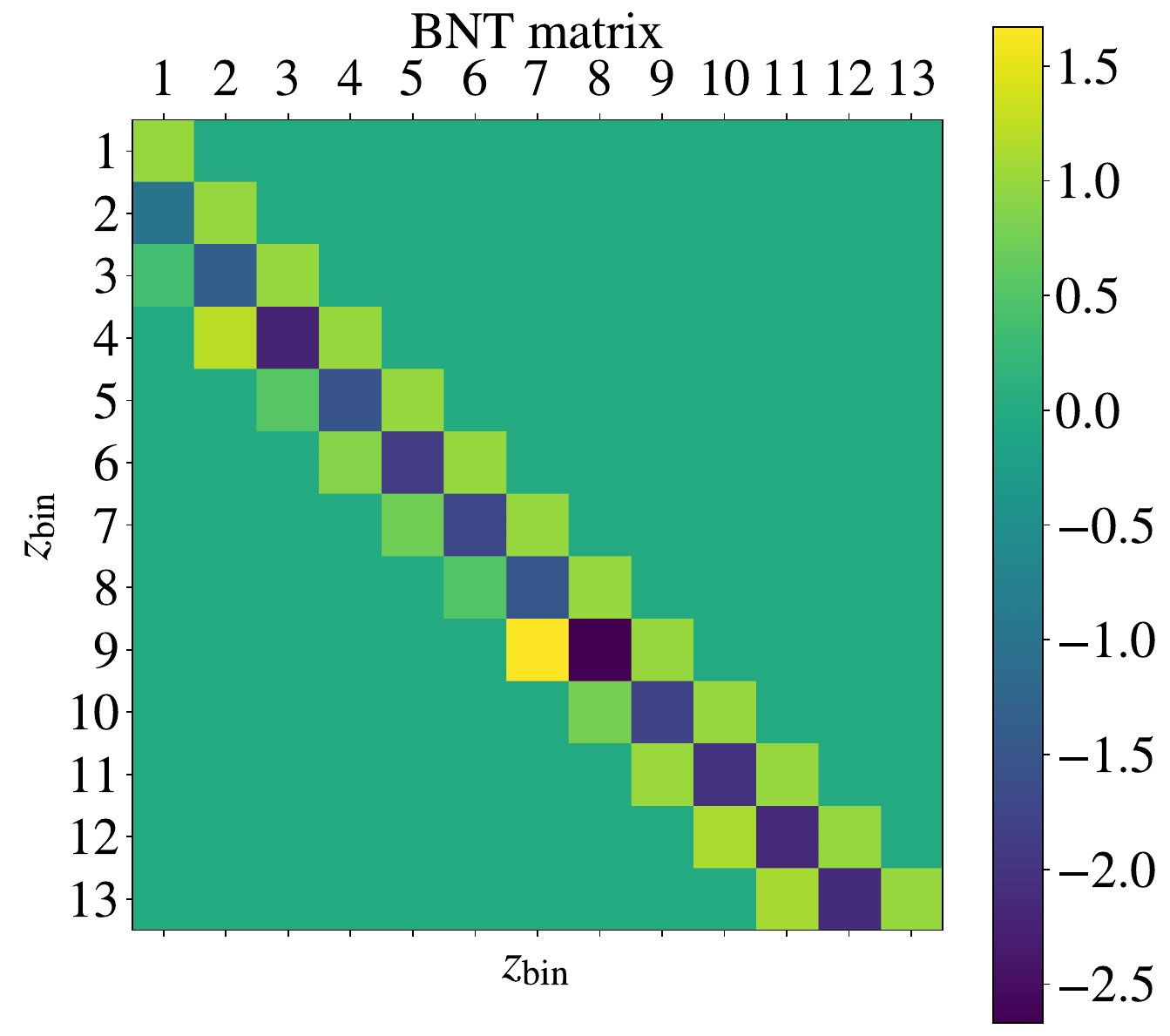}
    \caption{BNT matrix from the Flagship 2 $N$-body simulation.}
    \label{fig:bnt_matrix_fs2}
\end{figure}

We can further generalize this result to the case of a continuous distribution $n_i(z)$ instead of a set of planes, by defining \citep{Taylor2021_xcut} (in the flat case):
\begin{equation}\label{eq:Bi_bnt}
    B_i = \int_{0}^{z_{\rm H}} \diff z' \frac{n_i(z')}{\chi(z')} \; ,
\end{equation}
where $z_{\rm H}$ is the redshift of the horizon (or the maximum distance available in the survey). In this way, we finally get the system:
\begin{equation}\label{eq:bnt_continuous_eq_system}
    \left\{
    \begin{aligned}
        \sum_{j=i-2}^i \mathcal{M}_{ij}& =0 \\
        \sum_{j=i-2}^i \mathcal{M}_{ij} B_j&=0 \; .
    \end{aligned}
    \right.
\end{equation}
which again must be accompanied by the condition $\mathcal{M}_{ii} = 1$. From Eq.~\eqref{eq:Bi_bnt} we see that the BNT matrix is cosmology-dependent, since it involves an integral over the comoving distance; moreover, its computation requires an accurate characterization of the redshift distribution $n_i(z)$. We will nonetheless neglect this cosmology dependence, which \citet{Taylor2021_xcut} found to have little impact on the results. The BNT matrix is shown for the specifics under consideration in this Chapter in Fig.~\ref{fig:bnt_matrix_fs2}.
\\

Since the transform is linear, it can be applied to the kernels:
\begin{equation}
    \Tilde{\mathcal{K}}^{\rm L}_i = \mathcal{M}_{ij}\mathcal{K}^{\rm L}_j \; ,
\end{equation}
or to the $C(\ell)$s. For the 3$\times$2pt case, since the angular PS contain products of kernels, we have \citep{Taylor2021O_kcut_3x2pt}:
\begin{align}\label{eq:cl_BNT_3x2pt}
\begin{aligned}
    & \Tilde{C}^{\rm LL}_{ij}(\ell) = \mathcal{M}_{im} C^{\rm LL}_{mn}(\ell) (\mathcal{M}^T)_{nj} \\
    & \Tilde{C}^{\rm GL}_{ij}(\ell) = \delta^{\rm K}_{im} C^{\rm GL}_{mn}(\ell) (\mathcal{M}^T)_{nj} \\
    & \Tilde{C}^{\rm GG}_{ij}(\ell) = \delta^{\rm K}_{im} C^{\rm GG}_{mn}(\ell) \delta^{\rm K}_{nj} = C^{\rm GG}_{ij}(\ell) \; ,
\end{aligned}
\end{align}
where the Kronecker delta reflects the fact that the BNT is not applied to the galaxy kernels (which are already narrow). The angular PS transformed in this way have a very low sensitivity on the matter power spectrum above the chosen $k$ scale, which allows mitigating the potential bias due to incorrect modelling while retaining the large-scale information when cutting the angular scales. Additionally, the $k-\ell$ correspondence can be further improved by taking into account a larger number of $z$ bins \citep{Vazsonyi2021_kcut_HSC}.\\

Having defined the BNT matrix, we illustrate the procedure used to test the impact of these well-defined scale cuts, with a particular focus on the 3$\times$2pt FoM. This has been referred to \enquote{$k$-cut} analysis in \citet{Taylor2018_kcut}.
\section{Angular scale cuts}\label{sec:angular_scale_cuts}
\begin{figure}
    \centering
    \includegraphics[width=0.75\linewidth]{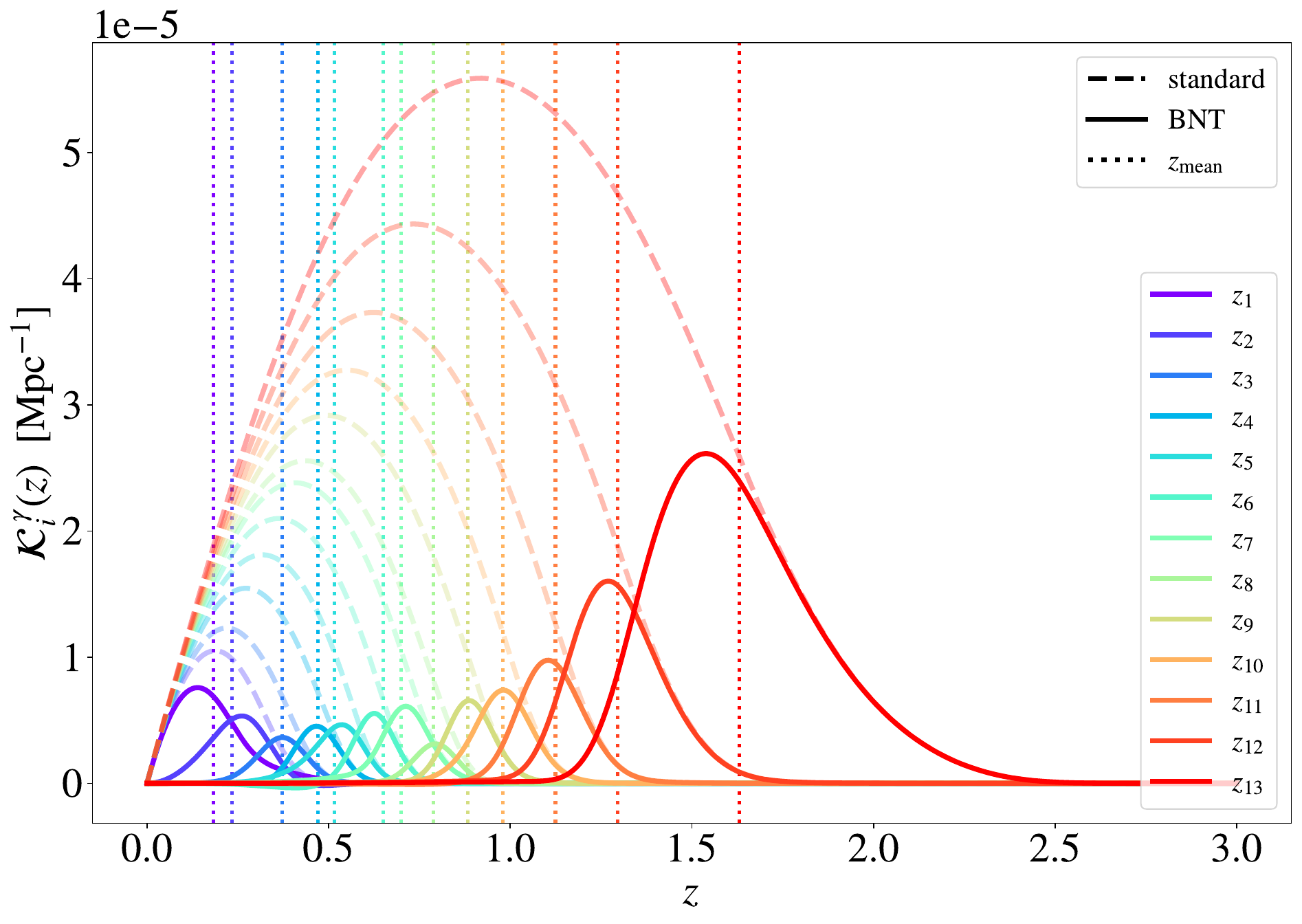}
    \caption{Standard (dashed lines) and BNT-transformed (solid lines) $\gamma$ kernels. The latter have much narrower support and are clearly separated, allowing the typical bin distance $\chi(z_{{\rm mean}, i})$ (dotted vertical lines) to be well-defined.}
    \label{fig:std_and_bnt_gamma_kernel}
\end{figure}

The above-mentioned Limber relation between physical and angular scale, after BNT-transforming the radial kernels, can be used as a way to define redshift-dependent angular scale cuts starting from a chosen $k_{\rm max}$. Inverting the Limber relation, the corresponding maximum multipole is 
\begin{equation}\label{eq:z_dep_ell_cuts_1d}
    \ell^A_{{\rm max}, i} = k_{\rm max}\chi(z^A_i) - 1/2 \; .
\end{equation}
The redshift $z_i$, the \enquote{typical distance} to the $i$-th bin, can be defined as the mean of the kernel (as usual, $A$ indicating the probe under consideration):
\begin{equation}
    z^A_i = \frac{\int \diff z \, z \Tilde{\mathcal{K}}^A_i(z)}{\int \diff z \, \Tilde{\mathcal{K}}^A_i(z)} \; .
\end{equation}
Figure~\ref{fig:std_and_bnt_gamma_kernel} shows the $z_i^{\rm L}$ values; note that we compute these from the BNT-transformed \textit{$\gamma$ kernel} (Eq~\ref{eq:w_gamma_ISTF}), excluding the IA kernel, which has a different expression (Eq.~\ref{eq:w_IA_ISTF}). This is not optimal, but the contribution of IA is subdominant \citep{Taylor2021O_kcut_3x2pt}. From the plot, it's clear how the BNT transform allows properly defining a typical distance for each kernel.
\\
We can now proceed to compute redshift-dependent $\ell$ cuts, for each tomographic combination; for the cross-bin and cross-probe case ($z_i \neq z_j$, or $A \neq B$) we choose take the more conservative (lower) $\ell$ cut \citep{Taylor2021O_kcut_3x2pt}:
\begin{equation}\label{eq:z_dep_ell_cuts_2d}
    \ell^{AB}_{{\rm max}, ij} = {\rm min}
    \{k_{\rm max}\chi(z^A_i) - 1/2,
    k_{\rm max}\chi(z^B_j) - 1/2\} \; .
\end{equation}
We can then proceed to \textit{cut} (remove) the elements of the (BNT-transformed) data vector where $C_{ij}^{AB}(\ell > \ell^{AB}_{{\rm max}, ij})$. Of course, the same elements have to be removed from the (BNT-transformed) covariance matrix as well. \\
We can BNT-transform the covariance matrix either by computing it from the BNT-transformed $C(\ell)$s (remembering to transform the noise spectra $N^{AB}_{ij}(\ell)$ as well), or by applying the BNT matrix to the covariance directly (cf. Appendix A of \citealt{Vazsonyi2021_kcut_HSC}):
\begin{equation}\label{eq:BNT_3x2pt_cov}
    \widetilde{{\rm Cov}}^{ABCD}_{abcd}( \ell, \ell') = \mathcal{X}^{AB}_{aebf} \mathcal{X}^{CD}_{cgdh} {\rm Cov}^{ABCD}_{efgh}( \ell, \ell') \; ,
\end{equation}
where 
\begin{equation}
\mathcal{X}_{AB}^{aebf}=
    \begin{cases}
          \mathcal{M}_{ae}\mathcal{M}_{bf}, & \text{if}\ AB={\rm LL} \\
          \delta^{\rm K}_{ae}\mathcal{M}_{bf}, & \text{if}\ AB={\rm GL} \\
          \delta^{\rm K}_{ae}\delta^{\rm K}_{bf}, & \text{if}\ AB={\rm GG} \\
    \end{cases}
\end{equation}
and the indices from $a$ to $h$ identify the redshift bin. We test that the two methods give consistent results to well below sub-percent agreement. We also validate our computation of the BNT by checking that the constraints from the FM analysis with and without applying the BNT, \textit{when no scale cuts are imposed}, agree to excellent precision. This is because, in the computation of the FM (Eq.~\ref{eq:fishmat}), the BNT matrix transforming the derivatives of the $C(\ell)$s cancels out with the inverse BNT matrices transforming the covariance matrix: indicating with $\vec{D}$ the vector of derivatives and taking as an example the WL probe, we have that
\begin{align}
    \begin{aligned}
        \Tilde{F} & = \tilde{\vec{D}}^{T} \widetilde{{\rm Cov}}^{-1} \Tilde{\vec{D}} \\
        & = (\mathcal{M}\vec{D})^{T} (\mathcal{M}^T{\rm Cov}\mathcal{M})^{-1} \mathcal{M}\vec{D} \\
        & = \vec{D}^{T}\mathcal{M}^T (\mathcal{M}^T)^{-1}{\rm Cov}^{-1}\mathcal{M}^{-1} \mathcal{M}\vec{D} \\
        & = \vec{D}^{T} {\rm Cov}^{-1} \vec{D} = F \; .
    \end{aligned}
\end{align}
In the above equations, we assumed the covariance and the vector of derivatives to have size $\mathcal{N}_{\rm b}$, but the same relation holds for every combination of multipoles and probes. We also leveraged the fact that the covariance is an invertible matrix and the (approximate) cosmology independence of the BNT matrix, which makes the operations of differentiation and BNT transform commutative. On the other hand, applying scale cuts does \textit{not} commute with the BNT transform: this is the reason why the constraints in the $k$-cut case can differ from the standard ones. \\

\begin{figure}
    \centering
    \includegraphics[width=1\textwidth]{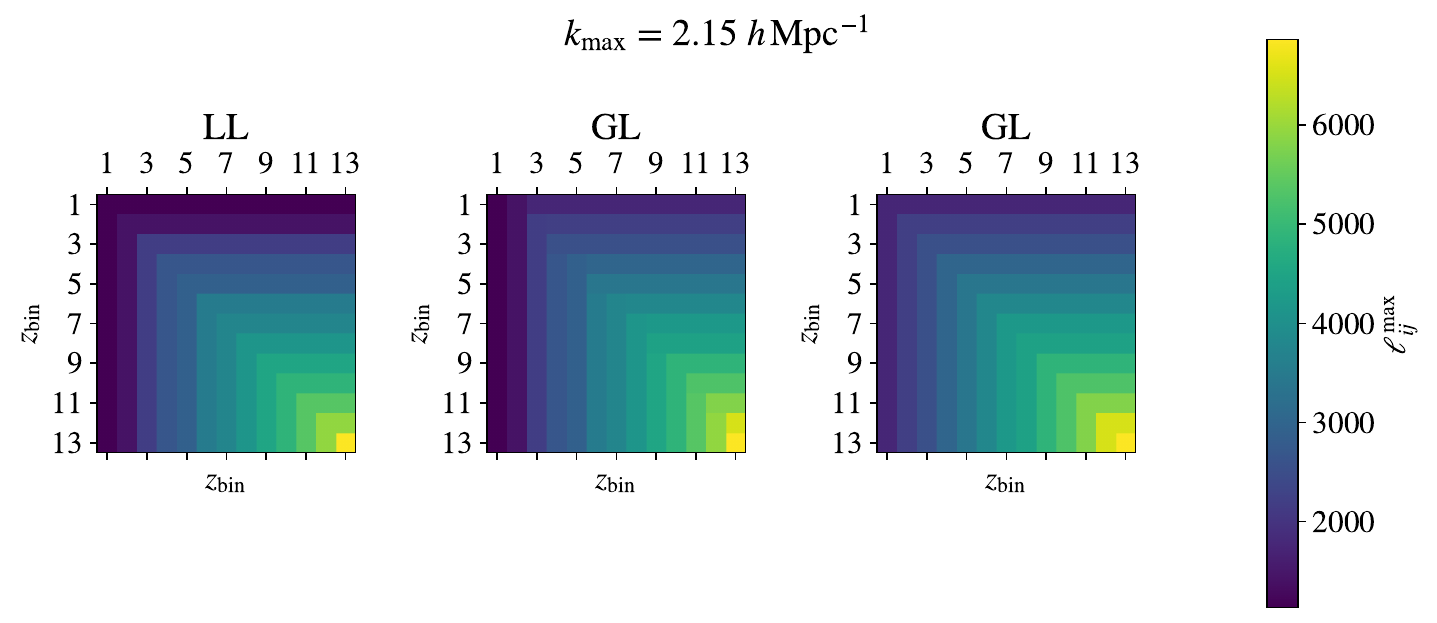}
    \caption{Redshift-dependent $\ell$ cuts for the different probe combinations, for the closest $\kmax$ to the threshold found to get a FoM of 400 in the reference case (see text). Note that the GL matrix is not symmetric under the exchange of redshift indices, as can be seen from the definition (Eq.~\ref{eq:z_dep_ell_cuts_2d}).}
    \label{fig:z_dependent_ell_cuts_kmax2.154435}
\end{figure}

To summarize, the steps taken for the $k$-cut procedure are the following:
\begin{enumerate}
    \item Starting from an $n_i(z)$, solve the system of equations~\eqref{eq:bnt_continuous_eq_system} to get the BNT matrix.
    \item Apply the BNT transform to the $\gamma$ kernels.
    \item Use the (narrow) BNT-transformed kernels to compute the mean redshifts $z^{\rm L}_i$. For GCph, we do the same using the standard galaxy kernels, to get $z^{\rm G}_i$.
    \item Choose a value for $k_{\rm max}$.
    \item Use the Limber relation through Eq.~\eqref{eq:z_dep_ell_cuts_2d} to translate the $k_{\rm max}$ into redshift- and probe-dependent $\ell$ cuts (see Fig.~\ref{fig:z_dependent_ell_cuts_kmax2.154435}).
    \item BNT-transform the covariance matrix through Eq.~\eqref{eq:BNT_3x2pt_cov} and the derivatives of the data vectors (in the same way as we transform the data vectors, i.e., through Eq.~\ref{eq:cl_BNT_3x2pt}), which are the two ingredients entering the computation of the FM.
    \item Cut the elements of the derivatives and covariance with $\ell^{AB}_{ij} > \ell^{AB}_{{\rm max}, ij}$.
    \item Compute the Fisher matrix and the $k$-cut parameter constraints (and FoM).
\end{enumerate}
The results of this procedure are presented in the next section.
\section{$k$-cut forecasts}\label{sec:k_cut_forecasts}
Having presented the formalism of the BNT transform and the actual algorithm used to derive the redshift- and probe-dependent angular scale cuts, we can proceed to apply it in our FM analysis, using this time the latest ingredients and prescriptions concerning the modelling of the observables and systematics for the \Euclid mission. As mentioned in the introduction, finding an optimal combination of these parameters is precisely the aim of the SPV exercise. This is necessarily a dynamic process since some of the survey specifications are being updated as the commissioning and performance verification phases of the mission unfold (see Fig.~\ref{fig:euclid_timeline}). \\

\begin{figure}
    \centering
    \includegraphics[width=0.7\linewidth]{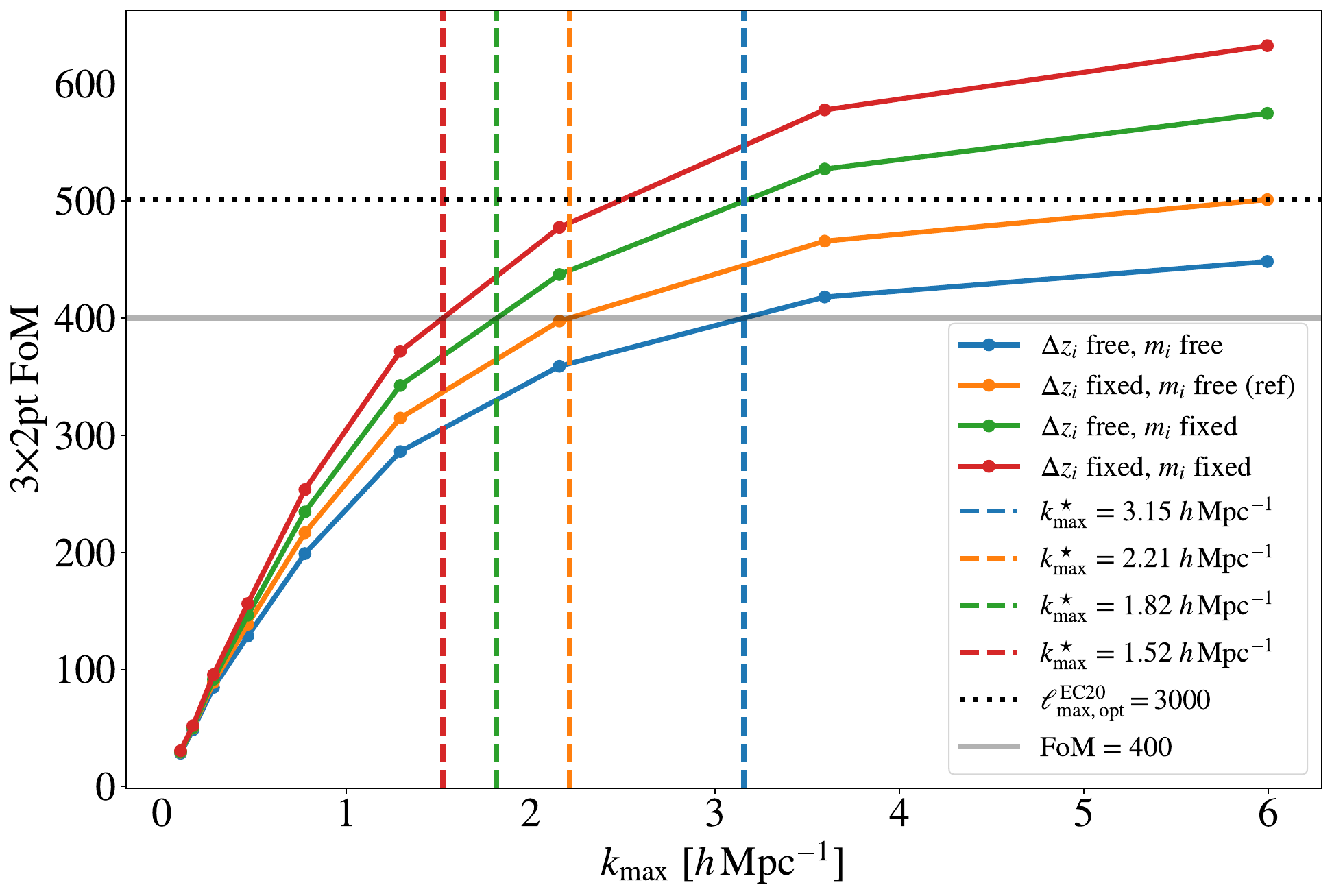}
    \caption{FoM vs. $k_{\rm max}$ for the 3$\times$2pt case (solid coloured lines), using \texttt{HMcode2020} with baryon corrections to model the nonlinear matter PS. The Red Book FoM (\ML{equals to 400,} grey horizontal solid line) and the optimistic-case FoM (\ML{equals to 501, }black horizontal dotted line) are also shown, allowing finding the minimum required $k_{\rm max}$ ($k_{\rm max}^\star$, colored vertical dashed lines) and to check the convergence of the results to the \enquote{hard-cut} case, respectively.}
    \label{fig:fom_hmcodebar_vs_kmax}
\end{figure}

\begin{figure}
    \centering
    \includegraphics[width=0.7\linewidth]{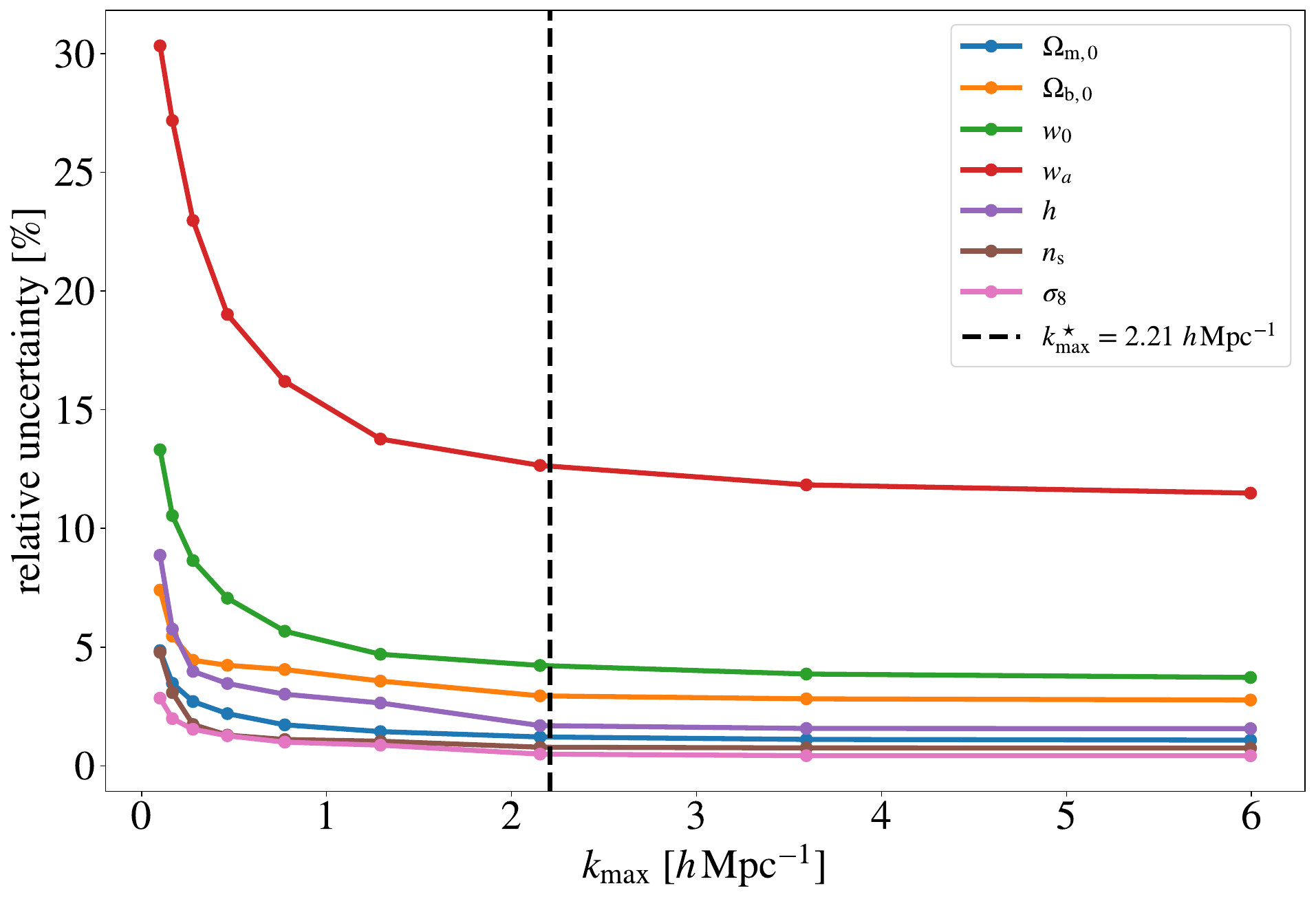}
    \caption{Relative 1$\sigma$ uncertainty of the different cosmological parameters vs. $k_{\rm max}$, for the reference case (\texttt{HMcode2020} with baryonic correction, $\Delta z_i$ fixed, $m_i$ free). The black vertical dashed line is the threshold cut $k_{\max}^\star$.}
    \label{fig:cosmo_params_vs_kmax}
\end{figure}

A number of elements differentiate the setup used here and the one described in the previous Chapters; these take into account the evolving specifications, and go in the direction of an increasingly realistic modelling. We list below the most notable changes:
\begin{itemize}
    \item We update the reference model for the nonlinear PS to a more recent recipe, \texttt{HMcode2020} \citep{Mead2021_hmcode2020}, which amongst the other improvements allows taking into account baryonic effects in the form of gas expulsion by AGN feedback.
    \item We cut the redshift distribution below $z=0.2$, due to the saturation of the detectors by luminous close sources.
    \item We include magnification bias in the GC angular PS (Sect.~\ref{sec:magnification_bias}).
    \item We fit the linear galaxy bias and magnification bias computed from the Flagship 2 $N$-body simulation with a third-order polynomial, resulting in 4 free parameters for each of these two systematics instead of $\mathcal{N}_{\rm b}$.
    \item We increase the number of redshift bins from 10 to 13, mainly to allow better tracking of the DE EoS time-dependence (hence a larger FoM). This has some drawbacks, such as a lower number of sources per bin, larger shot noise and number of bin-dependent nuisance parameters (shift in the mean of the redshift distribution $\Delta z_i$ -- Sect.~\ref{sec:z_distribution} -- and multiplicative shear bias -- Sect.~\ref{sec:mult_shear_bias}).
\end{itemize}
The forecasts obtained for the $k$-cut analysis with the reference setting described above are shown in Fig.~\ref{fig:fom_hmcodebar_vs_kmax} and Fig.~\ref{fig:cosmo_params_vs_kmax}.

In the first, we plot the FoM for the full 3$\times$2pt case as a function of $k_{\rm max}$, in 10 logarithmically-spaced steps from $10^{-1}$ to $10^{1} \, h\, {\rm Mpc}^{-1}$; it should be noted that in this analysis we maintain the (optimistic) \citetalias{ISTF2020} redshift-independent $\ell$ cuts. In fact, while the $k$-cut analysis allows properly taking into account the maximum scale at which we have reasonable belief in the \textit{modelling}, additional cuts may be imposed in angular space due to other sources of uncertainty, such as the PSF. Therefore, the asymptotic behaviour of the curves is due to the fact that, for increasing $k_{\rm max}$ values, increasingly more $\ell_{{\rm max}, ij}^{AB}$ values are above the redshift-independent, \enquote{hard} cut $\ell_{\rm max}=3000$, reducing to the optimistic case in the standard analysis (\ML{equals to 501, }dotted horizontal line, shown only for the reference case). In fact, we do not show the last of the 10 $k_{\rm max}$ values, since convergence is reached already at $k_{\rm max} \simeq 6 \, h\, {\rm Mpc}^{-1}$. The different curves correspond to increasingly optimistic setups, either fixing or marginalizing over the nuisance parameters $\Delta z_i$ and/or $m_i$; for the latter set of parameters, a Gaussian prior of width $\sigma_m = 5 \times 10^{-5}$ is imposed.
Now that the desired $k_{\rm max}$ is properly accounted for in the harmonic-space analysis, we can find, for example, the minimum scale cut $k_{\rm max}^{\star}$ for which the target FoM = 400 requirement defined in the \Euclid Red Book is met (albeit this is for the non-flat case, which we do not consider in this section). This value is shown, for the different setups, with colored vertical dashed lines, while the horizontal grey line denotes the target FoM. The reference case in the analysis is shown in orange, i.e., fixing only the mean $z$ shifts parameters, since we expect their prior to be quite narrow. Note that marginalizing over multiplicative shear bias has a larger impact than marginalizing over $\Delta z_i$, even when imposing the nominal prior (blue vs. green curves). The dispersion between the different curves increases with $k_{\rm max}$, showing how the uncertainty increase caused by marginalization over these nuisance parameters is less important than the one caused by taking aggressive scale cuts.

The whole set of cosmological parameters is shown instead in Fig.~\ref{fig:cosmo_params_vs_kmax}, this time only for the reference case. The $k_{\rm max}^{\star}$ requirement found for the FoM can be used to obtain the expected relative uncertainties, which apart from $w_a$ all lie in the range $[0.5, 4.2]\%$.
\\

\begin{figure}
    \centering
    \includegraphics[width=0.7\linewidth]{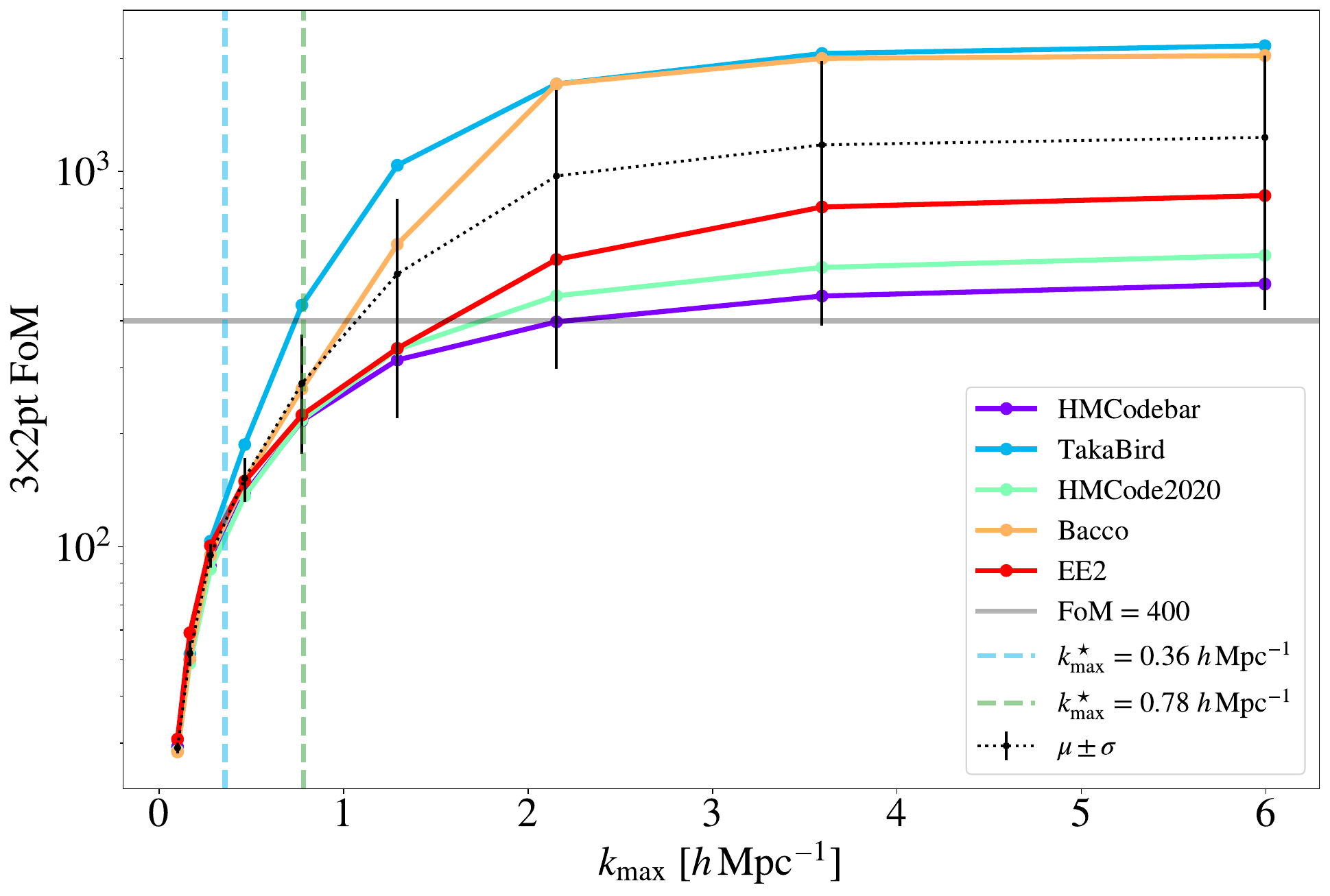}
    \caption{3$\times$2pt FoM as a function of $\kmax$ for the different nonlinear PS considered. The light green and light blue dashed vertical lines denote the maximum $\kmax^{\star}$ below which the dispersion between the different nonlinear recipes is lower than 10\%\ML{, respectively excluding and including the \texttt{TakaBird} recipe}. The black dotted line is the mean between the different values, with error bars given by the relative standard deviation. Note that these are made anisotropic by the logarithmic scale.}
    \label{fig:fom_vs_kmax_vs_pk}
\end{figure}

Another interesting possibility is to check the impact of different nonlinear prescriptions on the DE FoM. We explore five different options: the already introduced \texttt{TakaBird} and \texttt{HMcode2020} (with and without baryonic corrections), plus the emulators \texttt{Bacco} \citep{Arico2021_baccoemu} and \texttt{Euclid Emulator 2} \citep{Knabenhans2021_EE2}. The different results are shown in Fig.~\ref{fig:fom_vs_kmax_vs_pk}. \\
This time we gauge the spread between the FoM predicted by the different nonlinear prescriptions, with $\Delta z_i$ fixed and $m_i$ free, noting again that their difference vanishes for lower values of $k_{\max}$ (because they agree in the linear regime). The measure chosen to quantify this spread is simply the standard deviation $\sigma$ between the FoM of the different PS, at each $\kmax$. This dispersion is shown in the figure, along with their mean value ($\mu$). We can then find the $\kmax^{\star}$ values for which the relative scatter (defined as $\sigma/\mu$) drops below a certain threshold, in this case, 10\%. The resulting wavenumber, $\kmax^\star = 0.36 \, h \, {\rm Mpc}^{-1}$ is shown in light blue (dashed vertical line), and it is significantly lower than the minimum requirement found above to get to FoM =  400 (again, solid grey line). The requirement can be relaxed if we discard the \texttt{TakaBird} recipe, which is the most outdated of the ones considered (see \citealt{Smith2019_ngenhalofit} for some recent improvements to the original recipe); removing it from the sample we obtain the less stringent vertical dashed light green line, corresponding to $0.78 \, h \, {\rm Mpc}^{-1}$, still too severe to reach the target FoM (for which a FoM of 233 is reached instead). \\
We note that \citealt{Taylor2021O_kcut_3x2pt} finds for a similar analysis (albeit using \citetalias{ISTF2020} specifications) a threshold to reach the Red Book FoM of $\kmax^{\star} = 0.7 \, h \, {\rm Mpc}^{-1}$, which is in very good agreement with the value found here ($\kmax^{\star} = 0.698 \, h \, {\rm Mpc}^{-1}$) for the \texttt{TakaBird} case, which is supposedly the nonlinear PS used in by the authors -- being it the one chosen in \citetalias{ISTF2020}).

\begin{figure}
    \centering
    \includegraphics[width=\linewidth]{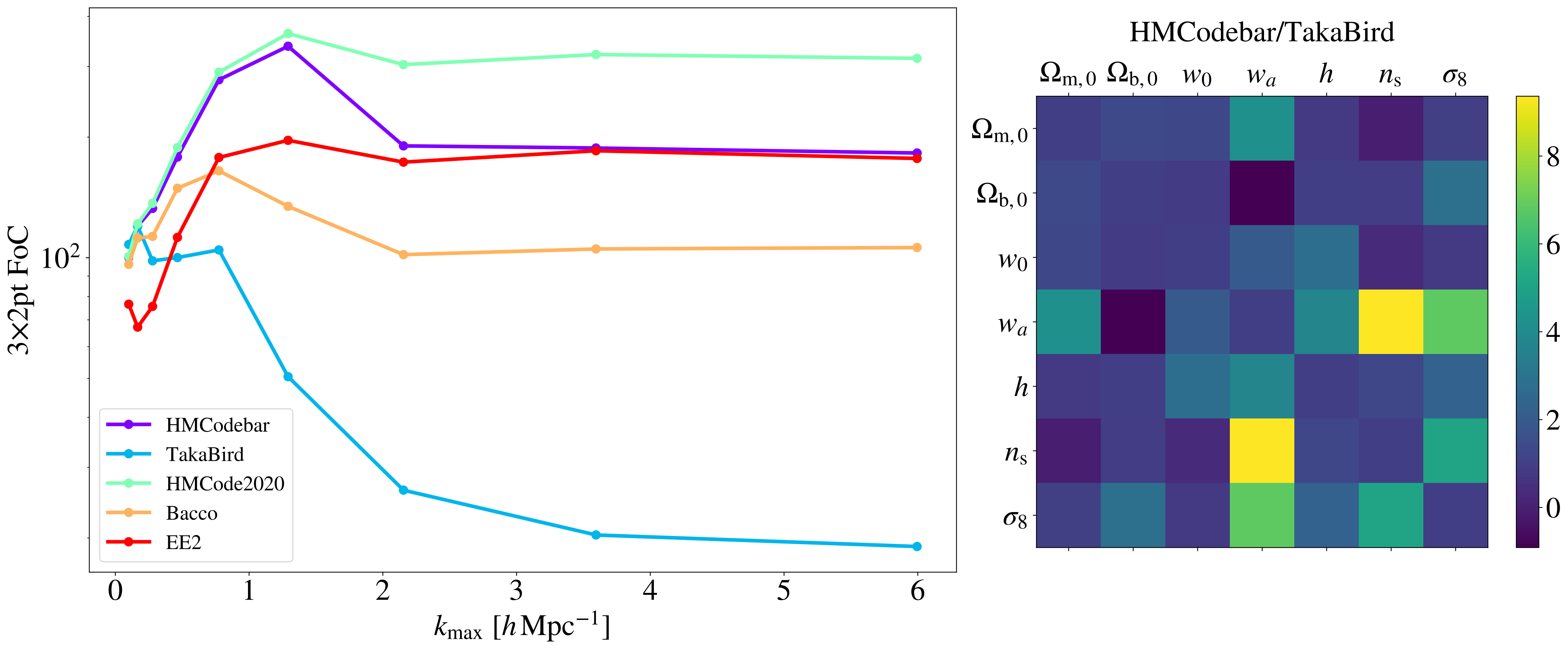}
    \caption{Left: Figure of Correlation as a function of $\kmax$ for the different nonlinear recipes. As seen above, the results tend to converge for lower values of the $k$-cut, because the models agree on linear scales. Right: Ratio between the correlation matrices obtained from \texttt{HMCode2020} including baryonic corrections and \texttt{TakaBird}.}
    \label{fig:pk_correlations_ppt}
\end{figure}

The difference between the curves in the figure is not necessarily representative, in principle, of the discrepancy between the power spectra themselves. Oftentimes the largest disagreement between the PS models and $N$-body simulations, hence the largest modelling uncertainty, is found precisely for the DE EoS parameters; this is the case for \citealt{Mead2021_hmcode2020}, for example. Moreover, the Figure of Merit also depends on the \textit{correlation} between the different parameters: in fact, to compute it we need to marginalize the FM over \textit{all} parameters but $w_0$ and $w_a$ -- see Eq.~\eqref{eq:fom}. Different recipes for the matter PS may give similar results as a function of $k$ given the same cosmology while displaying different parameters' covariance.\\
Figure~\ref{fig:pk_correlations_ppt} shows the ratio between the correlation matrices of the cosmological parameters obtained with \texttt{HMCode2020} including baryonic corrections and \texttt{TakaBird}, without scale cuts (the FMs having a sub-percent agreement with the case $\kmax \sim 6 \, \hMpc$). The correlation coefficient between $w_0$ and $w_a$ differs by a factor of two, and many other combinations have larger discrepancies. The correlation between different parameters can be summarized by the figure of correlation (FoC; \citealt{Casas2017_FoC}); a larger FoC indicates a higher correlation, which means a weaker constraining power. This can be defined as \citepalias{ISTF2020}:
\begin{equation}
    {\rm FoC} = \sqrt{{\rm det}({\rm Corr}^{-1})}
\end{equation}
A larger FoC will then, in general, correspond to a smaller FoM, and vice versa. The FoC is shown, in the context of the $k$-cuts analysis, in the left panel of Fig.~\ref{fig:pk_correlations_ppt}; the \texttt{Takabird} nonlinear power spectrum seems to underpredict the parameters' correlation, although, as can be seen from the curve of \texttt{Bacco}, this is not enough to explain for the larger FoM. As seen in Fig.~\ref{fig:fom_vs_kmax_vs_pk}, the spread between the curves tends to decrease when cutting nonlinear scales, since the recipes agree in the linear regime.\\

All of these results have been obtained through a FM analysis, hence they allow gauging the differences in \textit{constraining power} for the different cases presented. The other side of the coin is to assess the parameters \textit{bias}. This is most likely also going to decrease when applying $k$ cuts, both because of the larger uncertainties, as shown in Fig.~\ref{fig:cosmo_params_vs_kmax}, and because of the lower disagreement between the models themselves.\\
This requires a full MCMC analysis (see Sect.~\ref{sec:mcmc}) for each of the different scale cuts considered. We foresee this to be extremely numerically demanding, given the size of the data vector, the complexity of the model and the number of nuisance parameters. 

Another interesting possibility to retain more information is to use the $\ell_{{\rm max}, ij}^{AB}$ matrix to redefine the upper edge of the last $\ell$ bin, instead of cutting it altogether. Then, the PS would have to be recomputed either at that value -- i.e., $C^{AB}_{ij}(\ell_{{\rm max}, ij}^{AB})$, or at the centre of the newly defined, narrower bin. This would also require recomputing the covariance matrix (ideally, just the bins involved). \FB{Moreover, carrying out an analysis with a larger number of $\ell$ bins (or an unbinned one) would help in assessing the validity of taking wide bins, over the range of which large variations of relevant quantities (such as the comoving distance or the galaxy bias) occur}.  

Lastly, we note the potential of the BNT transform to mitigate the SSC, because the transformation of the highly-impacted lensing signal nulls the low-$z$ support for many of the redshift bins. As discussed in Sect.~\ref{sec:ssc_ref_results}, in this region the covariance of the background modes increases for $z \rightarrow 0$ because of the decrease in the effective volume. Moreover, removing nonlinear scales lessens the relative importance of non-Gaussian covariance. Preliminary results show this indeed to be the case; however, some numerical instabilities arise from the unstructured sparsity of the BNT-transformed GS covariance matrix. This induces larger uncertainties in its inversion, with the condition number being, e.g., $1.99\times 10^{18}$ at $\kmax = 2.15 \hMpc$, which results in an estimate for the precision of the matrix inversion of $4.42\times 10^2$ -- five orders of magnitude larger than what found in Sect.~\ref{sec:datavectors} for the non-BNT GS covariance. 

Investigating potential solutions to this problem, as well as developing the other points listed above, will constitute the next steps of our work. 
\section{Summary}\label{sec:bnt_conclusions}
The analysis presented in this Chapter, mainly based on the works of \citet{Bernardeau_2014} and \citet{Taylor2021O_kcut_3x2pt}, has shown the impact of the inclusion of scale cuts in the 3$\times$2pt forecast analysis. The procedure followed allows properly mapping such scale cuts to realistic, redshift- and probe-dependent $\ell$ cuts through BNT-transforming the cosmic shear signal. Small-scale modelling uncertainties, due to the limited understanding of nonlinear dark matter clustering and baryonic effects, can in this way be mitigated while retaining information from larger scales. 

Our $k$-cut analysis characterized different quantities as a function of the cut $\kmax$: the FoM when fixing different sets of nuisance parameters and the marginalized uncertainties on the cosmological ones, as well as the agreement between nonlinear models and their correlation.\\
The study takes into account a large set of nuisance parameters (${\cal N}_{\rm b}$ for multiplicative shear bias and shift in the mean $n_i(z)$, four for galaxy and magnification bias and two for intrinsic alignment, for a total of 36 parameters), a more realistic redshift distribution, a higher number of redshift bins, and more recent models (and tools) for the computation of the nonlinear matter PS. Some of these elements make the requirements on the FoM presented in \citet{laureijs2011euclid} difficult to meet, unless highly nonlinear scales ($\kmax \sim 2 \, \hMpc$) are included in the analysis. Our results show that this threshold is significantly above the scale at which the nonlinear models considered display consistent predictions ($\kmax \sim 0.4-0.8 \, \hMpc$ at the $10\%$ level) on the DE FoM. The bias on the best-fit parameters is another potential issue to be explored, and that can again be mitigated through the use of the BNT transform.

The BNT will also be used in the analysis of the soon-to-arrive \Euclid photometric data thanks to its development, to which I had the opportunity to participate, in the official code of the collaboration, \texttt{CLOE}.
\chapter*{Postface}
\addcontentsline{toc}{chapter}{Postface}
\lettrine[lines=2, findent=3pt, nindent=0pt]{T}{he} main aim of the work presented in this thesis has been to improve the realism and flexibility of the \Euclid forecast analysis, which allows propagating the impact of the different choices entering the modelling of the observables and survey specifications to the final uncertainties on the cosmological parameters. This makes it possible to optimize the survey to maximize the constraining power on the most important quantities of interest, such as the DE EoS parameters. Several different possibilities have been explored in this way, mainly within the context of the assessment of the impact of super-sample covariance and scale cuts, with a particular focus on the 3$\times$2pt analysis, which provides the majority of \Euclid constraining power.

We find the inclusion of SSC, computed with \texttt{PySSC} with the approximations presented in Chap.~\ref{chap:SSC} to be important for an accurate estimate of the precision of the measurements. We quantify its impact as the ratio between the 1$\sigma$ FM uncertainties including and neglecting this covariance term, and find it causes significant degradation of the WL constraints and the 3$\times$2pt FoM (up to 38\% in the \citetalias{ISTF2020} optimistic case). GCph is less affected because of the lower probe response and of the reduced contribution that the majority of redshift bins receive from the low-$z$ region, where the effective survey volume decreases and the covariance of the background modes increases. The results are in broad agreement with the recent works of \citet{Barreira2018cosmic_shear} and \citet{upham2021}, which have some differences in the setup. \\
The impact has been gauged for different survey settings, namely varying the number of redshift and multipole bins and broadening both the cosmological and nuisance parameters space by letting $\Omega_{\rm DE}$ and the multiplicative shear bias parameters free to vary. Variations in the redshift binning have almost no effect on the strength of SSC, while marginalization over multiplicative shear bias significantly reduces its importance for WL, at the cost of overall weaker constraints. This is because SSC is degenerate with the amplitude of the signal, as are multiplicative shear and galaxy bias parameters. We also quantified the requirements on the prior information to be added on galaxy bias and multiplicative shear bias to recover the Gaussian FoM, finding them to be quite stringent (at the percent-level on galaxy bias, when applying the fiducial prior with $\sigma_m = 5\times 10^{-5}$ on the multiplicative shear bias). \\

The second part of our work focused on the investigation of the BNT transform as a powerful tool to implement scale cuts, and in this way to mitigate modelling uncertainties which would potentially degrade and bias the final measurements. The framework used is flexible and does not introduce significant overhead to the analysis (apart from the necessity to loop over the $\kmax$ values and to apply the BNT transform), since the data vectors and covariance matrix are computed once and then cut in different ways. In this way, the requirement on the maximum scale to include in the analysis as a function of the desired FoM, and other interesting points such as the agreement between different nonlinear models, can be investigated.\\
This is likely to become a standard in the analysis of the upcoming photometric data, because of the significant theoretical uncertainties still affecting the modelling of the LSS observables on small scales.\\

Most of the tools and procedures developed to compute the forecasts presented here are ready to be applied to the analysis of the actual data, allowing us to finally witness the real power of the \Euclid mission.
\begin{acknowledgements}
\AckECol
\newline
\newline
{\large \textbf{Disclaimer}}\\
\textit{This thesis is based on, or contains non-public Euclid Consortium material or results that have not yet been endorsed by the Euclid Consortium.}
\end{acknowledgements}
\appendix
\chapter{High order bias from halo model} \label{sec:halomodel_appendix}
As described in Sect.~\ref{sec:higher_order_bias}, the second-order bias $b_{(2)}(z)$ has been estimated using the halo model. In the following Appendix, we provide further details on the input quantities, and how we set the relevant parameters. 

A key role is played by the halo mass function $\Phi_{\rm MF}(M, z)$, which we model as 
\begin{equation}
\Phi_{\rm MF}(M, z) = \frac{\bar\rho_{\rm m}}{M} f(\nu) \frac{\diff\ln{\sigma^{-1}}}{\diff M},
\label{eq:hmft10}
\end{equation}
with $M$ the halo mass, ${\bar\rho}_{\rm m}$ the mean matter density, $\nu= \delta_{\rm c}/\sigma(M, z)$, $\delta_{\rm c} = 1.686$ the critical overdensity for collapse, and $\sigma(M, z)$ the variance of linear perturbation smoothed with a top-hat filter of radius $R = \left[3M/(4 \pi \bar\rho_{\rm m})\right]^{1/3}$. We follow \citet{Tinker2010}, setting 
\begin{equation}
f(\nu) = {\cal N}_{\rm MF} \left [1 + \left ( \beta_{\rm MF} \nu \right )^{\, -2 \phi_{\rm MF}} \right ] \nu^{\, 2 \eta_{\rm MF}} \exp{\left ( - \gamma_{\rm MF} \nu^{\, 2}/2 \right )},
\label{eq:fnumf}
\end{equation}
where ${\cal N}_{\rm MF}$ is a normalisation constant, and the halo mass function fitting parameters $\beta_{\rm MF}, \eta_{\rm MF}, \gamma_{\rm MF}$ and $\phi_{\rm MF}$ -- not to be confused with $\Phi_{\rm MF}(M, z)$ -- scale with redshift as illustrated in Eqs.~(9--13) of the above-mentioned paper.

The other quantity needed is the average number of galaxies hosted by a halo of mass $M$ at redshift $z$. This is given by 
\begin{equation}
\langle N|M \rangle(M) = 
N_{\rm cen}(M) \left[1 + N_{\rm sat}(M)\right] \; ,
\label{eq:nmhod}
\end{equation}
where $N_{\rm cen}(M, z)$ and $N_{\rm sat}(M, z)$ account for the contributions of central and satellite galaxies, respectively. We model these terms as in \citet{White2011}
\begin{equation}
N_{\rm cen}(M) = \frac{1}{2} \left \{ 1 +  {\rm erfc}{\left [\frac{\ln{\big(M/M_{\rm cut}\big)}}{\sqrt{2} \sigma_c} \right ]} \right \} \; ,
\label{eq:ncen}
\end{equation}
\begin{equation}
N_{\rm sat}(M) = \left \{
\begin{array}{ll}
\displaystyle{0} & \displaystyle{M < \kappa_s M_{\rm cut}} \\
 & \\
\displaystyle{\left ( \frac{M - \kappa_s M_{\rm cut}}{M_1} \right)^{\alpha_s}} & \displaystyle{M \ge \kappa_s M_{\rm cut}}, \\
 \end{array}
\right . 
\label{eq:nsat}
\end{equation}
with fiducial parameter values
\begin{equation}
\{\logten{(M_{\rm cut}/M_\odot)}, \logten{(M_1/M_\odot)}, \sigma_c, \kappa_s, \alpha_s \} =
{13.04, 14.05, 0.94, 0.93, 0.97} \; ,
\end{equation}
$M_\odot$ being the mass of the Sun. These values give the best fit to the clustering of massive galaxies at $z \sim 0.5$ as measured from the first semester of BOSS data. It is, however, expected that they are redshift-dependent although the precise scaling with $z$ also depends on the galaxy population used as a tracer. We, therefore, adjust them so that the predicted galaxy bias matches, at each given redshift, our measured values from the Flagship simulation. Since, for each $z$, we have a single observable quantity, we can not fit all parameters. On the contrary, we fix all of them but $M_{\rm cut}$ to their fiducial values and use Eq.~\eqref{eq:bicalc} to compute the bias as a function of $M_{\rm cut}$. We then solve with respect to $M_{\rm cut}$ repeating this procedure for each redshift bin. We then linearly interpolate these values to get $M_{\rm cut}$ as a function of $z$, and use it to compute $b_{(2)}(z)$. Although quite crude, we have verified that changing the HOD parameter to be adjusted (e.g., using $\sigma_c$ or $M_1$) has a negligible impact on the predicted $R^{\rm gm}(\ell)$ and $R^{\rm gg}(\ell)$.

\chapter{Appendix B: Details of the code validation} \label{sec:validation_appendix}

In the following Appendix, we provide an overview of the steps undertaken to compare and validate the codes used in this work, and some of the lessons learnt in the process.

\begin{figure}
\centering
\includegraphics[width=.7\hsize]{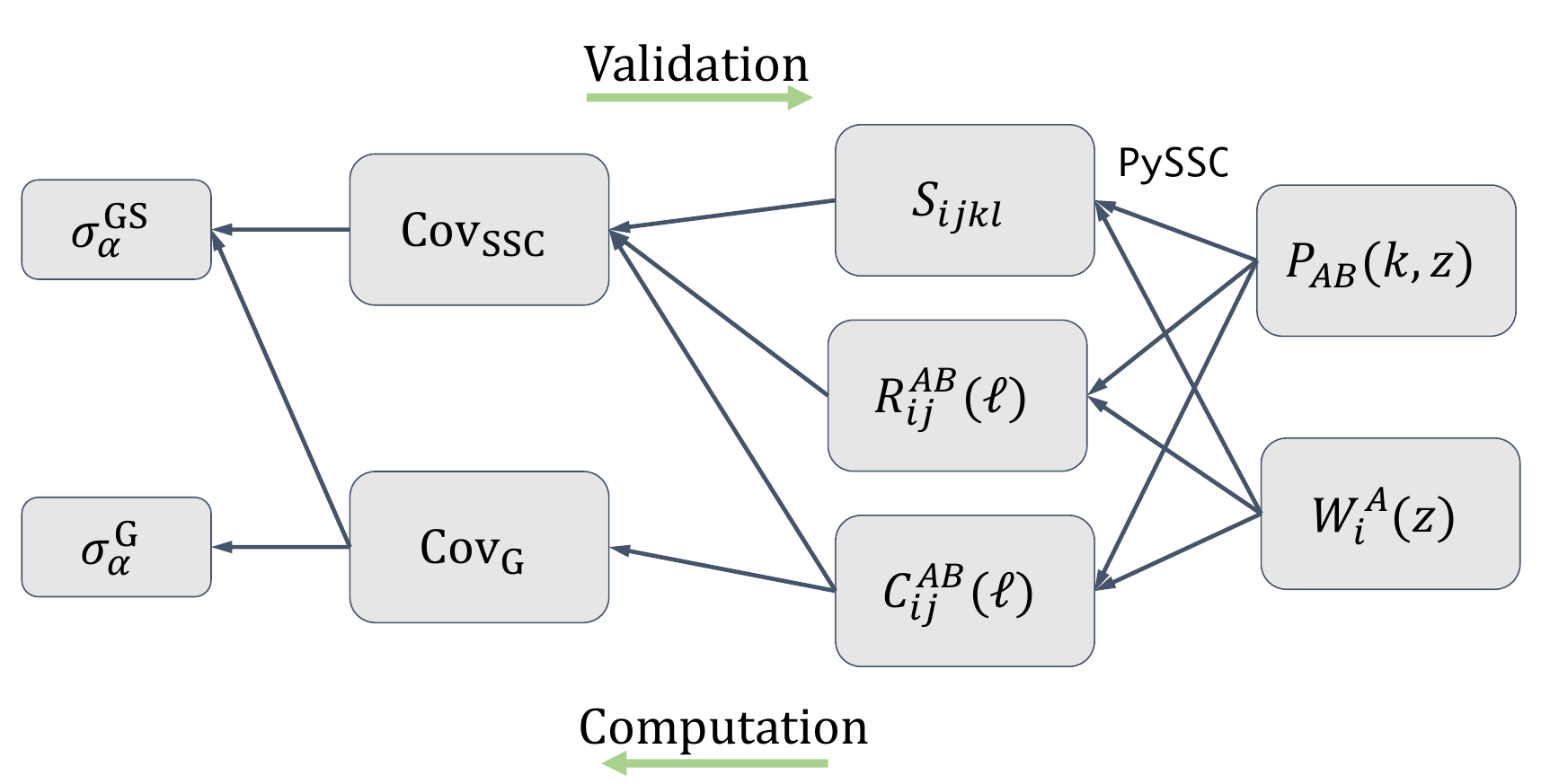}
\caption{Some of the most important elements examined in the comparison. The arrows show the ordering followed to produce the parameters constraints, which is opposite to the one followed to validate the code. The derivatives of the PS with respect to the cosmological parameters, entering the final step of the computation, are not shown.}
\label{fig:procedure}
\end{figure}

To compute and validate the results we adopt the scheme sketched in Fig.~\ref{fig:procedure}, which highlights the dependency of each main element of the forecast computation on the others. In particular, we have that:
\begin{enumerate}
    \item The $1\sigma$ constraints are obtained from the FM through Eq.~\eqref{eq:sigma_marg}, and the FM is built in turn from the (inverse) covariance matrix and the derivatives of the angular PS $C_{ij}^{AB}(\ell)$ as indicated in Eq.~\eqref{eq:fishmat}.
    \item The Gaussian covariance depends on the $C_{ij}^{AB}(\ell)$ through Eq.~\eqref{eq:covgauss} (and the noise PS, Eq.~\ref{eq:noiseps}). The SSC also depends on the $C_{ij}^{AB}(\ell)$, with the added contribution of the $\mathcal{R}_{ij}^{AB}(\ell)$ terms and the output of the \texttt{PySSC} module, the $S_{ijkl}$ matrix -- following Eq.~\eqref{eq:covssc_sijkl}. 
    \item The $C_{ij}^{AB}(\ell)$ are constructed by convolving the (nonlinear) matter PS with the lensing and galaxy weight functions, as in Eq.~\eqref{eq:cijdef_pyssc}. The $S_{ijkl}$ matrix also depends on the weight functions (see Eq.~\ref{eq:sijkl}), which are in fact the main external input needed by \texttt{PySSC}, and on the \textit{linear} matter PS through the $ \sigma^{2}(z_{1}, z_{2})$ term (Eq.~\ref{eq:sigma2_pyssc_4pi2}, modulo the $1/(4\pi)^2$ factor). It is to be noted, however, that \texttt{PySSC} computes this PS internally, needing only the specification of a dictionary of cosmological parameters with which to call the Boltzmann solver \texttt{CLASS} through the  \texttt{Python} wrapper \texttt{classy}. This means that we also have to make sure that the fiducial value of the parameters used to compute the PS of Eq.~\eqref{eq:cijdef_pyssc} are the same ones passed to \texttt{PySSC} (this time to compute the linear PS), in order to work with the same cosmology.
\end{enumerate}
While to compute the constraints we follow the scheme from right to left, starting from the basic ingredients to arrive at the final result, the general trend of the validation is the opposite: we begin by comparing the final results, then work our way back whenever we find disagreement.

We then start the comparison from the $\sigma_\alpha$. If a discrepancy larger than $10\%$ is found, we check the quantities they depend on, which in this case are the covariance matrices (see Eq.~\ref{eq:fishmat}). If these agree, we check the codes directly. If these disagree, we iterate the process by checking the subsequent element in the scheme (in this case the $S_{ijkl}$ matrix and the $C_{ij}^{AB}(\ell)$), until agreement is found. Essentially, this means that the disagreement in the outputs of the codes at each step can either come from the inputs, or from the codes themselves. Once the cause of the discrepancy is found and fixed, the computation is repeated and the process can start again.

The pipelines under comparison are both written in the $\texttt{Python}$ language. One of them requires as external inputs the weight functions, the angular PS $C^{AB}_{ij}(\ell)$ and their derivatives with respect to the cosmological parameters; whilst the other produces these through the use of \texttt{CosmoSIS}, and hence needs no external inputs but the vectors of fiducial cosmological and nuisance parameters.
For the reader wishing to repeat the validation, we list below some of the lessons learnt in the code comparison process.
\begin{itemize}
    \item \texttt{PySSC} needs as input the WL and GCph kernels of Eqs.~\eqref{eq:wildef} and \eqref{eq:wigdef}, as well as their argument, the redshift values. The code then uses this redshift array to perform the necessary integrals in $\diff V$ through Simpson's rule. The user is responsible for sampling the kernels on a sufficiently fine $z$ grid ($\mathcal{O}(10^4)$ values have been found to be sufficient in the present case) to make sure these integrals are performed accurately.
    \item The latest version of \texttt{PySSC} accepts a \texttt{convention} parameter. This specifies whether the kernels are in the form used in \citetalias{Lacasa_2019} (i.e., $K^A_i(z)$; \texttt{convention = 0}) or the one prescribed in \citetalias{ISTF2020} (i.e., $\mathcal{K}^A_i(z)$; \texttt{convention = 1}). The two differ by a $1/\chi^2(z)$ factor, as shown in Eq.~\eqref{eq:wfmatch}. Passing the kernels in the \citetalias{ISTF2020} form without changing the parameter's value from \texttt{0} -- the default -- to \texttt{1} will obviously yield incorrect results. 
    \item The ordering of the $S_{ijkl}$ matrix's elements depends on the ordering chosen when passing the input kernels to \texttt{PySSC}  -- whether $K_i^{\rm L}(z)$ first and $K_i^{\rm G}(z)$ second or vice versa. This must be kept in mind when implementing Eq.~\eqref{eq:covssc_sijkl}.
    \item The GCph constraints can show a discrepancy greater the 10\% for the dark energy equation of state parameters $w_0$ and $w_a$ even when the corresponding covariance is found to be in good agreement. This discrepancy is due to GCph being less numerically stable because of the lower constraining power compared to the other probes, and because the bias model considered has a strong degeneracy with $\sigma_8$, making the numerical derivatives unstable \citep[see e.g.,][]{Casas2023}. Since this is a known issue, not coming from the SSC computation, and the covariance matrices and angular PS show good agreement, we choose to overcome the problem by using, for GCph, one code to compute both sets of parameter constraints (that is, we run one FM evaluation code with as input the covariance matrices from both groups).
\end{itemize}
\chapter{Slowly varying response approximation for broad kernels}\label{sec:Sijkl_approximation}
\begin{figure}
    \centering
    \includegraphics[width=0.8\textwidth]{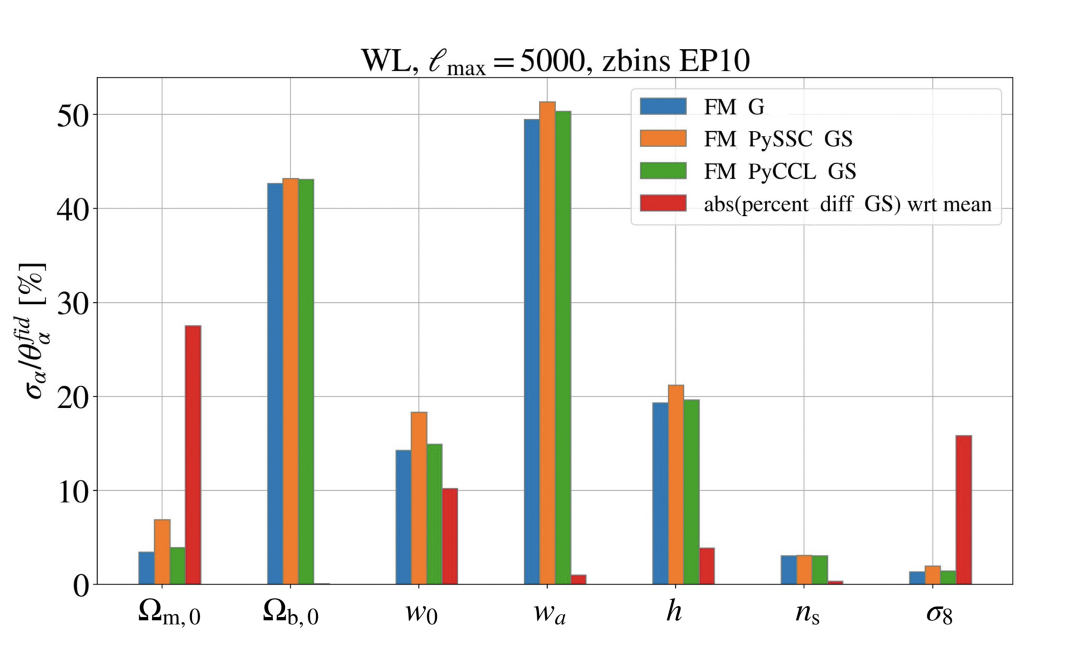}
    \caption{Gauss + SSC constraints for WL (10 equipopulated redshift bins), computed with \texttt{PySSC} and \texttt{PyCCL}. The percent difference with respect to the mean of the Gauss. + SSC constraints is shown in red; a marginally significant difference (10\% being the threshold for agreement considered in this work) is still present on the most affected parameters.}
    \label{fig:ccl_vs_pyssc}
\end{figure}

\FB{At the time of writing, all of the publicly available codes to compute the SSC term adopt some approximations. On one side, \texttt{PySSC} adopts the slowly varying response approximation, which has been tested, as mentioned in Sect.~\ref{sec:ssc_approx}, for GCph, obtaining good agreement against the full computation up to small angular scales, as well as for cluster number counts.\\ 
The two other codes are the Core Cosmology Library (\texttt{CCL} - \citealt{Chisari2019_CCL}) and \texttt{CosmoLike} \citep{Krause2017}, both of which implement the recipe given in the latter paper. This also adopts an approximation, dubbed the \enquote{KE approximation}: this consists of considering $\sigma^2(z_1, z_2)$ as a Dirac delta function at $z_1 = z_2$, so that the double redshift integral collapses into a single integral; this speeds up the code considerably, although not as much as the \citetalias{Lacasa_2019} approximation. \citet{Lima2018} argued the latter to underestimate the full SSC less than the KE approximation for cluster number counts, and in general for probes with narrow radial kernels. Besides its speed, \texttt{PySSC} was mainly chosen for the full control over the observables and the choice of the probe responses, which allowed us to use the measurements from the separate universe technique.\\
As mentioned in Sect.~\ref{sec:ssc_ref_results}, our results are in broad agreement with the ones from \cite{Barreira2018cosmic_shear} and \citet{upham2021} (computed with \texttt{CosmoLike}), which adopts the closest setup to the \Euclid settings used in this work. However, significant differences still persist; the use of pseudo-$C_\ell$s or of a reduced parameter space, just to name a few. An in-depth comparison between the different codes is currently undergoing: in Fig.~\ref{fig:ccl_vs_pyssc} we show the marginalised 1$\sigma$ uncertainties for the WL probe, for the Gaussian and Gaussian + SSC covariance matrix. The absolute value of the percent discrepancy w.r.t. the mean of the Gauss + SSC constraints computed with the two different codes is shown in red; in general the agreement is quite good, except for $\sigma_8$ and $\Omega_{{\rm m}, 0}$. To explore this issue more in detail, we are currently developing a new code to compute the SSC term without any of the above-mentioned approximations, the results of which will be shared in a forthcoming publication.}
\chapter{Multipole binning}\label{sec:appendix_binning}
We bin the $\ell$ space  according to the following procedure: the $\ell_k$ values, where $k = 1, ..., {\cal N}_\ell$, are the centers of ${\cal N}_\ell + 1$ logarithmically equispaced values, $\lambda_{k}$, which act as the edges of the ${\cal N}_\ell$ bins:
\begin{equation}
\ell_k = {\rm dex}\brackets{
    \paren{\lambda_{k}^{-} + \lambda_{k}^{+}}/2
    } \; ,
\label{eq: ellkxc}
\end{equation}
with ${\rm dex}(x) = 10^x$, $\paren{\lambda_{k}^{-}, \lambda_{k}^{+} } = \paren{\lambda_{k}, \lambda_{k + 1}}$, and
\begin{equation}
\lambda_k = \lambda_{\rm min}^{\rm XC} + (k - 1) (\lambda_{\rm max}^{\rm XC} - \lambda_{\rm min}^{\rm XC})/{\cal N}_{\ell} \; ,
\label{eq: lambdakxc}
\end{equation}
being
\begin{equation}
\left\{\lambda_{\rm min}^{\rm XC}, \lambda_{\rm max}^{\rm XC}\right\} = \left\{
\logten{(\ell_{\rm min}^{\rm XC})}, \, \logten{(\ell_{\rm max}^{\rm XC})}\right\} \; .
\label{eq: rangexc}
\end{equation}
In order to compute the Gaussian covariance, we also need the width of the bin, which will simply be
\begin{equation}
\Delta \ell_k = {\rm dex}(\lambda_{k + 1}) - {\rm dex}(\lambda_k) \; ,
\label{eq: deltaell} 
\end{equation}
so that $\Delta \ell_k$ is not the same for all bins, since the bins are logarithmically -- and not linearly -- equispaced.

\backmatter
\phantomsection
\bibliographystyle{bib_style}
\bibliography{main}

\end{document}